\documentstyle[eqsecnum,aps,epsf,
feynmf]{revtex}
\newcommand{\nc}{\newcommand}
\nc{\Gb}{\mbox{\boldmath $G$}}
\nc{\Gbs}{\mbox{\scriptsize \boldmath $G$}}
\nc{\scs}{\scriptstyle}
\nc{\beq}{\begin{eqnarray}}
\nc{\eeq}{\end{eqnarray}}
\nc{\la}{\label}
\nc{\r}{\ref}
\nc{\no}{\nonumber}
\nc{\ci}{\cite}
\nc{\trace}{{\rm Tr\,}}
\nc{\setval}{\fmfset{wiggly_len}{1.5mm}\fmfset{arrow_len}{1.5mm}\fmfset{arrow_ang}{13}\fmfset{dash_len}{1.5mm}\fmfpen{0.1mm}\fmfset{dot_size}{1thick}}
\nc{\dvertexx}[5]{\frac{\delta #1}{\rule[0pt]{0pt}{15pt}\delta
\parbox{10mm}{\centerline{
\begin{fmfgraph*}(4,4)
\fmfset{arrow_len}{1.5mm}
\fmfset{wiggly_len}{1.5mm}
\fmfset{dot_size}{1thick}
\fmfset{arrow_ang}{13}
\fmfpen{0.125mm}
\fmfleft{i2,i1}
\fmfright{o2,o1}
\fmf{plain}{i1,v1}
\fmf{plain}{v1,i2}
\fmf{plain}{v1,o1}
\fmf{plain}{o2,v1}
\fmfv{decor.size=0,label={\footnotesize #2},l.dist=0.5mm}{o1}
\fmfv{decor.size=0,label={\footnotesize #3},l.dist=0.5mm}{o2}
\fmfv{decor.size=0,label={\footnotesize #4},l.dist=0.5mm}{i2}
\fmfv{decor.size=0,label={\footnotesize #5},l.dist=0.5mm}{i1}
\fmfdot{v1}
\end{fmfgraph*}}}\rule[0pt]{0pt}{15pt}}}
\nc{\dvertex}[4]{\frac{\delta #1}{\rule[0pt]{0pt}{15pt}\delta
\parbox{10mm}{\centerline{
\begin{fmfgraph*}(4,3.464)
\fmfset{arrow_len}{1.5mm}
\fmfset{wiggly_len}{1.5mm}
\fmfset{dot_size}{1thick}
\fmfset{arrow_ang}{13}
\fmfpen{0.125mm}
\fmfforce{1w,0h}{v1}
\fmfforce{0w,0h}{v2}
\fmfforce{0.5w,1h}{v3}
\fmfforce{0.5w,0.2886h}{vm}
\fmf{plain}{v1,vm,v2}
\fmf{plain}{v3,vm}
\fmfv{decor.size=0,label={\footnotesize #3},l.dist=0.5mm}{v1}
\fmfv{decor.size=0,label={\footnotesize #4},l.dist=0.5mm}{v2}
\fmfv{decor.size=0,label={\footnotesize #2},l.dist=0.5mm}{v3}
\fmfdot{vm}
\end{fmfgraph*}}}\rule[0pt]{0pt}{15pt}}}
\nc{\fdphi}[2]{\frac{\delta #1}{\delta
\parbox{8mm}{\centerline{
\begin{fmfgraph*}(5,3)
\fmfpen{0.125mm}
\fmfset{wiggly_len}{1.5mm}
\fmfset{dot_size}{1thick}
\fmfleft{v1}
\fmfright{v2}
\fmf{boson}{v2,v1}
\fmfdot{v1}
\fmfv{decor.size=0,label={\footnotesize #2},l.dist=0.5mm}{v2}
\end{fmfgraph*}
}}}}
\nc{\cdphi}[2]{\frac{\delta #1}{\delta
\parbox{10mm}{\centerline{
\begin{fmfgraph*}(5,3)
\fmfpen{0.125mm}
\fmfleft{v1}
\fmfright{v2}
\fmf{plain}{v2,v1}
\fmfv{decor.size=0,label={\footnotesize #2},l.dist=0.5mm}{v2}
\fmfv{decor.shape=cross,decor.filled=shaded,decor.size=3thick}{v1}
\end{fmfgraph*}
}}}}
\nc{\dphi}[3]{\frac{\delta #1}{\delta
\parbox{10mm}{\centerline{
\begin{fmfgraph*}(5,3)
\fmfpen{0.125mm}
\fmfleft{v1}
\fmfright{v2}
\fmf{plain}{v2,v1}
\fmfv{decor.size=0,label={\footnotesize #2},l.dist=0.5mm}{v1}
\fmfv{decor.size=0,label={\footnotesize #3},l.dist=0.5mm}{v2}
\end{fmfgraph*}
}}}}
\nc{\ddphi}[5]{\frac{\delta^2 #1}{\delta
\parbox{10mm}{\centerline{
\begin{fmfgraph*}(5,3)
\fmfpen{0.125mm}
\fmfleft{v1}
\fmfright{v2}
\fmf{plain}{v2,v1}
\fmfv{decor.size=0,label={\footnotesize #2},l.dist=0.5mm}{v1}
\fmfv{decor.size=0,label={\footnotesize #3},l.dist=0.5mm}{v2}
\end{fmfgraph*}
}}\,\delta
\parbox{10mm}{\centerline{
\begin{fmfgraph*}(5,3)
\fmfpen{0.125mm}
\fmfleft{v1}
\fmfright{v2}
\fmf{plain}{v2,v1}
\fmfv{decor.size=0,label={\footnotesize #4},l.dist=0.5mm}{v1}
\fmfv{decor.size=0,label={\footnotesize #5},l.dist=0.5mm}{v2}
\end{fmfgraph*}
}}}}
\nc{\ddfphi}[4]{\frac{\delta^2 #1}{\delta
\parbox{10mm}{\centerline{
\begin{fmfgraph*}(5,3)
\fmfpen{0.125mm}
\fmfleft{v1}
\fmfright{v2}
\fmf{plain}{v2,v1}
\fmfv{decor.size=0,label={\footnotesize #2},l.dist=0.5mm}{v1}
\fmfv{decor.size=0,label={\footnotesize #3},l.dist=0.5mm}{v2}
\end{fmfgraph*}
}}\,\delta
\parbox{8mm}{\centerline{
\begin{fmfgraph*}(5,3)
\fmfpen{0.125mm}
\fmfset{dot_size}{1thick}
\fmfleft{v1}
\fmfright{v2}
\fmf{boson}{v2,v1}
\fmfdot{v1}
\fmfv{decor.size=0,label={\footnotesize #4},l.dist=0.5mm}{v2}
\end{fmfgraph*}
}}}}
\nc{\ddvertex}[6]{\frac{\delta^2 #1}{\rule[0pt]{0pt}{15pt}
\delta \parbox{10mm}{\centerline{
\begin{fmfgraph*}(4,3.464)
\fmfset{arrow_len}{1.5mm}
\fmfset{wiggly_len}{1.5mm}
\fmfset{dot_size}{1thick}
\fmfset{arrow_ang}{13}
\fmfpen{0.125mm}
\fmfforce{1w,0h}{v1}
\fmfforce{0w,0h}{v2}
\fmfforce{0.5w,1h}{v3}
\fmfforce{0.5w,0.2886h}{vm}
\fmf{plain}{v1,vm,v2}
\fmf{plain}{v3,vm}
\fmfv{decor.size=0,label={\footnotesize #3},l.dist=0.5mm}{v1}
\fmfv{decor.size=0,label={\footnotesize #4},l.dist=0.5mm}{v2}
\fmfv{decor.size=0,label={\footnotesize #2},l.dist=0.5mm}{v3}
\fmfdot{vm}
\end{fmfgraph*}}}
\,\,\,
\delta \parbox{10mm}{\centerline{
\begin{fmfgraph*}(5,3)
\fmfpen{0.125mm}
\fmfleft{v1}
\fmfright{v2}
\fmf{plain}{v2,v1}
\fmfv{decor.size=0,label={\footnotesize #5},l.dist=0.5mm}{v1}
\fmfv{decor.size=0,label={\footnotesize #6},l.dist=0.5mm}{v2}
\end{fmfgraph*}}}
}}
\nc{\sigmanew}[4]{\frac{\delta 
\parbox{8mm}{\centerline{
\begin{fmfgraph*}(5,3)
\setval
\fmfforce{0w,0.5h}{i1}
\fmfforce{1w,0.5h}{o1}
\fmfforce{0.5w,0.5h}{v1}
\fmf{double,width=0.25mm}{i1,v1}
\fmf{double,width=0.25mm}{v1,o1}
\fmfv{decor.size=0, label={\footnotesize #1}, l.dist=0.5mm, l.angle=-180}{i1}
\fmfv{decor.size=0, label={\footnotesize #2}, l.dist=0.5mm, l.angle=0}{o1}
\fmfv{decor.shape=circle,decor.filled=empty,decor.size=0.6mm}{v1}
\end{fmfgraph*}}} }{\delta
\parbox{10mm}{\centerline{
\begin{fmfgraph*}(5,3)
\fmfpen{0.125mm}
\fmfforce{0w,0.5h}{v1}
\fmfforce{1w,0.5h}{v2}
\fmf{double,width=0.25mm}{v2,v1}
\fmfv{decor.size=0,label={\footnotesize #3},l.dist=0.5mm}{v1}
\fmfv{decor.size=0,label={\footnotesize #4},l.dist=0.5mm}{v2}
\end{fmfgraph*} } } }}
\nc{\spvertex}[5]{\frac{\rule[0pt]{0pt}{15pt}\delta 
\parbox{10mm}{\centerline{
\begin{fmfgraph*}(4,5.464)
\fmfset{arrow_len}{1.5mm}
\fmfset{wiggly_len}{1.5mm}
\fmfset{dot_size}{1thick}
\fmfset{arrow_ang}{13}
\fmfpen{0.125mm}
\fmfforce{1w,0.366h}{v1}
\fmfforce{0w,0.366h}{v2}
\fmfforce{0.5w,1h}{v3}
\fmfforce{0.5w,0.549h}{vm}
\fmf{double,width=0.25mm}{v1,vm,v2}
\fmf{double,width=0.25mm}{v3,vm}
\fmfv{decor.shape=circle,decor.filled=empty,decor.size=1thick}{vm}
\fmfv{decor.size=0,label={\footnotesize #2},l.dist=0.5mm}{v1}
\fmfv{decor.size=0,label={\footnotesize #3},l.dist=0.5mm}{v2}
\fmfv{decor.size=0,label={\footnotesize #1},l.dist=0.5mm}{v3}
\end{fmfgraph*}}} \rule[0pt]{0pt}{15pt}}{\delta
\parbox{10mm}{\centerline{
\begin{fmfgraph*}(5,3)
\fmfpen{0.125mm}
\fmfleft{v1}
\fmfright{v2}
\fmf{double,width=0.25mm}{v2,v1}
\fmfv{decor.size=0,label={\footnotesize #4},l.dist=0.5mm}{v1}
\fmfv{decor.size=0,label={\footnotesize #5},l.dist=0.5mm}{v2}
\end{fmfgraph*}}}  }}
\nc{\ndvertexx}[5]{\frac{\delta #1}{\rule[0pt]{0pt}{15pt}\delta
\parbox{10mm}{\centerline{
\begin{fmfgraph*}(4,4)
\fmfset{arrow_len}{1.5mm}
\fmfset{wiggly_len}{1.5mm}
\fmfset{dot_size}{1thick}
\fmfset{arrow_ang}{13}
\fmfpen{0.125mm}
\fmfleft{i2,i1}
\fmfright{o2,o1}
\fmf{double,width=0.25mm}{i1,v1}
\fmf{double,width=0.25mm}{v1,i2}
\fmf{double,width=0.25mm}{v1,o1}
\fmf{double,width=0.25mm}{o2,v1}
\fmfv{decor.size=0,label={\footnotesize #2},l.dist=0.5mm}{o1}
\fmfv{decor.size=0,label={\footnotesize #3},l.dist=0.5mm}{o2}
\fmfv{decor.size=0,label={\footnotesize #4},l.dist=0.5mm}{i2}
\fmfv{decor.size=0,label={\footnotesize #5},l.dist=0.5mm}{i1}
\fmfdot{v1}
\end{fmfgraph*} }} \rule[0pt]{0pt}{15pt} }}
\nc{\ndvertex}[4]{\frac{\delta #1}{\rule[0pt]{0pt}{15pt}\delta
\parbox{10mm}{\centerline{
\begin{fmfgraph*}(4,3.464)
\fmfset{arrow_len}{1.5mm}
\fmfset{wiggly_len}{1.5mm}
\fmfset{dot_size}{1thick}
\fmfset{arrow_ang}{13}
\fmfpen{0.125mm}
\fmfforce{1w,0h}{v1}
\fmfforce{0w,0h}{v2}
\fmfforce{0.5w,1h}{v3}
\fmfforce{0.5w,0.2886h}{vm}
\fmf{double,width=0.25mm}{v1,vm,v2}
\fmf{double,width=0.25mm}{v3,vm}
\fmfv{decor.size=0,label={\footnotesize #3},l.dist=0.5mm}{v1}
\fmfv{decor.size=0,label={\footnotesize #4},l.dist=0.5mm}{v2}
\fmfv{decor.size=0,label={\footnotesize #2},l.dist=0.5mm}{v3}
\fmfdot{vm}
\end{fmfgraph*}}}\rule[0pt]{0pt}{15pt}}}
\nc{\nfdphi}[2]{\frac{\delta #1}{\delta
\parbox{8mm}{\centerline{
\begin{fmfgraph*}(5,3)
\fmfpen{0.125mm}
\fmfset{wiggly_len}{1.5mm}
\fmfset{dot_size}{1thick}
\fmfleft{v1}
\fmfright{v2}
\fmf{boson}{v2,v1}
\fmfdot{v1}
\fmfv{decor.size=0,label={\footnotesize #2},l.dist=0.5mm}{v2}
\end{fmfgraph*}
}}}}
\nc{\ncdphi}[2]{\frac{\delta #1}{\delta
\parbox{10mm}{\centerline{
\begin{fmfgraph*}(5,3)
\fmfpen{0.125mm}
\fmfleft{v1}
\fmfright{v2}
\fmf{double,width=0.25mm}{v2,v1}
\fmfv{decor.size=0,label={\footnotesize #2},l.dist=0.5mm}{v2}
\fmfv{decor.shape=cross,decor.filled=shaded,decor.size=3thick}{v1}
\end{fmfgraph*}
}}}}
\nc{\ndphi}[3]{\frac{\delta #1}{\delta
\parbox{10mm}{\centerline{
\begin{fmfgraph*}(5,3)
\fmfpen{0.125mm}
\fmfleft{v1}
\fmfright{v2}
\fmf{double,width=0.25mm}{v2,v1}
\fmfv{decor.size=0,label={\footnotesize #2},l.dist=0.5mm}{v1}
\fmfv{decor.size=0,label={\footnotesize #3},l.dist=0.5mm}{v2}
\end{fmfgraph*}
}}}}
\nc{\nddphi}[5]{\frac{\delta^2 #1}{\delta
\parbox{10mm}{\centerline{
\begin{fmfgraph*}(5,3)
\fmfpen{0.125mm}
\fmfleft{v1}
\fmfright{v2}
\fmf{double,width=0.25mm}{v2,v1}
\fmfv{decor.size=0,label={\footnotesize #2},l.dist=0.5mm}{v1}
\fmfv{decor.size=0,label={\footnotesize #3},l.dist=0.5mm}{v2}
\end{fmfgraph*}
}}\,\delta
\parbox{10mm}{\centerline{
\begin{fmfgraph*}(5,3)
\fmfpen{0.125mm}
\fmfleft{v1}
\fmfright{v2}
\fmf{double,width=0.25mm}{v2,v1}
\fmfv{decor.size=0,label={\footnotesize #4},l.dist=0.5mm}{v1}
\fmfv{decor.size=0,label={\footnotesize #5},l.dist=0.5mm}{v2}
\end{fmfgraph*}
}}}}
\nc{\nddfphi}[4]{\frac{\delta^2 #1}{\delta
\parbox{10mm}{\centerline{
\begin{fmfgraph*}(5,3)
\fmfpen{0.125mm}
\fmfleft{v1}
\fmfright{v2}
\fmf{double,width=0.25mm}{v2,v1}
\fmfv{decor.size=0,label={\footnotesize #2},l.dist=0.5mm}{v1}
\fmfv{decor.size=0,label={\footnotesize #3},l.dist=0.5mm}{v2}
\end{fmfgraph*}
}}\,\delta
\parbox{8mm}{\centerline{
\begin{fmfgraph*}(5,3)
\fmfpen{0.125mm}
\fmfset{dot_size}{1thick}
\fmfleft{v1}
\fmfright{v2}
\fmf{boson}{v2,v1}
\fmfdot{v1}
\fmfv{decor.size=0,label={\footnotesize #4},l.dist=0.5mm}{v2}
\end{fmfgraph*}
}}}}
\nc{\nddvertex}[6]{\frac{\delta^2 #1}{\rule[0pt]{0pt}{15pt}
\delta \parbox{10mm}{\centerline{
\begin{fmfgraph*}(4,3.464)
\fmfset{arrow_len}{1.5mm}
\fmfset{wiggly_len}{1.5mm}
\fmfset{dot_size}{1thick}
\fmfset{arrow_ang}{13}
\fmfpen{0.125mm}
\fmfforce{1w,0h}{v1}
\fmfforce{0w,0h}{v2}
\fmfforce{0.5w,1h}{v3}
\fmfforce{0.5w,0.2886h}{vm}
\fmf{double,width=0.25mm}{v1,vm,v2}
\fmf{double,width=0.25mm}{v3,vm}
\fmfv{decor.size=0,label={\footnotesize #3},l.dist=0.5mm}{v1}
\fmfv{decor.size=0,label={\footnotesize #4},l.dist=0.5mm}{v2}
\fmfv{decor.size=0,label={\footnotesize #2},l.dist=0.5mm}{v3}
\fmfdot{vm}
\end{fmfgraph*}}}
\,\,\,
\delta \parbox{10mm}{\centerline{
\begin{fmfgraph*}(5,3)
\fmfpen{0.125mm}
\fmfleft{v1}
\fmfright{v2}
\fmf{double,width=0.25mm}{v2,v1}
\fmfv{decor.size=0,label={\footnotesize #5},l.dist=0.5mm}{v1}
\fmfv{decor.size=0,label={\footnotesize #6},l.dist=0.5mm}{v2}
\end{fmfgraph*}}}
}}
\begin{document}
\setlength{\unitlength}{1mm}
\begin{fmffile}{fg1}
\title{Functional Differential Equations for the Free Energy and the Effective
Energy\\ in the Broken-Symmetry Phase
of $\phi^4$-Theory and Their Recursive Graphical Solution}
\author{Axel Pelster, Hagen Kleinert}
\address{Institut f\"ur Theoretische Physik, Freie Universit\"at Berlin, 
Arnimallee 14, 14195 Berlin, Germany}
\date{\today}
\maketitle
\begin{abstract}
Extending recent work on QED and the symmetric phase of 
the euclidean multicomponent scalar $\phi^4$-theory, we construct the
vacuum diagrams of
the free energy and the effective energy in the {\it ordered}
phase of $\phi^4$-theory.  By regarding them as
functionals of the free correlation function and the interaction vertices,
we  graphically solve
nonlinear functional differential equations, obtaining
loop by loop all
connected and one-particle irreducible vacuum diagrams with their proper
weights.
\end{abstract}
\section{Introduction}
Some time ago, one of us proposed a program for a
systematical construction of all Feynman diagrams of a field theory together
with their proper weigths by graphically solving a set of functional
differential equations \cite{Kleinert1,Kleinert2}. 
It relies on considering a Feynman diagram
as a functional of its graphical elements, i.e., its lines and vertices.
Functional derivatives with respect to these 
graphical elements are represented by
removing lines or vertices of a Feynman diagram
in all possible ways. With these graphical operations, the program proceeds
in four steps. First, a nonlinear functional differential equation
for the free energy is derived as a consequence of the field equations.
Subsequently, this functional differential equation is converted into a 
recursion relation for the loop expansion coefficients of the free energy.
From its graphical solution, the connected vacuum diagrams are constructed.
Finally, all diagrams of $n$-point functions are obtained
from removing lines or vertices from the connected vacuum diagrams.\\

This program was recently used
to systematically generate all 
connected Feynman diagrams of QED \cite{QED} and of the euclidean 
multicomponent scalar $\phi^4$ theory \cite{PHI4,Boris}. In the disordered,
symmetric phase of the latter theory, where the field expectation value 
vanishes, the energy functional consists only of even powers of the field.
To generate all connected diagrams of the 
$n$-point functions, it was sufficient
to work with the functional derivative with respect to the free correlation
function \cite{PHI4}. In the ordered phase, however, where the symmetry
is spontaneously broken by a non-zero field expectation value, the
situation is more complicated as the energy functional also contains 
odd powers of the field. To handle these, it is necessary to extend the
symmetric treatment by a second type of functional derivative.
This was first done in Ref.~\cite{Boris} using  functional derivatives with 
respect to both the free correlation function and the external current
by keeping the number of derivatives at a minimum. 
The procedure 
led to {\em two} coupled nonlinear graphical recursion relations 
for each the connected and the one-particle 
irreducible vacuum diagrams, respectively.
In this paper we show that all these vacuum
diagrams can be obtained from a {\em single} 
nonlinear graphical recursion relation which is derived via
functional derivatives with respect to both the free correlation function
and the $3$-vertex.
\section{Negative Free Energy}
Consider a self-interacting scalar field $\phi$ with $N$ components 
in $d$ euclidean dimensions whose
thermal fluctuations are controlled by the energy functional
\beq
\la{EF}
E [ \phi ] = E [ 0 ] - \int_1 J_1 \phi_1 + 
\frac{1}{2} \int_{12} G^{-1}_{12} \phi_1 \phi_2 
+ \frac{1}{6} \int_{123} K_{123}  \phi_1 \phi_2 \phi_3 
+ \frac{1}{24} \int_{1234} L_{1234}  \phi_1 \phi_2 \phi_3 \phi_4 \, .
\eeq
In this short-hand notation, the spatial and tensorial
arguments of the field $\phi$, the current $J$, 
the bilocal kernel $G^{-1}$, as well as the cubic and the quartic 
interactions $K$ and $L$ are indicated
by simple number indices, i.e.,
\beq
1 \equiv \{ x_1 , \alpha_1 \} \, , \,\, 
\int_1 \equiv \sum_{\alpha_1} \int d^d x_1 \, , \,\,
\phi_1 \equiv \phi_{\alpha_1} ( x_1 ) \, , \hspace*{4cm}\no \\
J_1 \equiv J_{\alpha_1} ( x_1 ) \, , \,\,
G^{-1}_{12} \equiv G^{-1}_{\alpha_1 , \alpha_2} ( x_1 , x_2 ) \, ,  \,\,
K_{123} \equiv K_{\alpha_1 , \alpha_2 , \alpha_3} 
( x_1 , x_2 , x_3 ) \, ,\,\,
L_{1234} \equiv L_{\alpha_1 , \alpha_2 , \alpha_3 , \alpha_4} 
( x_1 , x_2 , x_3 , x_4 ) \, .
\eeq
The kernel is a functional matrix $G^{-1}$, while $K$ and $L$ are functional 
tensors, all being symmetric in their respective
indices. The energy functional (\r{EF}) describes generically
$d$-dimensional euclidean $\phi^4$-theories. These are models for
a family of universality classes of continuous phase transitions,
such as the $O(N)$-symmetric $\phi^4$-theory, which serves to derive
the critical phenomena in dilute polymer solutions ($N=0$), Ising- and
Heisenberg-like magnets ($N=1,3$), and superfluids ($N=2$).
In the disordered phase above the critical point, where 
the system displays the full $O(N)$ symmetry and the
field expectation value vanishes, the energy functional (\r{EF}) consists
of even powers of the field and is specified by
\beq
E [ 0 ] & = & 0 \, , \no \\
J_{\alpha_1} ( x_1 ) & = & 0 \, , \no \\
G_{\alpha_1 , \alpha_2}^{-1} ( x_1 , x_2 ) & = &
\delta_{\alpha_1 , \alpha_2} \, \left( - \partial_{x_1}^2 + m^2 
\right) \delta ( x_1 - x_2 ) \, , \no \\ 
K_{\alpha_1,\alpha_2,\alpha_3} ( x_1 , x_2 , x_3 ) & = & 0 \, , \no\\
L_{\alpha_1,\alpha_2,\alpha_3,\alpha_4} ( x_1 , x_2 , x_3 , x_4 ) 
&=&  \frac{g}{3} \, \left\{ 
\delta_{\alpha_1 , \alpha_2} \delta_{\alpha_3 , \alpha_4} +
\delta_{\alpha_1 , \alpha_3} \delta_{\alpha_2 , \alpha_4} +
\delta_{\alpha_1 , \alpha_4} \delta_{\alpha_2 , \alpha_3} \right\} \,
\delta ( x_1 - x_2 ) \delta ( x_1 - x_3 ) \delta ( x_1 - x_4 ) \, . \la{PH1}
\eeq
The bare mass $m^2$ is proportional to the temperature distance
from the critical point, and $g$ is the coupling strength. In the
ordered phase below the critical point, where the symmetry is spontaneously
broken by a non-zero field expectation value, one has to
allow also for odd powers of the field. This situation is modelled by the
energy functional (\r{EF}), (\r{PH1})
if an additional shift of the field $\phi$ around
some background field $\chi$ is taken into account
according to the replacement $\phi \rightarrow \chi + \phi$ 
(compare Section 5.3 in the textbook \cite{Kleinert3}). 
Thus the energy functional (\r{EF}) is specified by
\beq
E [ 0 ] & = &  \frac{1}{2} \sum_{\alpha_1} \int d^d x_1 \chi_{\alpha_1} ( x_1 )
\left( - \partial_{x_1}^2 + m^2 \right) \chi_{\alpha_1} ( x_1 )
+ \frac{g}{24} \sum_{\alpha_1,\alpha_2} \int d^d x_1 
\chi_{\alpha_1}^2 ( x_1 ) \chi_{\alpha_2}^2 ( x_1 ) \, , \no \\
J_{\alpha_1} ( x_1 ) & = & - \left( - \partial_{x_1}^2 + m^2 \right)
\chi_{\alpha_1} ( x_1 ) - \frac{g}{6} \chi_{\alpha_1} ( x_1 ) 
\sum_{\alpha_2} \chi_{\alpha_2}^2 ( x_1 )  \, , \no \\
G_{\alpha_1 , \alpha_2}^{-1} ( x_1 , x_2 ) & = & \left\{ 
\delta_{\alpha_1 , \alpha_2} \left( - \partial_{x_1}^2 + m^2 + \frac{g}{6}
\sum_{\alpha_3} \chi_{\alpha_3}^2 ( x_1 ) \right) + \frac{g}{3} 
\chi_{\alpha_1} ( x_1 ) \chi_{\alpha_2} ( x_1 ) \right\} \delta ( x_1 -
x_2 )  \, , \no \\ 
K_{\alpha_1,\alpha_2,\alpha_3} ( x_1 , x_2 , x_3 ) & = &  \frac{g}{3} \left\{
\delta_{\alpha_1 , \alpha_2} \chi_{\alpha_3} ( x_1 ) +
\delta_{\alpha_1 , \alpha_3} \chi_{\alpha_2} ( x_1 ) +
\delta_{\alpha_2 , \alpha_3} \chi_{\alpha_1} ( x_1 ) \right\}
\delta ( x_1 - x_2 ) \delta ( x_1 - x_3 ) \, , \no \\
 L_{\alpha_1,\alpha_2,\alpha_3,\alpha_4} ( x_1 , x_2 , x_3 , x_4 ) 
&=&   \frac{g}{3} \, \left\{ 
\delta_{\alpha_1 , \alpha_2} \delta_{\alpha_3 , \alpha_4} +
\delta_{\alpha_1 , \alpha_3} \delta_{\alpha_2 , \alpha_4} +
\delta_{\alpha_1 , \alpha_4} \delta_{\alpha_2 , \alpha_3} \right\} \,
\delta ( x_1 - x_2 ) \delta ( x_1 - x_3 ) \delta ( x_1 - x_4 ) \, . \la{PH2}
\eeq
In the following, we shall leave $J$, $G^{-1}$, $K$, and $L$ completely 
general,  except for the symmetry with respect to their indices, and insert
the physical values (\r{PH1}) or (\r{PH2}) only at the end when we 
are looking at the disordered or the ordered phase, respectively. 
By doing so we regard
the energy (\r{EF}) as a functional of its arguments $J$, $G^{-1}$, $K$,
$L$, i.e.
\beq
E [ \phi ] = E [ \phi , J , G^{-1} , K , L ] \, ,
\eeq
so that the same 
functional dependences are inherited by all field-theoretic quantities
derived from it. In particular, we are interested in studying the
dependence of the partition function, which is determined as a functional
integral over a Boltzmann weight in natural units 
\beq
\la{PF}
Z [ J , G^{-1} , K , L ]
= \int {\cal D} {\bf \phi} \, e^{- E [ \phi , J , G^{-1} , K , L ] }  \, ,
\eeq
and its logarithm, the negative free energy
\beq
\la{FE}
W [ J , G^{-1} , K , L ]  = \ln Z [ J , G^{-1} , K , L ] \, .
\eeq
By performing a loop expansion of the partition function (\r{PF}),
the contributions to the negative free energy (\r{FE})
consist of all
connected vacuum diagrams constructed according to Feynman rules. 
A single dot represents the energy shift%
\beq
\la{ES}
\parbox{8mm}{\begin{center}
\begin{fmfgraph*}(5,5)
\setval
\fmfstraight
\fmfforce{1/2w,1/2h}{v1}
\fmfdot{v1}
\end{fmfgraph*}
\end{center}}
\equiv \hspace*{0.1cm} - E [ 0 ] \, ,
\eeq
a cross an integral over the current
\beq
\la{CR}
\parbox{8mm}{\begin{center}
\begin{fmfgraph*}(5,5)
\setval
\fmfstraight
\fmfforce{0w,1/2h}{v1}
\fmfforce{1w,1/2h}{v2}
\fmf{plain}{v1,v2}
\fmfv{decor.shape=cross,decor.filled=shaded,decor.size=3thick}{v1}
\end{fmfgraph*}
\end{center}}
\equiv \hspace*{1mm} \int_1 J_1
\eeq
and a line represents the free correlation function 
\beq
\la{PRO}
\parbox{12mm}{\begin{center}
\begin{fmfgraph*}(5,5)
\setval
\fmfstraight
\fmfleft{v1}
\fmfright{v2}
\fmf{plain}{v1,v2}
\fmflabel{${\scs 1}$}{v1}
\fmflabel{${\scs 2}$}{v2}
\end{fmfgraph*}
\end{center}}
\hspace*{2mm} \equiv \hspace*{1mm} G_{12} \, ,
\eeq
which is the functional inverse of the kernel $G^{-1}$ in the energy
functional (\r{EF}), defined by
\beq
\la{FP}
\int_{2} G_{12} \, G^{-1}_{23} = \delta_{13} \, .
\eeq
A 3-vertex represents an integral over the cubic interaction 
\beq
\la{3V}
\parbox{10mm}{\begin{center}
\begin{fmfgraph*}(5,4.33)
\setval
\fmfforce{1w,0h}{v1}
\fmfforce{0w,0h}{v2}
\fmfforce{0.5w,1h}{v3}
\fmfforce{0.5w,0.2886h}{vm}
\fmf{plain}{v1,vm,v2}
\fmf{plain}{v3,vm}
\fmfdot{vm}
\end{fmfgraph*}
\end{center}}
\equiv \hspace*{0.1cm} - \int_{123} K_{123}
\eeq
and a 4-vertex stands for an integral over the quartic interaction
\beq
\la{4V}
\parbox{17mm}{\begin{center}
\begin{fmfgraph*}(4,4)
\setval
\fmfstraight
\fmfforce{0w,0h}{o2}
\fmfforce{0w,1h}{i1}
\fmfforce{1w,0h}{o1}
\fmfforce{1w,1h}{i2}
\fmfforce{1/2w,1/2h}{v1}
\fmf{plain}{i1,v1}
\fmf{plain}{v1,i2}
\fmf{plain}{v1,o1}
\fmf{plain}{o2,v1}
\fmfdot{v1}
\end{fmfgraph*}
\end{center}} 
\equiv \hspace*{0.1cm} - \int_{1234} \, L_{1234} \, .
\eeq
If the cubic and the quartic interactions $K$ and $L$ in (\r{EF}) vanish,
the functional integral in (\r{PF}) is Gaussian and can be immediately
calculated to obtain for the negative free energy
\beq
\la{W0}
W^{(0)} [ J , G^{-1} , 0 , 0 ] =  - E [ 0 ] 
- \frac{1}{2} \mbox{Tr} \ln G^{-1} + \frac{1}{2} \int_{12} G_{12}
J_1 J_2 \, ,
\eeq
where the trace of the logarithm of the kernel is defined by the series
\cite[p.~16]{Kleinert3}
\beq
\la{LOG}
\mbox{Tr} \ln G^{-1} = \sum_{n = 1}^{\infty} \frac{(-1)^{n + 1}}{n}
\int_{1 \ldots n} \left\{ G^{-1}_{12} - \delta_{12} \right\} \cdots
\left\{ G^{-1}_{n1} - \delta_{n1} \right\} \, .
\eeq
The zeroth order contribution (\r{W0}) to 
the negative free energy will be graphically represented by
\beq
W^{(0)} = 
\parbox{8mm}{\begin{center}
\begin{fmfgraph*}(5,5)
\setval
\fmfstraight
\fmfforce{1/2w,1/2h}{v1}
\fmfdot{v1}
\end{fmfgraph*}
\end{center}}
+ \frac{1}{2}\;
\parbox{8mm}{\centerline{
\begin{fmfgraph}(5,5)
\setval
\fmfi{plain}{reverse fullcircle scaled 1w shifted (0.5w,0.5h)}
\end{fmfgraph} }} 
+ \frac{1}{2} \hspace*{0.1cm}
\parbox{8mm}{\begin{center}
\begin{fmfgraph*}(5,5)
\setval
\fmfstraight
\fmfforce{0w,1/2h}{v1}
\fmfforce{1w,1/2h}{v2}
\fmf{plain}{v1,v2}
\fmfv{decor.shape=cross,decor.filled=shaded,decor.size=3thick}{v1}
\fmfv{decor.shape=cross,decor.filled=shaded,decor.size=3thick}{v2}
\end{fmfgraph*}
\end{center}} \, .
\eeq
In order to find the connected vacuum diagrams of the negative free energy
together with their weights for non-vanishing cubic and quartic
interactions $K$ and $L$, we proceed as follows. We first introduce, 
in Subsection II.A,
functional derivatives with respect to the graphical
elements $J$, $G^{-1}$, $K$, $L$ of Feynman diagrams. With these we derive,
in Subsection II.B, a single nonlinear functional differential equation for
the negative free energy. This is converted into a recursion relation in
Subsection II.C which is solved graphically in Subsection II.D.
In Subsection II.E, we derive even simpler graphical
recursion relations for certain subsets of connected vacuum diagrams.
\subsection{Functional Derivatives}
Each Feynman diagram may be considered as a functional of the quantities
in (\r{EF}) characterizing the field theory,
i.e. of the current $J$, the kernel $G^{-1}$, and the interactions
$K$ and $L$. In this subsection we introduce functional derivatives
with respect to these, identify their associated
graphical operations, and study field-theoretic relations between them.
\subsubsection{Graphical Representation}
We start with studying the functional derivative with respect to the
current $J$, whose basic rule is
\beq
\la{DR1}
\frac{\delta J_2}{\delta J_1} = \delta_{12} \, .
\eeq
We represent this graphically by
extending the elements of Feynman diagrams by an open dot with two labeled
line ends representing the delta function:
\beq
{\scs 1}
\parbox{6mm}{\centerline{
\begin{fmfgraph}(4,3)
\setval
\fmfforce{0w,0.5h}{i1}
\fmfforce{1w,0.5h}{o1}
\fmfforce{0.5w,0.5h}{v1}
\fmf{plain}{i1,v1}
\fmf{plain}{v1,o1}
\fmfv{decor.size=0, label={\scs 1}, l.dist=1mm, l.angle=-180}{i1}
\fmfv{decor.size=0, label={\scs 2}, l.dist=1mm, l.angle=0}{o1}
\fmfv{decor.shape=circle,decor.filled=empty,decor.size=0.6mm}{v1}
\end{fmfgraph}}} {\scs 2}
= \quad \delta_{12} \, .
\eeq
Thus we can write
the differentiation (\r{DR1}) graphically as 
\beq
\cdphi{}{1} 
\parbox{20mm}{\centerline{
\begin{fmfgraph*}(6,3)
\setval
\fmfleft{v1}
\fmfright{v2}
\fmf{plain}{v1,v2}
\fmflabel{${\scs 2}$}{v2}
\fmfv{decor.shape=cross,decor.filled=shaded,decor.size=3thick}{v1}
\end{fmfgraph*}
}} = \hspace*{1mm}
{\scs 1}
\parbox{6mm}{\centerline{
\begin{fmfgraph}(4,3)
\setval
\fmfforce{0w,0.5h}{i1}
\fmfforce{1w,0.5h}{o1}
\fmfforce{0.5w,0.5h}{v1}
\fmf{plain}{i1,v1}
\fmf{plain}{v1,o1}
\fmfv{decor.size=0, label=${\scs 1}$, l.dist=1mm, l.angle=-180}{i1}
\fmfv{decor.size=0, label=${\scs 2}$, l.dist=1mm, l.angle=0}{o1}
\fmfv{decor.shape=circle,decor.filled=empty,decor.size=0.6mm}{v1}
\end{fmfgraph}}} {\scs 2} \, .
\eeq
Differentiating a cross with respect to the current replaces the cross by
the spatial index of the current.\\

Since $\phi$ is a real scalar field, the kernel $G^{-1}$ is a symmetric
functional matrix.  This property has to be taken into account when performing
functional derivatives with respect to the kernel $G^{-1}$, whose basic
rule is \cite{PHI4,Boris}
\beq
\la{DR2}
\frac{\delta G^{-1}_{12}}{\delta G^{-1}_{34}} = \frac{1}{2} \left\{ 
\delta_{13} \delta_{42} + \delta_{14} \delta_{32} \right\} \, .
\eeq
From the identity (\r{FP}) and the functional chain rule, 
we find the effect of this derivative
on the free propagator
\beq
\la{ACT}
- 2 \, \frac{\delta G_{12}}{\delta G^{-1}_{34}} = 
G_{13} G_{42} + G_{14} G_{32} \, .
\eeq
This has the graphical representation
\beq
- 2 \, \frac{\delta}{\delta G^{-1}_{34}}
\parbox{20mm}{\centerline{
\begin{fmfgraph*}(8,3)
\setval
\fmfleft{v1}
\fmfright{v2}
\fmf{plain}{v1,v2}
\fmflabel{${\scs 1}$}{v1}
\fmflabel{${\scs 2}$}{v2}
\end{fmfgraph*}
}}
=
\parbox{20mm}{\centerline{
\begin{fmfgraph*}(8,3)
\setval
\fmfleft{v1}
\fmfright{v2}
\fmf{plain}{v1,v2}
\fmflabel{${\scs 1}$}{v1}
\fmflabel{${\scs 3}$}{v2}
\end{fmfgraph*}
}}
\parbox{20mm}{\centerline{
\begin{fmfgraph*}(8,3)
\setval
\fmfleft{v1}
\fmfright{v2}
\fmf{plain}{v1,v2}
\fmflabel{${\scs 4}$}{v1}
\fmflabel{${\scs 2}$}{v2}
\end{fmfgraph*}
}}
+
\parbox{20mm}{\centerline{
\begin{fmfgraph*}(8,3)
\setval
\fmfleft{v1}
\fmfright{v2}
\fmf{plain}{v1,v2}
\fmflabel{${\scs 1}$}{v1}
\fmflabel{${\scs 4}$}{v2}
\end{fmfgraph*}
}}
\parbox{20mm}{\centerline{
\begin{fmfgraph*}(8,3)
\setval
\fmfleft{v1}
\fmfright{v2}
\fmf{plain}{v1,v2}
\fmflabel{${\scs 3}$}{v1}
\fmflabel{${\scs 2}$}{v2}
\end{fmfgraph*}
}} \, .
\eeq
Thus, differentiating a free correlation function
with respect to the kernel $G^{-1}$
amounts to cutting the associated line
into two pieces. 
The differentiation rule (\r{DR2}) ensures that the spatial indices 
of the kernel are symmetrically
attached to the newly created line ends
in the two possible ways. Differentiating a general
Feynman diagram with respect to $G^{-1}$, the product rule of functional
differentiation leads to diagrams in each of which one of the $2p$ lines
of the original Feynman diagram is cut.\\

We now study the graphical effect of functional derivatives
with respect to the free propagator $G$,
where the basic differentiation rule reads
\beq
\la{DR2b}
\frac{\delta G_{12}}{\delta G_{34}} = \frac{1}{2} \left\{ 
\delta_{13} \delta_{42} + \delta_{14} \delta_{32} \right\} \, .
\eeq
This can be written graphically as follows:
\beq
\dphi{}{3}{4} 
\parbox{20mm}{\centerline{
\begin{fmfgraph*}(8,3)
\setval
\fmfleft{v1}
\fmfright{v2}
\fmf{plain}{v1,v2}
\fmflabel{${\scs 1}$}{v1}
\fmflabel{${\scs 2}$}{v2}
\end{fmfgraph*}
}} = \frac{1}{2} \Bigg\{
{\scs 1}
\parbox{6mm}{\centerline{
\begin{fmfgraph}(4,3)
\setval
\fmfforce{0w,0.5h}{i1}
\fmfforce{1w,0.5h}{o1}
\fmfforce{0.5w,0.5h}{v1}
\fmf{plain}{i1,v1}
\fmf{plain}{v1,o1}
\fmfv{decor.size=0, label=${\scs 1}$, l.dist=1mm, l.angle=-180}{i1}
\fmfv{decor.size=0, label=${\scs 2}$, l.dist=1mm, l.angle=0}{o1}
\fmfv{decor.shape=circle,decor.filled=empty,decor.size=0.6mm}{v1}
\end{fmfgraph}}} {\scs 3} \quad
{\scs 4}
\parbox{6mm}{\centerline{
\begin{fmfgraph}(4,3)
\setval
\fmfforce{0w,0.5h}{i1}
\fmfforce{1w,0.5h}{o1}
\fmfforce{0.5w,0.5h}{v1}
\fmf{plain}{i1,v1}
\fmf{plain}{v1,o1}
\fmfv{decor.size=0, label=${\scs 1}$, l.dist=1mm, l.angle=-180}{i1}
\fmfv{decor.size=0, label=${\scs 2}$, l.dist=1mm, l.angle=0}{o1}
\fmfv{decor.shape=circle,decor.filled=empty,decor.size=0.6mm}{v1}
\end{fmfgraph}}} {\scs 2} \quad + \quad
{\scs 1}
\parbox{6mm}{\centerline{
\begin{fmfgraph}(4,3)
\setval
\fmfforce{0w,0.5h}{i1}
\fmfforce{1w,0.5h}{o1}
\fmfforce{0.5w,0.5h}{v1}
\fmf{plain}{i1,v1}
\fmf{plain}{v1,o1}
\fmfv{decor.size=0, label=${\scs 1}$, l.dist=1mm, l.angle=-180}{i1}
\fmfv{decor.size=0, label=${\scs 2}$, l.dist=1mm, l.angle=0}{o1}
\fmfv{decor.shape=circle,decor.filled=empty,decor.size=0.6mm}{v1}
\end{fmfgraph}}} {\scs 4} \quad
{\scs 3}
\parbox{6mm}{\centerline{
\begin{fmfgraph}(4,3)
\setval
\fmfforce{0w,0.5h}{i1}
\fmfforce{1w,0.5h}{o1}
\fmfforce{0.5w,0.5h}{v1}
\fmf{plain}{i1,v1}
\fmf{plain}{v1,o1}
\fmfv{decor.size=0, label=${\scs 1}$, l.dist=1mm, l.angle=-180}{i1}
\fmfv{decor.size=0, label=${\scs 2}$, l.dist=1mm, l.angle=0}{o1}
\fmfv{decor.shape=circle,decor.filled=empty,decor.size=0.6mm}{v1}
\end{fmfgraph}}} {\scs 2}
\Bigg\} \, .
\eeq
Thus differentiating a line with respect to the free correlation function
removes the line, leaving in a symmetrized way the spatial indices of
the free correlation function on the vertices to which the line was
connected.\\

As the interactions $K$ and $L$ are functional tensors which are symmetric
in their respective indices, their functional derivatives are
\beq
\frac{\delta K_{123}}{\delta K_{456}} & = & \frac{1}{6} \hspace*{1mm}\Bigg\{
\delta_{14} \delta_{25} \delta_{36}+  5 \hspace*{1mm} \mbox{perm.}
 \Bigg\} \, , \la{DR3}\\
\frac{\delta L_{1234}}{\delta L_{5678}} & = & \frac{1}{24} \hspace*{1mm}
 \Bigg\{ \delta_{15} \delta_{26} \delta_{37} \delta_{48}
+ 23 \hspace*{1mm} \mbox{perm.} \Bigg\} \, . \la{DR4}
\eeq
They have the graphical representations
\beq
\dvertex{}{5}{6}{4}
\parbox{15mm}{\centerline{
\begin{fmfgraph*}(5,4.33)
\setval
\fmfforce{1w,0h}{v1}
\fmfforce{0w,0h}{v2}
\fmfforce{0.5w,1h}{v3}
\fmfforce{0.5w,0.2886h}{vm}
\fmf{plain}{v1,vm,v2}
\fmf{plain}{v3,vm}
\fmfv{decor.size=0,label={\footnotesize 2},l.dist=0.5mm}{v1}
\fmfv{decor.size=0,label={\footnotesize 3},l.dist=0.5mm}{v2}
\fmfv{decor.size=0,label={\footnotesize 1},l.dist=0.5mm}{v3}
\fmfdot{vm}
\end{fmfgraph*}
}}& = & \hspace*{2mm}
\frac{1}{6} \hspace*{1mm}
{\displaystyle \left\{ \rule[-10pt]{0pt}{40pt}\hspace*{2mm}
\parbox{15mm}{\centerline{
\begin{fmfgraph*}(12,10.392)
\setval
\fmfforce{1w,0h}{v1}
\fmfforce{0.7w,0.2h}{v1b}
\fmfforce{0.85w,0.1h}{v1c}
\fmfforce{0w,0h}{v2}
\fmfforce{0.3w,0.2h}{v2b}
\fmfforce{0.15w,0.1h}{v2c}
\fmfforce{0.5w,1h}{v3}
\fmfforce{0.5w,0.639h}{v3b}
\fmfforce{0.5w,0.82h}{v3c}
\fmf{plain}{v3,v3c,v3b}
\fmf{plain}{v1,v1c,v1b}
\fmf{plain}{v2b,v2c,v2}
\fmfv{decor.size=0,label={\footnotesize 2},l.dist=0.5mm}{v1}
\fmfv{decor.size=0,label={\footnotesize 3},l.dist=0.5mm}{v2}
\fmfv{decor.size=0,label={\footnotesize 1},l.dist=0.5mm}{v3}
\fmfv{decor.size=0,label={\footnotesize 5},l.dist=0.5mm,l.angle=120}{v1b}
\fmfv{decor.size=0,label={\footnotesize 6},l.dist=0.5mm,l.angle=60}{v2b}
\fmfv{decor.size=0,label={\footnotesize 4},l.dist=0.5mm,l.angle=-90}{v3b}
\fmfv{decor.shape=circle,decor.filled=empty,decor.size=0.6mm}{v1c}
\fmfv{decor.shape=circle,decor.filled=empty,decor.size=0.6mm}{v2c}
\fmfv{decor.shape=circle,decor.filled=empty,decor.size=0.6mm}{v3c}
\end{fmfgraph*}
}} 
+ 5 \hspace*{1mm} \mbox{perm.} \right\} } \, , \\  
\dvertexx{}{6}{7}{8}{5}
\parbox{17mm}{\begin{center}
\begin{fmfgraph*}(4,4)
\setval
\fmfstraight
\fmfforce{0w,0h}{o2}
\fmfforce{0w,1h}{i1}
\fmfforce{1w,0h}{o1}
\fmfforce{1w,1h}{i2}
\fmfforce{1/2w,1/2h}{v1}
\fmf{plain}{i1,v1}
\fmf{plain}{v1,i2}
\fmf{plain}{v1,o1}
\fmf{plain}{o2,v1}
\fmfv{decor.shape=circle,decor.filled=empty,decor.size=1thick}{v1}
\fmfv{decor.size=0,label={\footnotesize 1},l.dist=0.5mm,l.angle=135}{i1}
\fmfv{decor.size=0,label={\footnotesize 2},l.dist=0.5mm,l.angle=45}{i2}
\fmfv{decor.size=0,label={\footnotesize 3},l.dist=0.5mm,l.angle=-45}{o1}
\fmfv{decor.size=0,label={\footnotesize 4},l.dist=0.5mm,l.angle=-135}{o2}
\end{fmfgraph*}
\end{center}}
& = & \hspace*{2mm}
\frac{1}{24} \hspace*{1mm}
{\displaystyle \left\{ \rule[-10pt]{0pt}{40pt}\hspace*{2mm}
\parbox{15mm}{\centerline{
\begin{fmfgraph*}(12,12)
\setval
\fmfforce{1w,1h}{v1}
\fmfforce{0.85w,0.85h}{v1b}
\fmfforce{0.7w,0.7h}{v1c}
\fmfforce{1w,0h}{v2}
\fmfforce{0.85w,0.15h}{v2b}
\fmfforce{0.7w,0.3h}{v2c}
\fmfforce{0w,0h}{v3}
\fmfforce{0.15w,0.15h}{v3b}
\fmfforce{0.3w,0.3h}{v3c}
\fmfforce{0w,1h}{v4}
\fmfforce{0.15w,0.85h}{v4b}
\fmfforce{0.3w,0.7h}{v4c}
\fmf{plain}{v1,v1c,v1b}
\fmf{plain}{v2b,v2c,v2}
\fmf{plain}{v3,v3c,v3b}
\fmf{plain}{v4,v4c,v4b}
\fmfv{decor.size=0,label={\footnotesize 2},l.dist=0.5mm}{v1}
\fmfv{decor.size=0,label={\footnotesize 3},l.dist=0.5mm}{v2}
\fmfv{decor.size=0,label={\footnotesize 4},l.dist=0.5mm}{v3}
\fmfv{decor.size=0,label={\footnotesize 1},l.dist=0.5mm}{v4}
\fmfv{decor.shape=circle,decor.filled=empty,decor.size=0.6mm}{v1b}
\fmfv{decor.shape=circle,decor.filled=empty,decor.size=0.6mm}{v2b}
\fmfv{decor.shape=circle,decor.filled=empty,decor.size=0.6mm}{v3b}
\fmfv{decor.shape=circle,decor.filled=empty,decor.size=0.6mm}{v4b}
\fmfv{decor.size=0,label={\footnotesize 6},l.dist=0.5mm,l.angle=-135}{v1c}
\fmfv{decor.size=0,label={\footnotesize 7},l.dist=0.5mm,l.angle=135}{v2c}
\fmfv{decor.size=0,label={\footnotesize 8},l.dist=0.5mm,l.angle=45}{v3c}
\fmfv{decor.size=0,label={\footnotesize 5},l.dist=0.5mm,l.angle=-45}{v4c}
\end{fmfgraph*}
}} 
+ 23 \hspace*{1mm} \mbox{perm.} \right\} }   \, .
\eeq
Thus, differentiating a 3- or a 4-vertex with respect to the cubic or
the quartic interaction removes this vertex, leaving in a symmetrized way
the spatial indices of the interaction on the line ends to which the
vertex was connected.
\subsubsection{Field-Theoretic Identities}
The functional derivative of the energy functional (\r{EF}) with respect
to the current $J_1$ gives the field $\phi_1$:
\beq
\phi_1  =  - \frac{\delta E [ \phi ]}{\delta J_1} \, . \la{SB1} 
\eeq
Products of fields, on the other hand, can be obtained in various ways from
functional derivatives of the energy functional (\r{EF})
\beq
\phi_1 \phi_2 & = & \frac{\delta^2 E [ \phi ]}{\delta J_1 \delta J_2} =
2 \, \frac{\delta E [ \phi ]}{\delta G^{-1}_{12}} \, , \la{SB2} \\
\phi_1 \phi_2 \phi_3 & = & - \frac{\delta^3 E [ \phi ]}{\delta J_1 
\delta J_2 \delta J_3}  = - 2 \, \frac{\delta^2 E [ \phi ]}{\delta G^{-1}_{12}
\delta J_3} = 6 \, \frac{\delta E [ \phi ]}{\delta K_{123}} \, , \la{SB3} \\
\phi_1 \phi_2 \phi_3 \phi_4 & = & \frac{\delta^4 E [ \phi ]}{\delta J_1 
\delta J_2 \delta J_3 \delta J_4} = 2 \, \frac{\delta^3 E [ \phi ]}{\delta 
G^{-1}_{12} \delta J_3 \delta J_4} = -6 \,
\frac{\delta^2 E [ \phi ]}{\delta K_{123} \delta J_4} = 4 \,
\frac{\delta^2 E [ \phi ]}{\delta G^{-1}_{12} \delta G^{-1}_{34}} = 
24 \, \frac{\delta E [ \phi ]}{\delta L_{1234}}\, , \la{SB4} 
\eeq
as follows from (\r{DR1}), (\r{DR2}), (\r{DR3}) and (\r{DR4}).
Applying these rules to the functional integral
of the $n$-point function
\beq
\la{NPF}
\Gb_{1 \cdots n} = \frac{1}{Z} \int {\cal D} 
\phi \hspace*{1mm} \phi_1 \cdots \phi_n \, e^{- E [ \phi ]} \, , 
\eeq
we obtain the expectation values of the field and its products
from functional
derivatives of the negative free energy $W$:
\beq
\Gb_1 & = & \frac{\delta W}{\delta J_1} \, , \la{CR1} \\
\Gb_{12} & = & -2
\, \frac{\delta W}{\delta G^{-1}_{12}} \, , \la{CR2} \\
\Gb_{123} & =  &
-6  \,\frac{\delta W}{\delta K_{123}} \, , \la{CR3} \\
\Gb_{1234} &=& 
- 24 \,\frac{\delta W}{\delta L_{1234}} 
\, .\la{CR4}
\eeq
Due to the non-uniqueness in (\r{SB2})-(\r{SB4})
there exist various compatibility relations
between the different functional derivatives, for instance
\beq
\frac{\delta W}{\delta G^{-1}_{12}} & =  & - \frac{1}{2}
\left \{ \frac{\delta^2 W}{\delta J_1 \delta J_2} 
+   \frac{\delta W}{\delta J_1} \frac{\delta W}{\delta J_2} \right\} \, , 
\la{CP1} \\
\frac{\delta W}{\delta K_{123}}& =  & \frac{1}{3}
\left\{  \frac{\delta^2 W}{\delta G^{-1}_{12} \delta J_3}
+ \frac{\delta W}{\delta G^{-1}_{12}}  \frac{\delta W}{\delta J_3} 
\right\} \, , \la{CP2} \\
\frac{\delta W}{\delta L_{1234}} & =  & - \frac{1}{6}
\left\{  \frac{\delta^2 W}{\delta G^{-1}_{12} \delta G^{-1}_{34}} 
+  \frac{\delta W}{\delta G^{-1}_{12}} \frac{\delta W}{\delta G^{-1}_{34}} 
\right\} \, .\la{CP3}
\eeq
These imply that there exist different 
ways of obtaining all diagrams of the
$n$-point functions from the connected vacuum diagrams. From (\r{CR2})
and (\r{CP1})
we read off, for instance, that the diagrams of the 
two-point function follow either from cutting a line or from removing two
crosses of the connected vacuum diagrams in all possible ways.\\

Using the basic compatibility relations (\r{CP1})--(\r{CP3}), we
obtain field-theoretic identities for the 
connected $n$-point functions which are defined as the $n$th functional
derivative of the negative free energy $W$ with respect to the current $J$:
\beq
\Gb_{1 \cdots n}^{\rm c} = \frac{\delta^n W}{\delta
J_1 \cdots \delta J_n} \, .
\eeq
In particuar, we obtain
\beq
\Gb_1^{\rm c} &=& \frac{\delta W}{\delta J_1} \,, \la{CPF1} \\
\Gb_{12}^{\rm c} &=& 
-2 \frac{\delta W}{\delta G^{-1}_{12}} -
\Gb_1^{\rm c} \Gb_2^{\rm c} \,, \la{CPF2} \\
\Gb_{123}^{\rm c} &=& -6 \frac{\delta W}{\delta K_{123}} 
- \Gb_{12}^{\rm c} \Gb_3^{\rm c}
- \Gb_{13}^{\rm c} \Gb_2^{\rm c}
- \Gb_{23}^{\rm c} \Gb_1^{\rm c}
- \Gb_1^{\rm c} \Gb_2^{\rm c} \Gb_3^{\rm c}
\,, \la{CPF3}
\eeq
and, correspondingly,
\beq
\Gb_{1234}^{\rm c} & =& 
- 24 \,\frac{\delta W}{\delta L_{1234}} 
- \Gb_{12}^{\rm c}  \Gb_{34}^{\rm c} 
- \Gb_{13}^{\rm c}  \Gb_{24}^{\rm c} 
- \Gb_{14}^{\rm c}  \Gb_{23}^{\rm c} 
- \Gb_{123}^{\rm c}  \Gb_4^{\rm c}
- \Gb_{124}^{\rm c}  \Gb_3^{\rm c}
- \Gb_{134}^{\rm c}  \Gb_2^{\rm c}
- \Gb_{234}^{\rm c}  \Gb_1^{\rm c} 
\no \\ && 
- \Gb_{12}^{\rm c} \Gb_3^{\rm c}\Gb_4^{\rm c}
- \Gb_{13}^{\rm c} \Gb_2^{\rm c}\Gb_4^{\rm c}
- \Gb_{14}^{\rm c} \Gb_2^{\rm c}\Gb_3^{\rm c}
- \Gb_{23}^{\rm c} \Gb_1^{\rm c}\Gb_4^{\rm c}
- \Gb_{24}^{\rm c} \Gb_1^{\rm c}\Gb_3^{\rm c}
- \Gb_{34}^{\rm c} \Gb_1^{\rm c}\Gb_2^{\rm c}
- \Gb_1^{\rm c} \Gb_2^{\rm c} \Gb_3^{\rm c}\Gb_4^{\rm c}
\, .\la{CPF4}
\eeq
From (\r{CPF2}) we read off, for example, that cutting a line 
of the connected vacuum diagrams in all possible ways leads to
disconnected pieces. These are removed by the term $\Gb^{\rm c}_1
\Gb^{\rm c}_2$, thus leading to the diagrams contributing to the 
connected two-point function $\Gb^{\rm c}_{12}$.
\end{fmffile}
\begin{fmffile}{fg2}
\subsection{Functional Differential Equation for $W = \ln Z$}
We derive a first functional differential equation for the negative free
energy $W$ by starting from the identity
\beq
\la{ID1}
\int {\cal D} \phi \, \frac{\delta}{\delta \phi_1} \, \left\{
\phi_2 \, e^{- E [ \phi ]} \right\} = 0 \, , 
\eeq
which follows by direct functional integration from the vanishing
of the exponential at infinite fields. Taking into account the explicit
form of the energy functional (\r{EF}), we perform the functional derivative
with respect to the field and obtain
\beq
\la{ID1b}
\int {\cal D} \phi \left\{ \delta_{12} + J_1 \phi_2 - \int_3 G^{-1}_{13}
\, \phi_2 \phi_3 - \frac{1}{2} \int_{34} K_{134} \, \phi_2 \phi_3 \phi_4 
- \frac{1}{6} \int_{345} L_{1345}\phi_2 \phi_3 \phi_4 \phi_5 \right\}
e^{- E [ \phi ]} = 0 \, .
\eeq
Substituting field products by functional derivatives according to
(\r{SB1})--(\r{SB4}), we
keep the number of these derivatives at a minimum and express
the resulting equation in terms of the partition function (\r{PF}):
\beq
\delta_{12} \, Z + J_1 \, \frac{\delta Z}{\delta J_2} + 2 \int_3 
G^{-1}_{13} \, \frac{\delta Z}{\delta G^{-1}_{23}} + 3
\int_{34} K_{134} \, \frac{\delta Z}{\delta K_{123}} + 4 
\int_{345} L_{1345}
\, \frac{\delta Z}{\delta L_{2345}} = 0 \, .
\eeq
Going over from $Z$ to $W = \ln Z$, we obtain the linear functional
differential equation 
\beq
\la{CHECK1}
\delta_{12} + J_1 \, \frac{\delta W}{\delta J_2} + 2 \int_3 
G^{-1}_{13} \, \frac{\delta W}{\delta G^{-1}_{23}} + 3
\int_{34} K_{134} \, \frac{\delta W}{\delta K_{123}} + 4 
\int_{345} L_{1345}
\, \frac{\delta W}{\delta L_{2345}} = 0 \, .
\eeq
Applying the compatibility relations (\r{CP2}) and (\r{CP3}), 
this linear functional differential
equation turns into the nonlinear one
\beq
&& \delta_{12}   + J_1 \, \frac{\delta W}{\delta J_2} + 2 \int_{3} 
G^{-1}_{13} \, \frac{\delta W}{\delta G^{-1}_{23}} \no \\
&& \hspace*{2cm} = - \int_{34} K_{134} \, 
\left\{ \frac{\delta^2 W}{\delta G^{-1}_{23} 
\delta J_4} + \frac{\delta W}{\delta G^{-1}_{23}} \, \frac{\delta W}{\delta
J_4} \right\} + \frac{2}{3} \int_{345} L_{1345} \, \left\{
\frac{\delta^2 W}{\delta G^{-1}_{23} \delta G^{-1}_{45}} + 
\frac{\delta W}{\delta G^{-1}_{23}} 
\frac{\delta W}{\delta G^{-1}_{45}} \right\} \, , 
\la{NL1}
\eeq
which is identical with Eq.~(56) in Ref. \cite{Boris}. 
In order to eliminate functional derivatives with respect to the current $J$,
we consider the second identity
\beq
\la{ID2}
\int {\cal D} \phi \frac{\delta}{\delta \phi_1} \, 
e^{- E [ \phi ]} = 0 \, , 
\eeq
which leads to 
\beq
\int {\cal D} \phi \left\{ - J_1 + \int_2 G^{-1}_{12} \phi_2 + \frac{1}{2}
\, \int_{23} K_{123} \, \phi_2 \phi_3 + \frac{1}{6} 
\int_{234} L_{1234} \, \phi_2 \phi_3 \phi_4 \right\} \,
e^{- E [ \phi ]} = 0 \, .
\eeq
Applying again the substitution rules (\r{SB1})--(\r{SB3}) while keeping the 
number of functional derivatives at a minimum,
and taking into account the partition function
(\r{PF}), we obtain for the negative free energy $W = \ln Z$:
\beq
\frac{\delta W}{\delta J_1} = \int_2 G_{12} J_2 + \int_{234} K_{234} G_{12}
\frac{\delta W}{\delta G^{-1}_{34}} + \int_{2345} L_{2345} G_{12} 
\frac{\delta W}{\delta K_{345}} \, .
\la{CC}
\eeq
Differentiating this further, we find
\beq
&&\frac{\delta^2 W}{\delta G^{-1}_{23} \delta J_1} = - \frac{1}{2}
\int_4 \left\{ G_{21} G_{34} + G_{24} G_{31} \right\} J_4 
- \frac{1}{2} \int_{456} K_{456} \left\{ 
G_{21} G_{34} + G_{24} G_{31} \right\} \frac{\delta W}{\delta G^{-1}_{56}}
\no \\
&& \hspace*{1cm} - \frac{1}{2} \int_{4567} L_{4567} \left\{ 
G_{21} G_{34} + G_{24} G_{31} \right\} \frac{\delta W}{\delta K_{567}}
+ \int_{456} K_{456} G_{14} 
\frac{\delta^2 W}{\delta G^{-1}_{23}\delta G^{-1}_{56}}
+ \int_{4567} L_{4567} G_{14} 
\frac{\delta^2 W}{\delta G^{-1}_{23}\delta K_{567}} \, .
\la{DD}
\eeq
This allows us to eliminate the functional derivatives with respect
to the current $J$ in Eq.~(\r{NL1}). The result is a single nonlinear
functional differential equation for the negative free energy $W$ 
which involves only
functional derivatives with respect to the kernel $G^{-1}_{12}$
and the cubic interaction $K_{123}$:
\beq
&&
\delta_{11} \int_1 
+ \int_{12} G_{12} J_1 J_2 
+ 2 \int_{12} G^{-1}_{12} 
\frac{\delta W}{\delta G^{-1}_{12}} 
= \int_{1234} K_{123} G_{12} G_{34} J_4 
- 2 \int_{1234} K_{123} G_{14} J_4 \frac{\delta W}{\delta G^{-1}_{23}} 
-\int_{12345} L_{1234} G_{15} J_5 \frac{\delta W}{\delta K_{234}} 
\no \\ && 
+ \int_{123456} K_{123} K_{456} G_{12} G_{34} \frac{\delta W}{\delta 
G^{-1}_{56}}
+ \int_{1234567} K_{123} L_{4567} G_{12} G_{34} \frac{\delta 
W}{\delta K_{567}}
+ \frac{2}{3} \int_{1234} L_{1234} \left\{ \frac{\delta^2 W}{\delta 
G^{-1}_{12} \delta G^{-1}_{34}} 
+ \frac{\delta W}{\delta G^{-1}_{12}} \frac{\delta W}{\delta G^{-1}_{34}} 
\right\} \no \\ &&
- \int_{1234567} K_{123} L_{4567} G_{14} \left\{ \frac{\delta ^2 W}{\delta 
G^{-1}_{23} \delta K_{567}} 
+ \frac{\delta W}{\delta G^{-1}_{23}} \frac{\delta W}{\delta K_{567}}
\right\}
- \int_{123456} K_{123} K_{456} G_{14} \left\{ \frac{\delta^2 W}{\delta 
G^{-1}_{23} \delta G^{-1}_{56}}  
+ \frac{\delta W}{\delta G^{-1}_{23}} 
\frac{\delta W}{\delta G^{-1}_{56}} \right\} 
\, . 
\la{NL2}
\eeq
Note that due to (\r{CP2}) the functional derivative with respect to 
the cubic interaction $K$ in (\r{CC}) is compatible with functional derivatives
with respect to the current $J$ and the kernel $G^{-1}$.
Inserting (\r{CP2}) in (\r{CC}) would lead to Eq.~(55) in Ref. \cite{Boris},
\beq
\frac{\delta W}{\delta J_1} = \int_2 G_{12} J_2 + \int_{234} K_{234} G_{12}
\frac{\delta W}{\delta G^{-1}_{34}} + \frac{1}{3} \int_{2345} L_{2345} G_{12} 
\left\{ \frac{\delta^2 W}{\delta G^{-1}_{12} \delta J_3} + 
\frac{\delta W}{\delta G^{-1}_{12}} \frac{\delta W}{\delta J_3} \right\}
\, ,
\la{CC2}
\eeq
such that functional derivatives with respect to the current $J$
in Eq.~(\r{NL1})
can no longer be eliminated. This line of approach has been 
pursued in Ref.~\cite{Boris} where
the two coupled nonlinear differential equations (\r{NL1})
and (\r{CC2}) for the negative free energy are used for deriving
all connected vacuum diagrams.
\subsection{Graphical Relation}
With the help of the functional chain rule, the first and second 
derivatives with respect to the kernel $G^{-1}$ are rewritten as
\beq
\la{NR1}
\frac{\delta}{\delta G^{-1}_{12}} = - \int_{34} G_{13} G_{24} 
\frac{\delta}{\delta G_{34}} 
\eeq
and
\beq
\label{NR2}
\frac{\delta^2}{\delta G^{-1}_{12} \delta G^{-1}_{34}} & = & \int_{5678}
G_{15} G_{26} G_{37} G_{48} \frac{\delta^2}{\delta G_{56} \delta G_{78}} 
\no \\
&&+ \frac{1}{2} \int_{56} \left\{ G_{13} G_{25} G_{46} +
G_{14} G_{25} G_{36} + G_{23} G_{15} G_{46} + G_{24} G_{15} G_{36} 
\right\} \frac{\delta}{\delta G_{56}} \, ,
\eeq
respectively. The functional differential equation (\r{NL2}) for  
$W$ takes the form
\beq
&&\delta_{11} \int_1 
+ \int_{12} G_{12} J_1 J_2 
- 2 \int_{12} G_{12} \frac{\delta W}{\delta G_{12}} 
= \int_{1234} K_{123} G_{12} G_{34} J_4
\no \\
&& \hspace*{5mm}
+ 2 \int_{123456} K_{123} G_{14} G_{25} G_{36} J_4 
\frac{\delta W}{\delta G_{56}} 
- \int_{12345} L_{1234} G_{15} J_5 \frac{\delta W}{\delta K_{234}} 
- 2 \int_{12345678} K_{123} K_{456} G_{14} G_{25} G_{37} G_{68} 
\frac{\delta W}{\delta G_{78}} 
\no \\
&&\hspace*{5mm}
- \int_{12345678} K_{123} K_{456} G_{12} G_{34} G_{57} G_{68} 
\frac{\delta W}{\delta G_{78}} 
- \int_{123456789\bar{1}}
K_{123} K_{456} G_{14} G_{27} G_{38} G_{59} G_{6\bar{1}} \left\{ 
\frac{\delta^2 W}{\delta G_{78} \delta G_{9\bar{1}}} + \frac{\delta W}{\delta
G_{78}} \frac{\delta W}{\delta G_{9\bar{1}}} \right\} 
\no \\
&&\hspace*{5mm}
+ \int_{1234567} K_{123} L_{4567} G_{12} G_{34}
\frac{\delta W}{\delta K_{567}} 
+ \int_{123456789} K_{123} L_{4567} G_{14} G_{27} G_{38} \left\{ 
\frac{\delta^2 W}{\delta K_{567} \delta G_{89}} + \frac{\delta W}{\delta 
K_{567}} \frac{\delta W}{\delta G_{89}}
\right\} 
\no \\
&&\hspace*{5mm}
+ \frac{4}{3} \int_{123456} L_{1234} G_{12} G_{35} G_{46} \frac{\delta 
W}{\delta G_{56}} 
+ \frac{2}{3} \int_{12345678} L_{1234} G_{15} G_{26} G_{37} G_{48} \left\{ 
\frac{\delta^2 W}{\delta G_{56} \delta G_{78}} 
+ \frac{\delta W}{\delta G_{56}} \frac{\delta W}{\delta G_{78}} \right\} \, .
\la{NL3}
\eeq
If the cubic and the quartic interactions $K$ and $L$ vanish, Eq.~(\r{NL3})
is solved by the zeroth order contribution to the negative free energy
(\r{W0}) which has the functional derivatives
\beq
\la{SIDE}
\frac{\delta W^{(0)}}{\delta J_1} = \int_2 G_{12} J_2 \, , \hspace*{4mm}
\frac{\delta W^{(0)}}{\delta G_{12}} = \frac{1}{2} G^{-1}_{12} +
\frac{1}{2} J_1 J_2 \, , \hspace*{4mm}
\frac{\delta^2 W^{(0)}}{\delta G_{12} \delta G_{34}} = 
- \frac{1}{4} \left\{ G^{-1}_{13} G^{-1}_{24} + G^{-1}_{14} G^{-1}_{23}
\right\} \, .
\eeq
For non-vanishing cubic and quartic interactions $K$ and $L$, the 
right-hand side in Eq.~(\r{NL3}) produces corrections to (\r{W0})
which we shall denote by $W^{({\rm int})}$. Thus
the negative free energy $W$ decomposes according to
\beq
\la{DEC}
W = W^{(0)} + W^{({\rm int})} \, .
\eeq
Inserting this into (\r{NL3}) and using (\ref{SIDE}),
we obtain the following functional differential equation for the interaction
negative free energy $W^{({\rm int})}$:
\beq
&&\int_{12} G_{12} \frac{\delta W^{({\rm int})}}{\delta G_{12}} 
= - \frac{1}{4} \int_{1234} L_{1234} G_{12} G_{34} 
+ \frac{1}{4} \int_{123456} K_{123} K_{456} G_{14} G_{25} G_{36} 
+ \frac{3}{8}\int_{123456} K_{123} K_{456} G_{14} G_{23} G_{56} 
\no \\ && \hspace*{5mm}
- \int_{1234} K_{123} G_{12} G_{34} J_4 
- \frac{1}{2} \int_{123456} L_{1234} G_{12} G_{35} G_{46} J_5 J_6 
+ \frac{1}{2} \int_{12345678} K_{123} K_{456} G_{14} G_{25} G_{37} 
G_{68} J_7 J_8 
\no \\ && \hspace*{5mm}
+ \frac{1}{2} \int_{12345678} K_{123} K_{456} G_{12} G_{34} G_{57} 
G_{68} J_7 J_8 
- \frac{1}{2} \int_{123456} K_{123} G_{14} G_{25} G_{36} J_4 J_5 J_6 
\no \\ &&\hspace*{5mm}
- \frac{1}{12} \int_{12345678} L_{1234} G_{15} G_{26} G_{37} G_{48} 
J_5 J_6 J_7 J_8 
+ \frac{1}{8} \int_{123456789\bar{1}} K_{123} K_{456} G_{14} G_{27} G_{38} 
G_{59} G_{6\bar{1}}J_7 J_8 J_9 J_{\bar{1}} 
\no \\ &&\hspace*{5mm}
- \int_{123456} L_{1234} G_{12} G_{35} G_{46} \frac{\delta 
W^{({\rm int})}}{\delta G_{56}} 
+ \int_{12345678} K_{123} K_{456} G_{14} G_{25} G_{37} G_{48} 
\frac{\delta W^{({\rm int})}}{\delta G_{78}}
\no \\ &&\hspace*{5mm}
+ \int_{12345678} K_{123} K_{456} G_{12} G_{34} G_{57} G_{68}
\frac{\delta W^{({\rm int})}}{\delta G_{78}} 
- \frac{3}{4} \int_{1234567} K_{123} L_{4567} G_{12} G_{34} 
\frac{\delta W^{({\rm int})}}{\delta K_{567}}
\no \\ &&\hspace*{5mm}
- \int_{123456} K_{123} G_{14} G_{25} G_{36} J_4 
\frac{\delta W^{({\rm int})}}{\delta G_{56}}
+ \frac{1}{2} \int_{12345} L_{1234} G_{15} J_5 
\frac{\delta W^{({\rm int})}}{\delta K_{234}} 
+\frac{1}{3} \int_{12345678} L_{1234} G_{15} G_{26} G_{37} G_{48} J_5 J_6
\frac{\delta W^{({\rm int})}}{\delta G_{78}}
\no \\ && \hspace*{5mm}
+ \frac{1}{2} \int_{123456789\bar{1}} K_{123} K_{456} G_{14} G_{27} G_{38}
G_{59} G_{6\bar{1}} J_7 J_8 \frac{\delta W^{({\rm int})}}{\delta G_{9\bar{1}}}
- \frac{1}{4} \int_{123456789} K_{123} L_{4567} G_{14} G_{27} G_{38}
J_7 J_8 \frac{\delta W^{({\rm int})}}{\delta K_{567}}
\no \\ &&  \hspace*{5mm}
- \frac{1}{3} \int_{12345678} L_{1234} G_{15} G_{26} G_{37} G_{48}
\left\{ \frac{\delta^2 W^{({\rm int})}}{\delta G_{56} \delta G_{78}} +
\frac{\delta W^{({\rm int})}}{\delta G_{56}}
\frac{\delta W^{({\rm int})}}{\delta G_{78}} \right\} 
\no \\ &&\hspace*{5mm}
+ \frac{1}{2} \int_{123456789\bar{1}} K_{123} K_{456} G_{14} G_{27} G_{38}
G_{59} G_{6\bar{1}} \left\{ \frac{\delta^2 W^{({\rm int})}}{\delta G_{78} 
\delta G_{9\bar{1}}} + \frac{\delta W^{({\rm int})}}{\delta G_{78}}
\frac{\delta W^{({\rm int})}}{\delta G_{9\bar{1}}} \right\} 
\no \\ &&\hspace*{5mm}
- \frac{1}{2} \int_{123456789} K_{123} L_{4567} G_{14} G_{27} G_{38} 
\left\{ \frac{\delta^2 W^{({\rm int})}}{\delta K_{567} 
\delta G_{89}} + \frac{\delta W^{({\rm int})}}{\delta K_{567}}
\frac{\delta W^{({\rm int})}}{\delta G_{89}} \right\} \, . \la{IT1}
\eeq
With the help of the graphical rules (\r{CR}), (\r{PRO}), (\r{3V}),
(\r{4V}), this can be written diagrammatically as follows:
\beq
&&
\parbox{5.5mm}{\begin{center}
\begin{fmfgraph*}(2.5,5)
\setval
\fmfstraight
\fmfforce{1w,0h}{v1}
\fmfforce{1w,1h}{v2}
\fmf{plain,left=1}{v1,v2}
\fmfv{decor.size=0, label=${\scs 2}$, l.dist=1mm, l.angle=0}{v1}
\fmfv{decor.size=0, label=${\scs 1}$, l.dist=1mm, l.angle=0}{v2}
\end{fmfgraph*}
\end{center}}
\hspace*{0.3cm} \dphi{W^{({\rm int})}}{1}{2} = 
\hspace*{1mm} \frac{1}{4}\hspace*{1mm}
\parbox{11mm}{\begin{center}
\begin{fmfgraph*}(8,4)
\setval
\fmfleft{i1}
\fmfright{o1}
\fmf{plain,left=1}{i1,v1,i1}
\fmf{plain,left=1}{o1,v1,o1}
\fmfdot{v1}
\end{fmfgraph*}\end{center}}
\hspace*{1mm}+ \frac{1}{4}\hspace*{1mm}
\parbox{7mm}{\begin{center}
\begin{fmfgraph*}(4,4)
\setval
\fmfforce{0w,0.5h}{v1}
\fmfforce{1w,0.5h}{v2}
\fmf{plain,left=1}{v1,v2,v1}
\fmf{plain}{v1,v2}
\fmfdot{v1,v2}
\end{fmfgraph*}\end{center}}
\hspace*{1mm}+ \frac{3}{8}\hspace*{1mm}
\parbox{15mm}{\begin{center}
\begin{fmfgraph*}(12,4)
\setval
\fmfforce{0w,0.5h}{v1}
\fmfforce{1/3w,0.5h}{v2}
\fmfforce{2/3w,0.5h}{v3}
\fmfforce{1w,0.5h}{v4}
\fmf{plain,left=1}{v1,v2,v1}
\fmf{plain,left=1}{v3,v4,v3}
\fmf{plain}{v2,v3}
\fmfdot{v2,v3}
\end{fmfgraph*}\end{center}} 
\hspace*{1mm}+ \hspace*{1mm}
\parbox{11mm}{\begin{center}
\begin{fmfgraph*}(8,4)
\setval
\fmfforce{0w,1/2h}{v1}
\fmfforce{1/2w,1/2h}{v2}
\fmfforce{1w,1/2h}{v3}
\fmf{plain,left=1}{v2,v3,v2}
\fmf{plain}{v1,v2}
\fmfdot{v2}
\fmfv{decor.shape=cross,decor.filled=shaded,decor.size=3thick}{v1}
\end{fmfgraph*}\end{center}}
\hspace*{1mm}+ \frac{1}{2}\hspace*{1mm}
\parbox{11mm}{\begin{center}
\begin{fmfgraph*}(8,4)
\setval
\fmfforce{0w,0h}{v1}
\fmfforce{1/2w,0h}{v2}
\fmfforce{1/2w,1h}{v3}
\fmfforce{1w,0h}{w1}
\fmf{plain,left=1}{v2,v3,v2}
\fmf{plain}{v1,w1}
\fmfdot{v2}
\fmfv{decor.shape=cross,decor.filled=shaded,decor.size=3thick}{v1}
\fmfv{decor.shape=cross,decor.filled=shaded,decor.size=3thick}{w1}
\end{fmfgraph*}\end{center}}
\hspace*{1mm}+ \frac{1}{2}\hspace*{1mm}
\parbox{15mm}{\begin{center}
\begin{fmfgraph*}(12,4)
\setval
\fmfforce{0w,1/2h}{v1}
\fmfforce{1/3w,1/2h}{v2}
\fmfforce{2/3w,1/2h}{v3}
\fmfforce{1w,1/2h}{w1}
\fmf{plain,left=1}{v2,v3,v2}
\fmf{plain}{v1,v2}
\fmf{plain}{v3,w1}
\fmfdot{v2,v3}
\fmfv{decor.shape=cross,decor.filled=shaded,decor.size=3thick}{v1}
\fmfv{decor.shape=cross,decor.filled=shaded,decor.size=3thick}{w1}
\end{fmfgraph*}\end{center}}
\hspace*{1mm}+ \frac{1}{2}\hspace*{1mm}
\parbox{11mm}{\begin{center}
\begin{fmfgraph*}(8,8)
\setval
\fmfforce{0w,0h}{v1}
\fmfforce{1/2w,0h}{v2}
\fmfforce{1/2w,1/2h}{v3}
\fmfforce{1/2w,1h}{v4}
\fmfforce{1w,0h}{w1}
\fmf{plain,left=1}{v3,v4,v3}
\fmf{plain}{v1,w1}
\fmf{plain}{v2,v3}
\fmfdot{v2,v3}
\fmfv{decor.shape=cross,decor.filled=shaded,decor.size=3thick}{v1}
\fmfv{decor.shape=cross,decor.filled=shaded,decor.size=3thick}{w1}
\end{fmfgraph*}\end{center}}
\no \\
&& 
\hspace*{5mm} + \frac{1}{2}\hspace*{1mm}
\parbox{11mm}{\begin{center}
\begin{fmfgraph*}(6.928,12)
\setval
\fmfforce{1/2w,5/6h}{v1}
\fmfforce{1w,1/4h}{w1}
\fmfforce{0w,1/4h}{u1}
\fmfforce{1/2w,1/2h}{v2}
\fmf{plain}{v2,v1}
\fmf{plain}{v2,w1}
\fmf{plain}{v2,u1}
\fmfdot{v2}
\fmfv{decor.shape=cross,decor.filled=shaded,decor.size=3thick}{v1}
\fmfv{decor.shape=cross,decor.filled=shaded,decor.size=3thick}{w1}
\fmfv{decor.shape=cross,decor.filled=shaded,decor.size=3thick}{u1}
\end{fmfgraph*}\end{center}}
\hspace*{1mm} + \frac{1}{12}\hspace*{1mm}
\parbox{11mm}{\begin{center}
\begin{fmfgraph*}(8,12)
\setval
\fmfforce{0w,1/2h}{v1}
\fmfforce{1w,1/2h}{w1}
\fmfforce{1/2w,1/6h}{u1}
\fmfforce{1/2w,5/6h}{x1}
\fmfforce{1/2w,1/2h}{v2}
\fmf{plain}{w1,v1}
\fmf{plain}{x1,u1}
\fmfdot{v2}
\fmfv{decor.shape=cross,decor.filled=shaded,decor.size=3thick}{u1}
\fmfv{decor.shape=cross,decor.filled=shaded,decor.size=3thick}{v1}
\fmfv{decor.shape=cross,decor.filled=shaded,decor.size=3thick}{w1}
\fmfv{decor.shape=cross,decor.filled=shaded,decor.size=3thick}{x1}
\end{fmfgraph*}\end{center}}
+\hspace*{1mm} \frac{1}{8}\hspace*{1mm}
\parbox{11mm}{\begin{center}
\begin{fmfgraph*}(8,4)
\setval
\fmfforce{0w,0h}{v1}
\fmfforce{1w,0h}{w1}
\fmfforce{0w,1h}{u1}
\fmfforce{1w,1h}{x1}
\fmfforce{1/2w,0h}{v2}
\fmfforce{1/2w,1h}{v3}
\fmf{plain}{w1,v1}
\fmf{plain}{v2,v3}
\fmf{plain}{x1,u1}
\fmfdot{v2,v3}
\fmfv{decor.shape=cross,decor.filled=shaded,decor.size=3thick}{u1}
\fmfv{decor.shape=cross,decor.filled=shaded,decor.size=3thick}{v1}
\fmfv{decor.shape=cross,decor.filled=shaded,decor.size=3thick}{w1}
\fmfv{decor.shape=cross,decor.filled=shaded,decor.size=3thick}{x1}
\end{fmfgraph*}\end{center}}
\hspace*{1mm}+\hspace*{1mm} 
\parbox{9mm}{\begin{center}
\begin{fmfgraph*}(6,4)
\setval
\fmfstraight
\fmfforce{0w,1/2h}{v1}
\fmfforce{4/6w,1/2h}{v2}
\fmfforce{1w,1h}{i2}
\fmfforce{1w,0h}{i1}
\fmf{plain}{i1,v2}
\fmf{plain}{v2,i2}
\fmf{plain,left}{v1,v2,v1}
\fmfdot{v2}
\fmfv{decor.size=0, label=${\scs 2}$, l.dist=1mm, l.angle=0}{i1}
\fmfv{decor.size=0, label=${\scs 1}$, l.dist=1mm, l.angle=0}{i2}
\end{fmfgraph*}
\end{center}}
\hspace*{3mm} \dphi{W^{({\rm int})}}{1}{2} 
\hspace*{1mm} + \hspace*{1mm}
\parbox{7mm}{\begin{center}
\begin{fmfgraph*}(6,4)
\setval
\fmfstraight
\fmfforce{1/3w,0h}{v1}
\fmfforce{1/3w,1h}{v2}
\fmfforce{1w,1h}{i2}
\fmfforce{1w,0h}{i1}
\fmf{plain}{i1,v1}
\fmf{plain}{v2,i2}
\fmf{plain,left}{v1,v2,v1}
\fmfdot{v2,v1}
\fmfv{decor.size=0, label=${\scs 2}$, l.dist=1mm, l.angle=0}{i1}
\fmfv{decor.size=0, label=${\scs 1}$, l.dist=1mm, l.angle=0}{i2}
\end{fmfgraph*}
\end{center}}
\hspace*{3mm} \dphi{W^{({\rm int})}}{1}{2}
\hspace*{1mm} + \hspace*{1mm}
\parbox{13mm}{\begin{center}
\begin{fmfgraph*}(10,4)
\setval
\fmfstraight
\fmfforce{0w,1/2h}{v1}
\fmfforce{4/10w,1/2h}{v2}
\fmfforce{8/10w,1/2h}{v3}
\fmfforce{1w,0h}{i2}
\fmfforce{1w,1h}{i1}
\fmf{plain,left}{v1,v2,v1}
\fmf{plain}{v3,v2}
\fmf{plain}{v3,i1}
\fmf{plain}{v3,i2}
\fmfdot{v2,v3}
\fmfv{decor.size=0, label=${\scs 2}$, l.dist=1mm, l.angle=0}{i2}
\fmfv{decor.size=0, label=${\scs 1}$, l.dist=1mm, l.angle=0}{i1}
\end{fmfgraph*}
\end{center}}
\hspace*{3mm} \dphi{W^{({\rm int})}}{1}{2}
\nonumber \\ &&
\hspace*{5mm} 
+ \frac{3}{4} \hspace*{1mm}
\parbox{13mm}{\begin{center}
\begin{fmfgraph*}(10,4)
\setval
\fmfstraight
\fmfforce{0w,1/2h}{v1}
\fmfforce{4/10w,1/2h}{v2}
\fmfforce{8/10w,1/2h}{v3}
\fmfforce{1w,-0.25h}{i1}
\fmfforce{1w,0.5h}{i2}
\fmfforce{1w,1.25h}{i3}
\fmf{plain,left}{v1,v2,v1}
\fmf{plain}{v3,v2}
\fmf{plain}{v3,i1}
\fmf{plain}{v3,i2}
\fmf{plain}{v3,i3}
\fmfdot{v2,v3}
\fmfv{decor.size=0, label=${\scs 1}$, l.dist=1mm, l.angle=0}{i3}
\fmfv{decor.size=0, label=${\scs 2}$, l.dist=1mm, l.angle=0}{i2}
\fmfv{decor.size=0, label=${\scs 3}$, l.dist=1mm, l.angle=0}{i1}
\end{fmfgraph*}
\end{center}}
\hspace*{3mm} \dvertex{W^{({\rm int})}}{1}{2}{3} \hspace*{1mm} +
\parbox{9mm}{\begin{center}
\begin{fmfgraph*}(6,4)
\setval
\fmfstraight
\fmfforce{0w,1/2h}{v1}
\fmfforce{4/6w,1/2h}{v2}
\fmfforce{1w,1h}{i1}
\fmfforce{1w,0h}{i2}
\fmf{plain}{v2,v1}
\fmf{plain}{v2,i1}
\fmf{plain}{v2,i2}
\fmfdot{v2}
\fmfv{decor.shape=cross,decor.filled=shaded,decor.size=3thick}{v1}
\fmfv{decor.size=0, label=${\scs 2}$, l.dist=1mm, l.angle=0}{i2}
\fmfv{decor.size=0, label=${\scs 1}$, l.dist=1mm, l.angle=0}{i1}
\end{fmfgraph*}
\end{center}}
\hspace*{3mm} \dphi{W^{({\rm int})}}{1}{2}
\hspace*{1mm} + \frac{1}{2} \hspace*{1mm}
\parbox{9mm}{\begin{center}
\begin{fmfgraph*}(6,4)
\setval
\fmfstraight
\fmfforce{0w,1/2h}{v1}
\fmfforce{4/6w,1/2h}{v2}
\fmfforce{1w,1.25h}{i1}
\fmfforce{1w,1/2h}{i2}
\fmfforce{1w,-0.25h}{i3}
\fmf{plain}{v2,v1}
\fmf{plain}{v2,i1}
\fmf{plain}{v2,i2}
\fmf{plain}{v2,i3}
\fmfdot{v2}
\fmfv{decor.shape=cross,decor.filled=shaded,decor.size=3thick}{v1}
\fmfv{decor.size=0, label=${\scs 3}$, l.dist=1mm, l.angle=0}{i3}
\fmfv{decor.size=0, label=${\scs 2}$, l.dist=1mm, l.angle=0}{i2}
\fmfv{decor.size=0, label=${\scs 1}$, l.dist=1mm, l.angle=0}{i1}
\end{fmfgraph*}
\end{center}}
\hspace*{3mm} \dvertex{W^{({\rm int})}}{1}{2}{3} 
\hspace*{1mm} + \frac{1}{3} \hspace*{1mm}
\parbox{9mm}{\begin{center}
\begin{fmfgraph*}(6,6)
\setval
\fmfstraight
\fmfforce{0w,1h}{v1}
\fmfforce{0w,0h}{v2}
\fmfforce{4/6w,1/2h}{v3}
\fmfforce{1w,5/6h}{i1}
\fmfforce{1w,1/6h}{i2}
\fmf{plain}{v3,v1}
\fmf{plain}{v3,v2}
\fmf{plain}{v3,i1}
\fmf{plain}{v3,i2}
\fmfdot{v3}
\fmfv{decor.shape=cross,decor.filled=shaded,decor.size=3thick}{v1}
\fmfv{decor.shape=cross,decor.filled=shaded,decor.size=3thick}{v2}
\fmfv{decor.size=0, label=${\scs 2}$, l.dist=1mm, l.angle=0}{i2}
\fmfv{decor.size=0, label=${\scs 1}$, l.dist=1mm, l.angle=0}{i1}
\end{fmfgraph*}
\end{center}}
\hspace*{3mm} \dphi{W^{({\rm int})}}{1}{2}
\hspace*{1mm} + \frac{1}{2} \hspace*{1mm}
\parbox{13mm}{\begin{center}
\begin{fmfgraph*}(10,6)
\setval
\fmfstraight
\fmfforce{0w,1h}{v1}
\fmfforce{0w,0h}{v2}
\fmfforce{4/10w,1/2h}{v3}
\fmfforce{8/10w,1/2h}{v4}
\fmfforce{1w,5/6h}{i1}
\fmfforce{1w,1/6h}{i2}
\fmf{plain}{v3,v1}
\fmf{plain}{v3,v2}
\fmf{plain}{v3,v4}
\fmf{plain}{v4,i1}
\fmf{plain}{v4,i2}
\fmfdot{v3,v4}
\fmfv{decor.shape=cross,decor.filled=shaded,decor.size=3thick}{v1}
\fmfv{decor.shape=cross,decor.filled=shaded,decor.size=3thick}{v2}
\fmfv{decor.size=0, label=${\scs 2}$, l.dist=1mm, l.angle=0}{i2}
\fmfv{decor.size=0, label=${\scs 1}$, l.dist=1mm, l.angle=0}{i1}
\end{fmfgraph*}
\end{center}}
\hspace*{3mm} \dphi{W^{({\rm int})}}{1}{2}
\no \\ &&
\hspace*{5mm} + \frac{1}{4} \hspace*{1mm}
\parbox{13mm}{\begin{center}
\begin{fmfgraph*}(10,6)
\setval
\fmfstraight
\fmfforce{0w,1h}{v1}
\fmfforce{0w,0h}{v2}
\fmfforce{4/10w,1/2h}{v3}
\fmfforce{8/10w,1/2h}{v4}
\fmfforce{1w,7/6h}{i1}
\fmfforce{1w,1/2h}{i2}
\fmfforce{1w,-1/6h}{i3}
\fmf{plain}{v3,v1}
\fmf{plain}{v3,v2}
\fmf{plain}{v3,v4}
\fmf{plain}{v4,i1}
\fmf{plain}{v4,i2}
\fmf{plain}{v4,i3}
\fmfdot{v3,v4}
\fmfv{decor.shape=cross,decor.filled=shaded,decor.size=3thick}{v1}
\fmfv{decor.shape=cross,decor.filled=shaded,decor.size=3thick}{v2}
\fmfv{decor.size=0, label=${\scs 3}$, l.dist=1mm, l.angle=0}{i3}
\fmfv{decor.size=0, label=${\scs 2}$, l.dist=1mm, l.angle=0}{i2}
\fmfv{decor.size=0, label=${\scs 1}$, l.dist=1mm, l.angle=0}{i1}
\end{fmfgraph*}
\end{center}}
\hspace*{3mm} \dvertex{W^{({\rm int})}}{1}{2}{3} \hspace*{1mm} 
+ \frac{1}{3} \hspace*{1mm}
\parbox{7mm}{\begin{center}
\begin{fmfgraph*}(3,3)
\setval
\fmfstraight
\fmfforce{0w,1/2h}{v1}
\fmfforce{1w,2h}{i1}
\fmfforce{1w,1h}{i2}
\fmfforce{1w,0h}{i3}
\fmfforce{1w,-1h}{i4}
\fmf{plain}{v1,i1}
\fmf{plain}{v1,i2}
\fmf{plain}{v1,i3}
\fmf{plain}{v1,i4}
\fmfdot{v1}
\fmfv{decor.size=0, label=${\scs 4}$, l.dist=1mm, l.angle=0}{i4}
\fmfv{decor.size=0, label=${\scs 3}$, l.dist=1mm, l.angle=0}{i3}
\fmfv{decor.size=0, label=${\scs 2}$, l.dist=1mm, l.angle=0}{i2}
\fmfv{decor.size=0, label=${\scs 1}$, l.dist=1mm, l.angle=0}{i1}
\end{fmfgraph*}
\end{center}}
\hspace*{3mm} \ddphi{W^{({\rm int})}}{1}{2}{3}{4}
\hspace*{1mm} + \frac{1}{2} \hspace*{1mm}
\parbox{7mm}{\begin{center}
\begin{fmfgraph*}(3,9)
\setval
\fmfstraight
\fmfforce{1w,1h}{o1}
\fmfforce{1w,2/3h}{o2}
\fmfforce{0w,5/6h}{v1}
\fmfforce{0w,1/6h}{v2}
\fmfforce{1w,1/3h}{i1}
\fmfforce{1w,0h}{i2}
\fmf{plain}{v1,v2}
\fmf{plain}{v1,o1}
\fmf{plain}{v1,o2}
\fmf{plain}{v2,i1}
\fmf{plain}{v2,i2}
\fmfdot{v1,v2}
\fmfv{decor.size=0, label=${\scs 4}$, l.dist=1mm, l.angle=0}{i2}
\fmfv{decor.size=0, label=${\scs 3}$, l.dist=1mm, l.angle=0}{i1}
\fmfv{decor.size=0, label=${\scs 2}$, l.dist=1mm, l.angle=0}{o2}
\fmfv{decor.size=0, label=${\scs 1}$, l.dist=1mm, l.angle=0}{o1}
\end{fmfgraph*}
\end{center}}
\hspace*{3mm} \ddphi{W^{({\rm int})}}{1}{2}{3}{4} 
\hspace*{1mm} + \frac{1}{2} \hspace*{1mm}
\parbox{7mm}{\begin{center}
\begin{fmfgraph*}(3,12)
\setval
\fmfstraight
\fmfforce{1w,1h}{o0}
\fmfforce{1w,3/4h}{o1}
\fmfforce{1w,1/2h}{o2}
\fmfforce{0w,3/4h}{v1}
\fmfforce{0w,1/8h}{v2}
\fmfforce{1w,1/4h}{i1}
\fmfforce{1w,0h}{i2}
\fmf{plain}{v1,v2}
\fmf{plain}{v1,o0}
\fmf{plain}{v1,o1}
\fmf{plain}{v1,o2}
\fmf{plain}{v2,i1}
\fmf{plain}{v2,i2}
\fmfdot{v1,v2}
\fmfv{decor.size=0, label=${\scs 5}$, l.dist=1mm, l.angle=0}{i2}
\fmfv{decor.size=0, label=${\scs 4}$, l.dist=1mm, l.angle=0}{i1}
\fmfv{decor.size=0, label=${\scs 3}$, l.dist=1mm, l.angle=0}{o2}
\fmfv{decor.size=0, label=${\scs 2}$, l.dist=1mm, l.angle=0}{o1}
\fmfv{decor.size=0, label=${\scs 1}$, l.dist=1mm, l.angle=0}{o0}
\end{fmfgraph*}
\end{center}}
\hspace*{3mm} \ddvertex{W^{({\rm int})}}{1}{2}{3}{4}{5}
\nonumber \\
&& \la{GRC}
\hspace*{5mm} + \frac{1}{3} \hspace*{1mm}
\hspace*{1mm} \dphi{W^{({\rm int})}}{1}{2}\hspace*{3mm}
\parbox{7mm}{\begin{center}
\begin{fmfgraph*}(4,4)
\setval
\fmfstraight
\fmfforce{0w,1h}{o1}
\fmfforce{0w,0h}{o2}
\fmfforce{1/2w,1/2h}{v1}
\fmfforce{1w,1h}{i1}
\fmfforce{1w,0h}{i2}
\fmf{plain}{v1,o1}
\fmf{plain}{v1,o2}
\fmf{plain}{v1,i1}
\fmf{plain}{v1,i2}
\fmfdot{v1}
\fmfv{decor.size=0, label=${\scs 4}$, l.dist=1mm, l.angle=0}{i2}
\fmfv{decor.size=0, label=${\scs 3}$, l.dist=1mm, l.angle=0}{i1}
\fmfv{decor.size=0, label=${\scs 2}$, l.dist=1mm, l.angle=-180}{o2}
\fmfv{decor.size=0, label=${\scs 1}$, l.dist=1mm, l.angle=-180}{o1}
\end{fmfgraph*}
\end{center}}
\hspace*{3mm} \dphi{W^{({\rm int})}}{3}{4} 
\hspace*{1mm} + \frac{1}{2} \hspace*{1mm}
\hspace*{1mm} \dphi{W^{({\rm int})}}{1}{2}\hspace*{3mm}
\parbox{11mm}{\begin{center}
\begin{fmfgraph*}(8,4)
\setval
\fmfstraight
\fmfforce{0w,1h}{o1}
\fmfforce{0w,0h}{o2}
\fmfforce{1/4w,1/2h}{v1}
\fmfforce{3/4w,1/2h}{v2}
\fmfforce{1w,1h}{i1}
\fmfforce{1w,0h}{i2}
\fmf{plain}{v1,v2}
\fmf{plain}{v1,o1}
\fmf{plain}{v1,o2}
\fmf{plain}{v2,i1}
\fmf{plain}{v2,i2}
\fmfdot{v1,v2}
\fmfv{decor.size=0, label=${\scs 4}$, l.dist=1mm, l.angle=0}{i2}
\fmfv{decor.size=0, label=${\scs 3}$, l.dist=1mm, l.angle=0}{i1}
\fmfv{decor.size=0, label=${\scs 2}$, l.dist=1mm, l.angle=-180}{o2}
\fmfv{decor.size=0, label=${\scs 1}$, l.dist=1mm, l.angle=-180}{o1}
\end{fmfgraph*}
\end{center}}
\hspace*{3mm} \dphi{W^{({\rm int})}}{3}{4}
\hspace*{1mm} + \frac{1}{2} \hspace*{1mm}
\hspace*{1mm} \dvertex{W^{({\rm int})}}{1}{2}{3}\hspace*{3mm}
\parbox{11mm}{\begin{center}
\begin{fmfgraph*}(8,4)
\setval
\fmfstraight
\fmfforce{0w,1.5h}{o1}
\fmfforce{0w,0.5h}{o2}
\fmfforce{0w,-0.5h}{o3}
\fmfforce{1/4w,1/2h}{v1}
\fmfforce{3/4w,1/2h}{v2}
\fmfforce{1w,1h}{i1}
\fmfforce{1w,0h}{i2}
\fmf{plain}{v1,v2}
\fmf{plain}{v1,o1}
\fmf{plain}{v1,o2}
\fmf{plain}{v1,o3}
\fmf{plain}{v2,i1}
\fmf{plain}{v2,i2}
\fmfdot{v1,v2}
\fmfv{decor.size=0, label=${\scs 5}$, l.dist=1mm, l.angle=0}{i2}
\fmfv{decor.size=0, label=${\scs 4}$, l.dist=1mm, l.angle=0}{i1}
\fmfv{decor.size=0, label=${\scs 3}$, l.dist=1mm, l.angle=-180}{o3}
\fmfv{decor.size=0, label=${\scs 2}$, l.dist=1mm, l.angle=-180}{o2}
\fmfv{decor.size=0, label=${\scs 1}$, l.dist=1mm, l.angle=-180}{o1}
\end{fmfgraph*}
\end{center}}
\hspace*{3mm} \dphi{W^{({\rm int})}}{4}{5} \hspace*{4mm} \, .
\eeq
\end{fmffile}
\begin{fmffile}{fg3}
The effect of the term on the left-hand side is to count the number of lines
of each connected vacuum diagram. Indeed, the functional derivative 
$\delta/\delta G_{12}$ removes successively the lines which are, subsequently,
reinserted by the operation $\int_{12} G_{12}$.
The right-hand side contains altogether 25 terms, 10 without 
$W^{({\rm int})}$, 12 linear in $W^{({\rm int})}$ and 3 bilinear
in $W^{({\rm int})}$. 
\subsection{Loopwise Recursive Graphical Solution}
Now we show how Eq.~(\r{GRC}) is solved graphically. To this end we expand
the interaction negative free energy $W^{({\rm int})}$ with respect to
the number $n$ of currents $J$  and the loop order $l$ 
\beq
\la{DECC}
W^{({\rm int})} = \sum_{n=0}^{\infty} \sum_{l=0}^{\infty} W^{(n,l)} \, .
\eeq
Here we can exclude the combinations $(n,l)\in \{ (0,0), (0,1), (1,0),
(2,0)\}$, as the corresponding expansion coefficients $W^{(n,l)}$ turn out to
be zero. With the help of (\r{DECC})
we convert Eq.~(\r{GRC}) into a recursive graphical solution
for the expansion coefficients $W^{(n,l)}$. As an example we consider 
the graphical recursion relation for the
current-free connected vacuum diagrams $W^{(0,l)}$.  
For $n=0$ and $l=2$, Eq.~(\r{GRC}) reduces to
\beq
\parbox{8mm}{\begin{center}
\begin{fmfgraph*}(2.5,5)
\setval
\fmfstraight
\fmfforce{1w,0h}{v1}
\fmfforce{1w,1h}{v2}
\fmf{plain,left=1}{v1,v2}
\fmfv{decor.size=0, label=${\scs 2}$, l.dist=1mm, l.angle=0}{v1}
\fmfv{decor.size=0, label=${\scs 1}$, l.dist=1mm, l.angle=0}{v2}
\end{fmfgraph*}
\end{center}}
\hspace*{0.3cm} \dphi{W^{(0,2)}}{1}{2} = 
\frac{1}{4}
\parbox{11mm}{\begin{center}
\begin{fmfgraph*}(8,4)
\setval
\fmfleft{i1}
\fmfright{o1}
\fmf{plain,left=1}{i1,v1,i1}
\fmf{plain,left=1}{o1,v1,o1}
\fmfdot{v1}
\end{fmfgraph*}\end{center}}
+ \frac{1}{4}
\parbox{7mm}{\begin{center}
\begin{fmfgraph*}(4,4)
\setval
\fmfforce{0w,0.5h}{v1}
\fmfforce{1w,0.5h}{v2}
\fmf{plain,left=1}{v1,v2,v1}
\fmf{plain}{v1,v2}
\fmfdot{v1,v2}
\end{fmfgraph*}\end{center}}
+ \frac{3}{8}
\parbox{15mm}{\begin{center}
\begin{fmfgraph*}(12,4)
\setval
\fmfforce{0w,0.5h}{v1}
\fmfforce{1/3w,0.5h}{v2}
\fmfforce{2/3w,0.5h}{v3}
\fmfforce{1w,0.5h}{v4}
\fmf{plain,left=1}{v1,v2,v1}
\fmf{plain,left=1}{v3,v4,v3}
\fmf{plain}{v2,v3}
\fmfdot{v2,v3}
\end{fmfgraph*}\end{center}} \, ,
\eeq
which is immediately solved by
\beq
W^{(0,2)} = 
\frac{1}{8}
\parbox{11mm}{\begin{center}
\begin{fmfgraph*}(8,4)
\setval
\fmfleft{i1}
\fmfright{o1}
\fmf{plain,left=1}{i1,v1,i1}
\fmf{plain,left=1}{o1,v1,o1}
\fmfdot{v1}
\end{fmfgraph*}\end{center}}
+ \frac{1}{12}
\parbox{7mm}{\begin{center}
\begin{fmfgraph*}(4,4)
\setval
\fmfforce{0w,0.5h}{v1}
\fmfforce{1w,0.5h}{v2}
\fmf{plain,left=1}{v1,v2,v1}
\fmf{plain}{v1,v2}
\fmfdot{v1,v2}
\end{fmfgraph*}\end{center}}
+ \frac{1}{8}
\parbox{15mm}{\begin{center}
\begin{fmfgraph*}(12,4)
\setval
\fmfforce{0w,0.5h}{v1}
\fmfforce{1/3w,0.5h}{v2}
\fmfforce{2/3w,0.5h}{v3}
\fmfforce{1w,0.5h}{v4}
\fmf{plain,left=1}{v1,v2,v1}
\fmf{plain,left=1}{v3,v4,v3}
\fmf{plain}{v2,v3}
\fmfdot{v2,v3}
\end{fmfgraph*}\end{center}} \, ,
\la{W2}
\eeq
as the first vacuum diagram contains 2 and the last two vacuum diagrams
3 lines. For $n=0$ and $l\ge3$ we obtain the graphical recursion relation
\beq
\parbox{5.5mm}{\begin{center}
\begin{fmfgraph*}(2.5,5)
\setval
\fmfstraight
\fmfforce{1w,0h}{v1}
\fmfforce{1w,1h}{v2}
\fmf{plain,left=1}{v1,v2}
\fmfv{decor.size=0, label=${\scs 2}$, l.dist=1mm, l.angle=0}{v1}
\fmfv{decor.size=0, label=${\scs 1}$, l.dist=1mm, l.angle=0}{v2}
\end{fmfgraph*}
\end{center}}
\hspace*{0.3cm} \dphi{W^{(0,l)}}{1}{2} &=& 
\hspace*{1mm} 
\parbox{9mm}{\begin{center}
\begin{fmfgraph*}(6,4)
\setval
\fmfstraight
\fmfforce{0w,1/2h}{v1}
\fmfforce{4/6w,1/2h}{v2}
\fmfforce{1w,1h}{i2}
\fmfforce{1w,0h}{i1}
\fmf{plain}{i1,v2}
\fmf{plain}{v2,i2}
\fmf{plain,left}{v1,v2,v1}
\fmfdot{v2}
\fmfv{decor.size=0, label=${\scs 2}$, l.dist=1mm, l.angle=0}{i1}
\fmfv{decor.size=0, label=${\scs 1}$, l.dist=1mm, l.angle=0}{i2}
\end{fmfgraph*}
\end{center}}
\hspace*{3mm} \dphi{W^{(0,l-1)}}{1}{2} 
\hspace*{1mm} + \hspace*{1mm}
\parbox{7mm}{\begin{center}
\begin{fmfgraph*}(6,4)
\setval
\fmfstraight
\fmfforce{1/3w,0h}{v1}
\fmfforce{1/3w,1h}{v2}
\fmfforce{1w,1h}{i2}
\fmfforce{1w,0h}{i1}
\fmf{plain}{i1,v1}
\fmf{plain}{v2,i2}
\fmf{plain,left}{v1,v2,v1}
\fmfdot{v2,v1}
\fmfv{decor.size=0, label=${\scs 2}$, l.dist=1mm, l.angle=0}{i1}
\fmfv{decor.size=0, label=${\scs 1}$, l.dist=1mm, l.angle=0}{i2}
\end{fmfgraph*}
\end{center}}
\hspace*{3mm} \dphi{W^{(0,l-1)}}{1}{2}
\hspace*{1mm} + \hspace*{1mm}
\parbox{13mm}{\begin{center}
\begin{fmfgraph*}(10,4)
\setval
\fmfstraight
\fmfforce{0w,1/2h}{v1}
\fmfforce{4/10w,1/2h}{v2}
\fmfforce{8/10w,1/2h}{v3}
\fmfforce{1w,0h}{i2}
\fmfforce{1w,1h}{i1}
\fmf{plain,left}{v1,v2,v1}
\fmf{plain}{v3,v2}
\fmf{plain}{v3,i1}
\fmf{plain}{v3,i2}
\fmfdot{v2,v3}
\fmfv{decor.size=0, label=${\scs 2}$, l.dist=1mm, l.angle=0}{i2}
\fmfv{decor.size=0, label=${\scs 1}$, l.dist=1mm, l.angle=0}{i1}
\end{fmfgraph*}
\end{center}}
\hspace*{3mm} \dphi{W^{(0,l-1)}}{1}{2}
\hspace*{1mm} + \frac{3}{4} \hspace*{1mm}
\parbox{13mm}{\begin{center}
\begin{fmfgraph*}(10,4)
\setval
\fmfstraight
\fmfforce{0w,1/2h}{v1}
\fmfforce{4/10w,1/2h}{v2}
\fmfforce{8/10w,1/2h}{v3}
\fmfforce{1w,-0.25h}{i1}
\fmfforce{1w,0.5h}{i2}
\fmfforce{1w,1.25h}{i3}
\fmf{plain,left}{v1,v2,v1}
\fmf{plain}{v3,v2}
\fmf{plain}{v3,i1}
\fmf{plain}{v3,i2}
\fmf{plain}{v3,i3}
\fmfdot{v2,v3}
\fmfv{decor.size=0, label=${\scs 1}$, l.dist=1mm, l.angle=0}{i3}
\fmfv{decor.size=0, label=${\scs 2}$, l.dist=1mm, l.angle=0}{i2}
\fmfv{decor.size=0, label=${\scs 3}$, l.dist=1mm, l.angle=0}{i1}
\end{fmfgraph*}
\end{center}}
\hspace*{3mm} \dvertex{W^{(0,l-1)}}{1}{2}{3} 
\nonumber \\
&&
+ \frac{1}{3} \hspace*{1mm}
\parbox{7mm}{\begin{center}
\begin{fmfgraph*}(3,3)
\setval
\fmfstraight
\fmfforce{0w,1/2h}{v1}
\fmfforce{1w,2h}{i1}
\fmfforce{1w,1h}{i2}
\fmfforce{1w,0h}{i3}
\fmfforce{1w,-1h}{i4}
\fmf{plain}{v1,i1}
\fmf{plain}{v1,i2}
\fmf{plain}{v1,i3}
\fmf{plain}{v1,i4}
\fmfdot{v1}
\fmfv{decor.size=0, label=${\scs 2}$, l.dist=1mm, l.angle=0}{i4}
\fmfv{decor.size=0, label=${\scs 1}$, l.dist=1mm, l.angle=0}{i3}
\fmfv{decor.size=0, label=${\scs 4}$, l.dist=1mm, l.angle=0}{i2}
\fmfv{decor.size=0, label=${\scs 3}$, l.dist=1mm, l.angle=0}{i1}
\end{fmfgraph*}
\end{center}}
\hspace*{3mm} \ddphi{W^{(0,l-1)}}{1}{2}{3}{4}
\hspace*{1mm} + \frac{1}{2} \hspace*{1mm}
\parbox{7mm}{\begin{center}
\begin{fmfgraph*}(3,9)
\setval
\fmfstraight
\fmfforce{1w,1h}{o1}
\fmfforce{1w,2/3h}{o2}
\fmfforce{0w,5/6h}{v1}
\fmfforce{0w,1/6h}{v2}
\fmfforce{1w,1/3h}{i1}
\fmfforce{1w,0h}{i2}
\fmf{plain}{v1,v2}
\fmf{plain}{v1,o1}
\fmf{plain}{v1,o2}
\fmf{plain}{v2,i1}
\fmf{plain}{v2,i2}
\fmfdot{v1,v2}
\fmfv{decor.size=0, label=${\scs 4}$, l.dist=1mm, l.angle=0}{i2}
\fmfv{decor.size=0, label=${\scs 3}$, l.dist=1mm, l.angle=0}{i1}
\fmfv{decor.size=0, label=${\scs 2}$, l.dist=1mm, l.angle=0}{o2}
\fmfv{decor.size=0, label=${\scs 1}$, l.dist=1mm, l.angle=0}{o1}
\end{fmfgraph*}
\end{center}}
\hspace*{3mm} \ddphi{W^{(0,l-1)}}{1}{2}{3}{4} 
\hspace*{1mm} + \frac{1}{2} \hspace*{1mm}
\parbox{7mm}{\begin{center}
\begin{fmfgraph*}(3,12)
\setval
\fmfstraight
\fmfforce{1w,1h}{o0}
\fmfforce{1w,3/4h}{o1}
\fmfforce{1w,1/2h}{o2}
\fmfforce{0w,3/4h}{v1}
\fmfforce{0w,1/8h}{v2}
\fmfforce{1w,1/4h}{i1}
\fmfforce{1w,0h}{i2}
\fmf{plain}{v1,v2}
\fmf{plain}{v1,o0}
\fmf{plain}{v1,o1}
\fmf{plain}{v1,o2}
\fmf{plain}{v2,i1}
\fmf{plain}{v2,i2}
\fmfdot{v1,v2}
\fmfv{decor.size=0, label=${\scs 5}$, l.dist=1mm, l.angle=0}{i2}
\fmfv{decor.size=0, label=${\scs 4}$, l.dist=1mm, l.angle=0}{i1}
\fmfv{decor.size=0, label=${\scs 3}$, l.dist=1mm, l.angle=0}{o2}
\fmfv{decor.size=0, label=${\scs 2}$, l.dist=1mm, l.angle=0}{o1}
\fmfv{decor.size=0, label=${\scs 1}$, l.dist=1mm, l.angle=0}{o0}
\end{fmfgraph*}
\end{center}}
\hspace*{3mm} \ddvertex{W^{(0,l-1)}}{1}{2}{3}{4}{5} 
\nonumber \\
&&
+  \sum_{l'=2}^{l-2} \Bigg\{ \hspace*{1mm} \frac{1}{3} 
\hspace*{1mm} \dphi{W^{(0,l')}}{1}{2}\hspace*{3mm}
\parbox{7mm}{\begin{center}
\begin{fmfgraph*}(4,4)
\setval
\fmfstraight
\fmfforce{0w,1h}{o1}
\fmfforce{0w,0h}{o2}
\fmfforce{1/2w,1/2h}{v1}
\fmfforce{1w,1h}{i1}
\fmfforce{1w,0h}{i2}
\fmf{plain}{v1,o1}
\fmf{plain}{v1,o2}
\fmf{plain}{v1,i1}
\fmf{plain}{v1,i2}
\fmfdot{v1}
\fmfv{decor.size=0, label=${\scs 4}$, l.dist=1mm, l.angle=0}{i2}
\fmfv{decor.size=0, label=${\scs 3}$, l.dist=1mm, l.angle=0}{i1}
\fmfv{decor.size=0, label=${\scs 2}$, l.dist=1mm, l.angle=-180}{o2}
\fmfv{decor.size=0, label=${\scs 1}$, l.dist=1mm, l.angle=-180}{o1}
\end{fmfgraph*}
\end{center}}
\hspace*{3mm} \dphi{W^{(0,l-l'-1)}}{3}{4} 
\hspace*{1mm} + \frac{1}{2} \hspace*{1mm} 
%
\hspace*{1mm} \dphi{W^{(0,l')}}{1}{2}\hspace*{3mm}
\parbox{11mm}{\begin{center}
\begin{fmfgraph*}(8,4)
\setval
\fmfstraight
\fmfforce{0w,1h}{o1}
\fmfforce{0w,0h}{o2}
\fmfforce{1/4w,1/2h}{v1}
\fmfforce{3/4w,1/2h}{v2}
\fmfforce{1w,1h}{i1}
\fmfforce{1w,0h}{i2}
\fmf{plain}{v1,v2}
\fmf{plain}{v1,o1}
\fmf{plain}{v1,o2}
\fmf{plain}{v2,i1}
\fmf{plain}{v2,i2}
\fmfdot{v1,v2}
\fmfv{decor.size=0, label=${\scs 4}$, l.dist=1mm, l.angle=0}{i2}
\fmfv{decor.size=0, label=${\scs 3}$, l.dist=1mm, l.angle=0}{i1}
\fmfv{decor.size=0, label=${\scs 2}$, l.dist=1mm, l.angle=-180}{o2}
\fmfv{decor.size=0, label=${\scs 1}$, l.dist=1mm, l.angle=-180}{o1}
\end{fmfgraph*}
\end{center}}
\hspace*{3mm} \dphi{W^{(0,l-l'-1)}}{3}{4} 
\no \\
&& 
+ \frac{1}{2} \hspace*{1mm} 
%
\hspace*{1mm} \dvertex{W^{(0,l')}}{1}{2}{3}\hspace*{3mm}
\parbox{11mm}{\begin{center}
\begin{fmfgraph*}(8,4)
\setval
\fmfstraight
\fmfforce{0w,1.5h}{o1}
\fmfforce{0w,0.5h}{o2}
\fmfforce{0w,-0.5h}{o3}
\fmfforce{1/4w,1/2h}{v1}
\fmfforce{3/4w,1/2h}{v2}
\fmfforce{1w,1h}{i1}
\fmfforce{1w,0h}{i2}
\fmf{plain}{v1,v2}
\fmf{plain}{v1,o1}
\fmf{plain}{v1,o2}
\fmf{plain}{v1,o3}
\fmf{plain}{v2,i1}
\fmf{plain}{v2,i2}
\fmfdot{v1,v2}
\fmfv{decor.size=0, label=${\scs 5}$, l.dist=1mm, l.angle=0}{i2}
\fmfv{decor.size=0, label=${\scs 4}$, l.dist=1mm, l.angle=0}{i1}
\fmfv{decor.size=0, label=${\scs 3}$, l.dist=1mm, l.angle=-180}{o3}
\fmfv{decor.size=0, label=${\scs 2}$, l.dist=1mm, l.angle=-180}{o2}
\fmfv{decor.size=0, label=${\scs 1}$, l.dist=1mm, l.angle=-180}{o1}
\end{fmfgraph*}
\end{center}}
\hspace*{3mm} \dphi{W^{(0,l-l'-1)}}{4}{5}  \Bigg\} \, .
\la{GNU}
\eeq
We observe that for either a vanishing cubic or quartic interaction the
graphical recursion relation (\r{GNU}) only involves the graphical
operation of removing lines.
Proceeding to the loop order $l=3$, we have to evaluate from the vacuum 
diagrams (\r{W2}) a one-line amputation
\beq
\la{NNR1}
\dphi{W^{(0,2)}}{1}{2} = \frac{1}{4} 
\parbox{9mm}{\begin{center}
\begin{fmfgraph*}(6,4)
\setval
\fmfstraight
\fmfforce{0w,1/2h}{v1}
\fmfforce{4/6w,1/2h}{v2}
\fmfforce{1w,1h}{i2}
\fmfforce{1w,0h}{i1}
\fmf{plain}{i1,v2}
\fmf{plain}{v2,i2}
\fmf{plain,left}{v1,v2,v1}
\fmfdot{v2}
\fmfv{decor.size=0, label=${\scs 2}$, l.dist=1mm, l.angle=0}{i1}
\fmfv{decor.size=0, label=${\scs 1}$, l.dist=1mm, l.angle=0}{i2}
\end{fmfgraph*}
\end{center}}
\hspace*{1mm} + \frac{1}{4} \hspace*{1mm}
\parbox{7mm}{\begin{center}
\begin{fmfgraph*}(6,4)
\setval
\fmfstraight
\fmfforce{1/3w,0h}{v1}
\fmfforce{1/3w,1h}{v2}
\fmfforce{1w,1h}{i2}
\fmfforce{1w,0h}{i1}
\fmf{plain}{i1,v1}
\fmf{plain}{v2,i2}
\fmf{plain,left}{v1,v2,v1}
\fmfdot{v2,v1}
\fmfv{decor.size=0, label=${\scs 2}$, l.dist=1mm, l.angle=0}{i1}
\fmfv{decor.size=0, label=${\scs 1}$, l.dist=1mm, l.angle=0}{i2}
\end{fmfgraph*}
\end{center}}
\hspace*{1mm} + \frac{1}{4} \hspace*{1mm}
\parbox{13mm}{\begin{center}
\begin{fmfgraph*}(10,4)
\setval
\fmfstraight
\fmfforce{0w,1/2h}{v1}
\fmfforce{4/10w,1/2h}{v2}
\fmfforce{8/10w,1/2h}{v3}
\fmfforce{1w,0h}{i2}
\fmfforce{1w,1h}{i1}
\fmf{plain,left}{v1,v2,v1}
\fmf{plain}{v3,v2}
\fmf{plain}{v3,i1}
\fmf{plain}{v3,i2}
\fmfdot{v2,v3}
\fmfv{decor.size=0, label=${\scs 2}$, l.dist=1mm, l.angle=0}{i2}
\fmfv{decor.size=0, label=${\scs 1}$, l.dist=1mm, l.angle=0}{i1}
\end{fmfgraph*}
\end{center}}
\hspace*{1mm} + \frac{1}{8} \hspace*{1mm}
\parbox{9mm}{\begin{center}
\begin{fmfgraph*}(6,10)
\setval
\fmfstraight
\fmfforce{0w,2/10h}{v1}
\fmfforce{4/6w,2/10h}{v2}
\fmfforce{1w,2/10h}{i1}
\fmfforce{0w,8/10h}{v3}
\fmfforce{4/6w,8/10h}{v4}
\fmfforce{1w,8/10h}{i2}
\fmf{plain,left}{v1,v2,v1}
\fmf{plain,left}{v3,v4,v3}
\fmf{plain}{v4,i2}
\fmf{plain}{v2,i1}
\fmfdot{v4,v2}
\fmfv{decor.size=0, label=${\scs 2}$, l.dist=1mm, l.angle=0}{i1}
\fmfv{decor.size=0, label=${\scs 1}$, l.dist=1mm, l.angle=0}{i2}
\end{fmfgraph*}
\end{center}} \hspace*{4mm} , 
\eeq
a two-line amputation
\beq
\la{NNR2}
\ddphi{W^{(0,2)}}{1}{2}{3}{4} = \hspace*{1mm}\frac{1}{4} \hspace*{2mm}
\parbox{7mm}{\begin{center}
\begin{fmfgraph*}(4,4)
\setval
\fmfstraight
\fmfforce{0w,1h}{o1}
\fmfforce{0w,0h}{o2}
\fmfforce{1/2w,1/2h}{v1}
\fmfforce{1w,1h}{i1}
\fmfforce{1w,0h}{i2}
\fmf{plain}{v1,o1}
\fmf{plain}{v1,o2}
\fmf{plain}{v1,i1}
\fmf{plain}{v1,i2}
\fmfdot{v1}
\fmfv{decor.size=0, label=${\scs 4}$, l.dist=1mm, l.angle=0}{i2}
\fmfv{decor.size=0, label=${\scs 3}$, l.dist=1mm, l.angle=0}{i1}
\fmfv{decor.size=0, label=${\scs 2}$, l.dist=1mm, l.angle=-180}{o2}
\fmfv{decor.size=0, label=${\scs 1}$, l.dist=1mm, l.angle=-180}{o1}
\end{fmfgraph*}
\end{center}}
\hspace*{2mm}
+ \hspace*{2mm}\frac{1}{4} \hspace*{2mm} {\displaystyle \left\{ \hspace*{2mm}
\parbox{11mm}{\begin{center}
\begin{fmfgraph*}(8,4)
\setval
\fmfstraight
\fmfforce{0w,1h}{o1}
\fmfforce{0w,0h}{o2}
\fmfforce{1/4w,1/2h}{v1}
\fmfforce{3/4w,1/2h}{v2}
\fmfforce{1w,1h}{i1}
\fmfforce{1w,0h}{i2}
\fmf{plain}{v1,v2}
\fmf{plain}{v1,o1}
\fmf{plain}{v1,o2}
\fmf{plain}{v2,i1}
\fmf{plain}{v2,i2}
\fmfdot{v1,v2}
\fmfv{decor.size=0, label=${\scs 4}$, l.dist=1mm, l.angle=0}{i2}
\fmfv{decor.size=0, label=${\scs 3}$, l.dist=1mm, l.angle=0}{i1}
\fmfv{decor.size=0, label=${\scs 2}$, l.dist=1mm, l.angle=-180}{o2}
\fmfv{decor.size=0, label=${\scs 1}$, l.dist=1mm, l.angle=-180}{o1}
\end{fmfgraph*}
\end{center}}
\hspace*{2mm} + 2\,\mbox{perm.} \right\} }
+ \hspace*{2mm}\frac{1}{8} 
{\displaystyle \left\{
\parbox{9mm}{\centerline{
\begin{fmfgraph*}(5,12.33)
\setval
\fmfforce{1w,0h}{v1}
\fmfforce{0w,0h}{v2}
\fmfforce{0.5w,4.33/12.33h}{v3}
\fmfforce{0.5w,1.25/12.33h}{vm}
\fmfforce{-1/5w,10.33/12.33h}{v4}
\fmfforce{3/5w,10.33/12.33h}{v5}
\fmfforce{1w,10.33/12.33h}{v6}
\fmf{plain}{v1,vm,v2}
\fmf{plain,left=1}{v4,v5,v4}
\fmf{plain}{v3,vm}
\fmf{plain}{v5,v6}
\fmfv{decor.size=0,label={\footnotesize 4},l.dist=0.5mm, l.angle=-30}{v1}
\fmfv{decor.size=0,label={\footnotesize 1},l.dist=0.5mm, l.angle=-150}{v2}
\fmfv{decor.size=0,label={\footnotesize 3},l.dist=0.5mm, l.angle=90}{v3}
\fmfv{decor.size=0,label={\footnotesize 2},l.dist=0.5mm, l.angle=0}{v6}
\fmfdot{vm,v5}
\end{fmfgraph*}}}
\hspace*{1mm} + \, 3 \, \mbox{perm.} \right\}}  \hspace*{4mm} , 
\eeq
a 3-vertex amputation
\beq
\la{NNR3}
\dvertex{W^{(0,2)}}{1}{2}{3} = 
\frac{1}{6} \hspace*{2mm}
\parbox{8mm}{\centerline{
\begin{fmfgraph*}(5,4.33)
\setval
\fmfforce{1w,0h}{v1}
\fmfforce{0w,0h}{v2}
\fmfforce{0.5w,1h}{v3}
\fmfforce{0.5w,0.2886h}{vm}
\fmf{plain}{v1,vm,v2}
\fmf{plain}{v3,vm}
\fmfv{decor.size=0,label={\footnotesize 2},l.dist=0.5mm}{v1}
\fmfv{decor.size=0,label={\footnotesize 3},l.dist=0.5mm}{v2}
\fmfv{decor.size=0,label={\footnotesize 1},l.dist=0.5mm}{v3}
\fmfdot{vm}
\end{fmfgraph*}}}
+ \hspace*{2mm} \frac{1}{12} 
{\displaystyle \left\{
\parbox{9mm}{\centerline{
\begin{fmfgraph*}(6,6)
\setval
\fmfforce{0w,1/3h}{v1}
\fmfforce{2/3w,1/3h}{v2}
\fmfforce{1w,1/3h}{v3}
\fmfforce{1/3w,1h}{v4}
\fmfforce{1w,1h}{v5}
\fmf{plain}{v4,v5}
\fmf{plain}{v2,v3}
\fmf{plain,left=1}{v1,v2,v1}
\fmfv{decor.size=0,label={\footnotesize 3},l.dist=0.5mm, l.angle=0}{v3}
\fmfv{decor.size=0,label={\footnotesize 1},l.dist=0.5mm, l.angle=-180}{v4}
\fmfv{decor.size=0,label={\footnotesize 2},l.dist=0.5mm, l.angle=0}{v5}
\fmfdot{v2}
\end{fmfgraph*}}}
\hspace*{2mm} + \, 2 \, \mbox{perm.} \right\} } \, ,
\eeq
and the amputation of one line and one 3-vertex
\beq
\la{NNR4}
\ddvertex{W^{(0,2)}}{1}{2}{3}{4}{5} & = &
\frac{1}{12}  \hspace*{2mm} 
{\displaystyle \left\{ \,
\parbox{8mm}{\centerline{
\begin{fmfgraph*}(5,18.33)
\setval
\fmfforce{1w,2/18.33h}{v1}
\fmfforce{0w,2/18.33h}{v2}
\fmfforce{0.5w,6.33/18.33h}{v3}
\fmfforce{0.5w,3.25/18.33h}{vm}
\fmfforce{0.5w,12.33/18.33h}{v4}
\fmfforce{0.5w,14.33/18.33h}{v5}
\fmfforce{0.5w,16.33/18.33h}{v6}
\fmf{plain}{v1,vm,v2}
\fmf{plain}{v3,vm}
\fmf{plain}{v4,v5}
\fmf{plain}{v6,v5}
\fmfv{decor.size=0,label={\footnotesize 2},l.dist=0.5mm, l.angle=-30}{v1}
\fmfv{decor.size=0,label={\footnotesize 3},l.dist=0.5mm, l.angle=-150}{v2}
\fmfv{decor.size=0,label={\footnotesize 5},l.dist=0.5mm, l.angle=90}{v3}
\fmfv{decor.size=0,label={\footnotesize 1},l.dist=0.5mm, l.angle=90}{v6}
\fmfv{decor.size=0,label={\footnotesize 4},l.dist=0.5mm, l.angle=-90}{v4}
\fmfv{decor.shape=circle,decor.filled=empty,decor.size=0.6mm}{v5}
\fmfdot{vm}
\end{fmfgraph*}}}
+ \,5\, \mbox{perm.} \right\} }
+ \hspace*{2mm} \frac{1}{24} {\displaystyle \left\{ \,
\parbox{16mm}{\centerline{
\begin{fmfgraph*}(13,6)
\setval
\fmfforce{2.5/13w,1/3h}{v1}
\fmfforce{6.5/13w,1/3h}{v2}
\fmfforce{8.5/13w,1/3h}{v3}
\fmfforce{0w,1h}{v4}
\fmfforce{2/13w,1h}{v5}
\fmfforce{4/13w,1h}{v6}
\fmfforce{9/13w,1h}{v7}
\fmfforce{11/13w,1h}{v8}
\fmfforce{13/13w,1h}{v9}
\fmf{plain}{v4,v5}
\fmf{plain}{v6,v5}
\fmf{plain}{v7,v8}
\fmf{plain}{v9,v8}
\fmf{plain}{v2,v3}
\fmf{plain,left=1}{v1,v2,v1}
\fmfv{decor.size=0,label={\footnotesize 1},l.dist=0.5mm, l.angle=-180}{v4}
\fmfv{decor.size=0,label={\footnotesize 5},l.dist=0.5mm, l.angle=0}{v6}
\fmfv{decor.size=0,label={\footnotesize 4},l.dist=0.5mm, l.angle=-180}{v7}
\fmfv{decor.size=0,label={\footnotesize 2},l.dist=0.5mm, l.angle=0}{v9}
\fmfv{decor.size=0,label={\footnotesize 3},l.dist=0.5mm, l.angle=0}{v3}
\fmfv{decor.shape=circle,decor.filled=empty,decor.size=0.6mm}{v5}
\fmfv{decor.shape=circle,decor.filled=empty,decor.size=0.6mm}{v8}
\fmfdot{v2}
\end{fmfgraph*}}} 
+ \, 5 \, \mbox{perm.} \right\}} \no \\
&& + \hspace*{2mm} \frac{1}{24} {\displaystyle \left\{
\parbox{16mm}{\centerline{
\begin{fmfgraph*}(13,6)
\setval
\fmfforce{2.5/13w,1/3h}{v1}
\fmfforce{6.5/13w,1/3h}{v2}
\fmfforce{8.5/13w,1/3h}{v3}
\fmfforce{0w,1h}{v4}
\fmfforce{2/13w,1h}{v5}
\fmfforce{4/13w,1h}{v6}
\fmfforce{9/13w,1h}{v7}
\fmfforce{11/13w,1h}{v8}
\fmfforce{13/13w,1h}{v9}
\fmf{plain}{v4,v5}
\fmf{plain}{v6,v5}
\fmf{plain}{v7,v8}
\fmf{plain}{v9,v8}
\fmf{plain}{v2,v3}
\fmf{plain,left=1}{v1,v2,v1}
\fmfv{decor.size=0,label={\footnotesize 1},l.dist=0.5mm, l.angle=-180}{v4}
\fmfv{decor.size=0,label={\footnotesize 4},l.dist=0.5mm, l.angle=0}{v6}
\fmfv{decor.size=0,label={\footnotesize 5},l.dist=0.5mm, l.angle=-180}{v7}
\fmfv{decor.size=0,label={\footnotesize 3},l.dist=0.5mm, l.angle=0}{v9}
\fmfv{decor.size=0,label={\footnotesize 2},l.dist=0.5mm, l.angle=0}{v3}
\fmfv{decor.shape=circle,decor.filled=empty,decor.size=0.6mm}{v5}
\fmfv{decor.shape=circle,decor.filled=empty,decor.size=0.6mm}{v8}
\fmfdot{v2}
\end{fmfgraph*}}}
+ \, 5 \mbox{perm.} \right\} }
\hspace*{2mm}+ \frac{1}{12} {\displaystyle \left\{
\parbox{9mm}{\centerline{
\begin{fmfgraph*}(5,8.33)
\setval
\fmfforce{1w,0h}{v1}
\fmfforce{0w,0h}{v2}
\fmfforce{0.5w,4.33/8.33h}{v3}
\fmfforce{0.5w,1.25/8.33h}{vm}
\fmfforce{0.5/5w,1h}{v4}
\fmfforce{4.5/5w,1h}{v5}
\fmf{plain}{v1,vm,v2}
\fmf{plain}{v4,v5}
\fmf{plain}{v3,vm}
\fmfv{decor.size=0,label={\footnotesize 4},l.dist=0.5mm, l.angle=-30}{v1}
\fmfv{decor.size=0,label={\footnotesize 5},l.dist=0.5mm, l.angle=-150}{v2}
\fmfv{decor.size=0,label={\footnotesize 3},l.dist=0.5mm, l.angle=90}{v3}
\fmfv{decor.size=0,label={\footnotesize 1},l.dist=0.5mm, l.angle=-180}{v4}
\fmfv{decor.size=0,label={\footnotesize 2},l.dist=0.5mm, l.angle=0}{v5}
\fmfdot{vm}
\end{fmfgraph*}}} 
+ \, 2 \, \mbox{perm.} \right\} } \hspace*{4mm} \, .
\eeq
Then we obtain from Eq.~(\r{GNU}) for $l=3$:
\beq
\parbox{5.5mm}{\begin{center}
\begin{fmfgraph*}(2.5,5)
\setval
\fmfstraight
\fmfforce{1w,0h}{v1}
\fmfforce{1w,1h}{v2}
\fmf{plain,left=1}{v1,v2}
\fmfv{decor.size=0, label=${\scs 2}$, l.dist=1mm, l.angle=0}{v1}
\fmfv{decor.size=0, label=${\scs 1}$, l.dist=1mm, l.angle=0}{v2}
\end{fmfgraph*}
\end{center}}
\hspace*{0.3cm} \dphi{W^{(0,3)}}{1}{2}  &=& 
\hspace*{1mm} \frac{1}{4}
\parbox{9mm}{\begin{center}
\begin{fmfgraph}(4,4)
\setval
\fmfforce{0w,0h}{v1}
\fmfforce{1w,0h}{v2}
\fmfforce{1w,1h}{v3}
\fmfforce{0w,1h}{v4}
\fmf{plain,right=1}{v1,v3,v1}
\fmf{plain}{v1,v3}
\fmf{plain}{v2,v4}
\fmfdot{v1,v2,v3,v4}
\end{fmfgraph}\end{center}} 
\hspace*{1mm} + \frac{3}{8} \hspace*{1mm}
\parbox{9mm}{\begin{center}
\begin{fmfgraph}(4,4)
\setval
\fmfforce{0w,0h}{v1}
\fmfforce{1w,0h}{v2}
\fmfforce{1w,1h}{v3}
\fmfforce{0w,1h}{v4}
\fmf{plain,right=1}{v1,v3,v1}
\fmf{plain,right=0.4}{v1,v4}
\fmf{plain,left=0.4}{v2,v3}
\fmfdot{v1,v2,v3,v4}
\end{fmfgraph}\end{center}} 
\hspace*{1mm} + \frac{3}{4}
\parbox{15mm}{\begin{center}
\begin{fmfgraph}(12,4)
\setval
\fmfforce{0w,1/2h}{v1}
\fmfforce{1/3w,1/2h}{v2}
\fmfforce{2/3w,1/2h}{v3}
\fmfforce{5/6w,0h}{v4}
\fmfforce{5/6w,1h}{v5}
\fmfforce{1w,1/2h}{v6}
\fmf{plain,left=1}{v1,v2,v1}
\fmf{plain,left=1}{v3,v6,v3}
\fmf{plain}{v2,v3}
\fmf{plain}{v4,v5}
\fmfdot{v2,v3,v4,v5}
\end{fmfgraph}\end{center}} 
\hspace*{1mm} + \frac{3}{8} \hspace*{1mm}
\parbox{23mm}{\begin{center}
\begin{fmfgraph}(20,4)
\setval
\fmfforce{0w,1/2h}{v1}
\fmfforce{1/5w,1/2h}{v2}
\fmfforce{2/5w,1/2h}{v3}
\fmfforce{3/5w,1/2h}{v4}
\fmfforce{4/5w,1/2h}{v5}
\fmfforce{1w,1/2h}{v6}
\fmf{plain,left=1}{v1,v2,v1}
\fmf{plain,left=1}{v3,v4,v3}
\fmf{plain,left=1}{v5,v6,v5}
\fmf{plain}{v2,v3}
\fmf{plain}{v4,v5}
\fmfdot{v2,v3,v4,v5}
\end{fmfgraph}\end{center}} 
\hspace*{1mm} + \frac{1}{8} \hspace*{1mm} 
\parbox{17mm}{\begin{center}
\begin{fmfgraph}(13.856,12)
\setval
\fmfforce{0w,0h}{v1}
\fmfforce{1/4w,1/6h}{v2}
\fmfforce{1/2w,1/3h}{v3}
\fmfforce{3/4w,1/6h}{v4}
\fmfforce{1w,0h}{v5}
\fmfforce{1/2w,2/3h}{v6}
\fmfforce{1/2w,1h}{v7}
\fmf{plain,left=1}{v1,v2,v1}
\fmf{plain,left=1}{v4,v5,v4}
\fmf{plain,left=1}{v6,v7,v6}
\fmf{plain}{v2,v3}
\fmf{plain}{v4,v3}
\fmf{plain}{v3,v6}
\fmfdot{v2,v3,v4,v6}
\end{fmfgraph}\end{center}} 
\hspace*{1mm} + \frac{5}{8} \hspace*{1mm} 
\parbox{9mm}{\begin{center}
\begin{fmfgraph}(6,6)
\setval
\fmfforce{0w,1/2h}{v1}
\fmfforce{1w,1/2h}{v2}
\fmfforce{1/2w,1h}{v3}
\fmf{plain,left=1}{v1,v2,v1}
\fmf{plain,left=0.4}{v3,v1}
\fmf{plain,left=0.4}{v2,v3}
\fmfdot{v1,v2,v3}
\end{fmfgraph}\end{center}}
\hspace*{1mm} + \frac{5}{8} \hspace*{1mm} 
\parbox{11mm}{\begin{center}
\begin{fmfgraph}(4,8)
\setval
\fmfforce{0w,1/4h}{v1}
\fmfforce{1w,1/4h}{v2}
\fmfforce{1/2w,1/2h}{v3}
\fmfforce{1/2w,1h}{v4}
\fmf{plain,left=1}{v1,v2,v1}
\fmf{plain,left=1}{v3,v4,v3}
\fmf{plain}{v1,v2}
\fmfdot{v2,v3,v1}
\end{fmfgraph}\end{center}} 
\nonumber \\
&& + \frac{5}{12} \hspace*{1mm} 
\parbox{15mm}{\begin{center}
\begin{fmfgraph}(12,4)
\setval
\fmfforce{0w,1/2h}{v1}
\fmfforce{1/3w,1/2h}{v2}
\fmfforce{2/3w,1/2h}{v3}
\fmfforce{1w,1/2h}{v4}
\fmf{plain,left=1}{v1,v2,v1}
\fmf{plain}{v2,v4}
\fmf{plain,left=1}{v3,v4,v3}
\fmfdot{v4,v2,v3}
\end{fmfgraph}\end{center}} 
\hspace*{1mm} + \frac{5}{8} \hspace*{1mm} 
\parbox{19mm}{\begin{center}
\begin{fmfgraph}(16,4)
\setval
\fmfforce{0w,1/2h}{v1}
\fmfforce{1/4w,1/2h}{v2}
\fmfforce{1/2w,1/2h}{v3}
\fmfforce{3/4w,1/2h}{v4}
\fmfforce{1w,1/2h}{v5}
\fmf{plain,left=1}{v1,v2,v1}
\fmf{plain,left=1}{v2,v3,v2}
\fmf{plain}{v3,v4}
\fmf{plain,left=1}{v4,v5,v4}
\fmfdot{v4,v2,v3}
\end{fmfgraph}\end{center}} 
\hspace*{1mm}+ \frac{5}{16} \hspace*{1mm}
\parbox{19mm}{\begin{center}
\begin{fmfgraph}(16,6)
\setval
\fmfforce{0w,1/3h}{v1}
\fmfforce{1/4w,1/3h}{v2}
\fmfforce{1/2w,1/3h}{v3}
\fmfforce{1/2w,1h}{v4}
\fmfforce{1w,1/3h}{v6}
\fmfforce{3/4w,1/3h}{v5}
\fmf{plain,left=1}{v1,v2,v1}
\fmf{plain}{v2,v5}
\fmf{plain,left=1}{v3,v4,v3}
\fmf{plain,left=1}{v5,v6,v5}
\fmfdot{v5,v2,v3}
\end{fmfgraph}\end{center}} 
\hspace*{1mm}+ \frac{1}{12} \hspace*{1mm}
\parbox{9mm}{\begin{center}
\begin{fmfgraph}(6,4)
\setval
\fmfforce{0w,0.5h}{v1}
\fmfforce{1w,0.5h}{v2}
\fmf{plain,left=1}{v1,v2,v1}
\fmf{plain,left=0.4}{v1,v2,v1}
\fmfdot{v1,v2}
\end{fmfgraph}\end{center}} 
\hspace*{1mm}+ \frac{1}{4} \hspace*{1mm}
\parbox{15mm}{\begin{center}
\begin{fmfgraph}(12,4)
\setval
\fmfleft{i1}
\fmfright{o1}
\fmf{plain,left=1}{i1,v1,i1}
\fmf{plain,left=1}{v1,v2,v1}
\fmf{plain,left=1}{o1,v2,o1}
\fmfdot{v1,v2}
\end{fmfgraph}\end{center}} \hspace*{2mm},
\eeq
which leads to the connected vacuum diagrams listed in Table I together
with the subsequent loop order $l=4$. In a similar way, the graphical
relation (\r{GRC}) is recursively iterated to construct the connected vacuum
diagrams which involve currents. Table II depicts the resulting
diagrams for the respective first two loop orders with $n=1,2,3,4$.\\

The topology of each connected diagram in Table I and II
can be characterized
by the 5 component vector $(S,D,T,F;N)$. Here $S,D,T,F$ denote
the number of self-, double, triple and fourfold
connections, whereas $N$ stands for the number of identical vertex
permutations where the $3$- and $4$-vertices as well as the currents
remain attached to the lines
emerging from them in the same way as before. The proper weights of the 
connected vacuum diagrams in the $\phi^3$-$\phi^4$-theory are then
given by the formula \cite{PHI4,Neu,Verena}
\beq
\la{W}
W = \frac{1}{2!^{S+D}\, 3!^T \, 4!^F \, N} \, .
\eeq
For higher orders, it becomes more and more difficult to
identify by inspection the number $N$ of identical vertex permutations.
A mnemonic rule states that the number $N$ of 
identical vertex permutations is given by twice the number of symmetry
axes, if the diagram is imagined 
in a suitable maximally symmetric way in some
higher dimensional space. A more systematic determination of $N$ is
possible by introducing a matrix notation for the diagrams as explained
in detail in Refs. \cite{PHI4,Verena}. 
\end{fmffile}
\begin{fmffile}{fg4}
\subsection{Simpler Recursion Relations}
The graphical relation (\r{GRC}) allows in principle to 
construct all connected vacuum diagrams contributing to the interaction
negative free energy $W^{({\rm int})}$. However, the iteration of (\r{GRC})
is in practice a tedious task, 
as quite often different terms lead to the same topological diagram so
that the corresponding contributions pile up to its proper weight.
Thus it would be advantageous to obtain simpler graphical  
relations for certain subsets of connected vacuum diagrams. To this
end we remember that Eq.~(\r{GRC}) is based on counting the lines.
In the following we aim at deriving graphical recursion relations which
rely on counting other graphical elements of diagrams such as the
currents, the 3- and the 4-vertices.
\subsubsection{Counting Currents}
Applying (\r{NR1}) and (\r{DEC}) to
Eq.~(\r{CC}), we immediately obtain the simple linear functional differential
equation
\beq
\int_1 J_1 \frac{\delta W^{({\rm int})}}{\delta J_1} &= &
- \frac{1}{2} \int_{1234} K_{123} G_{12} G_{34} J_4 
- \frac{1}{2} \int_{123456} K_{123} G_{14} G_{25} G_{36} J_4 J_5 J_6
\nonumber \\
&&- \frac{1}{2} \int_{123456} K_{123} G_{14} G_{25} G_{36} J_4 \frac{\delta
W^{({\rm int})}}{\delta G_{56}} 
+ \int_{12345} L_{1234} G_{15} J_5 \frac{\delta W^{({\rm int})}}{\delta
K_{234}} \hspace*{0.4cm} \la{CWW} ,
\eeq
where the term on the left-hand side counts the number of currents in each
connected vacuum diagram. It can be diagramatically written as follows
\beq
\parbox{8mm}{\begin{center}
\begin{fmfgraph*}(5,5)
\setval
\fmfstraight
\fmfforce{0w,1/2h}{v1}
\fmfforce{1w,1/2h}{v2}
\fmf{plain}{v1,v2}
\fmfv{decor.shape=cross,decor.filled=shaded,decor.size=3thick}{v1}
\fmfv{decor.size=0, label=${\scs 1}$, l.dist=1mm, l.angle=0}{v2}
\end{fmfgraph*}
\end{center}}
\hspace*{2mm} \cdphi{W^{({\rm int})}}{1} =  \frac{1}{2}
\parbox{11mm}{\begin{center}
\begin{fmfgraph*}(8,4)
\setval
\fmfforce{0w,1/2h}{v1}
\fmfforce{1/2w,1/2h}{v2}
\fmfforce{1w,1/2h}{v3}
\fmf{plain,left=1}{v2,v3,v2}
\fmf{plain}{v1,v2}
\fmfdot{v2}
\fmfv{decor.shape=cross,decor.filled=shaded,decor.size=3thick}{v1}
\end{fmfgraph*}\end{center}}
+ \frac{1}{2} 
\parbox{11mm}{\begin{center}
\begin{fmfgraph*}(6.928,12)
\setval
\fmfforce{1/2w,5/6h}{v1}
\fmfforce{1w,1/4h}{w1}
\fmfforce{0w,1/4h}{u1}
\fmfforce{1/2w,1/2h}{v2}
\fmf{plain}{v2,v1}
\fmf{plain}{v2,w1}
\fmf{plain}{v2,u1}
\fmfdot{v2}
\fmfv{decor.shape=cross,decor.filled=shaded,decor.size=3thick}{v1}
\fmfv{decor.shape=cross,decor.filled=shaded,decor.size=3thick}{w1}
\fmfv{decor.shape=cross,decor.filled=shaded,decor.size=3thick}{u1}
\end{fmfgraph*}\end{center}}
+ 
\parbox{9mm}{\begin{center}
\begin{fmfgraph*}(6,4)
\setval
\fmfstraight
\fmfforce{0w,1/2h}{v1}
\fmfforce{4/6w,1/2h}{v2}
\fmfforce{1w,1h}{i1}
\fmfforce{1w,0h}{i2}
\fmf{plain}{v2,v1}
\fmf{plain}{v2,i1}
\fmf{plain}{v2,i2}
\fmfdot{v2}
\fmfv{decor.shape=cross,decor.filled=shaded,decor.size=3thick}{v1}
\fmfv{decor.size=0, label=${\scs 2}$, l.dist=1mm, l.angle=0}{i2}
\fmfv{decor.size=0, label=${\scs 1}$, l.dist=1mm, l.angle=0}{i1}
\end{fmfgraph*}
\end{center}}
\hspace*{3mm} \dphi{W^{({\rm int})}}{1}{2}
+
\parbox{9mm}{\begin{center}
\begin{fmfgraph*}(6,4)
\setval
\fmfstraight
\fmfforce{0w,1/2h}{v1}
\fmfforce{4/6w,1/2h}{v2}
\fmfforce{1w,1.25h}{i1}
\fmfforce{1w,1/2h}{i2}
\fmfforce{1w,-0.25h}{i3}
\fmf{plain}{v2,v1}
\fmf{plain}{v2,i1}
\fmf{plain}{v2,i2}
\fmf{plain}{v2,i3}
\fmfdot{v2}
\fmfv{decor.shape=cross,decor.filled=shaded,decor.size=3thick}{v1}
\fmfv{decor.size=0, label=${\scs 3}$, l.dist=1mm, l.angle=0}{i3}
\fmfv{decor.size=0, label=${\scs 2}$, l.dist=1mm, l.angle=0}{i2}
\fmfv{decor.size=0, label=${\scs 1}$, l.dist=1mm, l.angle=0}{i1}
\end{fmfgraph*}
\end{center}}
\hspace*{3mm} \dvertex{W^{({\rm int})}}{1}{2}{3} \hspace*{4mm} ,
\la{CC1}
\eeq
where the right-hand side contains only 4 terms, 2 without 
$W^{({\rm int})}$ and 2 linear in $W^{({\rm int})}$. With the decomposition
(\r{DECC}), the graphical recursion relation which is derived from 
(\r{CC1}) reads for $n > 0$
\beq
W^{(n,l)} = \frac{1}{n} {\displaystyle \left\{ 
\parbox{9mm}{\begin{center}
\begin{fmfgraph*}(6,4)
\setval
\fmfstraight
\fmfforce{0w,1/2h}{v1}
\fmfforce{4/6w,1/2h}{v2}
\fmfforce{1w,1h}{i1}
\fmfforce{1w,0h}{i2}
\fmf{plain}{v2,v1}
\fmf{plain}{v2,i1}
\fmf{plain}{v2,i2}
\fmfdot{v2}
\fmfv{decor.shape=cross,decor.filled=shaded,decor.size=3thick}{v1}
\fmfv{decor.size=0, label=${\scs 2}$, l.dist=1mm, l.angle=0}{i2}
\fmfv{decor.size=0, label=${\scs 1}$, l.dist=1mm, l.angle=0}{i1}
\end{fmfgraph*}
\end{center}}
\hspace*{3mm} \dphi{W^{(n-1,l)}}{1}{2}
+
\parbox{9mm}{\begin{center}
\begin{fmfgraph*}(6,4)
\setval
\fmfstraight
\fmfforce{0w,1/2h}{v1}
\fmfforce{4/6w,1/2h}{v2}
\fmfforce{1w,1.25h}{i1}
\fmfforce{1w,1/2h}{i2}
\fmfforce{1w,-0.25h}{i3}
\fmf{plain}{v2,v1}
\fmf{plain}{v2,i1}
\fmf{plain}{v2,i2}
\fmf{plain}{v2,i3}
\fmfdot{v2}
\fmfv{decor.shape=cross,decor.filled=shaded,decor.size=3thick}{v1}
\fmfv{decor.size=0, label=${\scs 3}$, l.dist=1mm, l.angle=0}{i3}
\fmfv{decor.size=0, label=${\scs 2}$, l.dist=1mm, l.angle=0}{i2}
\fmfv{decor.size=0, label=${\scs 1}$, l.dist=1mm, l.angle=0}{i1}
\end{fmfgraph*}
\end{center}}
\hspace*{3mm} \dvertex{W^{(n-1,l)}}{1}{2}{3}
\right\} }
\eeq
which is iterated starting from
\beq
W^{(1,1)} = \frac{1}{2}
\parbox{11mm}{\begin{center}
\begin{fmfgraph*}(8,4)
\setval
\fmfforce{0w,1/2h}{v1}
\fmfforce{1/2w,1/2h}{v2}
\fmfforce{1w,1/2h}{v3}
\fmf{plain,left=1}{v2,v3,v2}
\fmf{plain}{v1,v2}
\fmfdot{v2}
\fmfv{decor.shape=cross,decor.filled=shaded,decor.size=3thick}{v1}
\end{fmfgraph*}\end{center}}
\, , \hspace*{1cm} W^{(3,0)} = \frac{1}{6}
\parbox{11mm}{\begin{center}
\begin{fmfgraph*}(6.928,12)
\setval
\fmfforce{1/2w,5/6h}{v1}
\fmfforce{1w,1/4h}{w1}
\fmfforce{0w,1/4h}{u1}
\fmfforce{1/2w,1/2h}{v2}
\fmf{plain}{v2,v1}
\fmf{plain}{v2,w1}
\fmf{plain}{v2,u1}
\fmfdot{v2}
\fmfv{decor.shape=cross,decor.filled=shaded,decor.size=3thick}{v1}
\fmfv{decor.shape=cross,decor.filled=shaded,decor.size=3thick}{w1}
\fmfv{decor.shape=cross,decor.filled=shaded,decor.size=3thick}{u1}
\end{fmfgraph*}\end{center}}
\eeq
and the current-free connected vacuum diagrams $W^{(0,l)}$. As a first
example, we consider the case $n=1$ and $l=2$, insert (\r{NNR1}), (\r{NNR3})
into 
\beq
W^{(1,2)} =  
\parbox{9mm}{\begin{center}
\begin{fmfgraph*}(6,4)
\setval
\fmfstraight
\fmfforce{0w,1/2h}{v1}
\fmfforce{4/6w,1/2h}{v2}
\fmfforce{1w,1h}{i1}
\fmfforce{1w,0h}{i2}
\fmf{plain}{v2,v1}
\fmf{plain}{v2,i1}
\fmf{plain}{v2,i2}
\fmfdot{v2}
\fmfv{decor.shape=cross,decor.filled=shaded,decor.size=3thick}{v1}
\fmfv{decor.size=0, label=${\scs 2}$, l.dist=1mm, l.angle=0}{i2}
\fmfv{decor.size=0, label=${\scs 1}$, l.dist=1mm, l.angle=0}{i1}
\end{fmfgraph*}
\end{center}}
\hspace*{3mm} \dphi{W^{(0,2)}}{1}{2}
+
\parbox{9mm}{\begin{center}
\begin{fmfgraph*}(6,4)
\setval
\fmfstraight
\fmfforce{0w,1/2h}{v1}
\fmfforce{4/6w,1/2h}{v2}
\fmfforce{1w,1.25h}{i1}
\fmfforce{1w,1/2h}{i2}
\fmfforce{1w,-0.25h}{i3}
\fmf{plain}{v2,v1}
\fmf{plain}{v2,i1}
\fmf{plain}{v2,i2}
\fmf{plain}{v2,i3}
\fmfdot{v2}
\fmfv{decor.shape=cross,decor.filled=shaded,decor.size=3thick}{v1}
\fmfv{decor.size=0, label=${\scs 3}$, l.dist=1mm, l.angle=0}{i3}
\fmfv{decor.size=0, label=${\scs 2}$, l.dist=1mm, l.angle=0}{i2}
\fmfv{decor.size=0, label=${\scs 1}$, l.dist=1mm, l.angle=0}{i1}
\end{fmfgraph*}
\end{center}}
\hspace*{3mm} \dvertex{W^{(0,2)}}{1}{2}{3}
\eeq
and thus obtain
\beq
W^{(1,2)} = \frac{1}{4} \hspace*{1mm}
\parbox{11mm}{\begin{center}
\begin{fmfgraph*}(8,4)
\setval
\fmfforce{0w,1/2h}{v1}
\fmfforce{1/2w,1/2h}{v2}
\fmfforce{1w,1/2h}{v3}
\fmfforce{3/4w,0h}{v4}
\fmfforce{3/4w,1h}{v5}
\fmf{plain,left=1}{v2,v3,v2}
\fmf{plain}{v1,v2}
\fmf{plain}{v4,v5}
\fmfdot{v2,v4,v5}
\fmfv{decor.shape=cross,decor.filled=shaded,decor.size=3thick}{v1}
\end{fmfgraph*}\end{center}}
\hspace*{1mm} + \frac{1}{4} \hspace*{1mm}
\parbox{19mm}{\begin{center}
\begin{fmfgraph*}(16,4)
\setval
\fmfforce{0w,1/2h}{v1}
\fmfforce{1/4w,1/2h}{v2}
\fmfforce{1/2w,1/2h}{v3}
\fmfforce{3/4w,1/2h}{v4}
\fmfforce{1w,1/2h}{v5}
\fmf{plain,left=1}{v2,v3,v2}
\fmf{plain,left=1}{v4,v5,v4}
\fmf{plain}{v1,v2}
\fmf{plain}{v3,v4}
\fmfdot{v2,v4,v3}
\fmfv{decor.shape=cross,decor.filled=shaded,decor.size=3thick}{v1}
\end{fmfgraph*}\end{center}}
\hspace*{1mm}+ \frac{1}{8} \hspace*{1mm}
\parbox{11mm}{\begin{center}
\begin{fmfgraph*}(8,13.856)
\setval
\fmfforce{0w,1/2h}{v1}
\fmfforce{1/2w,1/2h}{v2}
\fmfforce{3/4w,1/4h}{v3}
\fmfforce{3/4w,3/4h}{v4}
\fmfforce{1w,0h}{v5}
\fmfforce{1w,1h}{v6}
\fmf{plain,left=1}{v3,v5,v3}
\fmf{plain,left=1}{v4,v6,v4}
\fmf{plain}{v1,v2}
\fmf{plain}{v2,v3}
\fmf{plain}{v2,v4}
\fmfdot{v2,v4,v3}
\fmfv{decor.shape=cross,decor.filled=shaded,decor.size=3thick}{v1}
\end{fmfgraph*}\end{center}} 
\hspace*{1mm}+ \frac{1}{6} \hspace*{1mm}
\parbox{11mm}{\begin{center}
\begin{fmfgraph*}(8,4)
\setval
\fmfforce{0w,1/2h}{v1}
\fmfforce{1/2w,1/2h}{v2}
\fmfforce{1w,1/2h}{v3}
\fmf{plain,left=1}{v2,v3,v2}
\fmf{plain}{v1,v3}
\fmfdot{v2,v3}
\fmfv{decor.shape=cross,decor.filled=shaded,decor.size=3thick}{v1}
\end{fmfgraph*}\end{center}}
\hspace*{1mm}+ \frac{1}{4} \hspace*{1mm}
\parbox{15mm}{\begin{center}
\begin{fmfgraph*}(12,4)
\setval
\fmfforce{0w,1/2h}{v1}
\fmfforce{1/3w,1/2h}{v2}
\fmfforce{2/3w,1/2h}{v3}
\fmfforce{1w,1/2h}{v4}
\fmf{plain,left=1}{v2,v3,v2}
\fmf{plain,left=1}{v3,v4,v3}
\fmf{plain}{v1,v2}
\fmfdot{v2,v3}
\fmfv{decor.shape=cross,decor.filled=shaded,decor.size=3thick}{v1}
\end{fmfgraph*}\end{center}}
\hspace*{1mm}+ \frac{1}{4} \hspace*{1mm}
\parbox{15mm}{\begin{center}
\begin{fmfgraph*}(12,6)
\setval
\fmfforce{0w,1/3h}{v1}
\fmfforce{1/3w,1/3h}{v2}
\fmfforce{2/3w,1/3h}{v3}
\fmfforce{1w,1/3h}{v4}
\fmfforce{1/3w,1h}{v5}
\fmf{plain,left=1}{v2,v5,v2}
\fmf{plain,left=1}{v3,v4,v3}
\fmf{plain}{v1,v3}
\fmfdot{v2,v3}
\fmfv{decor.shape=cross,decor.filled=shaded,decor.size=3thick}{v1}
\end{fmfgraph*}\end{center}}
\hspace*{4mm} .
\eeq
In the second example we set $n=2$ and $l=1$, evaluate from $W^{(1,1)}$  
a line as well as a 3-vertex amputation
\beq
\dphi{W^{(1,1)}}{1}{2} = \frac{1}{2} \hspace*{1mm}
\parbox{9mm}{\begin{center}
\begin{fmfgraph*}(6,4)
\setval
\fmfstraight
\fmfforce{0w,1/2h}{v1}
\fmfforce{4/6w,1/2h}{v2}
\fmfforce{1w,1h}{i1}
\fmfforce{1w,0h}{i2}
\fmf{plain}{v2,v1}
\fmf{plain}{v2,i1}
\fmf{plain}{v2,i2}
\fmfdot{v2}
\fmfv{decor.shape=cross,decor.filled=shaded,decor.size=3thick}{v1}
\fmfv{decor.size=0, label=${\scs 2}$, l.dist=1mm, l.angle=0}{i2}
\fmfv{decor.size=0, label=${\scs 1}$, l.dist=1mm, l.angle=0}{i1}
\end{fmfgraph*}
\end{center}}
\hspace*{1mm}+\frac{1}{4} {\displaystyle \left\{
\parbox{9mm}{\centerline{
\begin{fmfgraph*}(6,6)
\setval
\fmfforce{0w,1/3h}{v1}
\fmfforce{2/3w,1/3h}{v2}
\fmfforce{1w,1/3h}{v3}
\fmfforce{1/3w,1h}{v4}
\fmfforce{1w,1h}{v5}
\fmf{plain}{v4,v5}
\fmf{plain}{v2,v3}
\fmf{plain,left=1}{v1,v2,v1}
\fmfv{decor.size=0,label={\footnotesize 2},l.dist=0.5mm, l.angle=0}{v3}
\fmfv{decor.shape=cross,decor.filled=shaded,decor.size=3thick}{v4}
\fmfv{decor.size=0,label={\footnotesize 1},l.dist=0.5mm, l.angle=0}{v5}
\fmfdot{v2}
\end{fmfgraph*}}} 
\hspace*{2mm}+ \, 1 \, \mbox{perm.} \right\} } \hspace*{3mm} \, , \hspace*{5mm}
\dvertex{W^{(1,1)}}{1}{2}{3} = \frac{1}{6} {\displaystyle \left\{
\parbox{8mm}{\centerline{
\begin{fmfgraph*}(4,4)
\setval
\fmfforce{0w,1h}{v1}
\fmfforce{1w,1h}{v2}
\fmfforce{0w,0h}{v3}
\fmfforce{1w,0h}{v4}
\fmf{plain}{v2,v1}
\fmf{plain}{v4,v3}
\fmfv{decor.shape=cross,decor.filled=shaded,decor.size=3thick}{v3}
\fmfv{decor.size=0,label={\footnotesize 1},l.dist=0.5mm, l.angle=-180}{v1}
\fmfv{decor.size=0,label={\footnotesize 2},l.dist=0.5mm, l.angle=0}{v2}
\fmfv{decor.size=0,label={\footnotesize 3},l.dist=0.5mm, l.angle=0}{v4}
\end{fmfgraph*}}}
\hspace*{2mm} + \, 2 \, \mbox{perm.} \right\} }
\eeq
and insert this into
\beq
W^{(n,l)} = \frac{1}{2} {\displaystyle \left\{ 
\parbox{9mm}{\begin{center}
\begin{fmfgraph*}(6,4)
\setval
\fmfstraight
\fmfforce{0w,1/2h}{v1}
\fmfforce{4/6w,1/2h}{v2}
\fmfforce{1w,1h}{i1}
\fmfforce{1w,0h}{i2}
\fmf{plain}{v2,v1}
\fmf{plain}{v2,i1}
\fmf{plain}{v2,i2}
\fmfdot{v2}
\fmfv{decor.shape=cross,decor.filled=shaded,decor.size=3thick}{v1}
\fmfv{decor.size=0, label=${\scs 2}$, l.dist=1mm, l.angle=0}{i2}
\fmfv{decor.size=0, label=${\scs 1}$, l.dist=1mm, l.angle=0}{i1}
\end{fmfgraph*}
\end{center}}
\hspace*{3mm} \dphi{W^{(1,1)}}{1}{2}
+
\parbox{9mm}{\begin{center}
\begin{fmfgraph*}(6,4)
\setval
\fmfstraight
\fmfforce{0w,1/2h}{v1}
\fmfforce{4/6w,1/2h}{v2}
\fmfforce{1w,1.25h}{i1}
\fmfforce{1w,1/2h}{i2}
\fmfforce{1w,-0.25h}{i3}
\fmf{plain}{v2,v1}
\fmf{plain}{v2,i1}
\fmf{plain}{v2,i2}
\fmf{plain}{v2,i3}
\fmfdot{v2}
\fmfv{decor.shape=cross,decor.filled=shaded,decor.size=3thick}{v1}
\fmfv{decor.size=0, label=${\scs 3}$, l.dist=1mm, l.angle=0}{i3}
\fmfv{decor.size=0, label=${\scs 2}$, l.dist=1mm, l.angle=0}{i2}
\fmfv{decor.size=0, label=${\scs 1}$, l.dist=1mm, l.angle=0}{i1}
\end{fmfgraph*}
\end{center}}
\hspace*{3mm} \dvertex{W^{(1,1)}}{1}{2}{3}
\right\} } \hspace*{4mm} ,
\eeq
so that we result in
\beq
W^{(2,1)} = \frac{1}{4} \hspace*{1mm}
\parbox{11mm}{\begin{center}
\begin{fmfgraph*}(8,4)
\setval
\fmfforce{0w,0h}{v1}
\fmfforce{1/2w,0h}{v2}
\fmfforce{1/2w,1h}{v3}
\fmfforce{1w,0h}{w1}
\fmf{plain,left=1}{v2,v3,v2}
\fmf{plain}{v1,w1}
\fmfdot{v2}
\fmfv{decor.shape=cross,decor.filled=shaded,decor.size=3thick}{v1}
\fmfv{decor.shape=cross,decor.filled=shaded,decor.size=3thick}{w1}
\end{fmfgraph*}\end{center}}
\hspace*{1mm} + \frac{1}{4} \hspace*{1mm}
\parbox{15mm}{\begin{center}
\begin{fmfgraph*}(12,4)
\setval
\fmfforce{0w,1/2h}{v1}
\fmfforce{1/3w,1/2h}{v2}
\fmfforce{2/3w,1/2h}{v3}
\fmfforce{1w,1/2h}{w1}
\fmf{plain,left=1}{v2,v3,v2}
\fmf{plain}{v1,v2}
\fmf{plain}{v3,w1}
\fmfdot{v2,v3}
\fmfv{decor.shape=cross,decor.filled=shaded,decor.size=3thick}{v1}
\fmfv{decor.shape=cross,decor.filled=shaded,decor.size=3thick}{w1}
\end{fmfgraph*}\end{center}}
\hspace*{1mm}+ \frac{1}{4} \hspace*{1mm}
\parbox{11mm}{\begin{center}
\begin{fmfgraph*}(8,8)
\setval
\fmfforce{0w,0h}{v1}
\fmfforce{1/2w,0h}{v2}
\fmfforce{1/2w,1/2h}{v3}
\fmfforce{1/2w,1h}{v4}
\fmfforce{1w,0h}{w1}
\fmf{plain,left=1}{v3,v4,v3}
\fmf{plain}{v1,w1}
\fmf{plain}{v2,v3}
\fmfdot{v2,v3}
\fmfv{decor.shape=cross,decor.filled=shaded,decor.size=3thick}{v1}
\fmfv{decor.shape=cross,decor.filled=shaded,decor.size=3thick}{w1}
\end{fmfgraph*}\end{center}} \hspace*{0.5cm} \, .
\eeq
\end{fmffile}
\begin{fmffile}{fg5}
\subsubsection{Counting 3-Vertices}
The compatibility relation (\r{CP2}) between 
functional derivatives with respect to the field expectation value 
$\Phi$, the kernel
$G^{-1}$ and the $3$-vertex $K$
leads together with (\r{CC}) and (\r{DD}) to 
\beq
& & \int_{123} K_{123} \frac{\delta W}{\delta K_{123}} = 
- \frac{1}{3} \int_{1234} K_{123} G_{12} G_{34} J_4 
+ \frac{1}{3} \int_{1234} K_{123} G_{14} J_4 
\frac{\delta W}{\delta G^{-1}_{23}} \no \\
& & \hspace*{1cm}
- \frac{1}{3} \int_{1234567} K_{123} K_{456} G_{12} G_{34} 
\frac{\delta W}{\delta G^{-1}_{56}} 
+ \frac{1}{3} \int_{123456} K_{123} K_{456} G_{34} \left\{ 
\frac{\delta^2 W}{\delta G^{-1}_{12} \delta G^{-1}_{56}} 
+ \frac{\delta W}{\delta G^{-1}_{12}} 
\frac{\delta W}{\delta G^{-1}_{56}}  \right\} \no \\
& & \hspace*{1cm}
- \frac{1}{3} \int_{1234567} K_{123} L_{4567} G_{12} G_{34} 
\frac{\delta W}{\delta K_{567}} 
+ \frac{1}{3} \int_{1234567} K_{123} L_{4567} G_{34} \left\{ 
\frac{\delta^2 W}{\delta G^{-1}_{12} \delta K_{567}} 
+ \frac{\delta W}{\delta G^{-1}_{12}} 
\frac{\delta W}{\delta K_{567}}  \right\} \hspace*{4mm} .
\eeq
Applying (\r{NR1}), (\r{NR2}) and (\r{DEC}), we obtain
a functional differential equation which is based on counting $3$-vertices:
\beq
&&\int_{123} K_{123} \frac{\delta W^{({\rm int})}}{\delta K_{123}} 
= 
\frac{1}{6} \int_{123456} K_{123} K_{456} G_{14} G_{25} G_{36} 
+ \frac{1}{4}\int_{123456} K_{123} K_{456} G_{14} G_{23} G_{56} 
- \frac{1}{2} \int_{1234} K_{123} G_{12} G_{34} J_4 
\no \\ && \hspace*{5mm}
+ \frac{1}{3} \int_{12345678} K_{123} K_{456} G_{14} G_{25} G_{37} 
G_{68} J_7 J_8 
+ \frac{1}{3} \int_{12345678} K_{123} K_{456} G_{12} G_{34} G_{57} 
G_{68} J_7 J_8 
\no \\ &&\hspace*{5mm}
- \frac{1}{6} \int_{123456} K_{123} G_{14} G_{25} G_{36} J_4 J_5 J_6 
+ \frac{1}{12} \int_{123456789\bar{1}} K_{123} K_{456} G_{14} G_{27} G_{38} 
G_{59} G_{6\bar{1}}J_7 J_8 J_9 J_{\bar{1}} 
\no \\ &&\hspace*{5mm}
+ \frac{2}{3}
\int_{12345678} K_{123} K_{456} G_{14} G_{25} G_{37} G_{68} 
\frac{\delta W^{({\rm int})}}{\delta G_{78}}
+ \frac{2}{3}
\int_{12345678} K_{123} K_{456} G_{12} G_{34} G_{57} G_{68}
\frac{\delta W^{({\rm int})}}{\delta G_{78}} 
\no \\ &&\hspace*{5mm}
- \frac{1}{2} \int_{1234567} K_{123} L_{4567} G_{12} G_{34} 
\frac{\delta W^{({\rm int})}}{\delta K_{567}}
- \frac{1}{3} \int_{123456} K_{123} G_{14} G_{25} G_{36} J_4 
\frac{\delta W^{({\rm int})}}{\delta G_{56}}
\no \\&& \hspace*{5mm}
+ \frac{1}{3} \int_{123456789\bar{1}} K_{123} K_{456} G_{14} G_{27} G_{38}
G_{59} G_{6\bar{1}} J_7 J_8 \frac{\delta W^{({\rm int})}}{\delta G_{9\bar{1}}}
- \frac{1}{6} \int_{123456789} K_{123} L_{4567} G_{14} G_{27} G_{38}
J_7 J_8 \frac{\delta W^{({\rm int})}}{\delta K_{567}}
\no \\ && \hspace*{5mm}
+ \frac{1}{3} \int_{123456789\bar{1}} K_{123} K_{456} G_{14} G_{27} G_{38}
G_{59} G_{6\bar{1}} \left\{ \frac{\delta^2 W^{({\rm int})}}{\delta G_{78} 
\delta G_{9\bar{1}}} + \frac{\delta W^{({\rm int})}}{\delta G_{78}}
\frac{\delta W^{({\rm int})}}{\delta G_{9\bar{1}}} \right\} 
\no \\ && \hspace*{5mm}
- \frac{1}{3} \int_{123456789} K_{123} L_{4567} G_{14} G_{27} G_{38} 
\left\{ \frac{\delta^2 W^{({\rm int})}}{\delta K_{567} 
\delta G_{89}} + \frac{\delta W^{({\rm int})}}{\delta K_{567}}
\frac{\delta W^{({\rm int})}}{\delta G_{89}} \right\} \, . 
\eeq
The corresponding graphical representation reads
\beq
&&
\parbox{5mm}{\begin{center}
\begin{fmfgraph*}(2,4)
\setval
\fmfstraight
\fmfforce{0w,1/2h}{v1}
\fmfforce{1w,1/2h}{v2}
\fmfforce{1w,1.25h}{v3}
\fmfforce{1w,-0.25h}{v4}
\fmf{plain}{v1,v2}
\fmf{plain}{v1,v3}
\fmf{plain}{v1,v4}
\fmfv{decor.size=0, label=${\scs 2}$, l.dist=1mm, l.angle=0}{v2}
\fmfv{decor.size=0, label=${\scs 1}$, l.dist=1mm, l.angle=0}{v3}
\fmfv{decor.size=0, label=${\scs 3}$, l.dist=1mm, l.angle=0}{v4}
\fmfdot{v1}
\end{fmfgraph*}
\end{center}}
\hspace*{0.3cm} \dvertex{W^{({\rm int})}}{1}{2}{3} = 
\hspace*{1mm} \frac{1}{6}\hspace*{1mm}
\parbox{7mm}{\begin{center}
\begin{fmfgraph*}(4,4)
\setval
\fmfforce{0w,0.5h}{v1}
\fmfforce{1w,0.5h}{v2}
\fmf{plain,left=1}{v1,v2,v1}
\fmf{plain}{v1,v2}
\fmfdot{v1,v2}
\end{fmfgraph*}\end{center}}
\hspace*{1mm}+ \frac{1}{4}\hspace*{1mm}
\parbox{15mm}{\begin{center}
\begin{fmfgraph*}(12,4)
\setval
\fmfforce{0w,0.5h}{v1}
\fmfforce{1/3w,0.5h}{v2}
\fmfforce{2/3w,0.5h}{v3}
\fmfforce{1w,0.5h}{v4}
\fmf{plain,left=1}{v1,v2,v1}
\fmf{plain,left=1}{v3,v4,v3}
\fmf{plain}{v2,v3}
\fmfdot{v2,v3}
\end{fmfgraph*}\end{center}} 
\hspace*{1mm}+ \frac{1}{2} \hspace*{1mm}
\parbox{11mm}{\begin{center}
\begin{fmfgraph*}(8,4)
\setval
\fmfforce{0w,1/2h}{v1}
\fmfforce{1/2w,1/2h}{v2}
\fmfforce{1w,1/2h}{v3}
\fmf{plain,left=1}{v2,v3,v2}
\fmf{plain}{v1,v2}
\fmfdot{v2}
\fmfv{decor.shape=cross,decor.filled=shaded,decor.size=3thick}{v1}
\end{fmfgraph*}\end{center}}
\hspace*{1mm}+ \frac{1}{3}\hspace*{1mm}
\parbox{15mm}{\begin{center}
\begin{fmfgraph*}(12,4)
\setval
\fmfforce{0w,1/2h}{v1}
\fmfforce{1/3w,1/2h}{v2}
\fmfforce{2/3w,1/2h}{v3}
\fmfforce{1w,1/2h}{w1}
\fmf{plain,left=1}{v2,v3,v2}
\fmf{plain}{v1,v2}
\fmf{plain}{v3,w1}
\fmfdot{v2,v3}
\fmfv{decor.shape=cross,decor.filled=shaded,decor.size=3thick}{v1}
\fmfv{decor.shape=cross,decor.filled=shaded,decor.size=3thick}{w1}
\end{fmfgraph*}\end{center}}
\hspace*{1mm}+ \frac{1}{3}\hspace*{1mm}
\parbox{11mm}{\begin{center}
\begin{fmfgraph*}(8,8)
\setval
\fmfforce{0w,0h}{v1}
\fmfforce{1/2w,0h}{v2}
\fmfforce{1/2w,1/2h}{v3}
\fmfforce{1/2w,1h}{v4}
\fmfforce{1w,0h}{w1}
\fmf{plain,left=1}{v3,v4,v3}
\fmf{plain}{v1,w1}
\fmf{plain}{v2,v3}
\fmfdot{v2,v3}
\fmfv{decor.shape=cross,decor.filled=shaded,decor.size=3thick}{v1}
\fmfv{decor.shape=cross,decor.filled=shaded,decor.size=3thick}{w1}
\end{fmfgraph*}\end{center}}
\hspace*{1mm} + \frac{1}{6}\hspace*{1mm}
\parbox{11mm}{\begin{center}
\begin{fmfgraph*}(6.928,12)
\setval
\fmfforce{1/2w,5/6h}{v1}
\fmfforce{1w,1/4h}{w1}
\fmfforce{0w,1/4h}{u1}
\fmfforce{1/2w,1/2h}{v2}
\fmf{plain}{v2,v1}
\fmf{plain}{v2,w1}
\fmf{plain}{v2,u1}
\fmfdot{v2}
\fmfv{decor.shape=cross,decor.filled=shaded,decor.size=3thick}{v1}
\fmfv{decor.shape=cross,decor.filled=shaded,decor.size=3thick}{w1}
\fmfv{decor.shape=cross,decor.filled=shaded,decor.size=3thick}{u1}
\end{fmfgraph*}\end{center}}
\hspace*{1mm} + \frac{1}{12} 
\parbox{11mm}{\begin{center}
\begin{fmfgraph*}(8,4)
\setval
\fmfforce{0w,0h}{v1}
\fmfforce{1w,0h}{w1}
\fmfforce{0w,1h}{u1}
\fmfforce{1w,1h}{x1}
\fmfforce{1/2w,0h}{v2}
\fmfforce{1/2w,1h}{v3}
\fmf{plain}{w1,v1}
\fmf{plain}{v2,v3}
\fmf{plain}{x1,u1}
\fmfdot{v2,v3}
\fmfv{decor.shape=cross,decor.filled=shaded,decor.size=3thick}{u1}
\fmfv{decor.shape=cross,decor.filled=shaded,decor.size=3thick}{v1}
\fmfv{decor.shape=cross,decor.filled=shaded,decor.size=3thick}{w1}
\fmfv{decor.shape=cross,decor.filled=shaded,decor.size=3thick}{x1}
\end{fmfgraph*}\end{center}}
\no \\
&& 
\hspace*{5mm} + \frac{2}{3} \hspace*{1mm}
\parbox{7mm}{\begin{center}
\begin{fmfgraph*}(6,4)
\setval
\fmfstraight
\fmfforce{1/3w,0h}{v1}
\fmfforce{1/3w,1h}{v2}
\fmfforce{1w,1h}{i2}
\fmfforce{1w,0h}{i1}
\fmf{plain}{i1,v1}
\fmf{plain}{v2,i2}
\fmf{plain,left}{v1,v2,v1}
\fmfdot{v2,v1}
\fmfv{decor.size=0, label=${\scs 2}$, l.dist=1mm, l.angle=0}{i1}
\fmfv{decor.size=0, label=${\scs 1}$, l.dist=1mm, l.angle=0}{i2}
\end{fmfgraph*}
\end{center}}
\hspace*{3mm} \dphi{W^{({\rm int})}}{1}{2}
\hspace*{1mm} + \frac{2}{3} \hspace*{1mm}
\parbox{13mm}{\begin{center}
\begin{fmfgraph*}(10,4)
\setval
\fmfstraight
\fmfforce{0w,1/2h}{v1}
\fmfforce{4/10w,1/2h}{v2}
\fmfforce{8/10w,1/2h}{v3}
\fmfforce{1w,0h}{i2}
\fmfforce{1w,1h}{i1}
\fmf{plain,left}{v1,v2,v1}
\fmf{plain}{v3,v2}
\fmf{plain}{v3,i1}
\fmf{plain}{v3,i2}
\fmfdot{v2,v3}
\fmfv{decor.size=0, label=${\scs 2}$, l.dist=1mm, l.angle=0}{i2}
\fmfv{decor.size=0, label=${\scs 1}$, l.dist=1mm, l.angle=0}{i1}
\end{fmfgraph*}
\end{center}}
\hspace*{3mm} \dphi{W^{({\rm int})}}{1}{2}
\hspace*{1mm} + \frac{1}{2} \hspace*{1mm}
\parbox{13mm}{\begin{center}
\begin{fmfgraph*}(10,4)
\setval
\fmfstraight
\fmfforce{0w,1/2h}{v1}
\fmfforce{4/10w,1/2h}{v2}
\fmfforce{8/10w,1/2h}{v3}
\fmfforce{1w,-0.25h}{i1}
\fmfforce{1w,0.5h}{i2}
\fmfforce{1w,1.25h}{i3}
\fmf{plain,left}{v1,v2,v1}
\fmf{plain}{v3,v2}
\fmf{plain}{v3,i1}
\fmf{plain}{v3,i2}
\fmf{plain}{v3,i3}
\fmfdot{v2,v3}
\fmfv{decor.size=0, label=${\scs 1}$, l.dist=1mm, l.angle=0}{i3}
\fmfv{decor.size=0, label=${\scs 2}$, l.dist=1mm, l.angle=0}{i2}
\fmfv{decor.size=0, label=${\scs 3}$, l.dist=1mm, l.angle=0}{i1}
\end{fmfgraph*}
\end{center}}
\hspace*{3mm} \dvertex{W^{({\rm int})}}{1}{2}{3} 
\hspace*{1mm} + \frac{1}{3} \hspace*{1mm}
\parbox{9mm}{\begin{center}
\begin{fmfgraph*}(6,4)
\setval
\fmfstraight
\fmfforce{0w,1/2h}{v1}
\fmfforce{4/6w,1/2h}{v2}
\fmfforce{1w,1h}{i1}
\fmfforce{1w,0h}{i2}
\fmf{plain}{v2,v1}
\fmf{plain}{v2,i1}
\fmf{plain}{v2,i2}
\fmfdot{v2}
\fmfv{decor.shape=cross,decor.filled=shaded,decor.size=3thick}{v1}
\fmfv{decor.size=0, label=${\scs 2}$, l.dist=1mm, l.angle=0}{i2}
\fmfv{decor.size=0, label=${\scs 1}$, l.dist=1mm, l.angle=0}{i1}
\end{fmfgraph*}
\end{center}}
\hspace*{3mm} \dphi{W^{({\rm int})}}{1}{2}
\no \\ &&
\hspace*{5mm} + \frac{1}{3} \hspace*{1mm}
\parbox{13mm}{\begin{center}
\begin{fmfgraph*}(10,6)
\setval
\fmfstraight
\fmfforce{0w,1h}{v1}
\fmfforce{0w,0h}{v2}
\fmfforce{4/10w,1/2h}{v3}
\fmfforce{8/10w,1/2h}{v4}
\fmfforce{1w,5/6h}{i1}
\fmfforce{1w,1/6h}{i2}
\fmf{plain}{v3,v1}
\fmf{plain}{v3,v2}
\fmf{plain}{v3,v4}
\fmf{plain}{v4,i1}
\fmf{plain}{v4,i2}
\fmfdot{v3,v4}
\fmfv{decor.shape=cross,decor.filled=shaded,decor.size=3thick}{v1}
\fmfv{decor.shape=cross,decor.filled=shaded,decor.size=3thick}{v2}
\fmfv{decor.size=0, label=${\scs 2}$, l.dist=1mm, l.angle=0}{i2}
\fmfv{decor.size=0, label=${\scs 1}$, l.dist=1mm, l.angle=0}{i1}
\end{fmfgraph*}
\end{center}}
\hspace*{3mm} \dphi{W^{({\rm int})}}{1}{2}
\hspace*{1mm} + \frac{1}{6} \hspace*{1mm}
\parbox{13mm}{\begin{center}
\begin{fmfgraph*}(10,6)
\setval
\fmfstraight
\fmfforce{0w,1h}{v1}
\fmfforce{0w,0h}{v2}
\fmfforce{4/10w,1/2h}{v3}
\fmfforce{8/10w,1/2h}{v4}
\fmfforce{1w,7/6h}{i1}
\fmfforce{1w,1/2h}{i2}
\fmfforce{1w,-1/6h}{i3}
\fmf{plain}{v3,v1}
\fmf{plain}{v3,v2}
\fmf{plain}{v3,v4}
\fmf{plain}{v4,i1}
\fmf{plain}{v4,i2}
\fmf{plain}{v4,i3}
\fmfdot{v3,v4}
\fmfv{decor.shape=cross,decor.filled=shaded,decor.size=3thick}{v1}
\fmfv{decor.shape=cross,decor.filled=shaded,decor.size=3thick}{v2}
\fmfv{decor.size=0, label=${\scs 3}$, l.dist=1mm, l.angle=0}{i3}
\fmfv{decor.size=0, label=${\scs 2}$, l.dist=1mm, l.angle=0}{i2}
\fmfv{decor.size=0, label=${\scs 1}$, l.dist=1mm, l.angle=0}{i1}
\end{fmfgraph*}
\end{center}}
\hspace*{3mm} \dvertex{W^{({\rm int})}}{1}{2}{3}
\hspace*{1mm} + \frac{1}{3} \hspace*{1mm}
\parbox{7mm}{\begin{center}
\begin{fmfgraph*}(3,9)
\setval
\fmfstraight
\fmfforce{1w,1h}{o1}
\fmfforce{1w,2/3h}{o2}
\fmfforce{0w,5/6h}{v1}
\fmfforce{0w,1/6h}{v2}
\fmfforce{1w,1/3h}{i1}
\fmfforce{1w,0h}{i2}
\fmf{plain}{v1,v2}
\fmf{plain}{v1,o1}
\fmf{plain}{v1,o2}
\fmf{plain}{v2,i1}
\fmf{plain}{v2,i2}
\fmfdot{v1,v2}
\fmfv{decor.size=0, label=${\scs 4}$, l.dist=1mm, l.angle=0}{i2}
\fmfv{decor.size=0, label=${\scs 3}$, l.dist=1mm, l.angle=0}{i1}
\fmfv{decor.size=0, label=${\scs 2}$, l.dist=1mm, l.angle=0}{o2}
\fmfv{decor.size=0, label=${\scs 1}$, l.dist=1mm, l.angle=0}{o1}
\end{fmfgraph*}
\end{center}}
\hspace*{3mm} \ddphi{W^{({\rm int})}}{1}{2}{3}{4} 
\hspace*{1mm} + \frac{1}{3} \hspace*{1mm}
\parbox{7mm}{\begin{center}
\begin{fmfgraph*}(3,12)
\setval
\fmfstraight
\fmfforce{1w,1h}{o0}
\fmfforce{1w,3/4h}{o1}
\fmfforce{1w,1/2h}{o2}
\fmfforce{0w,3/4h}{v1}
\fmfforce{0w,1/8h}{v2}
\fmfforce{1w,1/4h}{i1}
\fmfforce{1w,0h}{i2}
\fmf{plain}{v1,v2}
\fmf{plain}{v1,o0}
\fmf{plain}{v1,o1}
\fmf{plain}{v1,o2}
\fmf{plain}{v2,i1}
\fmf{plain}{v2,i2}
\fmfdot{v1,v2}
\fmfv{decor.size=0, label=${\scs 5}$, l.dist=1mm, l.angle=0}{i2}
\fmfv{decor.size=0, label=${\scs 4}$, l.dist=1mm, l.angle=0}{i1}
\fmfv{decor.size=0, label=${\scs 3}$, l.dist=1mm, l.angle=0}{o2}
\fmfv{decor.size=0, label=${\scs 2}$, l.dist=1mm, l.angle=0}{o1}
\fmfv{decor.size=0, label=${\scs 1}$, l.dist=1mm, l.angle=0}{o0}
\end{fmfgraph*}
\end{center}}
\hspace*{3mm} \ddvertex{W^{({\rm int})}}{1}{2}{3}{4}{5}
\nonumber \\
&& 
\hspace*{5mm}  + \frac{1}{3} \hspace*{1mm}
\hspace*{1mm} \dphi{W^{({\rm int})}}{1}{2}\hspace*{3mm}
\parbox{11mm}{\begin{center}
\begin{fmfgraph*}(8,4)
\setval
\fmfstraight
\fmfforce{0w,1h}{o1}
\fmfforce{0w,0h}{o2}
\fmfforce{1/4w,1/2h}{v1}
\fmfforce{3/4w,1/2h}{v2}
\fmfforce{1w,1h}{i1}
\fmfforce{1w,0h}{i2}
\fmf{plain}{v1,v2}
\fmf{plain}{v1,o1}
\fmf{plain}{v1,o2}
\fmf{plain}{v2,i1}
\fmf{plain}{v2,i2}
\fmfdot{v1,v2}
\fmfv{decor.size=0, label=${\scs 4}$, l.dist=1mm, l.angle=0}{i2}
\fmfv{decor.size=0, label=${\scs 3}$, l.dist=1mm, l.angle=0}{i1}
\fmfv{decor.size=0, label=${\scs 2}$, l.dist=1mm, l.angle=-180}{o2}
\fmfv{decor.size=0, label=${\scs 1}$, l.dist=1mm, l.angle=-180}{o1}
\end{fmfgraph*}
\end{center}}
\hspace*{3mm} \dphi{W^{({\rm int})}}{3}{4}
\hspace*{1mm} + \frac{1}{3} \hspace*{1mm}
\hspace*{1mm} \dvertex{W^{({\rm int})}}{1}{2}{3}\hspace*{3mm}
\parbox{11mm}{\begin{center}
\begin{fmfgraph*}(8,4)
\setval
\fmfstraight
\fmfforce{0w,1.5h}{o1}
\fmfforce{0w,0.5h}{o2}
\fmfforce{0w,-0.5h}{o3}
\fmfforce{1/4w,1/2h}{v1}
\fmfforce{3/4w,1/2h}{v2}
\fmfforce{1w,1h}{i1}
\fmfforce{1w,0h}{i2}
\fmf{plain}{v1,v2}
\fmf{plain}{v1,o1}
\fmf{plain}{v1,o2}
\fmf{plain}{v1,o3}
\fmf{plain}{v2,i1}
\fmf{plain}{v2,i2}
\fmfdot{v1,v2}
\fmfv{decor.size=0, label=${\scs 5}$, l.dist=1mm, l.angle=0}{i2}
\fmfv{decor.size=0, label=${\scs 4}$, l.dist=1mm, l.angle=0}{i1}
\fmfv{decor.size=0, label=${\scs 3}$, l.dist=1mm, l.angle=-180}{o3}
\fmfv{decor.size=0, label=${\scs 2}$, l.dist=1mm, l.angle=-180}{o2}
\fmfv{decor.size=0, label=${\scs 1}$, l.dist=1mm, l.angle=-180}{o1}
\end{fmfgraph*}
\end{center}}
\hspace*{3mm} \dphi{W^{({\rm int})}}{4}{5} \hspace*{4mm} \, .
\la{GRC2}
\eeq
The right-hand side consists of only 17 out of 25 terms from Eq.~(\r{GRC}),
7 without $W^{({\rm int})}$, 8 linear in $W^{({\rm int})}$ and 2
bilinear in  $W^{({\rm int})}$. Therefore the recursive
iteration of the graphical
relation (\r{GRC2}) is simpler than (\r{GRC}). To demonstrate 
this, we perform the decomposition (\r{DECC}) and restrict ourselves again
to the current-free connected vacuum diagrams $W^{(0,l)}$. For $n=0$ and
$l=2$ Eq.~(\r{GRC2}) reduces to 
\beq
\parbox{5mm}{\begin{center}
\begin{fmfgraph*}(2,4)
\setval
\fmfstraight
\fmfforce{0w,1/2h}{v1}
\fmfforce{1w,1/2h}{v2}
\fmfforce{1w,1.25h}{v3}
\fmfforce{1w,-0.25h}{v4}
\fmf{plain}{v1,v2}
\fmf{plain}{v1,v3}
\fmf{plain}{v1,v4}
\fmfv{decor.size=0, label=${\scs 2}$, l.dist=1mm, l.angle=0}{v2}
\fmfv{decor.size=0, label=${\scs 1}$, l.dist=1mm, l.angle=0}{v3}
\fmfv{decor.size=0, label=${\scs 3}$, l.dist=1mm, l.angle=0}{v4}
\fmfdot{v1}
\end{fmfgraph*}
\end{center}}
\hspace*{0.3cm} \dvertex{W^{(0,2)}}{1}{2}{3} = 
\hspace*{1mm} \frac{1}{6}\hspace*{1mm}
\parbox{7mm}{\begin{center}
\begin{fmfgraph*}(4,4)
\setval
\fmfforce{0w,0.5h}{v1}
\fmfforce{1w,0.5h}{v2}
\fmf{plain,left=1}{v1,v2,v1}
\fmf{plain}{v1,v2}
\fmfdot{v1,v2}
\end{fmfgraph*}\end{center}}
\hspace*{1mm}+ \frac{1}{4}\hspace*{1mm}
\parbox{15mm}{\begin{center}
\begin{fmfgraph*}(12,4)
\setval
\fmfforce{0w,0.5h}{v1}
\fmfforce{1/3w,0.5h}{v2}
\fmfforce{2/3w,0.5h}{v3}
\fmfforce{1w,0.5h}{v4}
\fmf{plain,left=1}{v1,v2,v1}
\fmf{plain,left=1}{v3,v4,v3}
\fmf{plain}{v2,v3}
\fmfdot{v2,v3}
\end{fmfgraph*}\end{center}} \hspace*{4mm} ,
\la{VGR1}
\eeq
whereas for $n=0$ and $l\ge3$ we obtain the graphical recursion relation
\end{fmffile}
\begin{fmffile}{fg6}
\beq
\parbox{5mm}{\begin{center}
\begin{fmfgraph*}(2,4)
\setval
\fmfstraight
\fmfforce{0w,1/2h}{v1}
\fmfforce{1w,1/2h}{v2}
\fmfforce{1w,1.25h}{v3}
\fmfforce{1w,-0.25h}{v4}
\fmf{plain}{v1,v2}
\fmf{plain}{v1,v3}
\fmf{plain}{v1,v4}
\fmfv{decor.size=0, label=${\scs 2}$, l.dist=1mm, l.angle=0}{v2}
\fmfv{decor.size=0, label=${\scs 1}$, l.dist=1mm, l.angle=0}{v3}
\fmfv{decor.size=0, label=${\scs 3}$, l.dist=1mm, l.angle=0}{v4}
\fmfdot{v1}
\end{fmfgraph*}
\end{center}}
\hspace*{0.3cm} \dvertex{W^{(0,l)}}{1}{2}{3} &=& 
\frac{2}{3} \hspace*{1mm}
\parbox{7mm}{\begin{center}
\begin{fmfgraph*}(6,4)
\setval
\fmfstraight
\fmfforce{1/3w,0h}{v1}
\fmfforce{1/3w,1h}{v2}
\fmfforce{1w,1h}{i2}
\fmfforce{1w,0h}{i1}
\fmf{plain}{i1,v1}
\fmf{plain}{v2,i2}
\fmf{plain,left}{v1,v2,v1}
\fmfdot{v2,v1}
\fmfv{decor.size=0, label=${\scs 2}$, l.dist=1mm, l.angle=0}{i1}
\fmfv{decor.size=0, label=${\scs 1}$, l.dist=1mm, l.angle=0}{i2}
\end{fmfgraph*}
\end{center}}
\hspace*{3mm} \dphi{W^{(0,l-1)}}{1}{2}
\hspace*{1mm} + \frac{2}{3} \hspace*{1mm}
\parbox{13mm}{\begin{center}
\begin{fmfgraph*}(10,4)
\setval
\fmfstraight
\fmfforce{0w,1/2h}{v1}
\fmfforce{4/10w,1/2h}{v2}
\fmfforce{8/10w,1/2h}{v3}
\fmfforce{1w,0h}{i2}
\fmfforce{1w,1h}{i1}
\fmf{plain,left}{v1,v2,v1}
\fmf{plain}{v3,v2}
\fmf{plain}{v3,i1}
\fmf{plain}{v3,i2}
\fmfdot{v2,v3}
\fmfv{decor.size=0, label=${\scs 2}$, l.dist=1mm, l.angle=0}{i2}
\fmfv{decor.size=0, label=${\scs 1}$, l.dist=1mm, l.angle=0}{i1}
\end{fmfgraph*}
\end{center}}
\hspace*{3mm} \dphi{W^{(0,l-1)}}{1}{2}
\hspace*{1mm} 
+ \frac{1}{2} \hspace*{1mm}
\parbox{13mm}{\begin{center}
\begin{fmfgraph*}(10,4)
\setval
\fmfstraight
\fmfforce{0w,1/2h}{v1}
\fmfforce{4/10w,1/2h}{v2}
\fmfforce{8/10w,1/2h}{v3}
\fmfforce{1w,-0.25h}{i1}
\fmfforce{1w,0.5h}{i2}
\fmfforce{1w,1.25h}{i3}
\fmf{plain,left}{v1,v2,v1}
\fmf{plain}{v3,v2}
\fmf{plain}{v3,i1}
\fmf{plain}{v3,i2}
\fmf{plain}{v3,i3}
\fmfdot{v2,v3}
\fmfv{decor.size=0, label=${\scs 1}$, l.dist=1mm, l.angle=0}{i3}
\fmfv{decor.size=0, label=${\scs 2}$, l.dist=1mm, l.angle=0}{i2}
\fmfv{decor.size=0, label=${\scs 3}$, l.dist=1mm, l.angle=0}{i1}
\end{fmfgraph*}
\end{center}}
\hspace*{3mm} \dvertex{W^{(0,l-1)}}{1}{2}{3} 
\no \\ &&
\hspace*{5mm}  + \frac{1}{3} \hspace*{1mm}
\parbox{7mm}{\begin{center}
\begin{fmfgraph*}(3,9)
\setval
\fmfstraight
\fmfforce{1w,1h}{o1}
\fmfforce{1w,2/3h}{o2}
\fmfforce{0w,5/6h}{v1}
\fmfforce{0w,1/6h}{v2}
\fmfforce{1w,1/3h}{i1}
\fmfforce{1w,0h}{i2}
\fmf{plain}{v1,v2}
\fmf{plain}{v1,o1}
\fmf{plain}{v1,o2}
\fmf{plain}{v2,i1}
\fmf{plain}{v2,i2}
\fmfdot{v1,v2}
\fmfv{decor.size=0, label=${\scs 4}$, l.dist=1mm, l.angle=0}{i2}
\fmfv{decor.size=0, label=${\scs 3}$, l.dist=1mm, l.angle=0}{i1}
\fmfv{decor.size=0, label=${\scs 2}$, l.dist=1mm, l.angle=0}{o2}
\fmfv{decor.size=0, label=${\scs 1}$, l.dist=1mm, l.angle=0}{o1}
\end{fmfgraph*}
\end{center}}
\hspace*{3mm} \ddphi{W^{(0,l-1)}}{1}{2}{3}{4} 
\hspace*{1mm} + \frac{1}{3} \hspace*{1mm}
\parbox{7mm}{\begin{center}
\begin{fmfgraph*}(3,12)
\setval
\fmfstraight
\fmfforce{1w,1h}{o0}
\fmfforce{1w,3/4h}{o1}
\fmfforce{1w,1/2h}{o2}
\fmfforce{0w,3/4h}{v1}
\fmfforce{0w,1/8h}{v2}
\fmfforce{1w,1/4h}{i1}
\fmfforce{1w,0h}{i2}
\fmf{plain}{v1,v2}
\fmf{plain}{v1,o0}
\fmf{plain}{v1,o1}
\fmf{plain}{v1,o2}
\fmf{plain}{v2,i1}
\fmf{plain}{v2,i2}
\fmfdot{v1,v2}
\fmfv{decor.size=0, label=${\scs 5}$, l.dist=1mm, l.angle=0}{i2}
\fmfv{decor.size=0, label=${\scs 4}$, l.dist=1mm, l.angle=0}{i1}
\fmfv{decor.size=0, label=${\scs 3}$, l.dist=1mm, l.angle=0}{o2}
\fmfv{decor.size=0, label=${\scs 2}$, l.dist=1mm, l.angle=0}{o1}
\fmfv{decor.size=0, label=${\scs 1}$, l.dist=1mm, l.angle=0}{o0}
\end{fmfgraph*}
\end{center}}
\hspace*{3mm} \ddvertex{W^{(0,l-1)}}{1}{2}{3}{4}{5}
\nonumber \\
&& 
\hspace*{5mm}  + \sum_{l'=2}^{l-2}\, \left\{ \frac{1}{3} 
\hspace*{1mm} \dphi{W^{(0,l')}}{1}{2}\hspace*{3mm}
\parbox{11mm}{\begin{center}
\begin{fmfgraph*}(8,4)
\setval
\fmfstraight
\fmfforce{0w,1h}{o1}
\fmfforce{0w,0h}{o2}
\fmfforce{1/4w,1/2h}{v1}
\fmfforce{3/4w,1/2h}{v2}
\fmfforce{1w,1h}{i1}
\fmfforce{1w,0h}{i2}
\fmf{plain}{v1,v2}
\fmf{plain}{v1,o1}
\fmf{plain}{v1,o2}
\fmf{plain}{v2,i1}
\fmf{plain}{v2,i2}
\fmfdot{v1,v2}
\fmfv{decor.size=0, label=${\scs 4}$, l.dist=1mm, l.angle=0}{i2}
\fmfv{decor.size=0, label=${\scs 3}$, l.dist=1mm, l.angle=0}{i1}
\fmfv{decor.size=0, label=${\scs 2}$, l.dist=1mm, l.angle=-180}{o2}
\fmfv{decor.size=0, label=${\scs 1}$, l.dist=1mm, l.angle=-180}{o1}
\end{fmfgraph*}
\end{center}}
\hspace*{3mm} \dphi{W^{(0,l-l'-1)}}{3}{4}
\hspace*{1mm} + \frac{1}{3} \hspace*{1mm}
\hspace*{1mm} \dvertex{W^{(0,l')}}{1}{2}{3}\hspace*{3mm}
\parbox{11mm}{\begin{center}
\begin{fmfgraph*}(8,4)
\setval
\fmfstraight
\fmfforce{0w,1.5h}{o1}
\fmfforce{0w,0.5h}{o2}
\fmfforce{0w,-0.5h}{o3}
\fmfforce{1/4w,1/2h}{v1}
\fmfforce{3/4w,1/2h}{v2}
\fmfforce{1w,1h}{i1}
\fmfforce{1w,0h}{i2}
\fmf{plain}{v1,v2}
\fmf{plain}{v1,o1}
\fmf{plain}{v1,o2}
\fmf{plain}{v1,o3}
\fmf{plain}{v2,i1}
\fmf{plain}{v2,i2}
\fmfdot{v1,v2}
\fmfv{decor.size=0, label=${\scs 5}$, l.dist=1mm, l.angle=0}{i2}
\fmfv{decor.size=0, label=${\scs 4}$, l.dist=1mm, l.angle=0}{i1}
\fmfv{decor.size=0, label=${\scs 3}$, l.dist=1mm, l.angle=-180}{o3}
\fmfv{decor.size=0, label=${\scs 2}$, l.dist=1mm, l.angle=-180}{o2}
\fmfv{decor.size=0, label=${\scs 1}$, l.dist=1mm, l.angle=-180}{o1}
\end{fmfgraph*}
\end{center}}
\hspace*{3mm} \dphi{W^{(0,l-l'-1)}}{4}{5} 
\right\} \hspace*{4mm} \, .
\la{VGR2}
\eeq
Integrating (\r{VGR1}) and (\r{VGR2}), we have to take into account as a
yet undetermined integration constant all those connected vacuum diagrams
$\tilde{W}^{(0,l)}$ which only consist of $4$-vertices. A comparison
with (\r{GNU}) shows that those diagrams follow from the graphical
recursion relation
\beq
\tilde{W}^{(0,l)} = \frac{1}{2(l-1)} \left\{ 
\hspace*{1mm} 
\parbox{9mm}{\begin{center}
\begin{fmfgraph*}(6,4)
\setval
\fmfstraight
\fmfforce{0w,1/2h}{v1}
\fmfforce{4/6w,1/2h}{v2}
\fmfforce{1w,1h}{i2}
\fmfforce{1w,0h}{i1}
\fmf{plain}{i1,v2}
\fmf{plain}{v2,i2}
\fmf{plain,left}{v1,v2,v1}
\fmfdot{v2}
\fmfv{decor.size=0, label=${\scs 2}$, l.dist=1mm, l.angle=0}{i1}
\fmfv{decor.size=0, label=${\scs 1}$, l.dist=1mm, l.angle=0}{i2}
\end{fmfgraph*}
\end{center}}
\hspace*{3mm} \dphi{\tilde{W}^{(0,l-1)}}{1}{2} 
\hspace*{1mm} + \frac{1}{3} \hspace*{1mm}
\parbox{7mm}{\begin{center}
\begin{fmfgraph*}(3,3)
\setval
\fmfstraight
\fmfforce{0w,1/2h}{v1}
\fmfforce{1w,2h}{i1}
\fmfforce{1w,1h}{i2}
\fmfforce{1w,0h}{i3}
\fmfforce{1w,-1h}{i4}
\fmf{plain}{v1,i1}
\fmf{plain}{v1,i2}
\fmf{plain}{v1,i3}
\fmf{plain}{v1,i4}
\fmfdot{v1}
\fmfv{decor.size=0, label=${\scs 4}$, l.dist=1mm, l.angle=0}{i4}
\fmfv{decor.size=0, label=${\scs 3}$, l.dist=1mm, l.angle=0}{i3}
\fmfv{decor.size=0, label=${\scs 2}$, l.dist=1mm, l.angle=0}{i2}
\fmfv{decor.size=0, label=${\scs 1}$, l.dist=1mm, l.angle=0}{i1}
\end{fmfgraph*}
\end{center}}
\hspace*{3mm} \ddphi{\tilde{W}^{(0,l-1)}}{1}{2}{3}{4}
\hspace*{1mm} +  \sum_{l'=2}^{l-2} \hspace*{1mm} \frac{1}{3} 
\hspace*{1mm} \dphi{\tilde{W}^{(0,l')}}{1}{2}\hspace*{3mm}
\parbox{7mm}{\begin{center}
\begin{fmfgraph*}(4,4)
\setval
\fmfstraight
\fmfforce{0w,1h}{o1}
\fmfforce{0w,0h}{o2}
\fmfforce{1/2w,1/2h}{v1}
\fmfforce{1w,1h}{i1}
\fmfforce{1w,0h}{i2}
\fmf{plain}{v1,o1}
\fmf{plain}{v1,o2}
\fmf{plain}{v1,i1}
\fmf{plain}{v1,i2}
\fmfdot{v1}
\fmfv{decor.size=0, label=${\scs 4}$, l.dist=1mm, l.angle=0}{i2}
\fmfv{decor.size=0, label=${\scs 3}$, l.dist=1mm, l.angle=0}{i1}
\fmfv{decor.size=0, label=${\scs 2}$, l.dist=1mm, l.angle=-180}{o2}
\fmfv{decor.size=0, label=${\scs 1}$, l.dist=1mm, l.angle=-180}{o1}
\end{fmfgraph*}
\end{center}}
\hspace*{3mm} \dphi{\tilde{W}^{(0,l-l'-1)}}{3}{4} \right\} \hspace*{1mm} ,
\la{TI}
\eeq
which is to be iterated starting from
\beq
\la{TIS}
\tilde{W}^{(0,2)} = 
\frac{1}{8}
\parbox{11mm}{\begin{center}
\begin{fmfgraph*}(8,4)
\setval
\fmfleft{i1}
\fmfright{o1}
\fmf{plain,left=1}{i1,v1,i1}
\fmf{plain,left=1}{o1,v1,o1}
\fmfdot{v1}
\end{fmfgraph*}\end{center}}
\eeq
and has been already discussed in \cite{PHI4}. In the first iteration
step we evaluate the amputation of one or two lines from (\r{TIS})
\beq
\dphi{\tilde{W}^{(0,2)}}{1}{2} = \frac{1}{4} \hspace*{2mm}
\parbox{9mm}{\begin{center}
\begin{fmfgraph*}(6,4)
\setval
\fmfstraight
\fmfforce{0w,1/2h}{v1}
\fmfforce{4/6w,1/2h}{v2}
\fmfforce{1w,1h}{i2}
\fmfforce{1w,0h}{i1}
\fmf{plain}{i1,v2}
\fmf{plain}{v2,i2}
\fmf{plain,left}{v1,v2,v1}
\fmfdot{v2}
\fmfv{decor.size=0, label=${\scs 2}$, l.dist=1mm, l.angle=0}{i1}
\fmfv{decor.size=0, label=${\scs 1}$, l.dist=1mm, l.angle=0}{i2}
\end{fmfgraph*}
\end{center}}
\hspace*{2mm} , \hspace*{1cm}
\ddphi{\tilde{W}^{(0,2)}}{1}{2}{3}{4} = \frac{1}{4} \hspace*{2mm}
\parbox{7mm}{\begin{center}
\begin{fmfgraph*}(4,4)
\setval
\fmfstraight
\fmfforce{0w,1h}{o1}
\fmfforce{0w,0h}{o2}
\fmfforce{1/2w,1/2h}{v1}
\fmfforce{1w,1h}{i1}
\fmfforce{1w,0h}{i2}
\fmf{plain}{v1,o1}
\fmf{plain}{v1,o2}
\fmf{plain}{v1,i1}
\fmf{plain}{v1,i2}
\fmfdot{v1}
\fmfv{decor.size=0, label=${\scs 4}$, l.dist=1mm, l.angle=0}{i2}
\fmfv{decor.size=0, label=${\scs 3}$, l.dist=1mm, l.angle=0}{i1}
\fmfv{decor.size=0, label=${\scs 2}$, l.dist=1mm, l.angle=-180}{o2}
\fmfv{decor.size=0, label=${\scs 1}$, l.dist=1mm, l.angle=-180}{o1}
\end{fmfgraph*}
\end{center}}
\eeq
and insert this into (\r{TI}) for $l=3$
\beq
\tilde{W}^{(0,3)} = 
\frac{1}{4} \left\{ 
\hspace*{1mm} 
\parbox{9mm}{\begin{center}
\begin{fmfgraph*}(6,4)
\setval
\fmfstraight
\fmfforce{0w,1/2h}{v1}
\fmfforce{4/6w,1/2h}{v2}
\fmfforce{1w,1h}{i2}
\fmfforce{1w,0h}{i1}
\fmf{plain}{i1,v2}
\fmf{plain}{v2,i2}
\fmf{plain,left}{v1,v2,v1}
\fmfdot{v2}
\fmfv{decor.size=0, label=${\scs 2}$, l.dist=1mm, l.angle=0}{i1}
\fmfv{decor.size=0, label=${\scs 1}$, l.dist=1mm, l.angle=0}{i2}
\end{fmfgraph*}
\end{center}}
\hspace*{3mm} \dphi{\tilde{W}^{(0,2)}}{1}{2} 
\hspace*{1mm} + \frac{1}{3} \hspace*{1mm}
\parbox{7mm}{\begin{center}
\begin{fmfgraph*}(3,3)
\setval
\fmfstraight
\fmfforce{0w,1/2h}{v1}
\fmfforce{1w,2h}{i1}
\fmfforce{1w,1h}{i2}
\fmfforce{1w,0h}{i3}
\fmfforce{1w,-1h}{i4}
\fmf{plain}{v1,i1}
\fmf{plain}{v1,i2}
\fmf{plain}{v1,i3}
\fmf{plain}{v1,i4}
\fmfdot{v1}
\fmfv{decor.size=0, label=${\scs 4}$, l.dist=1mm, l.angle=0}{i4}
\fmfv{decor.size=0, label=${\scs 3}$, l.dist=1mm, l.angle=0}{i3}
\fmfv{decor.size=0, label=${\scs 2}$, l.dist=1mm, l.angle=0}{i2}
\fmfv{decor.size=0, label=${\scs 1}$, l.dist=1mm, l.angle=0}{i1}
\end{fmfgraph*}
\end{center}}
\hspace*{3mm} \ddphi{\tilde{W}^{(0,2)}}{1}{2}{3}{4} \right\}
\eeq
to obtain the three-loop result
\beq
\la{TWOL}
\tilde{W}^{(0,3)} = \frac{1}{48}\hspace*{1mm}
\parbox{9mm}{\begin{center}
\begin{fmfgraph}(6,4)
\setval
\fmfforce{0w,0.5h}{v1}
\fmfforce{1w,0.5h}{v2}
\fmf{plain,left=1}{v1,v2,v1}
\fmf{plain,left=0.4}{v1,v2,v1}
\fmfdot{v1,v2}
\end{fmfgraph}\end{center}} 
\hspace*{1mm}+ \frac{1}{16}\hspace*{1mm}
\parbox{15mm}{\begin{center}
\begin{fmfgraph}(12,4)
\setval
\fmfleft{i1}
\fmfright{o1}
\fmf{plain,left=1}{i1,v1,i1}
\fmf{plain,left=1}{v1,v2,v1}
\fmf{plain,left=1}{o1,v2,o1}
\fmfdot{v1,v2}
\end{fmfgraph}\end{center}} \hspace*{4mm}.
\eeq
We note that the solution of (\r{TI}), (\r{TIS}) has already been determined
in Ref.~\cite{PHI4} up to the loop order $l=5$
and by a MATHEMATICA program
in Ref. \cite{CODE} up to the loop order $l=7$.
Thus we assume
from now on that the connected vacuum diagrams $\tilde{W}^{(0,l)}$ above
the critical point are known and construct with them the connected vacuum
diagrams $W^{(0,l)}$ below the critical point. Integrating (\r{VGR1}) 
with the integration constant (\r{TIS}) leads, indeed, to the correct 
two-loop result (\r{W2}). Subsequently we insert (\r{NNR1})--(\r{NNR4}) 
into (\r{VGR2}) with $l=3$ and obtain
\beq
\parbox{5mm}{\begin{center}
\begin{fmfgraph*}(2,4)
\setval
\fmfstraight
\fmfforce{0w,1/2h}{v1}
\fmfforce{1w,1/2h}{v2}
\fmfforce{1w,1.25h}{v3}
\fmfforce{1w,-0.25h}{v4}
\fmf{plain}{v1,v2}
\fmf{plain}{v1,v3}
\fmf{plain}{v1,v4}
\fmfv{decor.size=0, label=${\scs 2}$, l.dist=1mm, l.angle=0}{v2}
\fmfv{decor.size=0, label=${\scs 1}$, l.dist=1mm, l.angle=0}{v3}
\fmfv{decor.size=0, label=${\scs 3}$, l.dist=1mm, l.angle=0}{v4}
\fmfdot{v1}
\end{fmfgraph*}
\end{center}}
\hspace*{0.3cm} \dvertex{W^{(0,3)}}{1}{2}{3} &=& 
\hspace*{1mm} \frac{1}{6}
\parbox{9mm}{\begin{center}
\begin{fmfgraph}(4,4)
\setval
\fmfforce{0w,0h}{v1}
\fmfforce{1w,0h}{v2}
\fmfforce{1w,1h}{v3}
\fmfforce{0w,1h}{v4}
\fmf{plain,right=1}{v1,v3,v1}
\fmf{plain}{v1,v3}
\fmf{plain}{v2,v4}
\fmfdot{v1,v2,v3,v4}
\end{fmfgraph}\end{center}} 
\hspace*{1mm} + \frac{1}{4} \hspace*{1mm}
\parbox{9mm}{\begin{center}
\begin{fmfgraph}(4,4)
\setval
\fmfforce{0w,0h}{v1}
\fmfforce{1w,0h}{v2}
\fmfforce{1w,1h}{v3}
\fmfforce{0w,1h}{v4}
\fmf{plain,right=1}{v1,v3,v1}
\fmf{plain,right=0.4}{v1,v4}
\fmf{plain,left=0.4}{v2,v3}
\fmfdot{v1,v2,v3,v4}
\end{fmfgraph}\end{center}} 
\hspace*{1mm} + \frac{1}{2}
\parbox{15mm}{\begin{center}
\begin{fmfgraph}(12,4)
\setval
\fmfforce{0w,1/2h}{v1}
\fmfforce{1/3w,1/2h}{v2}
\fmfforce{2/3w,1/2h}{v3}
\fmfforce{5/6w,0h}{v4}
\fmfforce{5/6w,1h}{v5}
\fmfforce{1w,1/2h}{v6}
\fmf{plain,left=1}{v1,v2,v1}
\fmf{plain,left=1}{v3,v6,v3}
\fmf{plain}{v2,v3}
\fmf{plain}{v4,v5}
\fmfdot{v2,v3,v4,v5}
\end{fmfgraph}\end{center}} 
\hspace*{1mm} + \frac{1}{4} \hspace*{1mm}
\parbox{23mm}{\begin{center}
\begin{fmfgraph}(20,4)
\setval
\fmfforce{0w,1/2h}{v1}
\fmfforce{1/5w,1/2h}{v2}
\fmfforce{2/5w,1/2h}{v3}
\fmfforce{3/5w,1/2h}{v4}
\fmfforce{4/5w,1/2h}{v5}
\fmfforce{1w,1/2h}{v6}
\fmf{plain,left=1}{v1,v2,v1}
\fmf{plain,left=1}{v3,v4,v3}
\fmf{plain,left=1}{v5,v6,v5}
\fmf{plain}{v2,v3}
\fmf{plain}{v4,v5}
\fmfdot{v2,v3,v4,v5}
\end{fmfgraph}\end{center}} 
\hspace*{1mm} + \frac{1}{12} \hspace*{1mm} 
\parbox{17mm}{\begin{center}
\begin{fmfgraph}(13.856,12)
\setval
\fmfforce{0w,0h}{v1}
\fmfforce{1/4w,1/6h}{v2}
\fmfforce{1/2w,1/3h}{v3}
\fmfforce{3/4w,1/6h}{v4}
\fmfforce{1w,0h}{v5}
\fmfforce{1/2w,2/3h}{v6}
\fmfforce{1/2w,1h}{v7}
\fmf{plain,left=1}{v1,v2,v1}
\fmf{plain,left=1}{v4,v5,v4}
\fmf{plain,left=1}{v6,v7,v6}
\fmf{plain}{v2,v3}
\fmf{plain}{v4,v3}
\fmf{plain}{v3,v6}
\fmfdot{v2,v3,v4,v6}
\end{fmfgraph}\end{center}} 
\hspace*{1mm} + \frac{1}{4} \hspace*{1mm} 
\parbox{9mm}{\begin{center}
\begin{fmfgraph}(6,6)
\setval
\fmfforce{0w,1/2h}{v1}
\fmfforce{1w,1/2h}{v2}
\fmfforce{1/2w,1h}{v3}
\fmf{plain,left=1}{v1,v2,v1}
\fmf{plain,left=0.4}{v3,v1}
\fmf{plain,left=0.4}{v2,v3}
\fmfdot{v1,v2,v3}
\end{fmfgraph}\end{center}}
\nonumber \\
&& 
+ \frac{1}{4} \hspace*{1mm} 
\parbox{11mm}{\begin{center}
\begin{fmfgraph}(4,8)
\setval
\fmfforce{0w,1/4h}{v1}
\fmfforce{1w,1/4h}{v2}
\fmfforce{1/2w,1/2h}{v3}
\fmfforce{1/2w,1h}{v4}
\fmf{plain,left=1}{v1,v2,v1}
\fmf{plain,left=1}{v3,v4,v3}
\fmf{plain}{v1,v2}
\fmfdot{v2,v3,v1}
\end{fmfgraph}\end{center}} 
\hspace*{1mm} + \frac{1}{6} \hspace*{1mm} 
\parbox{15mm}{\begin{center}
\begin{fmfgraph}(12,4)
\setval
\fmfforce{0w,1/2h}{v1}
\fmfforce{1/3w,1/2h}{v2}
\fmfforce{2/3w,1/2h}{v3}
\fmfforce{1w,1/2h}{v4}
\fmf{plain,left=1}{v1,v2,v1}
\fmf{plain}{v2,v4}
\fmf{plain,left=1}{v3,v4,v3}
\fmfdot{v4,v2,v3}
\end{fmfgraph}\end{center}} 
\hspace*{1mm} + \frac{1}{4} \hspace*{1mm} 
\parbox{19mm}{\begin{center}
\begin{fmfgraph}(16,4)
\setval
\fmfforce{0w,1/2h}{v1}
\fmfforce{1/4w,1/2h}{v2}
\fmfforce{1/2w,1/2h}{v3}
\fmfforce{3/4w,1/2h}{v4}
\fmfforce{1w,1/2h}{v5}
\fmf{plain,left=1}{v1,v2,v1}
\fmf{plain,left=1}{v2,v3,v2}
\fmf{plain}{v3,v4}
\fmf{plain,left=1}{v4,v5,v4}
\fmfdot{v4,v2,v3}
\end{fmfgraph}\end{center}} 
\hspace*{1mm}+ \frac{1}{8} \hspace*{1mm}
\parbox{19mm}{\begin{center}
\begin{fmfgraph}(16,6)
\setval
\fmfforce{0w,1/3h}{v1}
\fmfforce{1/4w,1/3h}{v2}
\fmfforce{1/2w,1/3h}{v3}
\fmfforce{1/2w,1h}{v4}
\fmfforce{1w,1/3h}{v6}
\fmfforce{3/4w,1/3h}{v5}
\fmf{plain,left=1}{v1,v2,v1}
\fmf{plain}{v2,v5}
\fmf{plain,left=1}{v3,v4,v3}
\fmf{plain,left=1}{v5,v6,v5}
\fmfdot{v5,v2,v3}
\end{fmfgraph}\end{center}} \hspace*{2mm},
\eeq
Integrating this with the integration constant (\r{TWOL}), we rederive
the three-loop result listed in Table I.
\subsubsection{Counting $4$-Vertices}
The compatibility relation (\r{CP3}) between 
functional derivatives with respect to the kernel $G^{-1}$ 
and the $4$-vertex $L$ leads to
\beq
\int_{1234} L_{1234} \frac{\delta W}{\delta L_{1234}} = 
- \frac{1}{6} \int_{1234} L_{1234} \frac{\delta^2 W}{\delta G^{-1}_{12}
\delta G^{-1}_{34}} - \frac{1}{6}\int_{1234} L_{1234} 
\frac{\delta W}{\delta G^{-1}_{12}}
\frac{\delta W}{\delta G^{-1}_{34}} \, .
\eeq
Applying (\r{NR1}), (\r{NR2}) and (\r{DEC}), we obtain
a functional differential equation which is based on counting $4$-vertices
\beq
&&\int_{1234} L_{1234} \frac{\delta W^{({\rm int})}}{\delta L_{1234}}=
- \frac{1}{8} \int_{1234} L_{1234} G_{12} G_{34} 
- \frac{1}{4} \int_{123456} L_{1234} G_{12} G_{35} G_{46} J_5 J_6 
\no \\ && \hspace*{1.5cm}
- \frac{1}{24} \int_{12345678} L_{1234} G_{15} G_{26} G_{37} G_{48}
J_5 J_6 J_7 J_8 
- \frac{1}{2} \int_{123456} L_{1234} G_{12} G_{35} G_{46}
\frac{\delta W^{({\rm int})}}{\delta G_{56}}
\no \\ && \hspace*{1.5cm}
- \frac{1}{6} \int_{12345678} L_{1234} G_{15} G_{26} G_{37} G_{48} J_5 
J_6  \frac{\delta W^{({\rm int})}}{\delta G_{78}}
- \frac{1}{6} \int_{12345678} L_{1234} G_{15} G_{26} G_{37} G_{48}
\frac{\delta^2 W^{({\rm int})}}{\delta G_{56} \delta G_{78}} 
\no \\ &&\hspace*{1.5cm}
- \frac{1}{6} \int_{12345678} L_{1234} G_{15} G_{26} G_{37} G_{48}
\frac{\delta W^{({\rm int})}}{\delta G_{56}}
\frac{\delta W^{({\rm int})}}{\delta G_{78}} \, .
\eeq
Its graphical representation reads
\beq
\parbox{7mm}{\begin{center}
\begin{fmfgraph*}(3,3)
\setval
\fmfstraight
\fmfforce{0w,1/2h}{v1}
\fmfforce{1w,2h}{i1}
\fmfforce{1w,1h}{i2}
\fmfforce{1w,0h}{i3}
\fmfforce{1w,-1h}{i4}
\fmf{plain}{v1,i1}
\fmf{plain}{v1,i2}
\fmf{plain}{v1,i3}
\fmf{plain}{v1,i4}
\fmfdot{v1}
\fmfv{decor.size=0, label=${\scs 4}$, l.dist=1mm, l.angle=0}{i4}
\fmfv{decor.size=0, label=${\scs 3}$, l.dist=1mm, l.angle=0}{i3}
\fmfv{decor.size=0, label=${\scs 2}$, l.dist=1mm, l.angle=0}{i2}
\fmfv{decor.size=0, label=${\scs 1}$, l.dist=1mm, l.angle=0}{i1}
\end{fmfgraph*}
\end{center}}
\hspace*{2mm} \dvertexx{ W^{({\rm int})}}{2}{3}{4}{1}\hspace*{1mm} &= &
\hspace*{1mm} \frac{1}{8}\hspace*{1mm}
\parbox{11mm}{\begin{center}
\begin{fmfgraph*}(8,4)
\setval
\fmfleft{i1}
\fmfright{o1}
\fmf{plain,left=1}{i1,v1,i1}
\fmf{plain,left=1}{o1,v1,o1}
\fmfdot{v1}
\end{fmfgraph*}\end{center}}
\hspace*{1mm}+ \frac{1}{4}\hspace*{1mm}
\parbox{11mm}{\begin{center}
\begin{fmfgraph*}(8,4)
\setval
\fmfforce{0w,0h}{v1}
\fmfforce{1/2w,0h}{v2}
\fmfforce{1/2w,1h}{v3}
\fmfforce{1w,0h}{w1}
\fmf{plain,left=1}{v2,v3,v2}
\fmf{plain}{v1,w1}
\fmfdot{v2}
\fmfv{decor.shape=cross,decor.filled=shaded,decor.size=3thick}{v1}
\fmfv{decor.shape=cross,decor.filled=shaded,decor.size=3thick}{w1}
\end{fmfgraph*}\end{center}}
\hspace*{1mm} + \frac{1}{24}\hspace*{1mm}
\parbox{11mm}{\begin{center}
\begin{fmfgraph*}(8,12)
\setval
\fmfforce{0w,1/2h}{v1}
\fmfforce{1w,1/2h}{w1}
\fmfforce{1/2w,1/6h}{u1}
\fmfforce{1/2w,5/6h}{x1}
\fmfforce{1/2w,1/2h}{v2}
\fmf{plain}{w1,v1}
\fmf{plain}{x1,u1}
\fmfdot{v2}
\fmfv{decor.shape=cross,decor.filled=shaded,decor.size=3thick}{u1}
\fmfv{decor.shape=cross,decor.filled=shaded,decor.size=3thick}{v1}
\fmfv{decor.shape=cross,decor.filled=shaded,decor.size=3thick}{w1}
\fmfv{decor.shape=cross,decor.filled=shaded,decor.size=3thick}{x1}
\end{fmfgraph*}\end{center}}
\hspace*{1mm}+ \frac{1}{2} \hspace*{1mm} 
\parbox{9mm}{\begin{center}
\begin{fmfgraph*}(6,4)
\setval
\fmfstraight
\fmfforce{0w,1/2h}{v1}
\fmfforce{4/6w,1/2h}{v2}
\fmfforce{1w,1h}{i2}
\fmfforce{1w,0h}{i1}
\fmf{plain}{i1,v2}
\fmf{plain}{v2,i2}
\fmf{plain,left}{v1,v2,v1}
\fmfdot{v2}
\fmfv{decor.size=0, label=${\scs 2}$, l.dist=1mm, l.angle=0}{i1}
\fmfv{decor.size=0, label=${\scs 1}$, l.dist=1mm, l.angle=0}{i2}
\end{fmfgraph*}
\end{center}}
\hspace*{3mm} \dphi{W^{({\rm int})}}{1}{2} 
\hspace*{1mm} + \frac{1}{6} \hspace*{1mm}
\parbox{9mm}{\begin{center}
\begin{fmfgraph*}(6,6)
\setval
\fmfstraight
\fmfforce{0w,1h}{v1}
\fmfforce{0w,0h}{v2}
\fmfforce{4/6w,1/2h}{v3}
\fmfforce{1w,5/6h}{i1}
\fmfforce{1w,1/6h}{i2}
\fmf{plain}{v3,v1}
\fmf{plain}{v3,v2}
\fmf{plain}{v3,i1}
\fmf{plain}{v3,i2}
\fmfdot{v3}
\fmfv{decor.shape=cross,decor.filled=shaded,decor.size=3thick}{v1}
\fmfv{decor.shape=cross,decor.filled=shaded,decor.size=3thick}{v2}
\fmfv{decor.size=0, label=${\scs 2}$, l.dist=1mm, l.angle=0}{i2}
\fmfv{decor.size=0, label=${\scs 1}$, l.dist=1mm, l.angle=0}{i1}
\end{fmfgraph*}
\end{center}}
\hspace*{3mm} \dphi{W^{({\rm int})}}{1}{2}
\no \\ 
&& 
+ \frac{1}{6} \hspace*{1mm}
\parbox{7mm}{\begin{center}
\begin{fmfgraph*}(3,3)
\setval
\fmfstraight
\fmfforce{0w,1/2h}{v1}
\fmfforce{1w,2h}{i1}
\fmfforce{1w,1h}{i2}
\fmfforce{1w,0h}{i3}
\fmfforce{1w,-1h}{i4}
\fmf{plain}{v1,i1}
\fmf{plain}{v1,i2}
\fmf{plain}{v1,i3}
\fmf{plain}{v1,i4}
\fmfdot{v1}
\fmfv{decor.size=0, label=${\scs 4}$, l.dist=1mm, l.angle=0}{i4}
\fmfv{decor.size=0, label=${\scs 3}$, l.dist=1mm, l.angle=0}{i3}
\fmfv{decor.size=0, label=${\scs 2}$, l.dist=1mm, l.angle=0}{i2}
\fmfv{decor.size=0, label=${\scs 1}$, l.dist=1mm, l.angle=0}{i1}
\end{fmfgraph*}
\end{center}}
\hspace*{3mm} \ddphi{W^{({\rm int})}}{1}{2}{3}{4}
\hspace*{1mm} + \frac{1}{6} \hspace*{1mm}
\hspace*{1mm} \dphi{W^{({\rm int})}}{1}{2}\hspace*{3mm}
\parbox{7mm}{\begin{center}
\begin{fmfgraph*}(4,4)
\setval
\fmfstraight
\fmfforce{0w,1h}{o1}
\fmfforce{0w,0h}{o2}
\fmfforce{1/2w,1/2h}{v1}
\fmfforce{1w,1h}{i1}
\fmfforce{1w,0h}{i2}
\fmf{plain}{v1,o1}
\fmf{plain}{v1,o2}
\fmf{plain}{v1,i1}
\fmf{plain}{v1,i2}
\fmfdot{v1}
\fmfv{decor.size=0, label=${\scs 4}$, l.dist=1mm, l.angle=0}{i2}
\fmfv{decor.size=0, label=${\scs 3}$, l.dist=1mm, l.angle=0}{i1}
\fmfv{decor.size=0, label=${\scs 2}$, l.dist=1mm, l.angle=-180}{o2}
\fmfv{decor.size=0, label=${\scs 1}$, l.dist=1mm, l.angle=-180}{o1}
\end{fmfgraph*}
\end{center}}
\hspace*{3mm} \dphi{W^{({\rm int})}}{3}{4}  
\hspace*{4mm} \, . \la{GRC4}
\eeq
The right-hand side consists of $7$ terms in Eq.~(\r{GRC}) which 
are missing in (\r{GRC2}), 3 without $W^{({\rm int})}$, 3 linear
in  $W^{({\rm int})}$ and 1 bilinear in $W^{({\rm int})}$. Iterating
the graphical relation (\r{GRC4}) for the current-free
connected vacuum diagrams $W^{(0,l)}$ is even simpler than (\r{GRC2}).
For $n=0$ and $l=2$ Eq.~(\ref{GRC}) reduces to
\beq
\parbox{7mm}{\begin{center}
\begin{fmfgraph*}(3,3)
\setval
\fmfstraight
\fmfforce{0w,1/2h}{v1}
\fmfforce{1w,2h}{i1}
\fmfforce{1w,1h}{i2}
\fmfforce{1w,0h}{i3}
\fmfforce{1w,-1h}{i4}
\fmf{plain}{v1,i1}
\fmf{plain}{v1,i2}
\fmf{plain}{v1,i3}
\fmf{plain}{v1,i4}
\fmfdot{v1}
\fmfv{decor.size=0, label=${\scs 4}$, l.dist=1mm, l.angle=0}{i4}
\fmfv{decor.size=0, label=${\scs 3}$, l.dist=1mm, l.angle=0}{i3}
\fmfv{decor.size=0, label=${\scs 2}$, l.dist=1mm, l.angle=0}{i2}
\fmfv{decor.size=0, label=${\scs 1}$, l.dist=1mm, l.angle=0}{i1}
\end{fmfgraph*}
\end{center}}
\hspace*{2mm} \dvertexx{ W^{(0,2)}}{2}{3}{4}{1}\hspace*{1mm} = 
\hspace*{1mm} \frac{1}{8}\hspace*{1mm}
\parbox{11mm}{\begin{center}
\begin{fmfgraph*}(8,4)
\setval
\fmfleft{i1}
\fmfright{o1}
\fmf{plain,left=1}{i1,v1,i1}
\fmf{plain,left=1}{o1,v1,o1}
\fmfdot{v1}
\end{fmfgraph*}\end{center}} \la{GV1} \, ,
\eeq
whereas for $n=0$ and $l \ge 3$ we obtain the graphical recursion relation
\beq
\parbox{7mm}{\begin{center}
\begin{fmfgraph*}(3,3)
\setval
\fmfstraight
\fmfforce{0w,1/2h}{v1}
\fmfforce{1w,2h}{i1}
\fmfforce{1w,1h}{i2}
\fmfforce{1w,0h}{i3}
\fmfforce{1w,-1h}{i4}
\fmf{plain}{v1,i1}
\fmf{plain}{v1,i2}
\fmf{plain}{v1,i3}
\fmf{plain}{v1,i4}
\fmfdot{v1}
\fmfv{decor.size=0, label=${\scs 4}$, l.dist=1mm, l.angle=0}{i4}
\fmfv{decor.size=0, label=${\scs 3}$, l.dist=1mm, l.angle=0}{i3}
\fmfv{decor.size=0, label=${\scs 2}$, l.dist=1mm, l.angle=0}{i2}
\fmfv{decor.size=0, label=${\scs 1}$, l.dist=1mm, l.angle=0}{i1}
\end{fmfgraph*}
\end{center}}
\hspace*{2mm} \dvertexx{ W^{(0,l)}}{2}{3}{4}{1}\hspace*{1mm} = \hspace*{1mm}
\frac{1}{2} \hspace*{1mm} 
\parbox{9mm}{\begin{center}
\begin{fmfgraph*}(6,4)
\setval
\fmfstraight
\fmfforce{0w,1/2h}{v1}
\fmfforce{4/6w,1/2h}{v2}
\fmfforce{1w,1h}{i2}
\fmfforce{1w,0h}{i1}
\fmf{plain}{i1,v2}
\fmf{plain}{v2,i2}
\fmf{plain,left}{v1,v2,v1}
\fmfdot{v2}
\fmfv{decor.size=0, label=${\scs 2}$, l.dist=1mm, l.angle=0}{i1}
\fmfv{decor.size=0, label=${\scs 1}$, l.dist=1mm, l.angle=0}{i2}
\end{fmfgraph*}
\end{center}}
\hspace*{3mm} \dphi{W^{(0,l-1)}}{1}{2} 
\hspace*{1mm}  + \frac{1}{6} \hspace*{1mm}
\parbox{7mm}{\begin{center}
\begin{fmfgraph*}(3,3)
\setval
\fmfstraight
\fmfforce{0w,1/2h}{v1}
\fmfforce{1w,2h}{i1}
\fmfforce{1w,1h}{i2}
\fmfforce{1w,0h}{i3}
\fmfforce{1w,-1h}{i4}
\fmf{plain}{v1,i1}
\fmf{plain}{v1,i2}
\fmf{plain}{v1,i3}
\fmf{plain}{v1,i4}
\fmfdot{v1}
\fmfv{decor.size=0, label=${\scs 4}$, l.dist=1mm, l.angle=0}{i4}
\fmfv{decor.size=0, label=${\scs 3}$, l.dist=1mm, l.angle=0}{i3}
\fmfv{decor.size=0, label=${\scs 2}$, l.dist=1mm, l.angle=0}{i2}
\fmfv{decor.size=0, label=${\scs 1}$, l.dist=1mm, l.angle=0}{i1}
\end{fmfgraph*}
\end{center}}
\hspace*{3mm} \ddphi{W^{(0,l-1)}}{1}{2}{3}{4}
\hspace*{1mm} + \sum_{l'=2}^{l-2} \frac{1}{6} \hspace*{1mm}
\hspace*{1mm} \dphi{W^{(0,l')}}{1}{2}\hspace*{3mm}
\parbox{7mm}{\begin{center}
\begin{fmfgraph*}(4,4)
\setval
\fmfstraight
\fmfforce{0w,1h}{o1}
\fmfforce{0w,0h}{o2}
\fmfforce{1/2w,1/2h}{v1}
\fmfforce{1w,1h}{i1}
\fmfforce{1w,0h}{i2}
\fmf{plain}{v1,o1}
\fmf{plain}{v1,o2}
\fmf{plain}{v1,i1}
\fmf{plain}{v1,i2}
\fmfdot{v1}
\fmfv{decor.size=0, label=${\scs 4}$, l.dist=1mm, l.angle=0}{i2}
\fmfv{decor.size=0, label=${\scs 3}$, l.dist=1mm, l.angle=0}{i1}
\fmfv{decor.size=0, label=${\scs 2}$, l.dist=1mm, l.angle=-180}{o2}
\fmfv{decor.size=0, label=${\scs 1}$, l.dist=1mm, l.angle=-180}{o1}
\end{fmfgraph*}
\end{center}}
\hspace*{3mm} \dphi{W^{(0,l-l'-1)}}{3}{4} \, . \la{GV2}
\eeq
Integrating (\r{GV1}) and (\r{GV2}), we have to take into account
as a yet undetermined integration constant all those connected vacuum
diagrams $\tilde{W}^{(0,l)}$ which only consist of $3$-vertices. A comparison
with (\r{GNU}) shows that those diagrams follow from the graphical 
recursion relation
\beq
\tilde{W}^{(0,l)}\hspace*{1mm}& =&\hspace*{1mm} \frac{1}{3(l-1)} \, \left\{ 
\parbox{7mm}{\begin{center}
\begin{fmfgraph*}(6,4)
\setval
\fmfstraight
\fmfforce{1/3w,0h}{v1}
\fmfforce{1/3w,1h}{v2}
\fmfforce{1w,1h}{i2}
\fmfforce{1w,0h}{i1}
\fmf{plain}{i1,v1}
\fmf{plain}{v2,i2}
\fmf{plain,left}{v1,v2,v1}
\fmfdot{v2,v1}
\fmfv{decor.size=0, label=${\scs 2}$, l.dist=1mm, l.angle=0}{i1}
\fmfv{decor.size=0, label=${\scs 1}$, l.dist=1mm, l.angle=0}{i2}
\end{fmfgraph*}
\end{center}}
\hspace*{3mm} \dphi{\tilde{W}^{(0,l-1)}}{1}{2}
\hspace*{1mm} + \hspace*{1mm}
\parbox{13mm}{\begin{center}
\begin{fmfgraph*}(10,4)
\setval
\fmfstraight
\fmfforce{0w,1/2h}{v1}
\fmfforce{4/10w,1/2h}{v2}
\fmfforce{8/10w,1/2h}{v3}
\fmfforce{1w,0h}{i2}
\fmfforce{1w,1h}{i1}
\fmf{plain,left}{v1,v2,v1}
\fmf{plain}{v3,v2}
\fmf{plain}{v3,i1}
\fmf{plain}{v3,i2}
\fmfdot{v2,v3}
\fmfv{decor.size=0, label=${\scs 2}$, l.dist=1mm, l.angle=0}{i2}
\fmfv{decor.size=0, label=${\scs 1}$, l.dist=1mm, l.angle=0}{i1}
\end{fmfgraph*}
\end{center}}
\hspace*{3mm} \dphi{\tilde{W}^{(0,l-1)}}{1}{2} \right. \no \\
&& \left.
+ \frac{1}{2} \hspace*{1mm}
\parbox{7mm}{\begin{center}
\begin{fmfgraph*}(3,9)
\setval
\fmfstraight
\fmfforce{1w,1h}{o1}
\fmfforce{1w,2/3h}{o2}
\fmfforce{0w,5/6h}{v1}
\fmfforce{0w,1/6h}{v2}
\fmfforce{1w,1/3h}{i1}
\fmfforce{1w,0h}{i2}
\fmf{plain}{v1,v2}
\fmf{plain}{v1,o1}
\fmf{plain}{v1,o2}
\fmf{plain}{v2,i1}
\fmf{plain}{v2,i2}
\fmfdot{v1,v2}
\fmfv{decor.size=0, label=${\scs 4}$, l.dist=1mm, l.angle=0}{i2}
\fmfv{decor.size=0, label=${\scs 3}$, l.dist=1mm, l.angle=0}{i1}
\fmfv{decor.size=0, label=${\scs 2}$, l.dist=1mm, l.angle=0}{o2}
\fmfv{decor.size=0, label=${\scs 1}$, l.dist=1mm, l.angle=0}{o1}
\end{fmfgraph*}
\end{center}}
\hspace*{3mm} \ddphi{\tilde{W}^{(0,l-1)}}{1}{2}{3}{4} 
\hspace*{1mm} + \sum_{l'=2}^{l-2} \frac{1}{2} \hspace*{1mm}
\hspace*{1mm} \dphi{\tilde{W}^{(0,l')}}{1}{2}\hspace*{3mm}
\parbox{11mm}{\begin{center}
\begin{fmfgraph*}(8,4)
\setval
\fmfstraight
\fmfforce{0w,1h}{o1}
\fmfforce{0w,0h}{o2}
\fmfforce{1/4w,1/2h}{v1}
\fmfforce{3/4w,1/2h}{v2}
\fmfforce{1w,1h}{i1}
\fmfforce{1w,0h}{i2}
\fmf{plain}{v1,v2}
\fmf{plain}{v1,o1}
\fmf{plain}{v1,o2}
\fmf{plain}{v2,i1}
\fmf{plain}{v2,i2}
\fmfdot{v1,v2}
\fmfv{decor.size=0, label=${\scs 4}$, l.dist=1mm, l.angle=0}{i2}
\fmfv{decor.size=0, label=${\scs 3}$, l.dist=1mm, l.angle=0}{i1}
\fmfv{decor.size=0, label=${\scs 2}$, l.dist=1mm, l.angle=-180}{o2}
\fmfv{decor.size=0, label=${\scs 1}$, l.dist=1mm, l.angle=-180}{o1}
\end{fmfgraph*}
\end{center}}
\hspace*{3mm} \dphi{\tilde{W}^{(0,l-l'-1)}}{3}{4} \right\} \hspace*{4mm},
\la{YTY}
\eeq
which is to be iterated starting from
\beq
\tilde{W}^{(0,2)}\hspace*{1mm} =  \hspace*{1mm}\frac{1}{12}\hspace*{1mm}
\parbox{7mm}{\begin{center}
\begin{fmfgraph*}(4,4)
\setval
\fmfforce{0w,0.5h}{v1}
\fmfforce{1w,0.5h}{v2}
\fmf{plain,left=1}{v1,v2,v1}
\fmf{plain}{v1,v2}
\fmfdot{v1,v2}
\end{fmfgraph*}\end{center}}
\hspace*{1mm}+ \frac{1}{8}\hspace*{1mm}
\parbox{15mm}{\begin{center}
\begin{fmfgraph*}(12,4)
\setval
\fmfforce{0w,0.5h}{v1}
\fmfforce{1/3w,0.5h}{v2}
\fmfforce{2/3w,0.5h}{v3}
\fmfforce{1w,0.5h}{v4}
\fmf{plain,left=1}{v1,v2,v1}
\fmf{plain,left=1}{v3,v4,v3}
\fmf{plain}{v2,v3}
\fmfdot{v2,v3}
\end{fmfgraph*}\end{center}}  \hspace*{4mm} . \la{S}
\eeq
In the first iteration step we have to evaluate the amputation of one
or two lines from (\r{S})
\beq
\la{NNNR1}
\dphi{\tilde{W}^{(0,2)}}{1}{2} &=&  \frac{1}{4} \hspace*{1mm}
\parbox{9mm}{\begin{center}
\begin{fmfgraph*}(6,4)
\setval
\fmfstraight
\fmfforce{1/3w,0h}{v1}
\fmfforce{1/3w,1h}{v2}
\fmfforce{1w,1h}{i2}
\fmfforce{1w,0h}{i1}
\fmf{plain}{i1,v1}
\fmf{plain}{v2,i2}
\fmf{plain,left}{v1,v2,v1}
\fmfdot{v2,v1}
\fmfv{decor.size=0, label=${\scs 2}$, l.dist=1mm, l.angle=0}{i1}
\fmfv{decor.size=0, label=${\scs 1}$, l.dist=1mm, l.angle=0}{i2}
\end{fmfgraph*}
\end{center}}
\hspace*{1mm} + \frac{1}{4} \hspace*{1mm}
\parbox{13mm}{\begin{center}
\begin{fmfgraph*}(10,4)
\setval
\fmfstraight
\fmfforce{0w,1/2h}{v1}
\fmfforce{4/10w,1/2h}{v2}
\fmfforce{8/10w,1/2h}{v3}
\fmfforce{1w,0h}{i2}
\fmfforce{1w,1h}{i1}
\fmf{plain,left}{v1,v2,v1}
\fmf{plain}{v3,v2}
\fmf{plain}{v3,i1}
\fmf{plain}{v3,i2}
\fmfdot{v2,v3}
\fmfv{decor.size=0, label=${\scs 2}$, l.dist=1mm, l.angle=0}{i2}
\fmfv{decor.size=0, label=${\scs 1}$, l.dist=1mm, l.angle=0}{i1}
\end{fmfgraph*}
\end{center}}
\hspace*{1mm} + \frac{1}{8} \hspace*{1mm}
\parbox{9mm}{\begin{center}
\begin{fmfgraph*}(6,10)
\setval
\fmfstraight
\fmfforce{0w,2/10h}{v1}
\fmfforce{4/6w,2/10h}{v2}
\fmfforce{1w,2/10h}{i1}
\fmfforce{0w,8/10h}{v3}
\fmfforce{4/6w,8/10h}{v4}
\fmfforce{1w,8/10h}{i2}
\fmf{plain,left}{v1,v2,v1}
\fmf{plain,left}{v3,v4,v3}
\fmf{plain}{v4,i2}
\fmf{plain}{v2,i1}
\fmfdot{v4,v2}
\fmfv{decor.size=0, label=${\scs 2}$, l.dist=1mm, l.angle=0}{i1}
\fmfv{decor.size=0, label=${\scs 1}$, l.dist=1mm, l.angle=0}{i2}
\end{fmfgraph*}
\end{center}} \hspace*{4mm}  \no \\
\la{NNNR2}
\ddphi{\tilde{W}^{(0,2)}}{1}{2}{3}{4} &=&  \frac{1}{4} \hspace*{2mm}
{\displaystyle \left\{ \hspace*{2mm}
\parbox{11mm}{\begin{center}
\begin{fmfgraph*}(8,4)
\setval
\fmfstraight
\fmfforce{0w,1h}{o1}
\fmfforce{0w,0h}{o2}
\fmfforce{1/4w,1/2h}{v1}
\fmfforce{3/4w,1/2h}{v2}
\fmfforce{1w,1h}{i1}
\fmfforce{1w,0h}{i2}
\fmf{plain}{v1,v2}
\fmf{plain}{v1,o1}
\fmf{plain}{v1,o2}
\fmf{plain}{v2,i1}
\fmf{plain}{v2,i2}
\fmfdot{v1,v2}
\fmfv{decor.size=0, label=${\scs 4}$, l.dist=1mm, l.angle=0}{i2}
\fmfv{decor.size=0, label=${\scs 3}$, l.dist=1mm, l.angle=0}{i1}
\fmfv{decor.size=0, label=${\scs 2}$, l.dist=1mm, l.angle=-180}{o2}
\fmfv{decor.size=0, label=${\scs 1}$, l.dist=1mm, l.angle=-180}{o1}
\end{fmfgraph*}
\end{center}}
\hspace*{2mm}+ 2 \, \mbox{perm.} \right\} }
\hspace*{2mm}+ \hspace*{2mm}\frac{1}{8}\hspace*{1mm} 
{\displaystyle \left\{
\parbox{9mm}{\centerline{
\begin{fmfgraph*}(5,12.33)
\setval
\fmfforce{1w,0h}{v1}
\fmfforce{0w,0h}{v2}
\fmfforce{0.5w,4.33/12.33h}{v3}
\fmfforce{0.5w,1.25/12.33h}{vm}
\fmfforce{-1/5w,10.33/12.33h}{v4}
\fmfforce{3/5w,10.33/12.33h}{v5}
\fmfforce{1w,10.33/12.33h}{v6}
\fmf{plain}{v1,vm,v2}
\fmf{plain,left=1}{v4,v5,v4}
\fmf{plain}{v3,vm}
\fmf{plain}{v5,v6}
\fmfv{decor.size=0,label={\footnotesize 3},l.dist=0.5mm, l.angle=-30}{v1}
\fmfv{decor.size=0,label={\footnotesize 4},l.dist=0.5mm, l.angle=-150}{v2}
\fmfv{decor.size=0,label={\footnotesize 2},l.dist=0.5mm, l.angle=90}{v3}
\fmfv{decor.size=0,label={\footnotesize 1},l.dist=0.5mm, l.angle=0}{v6}
\fmfdot{vm,v5}
\end{fmfgraph*}}}
+ \, 3 \, \mbox{perm.} \right\} } \hspace*{4mm} , 
\eeq
and insert this into (\r{YTY}) for $l=3$
\beq
\tilde{W}^{(0,3)}\hspace*{1mm} =\hspace*{1mm} \frac{1}{6} \, 
{\displaystyle \left\{ 
\parbox{7mm}{\begin{center}
\begin{fmfgraph*}(6,4)
\setval
\fmfstraight
\fmfforce{1/3w,0h}{v1}
\fmfforce{1/3w,1h}{v2}
\fmfforce{1w,1h}{i2}
\fmfforce{1w,0h}{i1}
\fmf{plain}{i1,v1}
\fmf{plain}{v2,i2}
\fmf{plain,left}{v1,v2,v1}
\fmfdot{v2,v1}
\fmfv{decor.size=0, label=${\scs 2}$, l.dist=1mm, l.angle=0}{i1}
\fmfv{decor.size=0, label=${\scs 1}$, l.dist=1mm, l.angle=0}{i2}
\end{fmfgraph*}
\end{center}}
\hspace*{3mm} \dphi{\tilde{W}^{(0,2)}}{1}{2}
\hspace*{1mm} + \hspace*{1mm}
\parbox{13mm}{\begin{center}
\begin{fmfgraph*}(10,4)
\setval
\fmfstraight
\fmfforce{0w,1/2h}{v1}
\fmfforce{4/10w,1/2h}{v2}
\fmfforce{8/10w,1/2h}{v3}
\fmfforce{1w,0h}{i2}
\fmfforce{1w,1h}{i1}
\fmf{plain,left}{v1,v2,v1}
\fmf{plain}{v3,v2}
\fmf{plain}{v3,i1}
\fmf{plain}{v3,i2}
\fmfdot{v2,v3}
\fmfv{decor.size=0, label=${\scs 2}$, l.dist=1mm, l.angle=0}{i2}
\fmfv{decor.size=0, label=${\scs 1}$, l.dist=1mm, l.angle=0}{i1}
\end{fmfgraph*}
\end{center}}
\hspace*{3mm} \dphi{\tilde{W}^{(0,2)}}{1}{2} 
+ \frac{1}{2} \hspace*{1mm}
\parbox{7mm}{\begin{center}
\begin{fmfgraph*}(3,9)
\setval
\fmfstraight
\fmfforce{1w,1h}{o1}
\fmfforce{1w,2/3h}{o2}
\fmfforce{0w,5/6h}{v1}
\fmfforce{0w,1/6h}{v2}
\fmfforce{1w,1/3h}{i1}
\fmfforce{1w,0h}{i2}
\fmf{plain}{v1,v2}
\fmf{plain}{v1,o1}
\fmf{plain}{v1,o2}
\fmf{plain}{v2,i1}
\fmf{plain}{v2,i2}
\fmfdot{v1,v2}
\fmfv{decor.size=0, label=${\scs 4}$, l.dist=1mm, l.angle=0}{i2}
\fmfv{decor.size=0, label=${\scs 3}$, l.dist=1mm, l.angle=0}{i1}
\fmfv{decor.size=0, label=${\scs 2}$, l.dist=1mm, l.angle=0}{o2}
\fmfv{decor.size=0, label=${\scs 1}$, l.dist=1mm, l.angle=0}{o1}
\end{fmfgraph*}
\end{center}}
\hspace*{3mm} \ddphi{\tilde{W}^{(0,2)}}{1}{2}{3}{4} \right\} }\hspace*{4mm}  
\la{YT}
\eeq
to obtain
\beq
\tilde{W}^{(0,3)} = \frac{1}{24}
\parbox{9mm}{\begin{center}
\begin{fmfgraph}(4,4)
\setval
\fmfforce{0w,0h}{v1}
\fmfforce{1w,0h}{v2}
\fmfforce{1w,1h}{v3}
\fmfforce{0w,1h}{v4}
\fmf{plain,right=1}{v1,v3,v1}
\fmf{plain}{v1,v3}
\fmf{plain}{v2,v4}
\fmfdot{v1,v2,v3,v4}
\end{fmfgraph}\end{center}} 
\hspace*{1mm} + \frac{1}{16} \hspace*{1mm}
\parbox{9mm}{\begin{center}
\begin{fmfgraph}(4,4)
\setval
\fmfforce{0w,0h}{v1}
\fmfforce{1w,0h}{v2}
\fmfforce{1w,1h}{v3}
\fmfforce{0w,1h}{v4}
\fmf{plain,right=1}{v1,v3,v1}
\fmf{plain,right=0.4}{v1,v4}
\fmf{plain,left=0.4}{v2,v3}
\fmfdot{v1,v2,v3,v4}
\end{fmfgraph}\end{center}} 
\hspace*{1mm} + \frac{1}{8}
\parbox{15mm}{\begin{center}
\begin{fmfgraph}(12,4)
\setval
\fmfforce{0w,1/2h}{v1}
\fmfforce{1/3w,1/2h}{v2}
\fmfforce{2/3w,1/2h}{v3}
\fmfforce{5/6w,0h}{v4}
\fmfforce{5/6w,1h}{v5}
\fmfforce{1w,1/2h}{v6}
\fmf{plain,left=1}{v1,v2,v1}
\fmf{plain,left=1}{v3,v6,v3}
\fmf{plain}{v2,v3}
\fmf{plain}{v4,v5}
\fmfdot{v2,v3,v4,v5}
\end{fmfgraph}\end{center}} 
\hspace*{1mm} + \frac{1}{16} \hspace*{1mm}
\parbox{23mm}{\begin{center}
\begin{fmfgraph}(20,4)
\setval
\fmfforce{0w,1/2h}{v1}
\fmfforce{1/5w,1/2h}{v2}
\fmfforce{2/5w,1/2h}{v3}
\fmfforce{3/5w,1/2h}{v4}
\fmfforce{4/5w,1/2h}{v5}
\fmfforce{1w,1/2h}{v6}
\fmf{plain,left=1}{v1,v2,v1}
\fmf{plain,left=1}{v3,v4,v3}
\fmf{plain,left=1}{v5,v6,v5}
\fmf{plain}{v2,v3}
\fmf{plain}{v4,v5}
\fmfdot{v2,v3,v4,v5}
\end{fmfgraph}\end{center}} 
\hspace*{1mm} + \frac{1}{48} \hspace*{1mm} 
\parbox{17mm}{\begin{center}
\begin{fmfgraph}(13.856,12)
\setval
\fmfforce{0w,0h}{v1}
\fmfforce{1/4w,1/6h}{v2}
\fmfforce{1/2w,1/3h}{v3}
\fmfforce{3/4w,1/6h}{v4}
\fmfforce{1w,0h}{v5}
\fmfforce{1/2w,2/3h}{v6}
\fmfforce{1/2w,1h}{v7}
\fmf{plain,left=1}{v1,v2,v1}
\fmf{plain,left=1}{v4,v5,v4}
\fmf{plain,left=1}{v6,v7,v6}
\fmf{plain}{v2,v3}
\fmf{plain}{v4,v3}
\fmf{plain}{v3,v6}
\fmfdot{v2,v3,v4,v6}
\end{fmfgraph}\end{center}} \hspace*{4mm} .\la{ITC}
\eeq
Integrating (\r{GV1}) with the integration constant (\r{S}) leads,
indeed, to the correct two-loop result (\r{W2}). Subsequently we  
insert (\r{NNR1})--(\r{NNR4}) into (\r{GV2}) with $l=3$ and obtain
\beq
\parbox{7mm}{\begin{center}
\begin{fmfgraph*}(3,3)
\setval
\fmfstraight
\fmfforce{0w,1/2h}{v1}
\fmfforce{1w,2h}{i1}
\fmfforce{1w,1h}{i2}
\fmfforce{1w,0h}{i3}
\fmfforce{1w,-1h}{i4}
\fmf{plain}{v1,i1}
\fmf{plain}{v1,i2}
\fmf{plain}{v1,i3}
\fmf{plain}{v1,i4}
\fmfdot{v1}
\fmfv{decor.size=0, label=${\scs 4}$, l.dist=1mm, l.angle=0}{i4}
\fmfv{decor.size=0, label=${\scs 3}$, l.dist=1mm, l.angle=0}{i3}
\fmfv{decor.size=0, label=${\scs 2}$, l.dist=1mm, l.angle=0}{i2}
\fmfv{decor.size=0, label=${\scs 1}$, l.dist=1mm, l.angle=0}{i1}
\end{fmfgraph*}
\end{center}}
\hspace*{2mm} \dvertexx{ W^{(0,3)}}{2}{3}{4}{1} & = &
\frac{1}{24}\hspace*{1mm}
\parbox{9mm}{\begin{center}
\begin{fmfgraph}(6,4)
\setval
\fmfforce{0w,0.5h}{v1}
\fmfforce{1w,0.5h}{v2}
\fmf{plain,left=1}{v1,v2,v1}
\fmf{plain,left=0.4}{v1,v2,v1}
\fmfdot{v1,v2}
\end{fmfgraph}\end{center}} 
\hspace*{1mm}+ \frac{1}{8}\hspace*{1mm}
\parbox{15mm}{\begin{center}
\begin{fmfgraph}(12,4)
\setval
\fmfleft{i1}
\fmfright{o1}
\fmf{plain,left=1}{i1,v1,i1}
\fmf{plain,left=1}{v1,v2,v1}
\fmf{plain,left=1}{o1,v2,o1}
\fmfdot{v1,v2}
\end{fmfgraph}\end{center}} \no \\
&& +\frac{1}{8} \hspace*{1mm} 
\parbox{9mm}{\begin{center}
\begin{fmfgraph}(6,6)
\setval
\fmfforce{0w,1/2h}{v1}
\fmfforce{1w,1/2h}{v2}
\fmfforce{1/2w,1h}{v3}
\fmf{plain,left=1}{v1,v2,v1}
\fmf{plain,left=0.4}{v3,v1}
\fmf{plain,left=0.4}{v2,v3}
\fmfdot{v1,v2,v3}
\end{fmfgraph}\end{center}}
\hspace*{1mm}+ \frac{1}{8} \hspace*{1mm} 
\parbox{11mm}{\begin{center}
\begin{fmfgraph}(4,8)
\setval
\fmfforce{0w,1/4h}{v1}
\fmfforce{1w,1/4h}{v2}
\fmfforce{1/2w,1/2h}{v3}
\fmfforce{1/2w,1h}{v4}
\fmf{plain,left=1}{v1,v2,v1}
\fmf{plain,left=1}{v3,v4,v3}
\fmf{plain}{v1,v2}
\fmfdot{v2,v3,v1}
\end{fmfgraph}\end{center}} 
\hspace*{1mm} + \frac{1}{12} \hspace*{1mm} 
\parbox{15mm}{\begin{center}
\begin{fmfgraph}(12,4)
\setval
\fmfforce{0w,1/2h}{v1}
\fmfforce{1/3w,1/2h}{v2}
\fmfforce{2/3w,1/2h}{v3}
\fmfforce{1w,1/2h}{v4}
\fmf{plain,left=1}{v1,v2,v1}
\fmf{plain}{v2,v4}
\fmf{plain,left=1}{v3,v4,v3}
\fmfdot{v4,v2,v3}
\end{fmfgraph}\end{center}} 
\hspace*{1mm} + \frac{1}{8} \hspace*{1mm} 
\parbox{19mm}{\begin{center}
\begin{fmfgraph}(16,4)
\setval
\fmfforce{0w,1/2h}{v1}
\fmfforce{1/4w,1/2h}{v2}
\fmfforce{1/2w,1/2h}{v3}
\fmfforce{3/4w,1/2h}{v4}
\fmfforce{1w,1/2h}{v5}
\fmf{plain,left=1}{v1,v2,v1}
\fmf{plain,left=1}{v2,v3,v2}
\fmf{plain}{v3,v4}
\fmf{plain,left=1}{v4,v5,v4}
\fmfdot{v4,v2,v3}
\end{fmfgraph}\end{center}} 
\hspace*{1mm}+ \frac{1}{16} \hspace*{1mm}
\parbox{19mm}{\begin{center}
\begin{fmfgraph}(16,6)
\setval
\fmfforce{0w,1/3h}{v1}
\fmfforce{1/4w,1/3h}{v2}
\fmfforce{1/2w,1/3h}{v3}
\fmfforce{1/2w,1h}{v4}
\fmfforce{1w,1/3h}{v6}
\fmfforce{3/4w,1/3h}{v5}
\fmf{plain,left=1}{v1,v2,v1}
\fmf{plain}{v2,v5}
\fmf{plain,left=1}{v3,v4,v3}
\fmf{plain,left=1}{v5,v6,v5}
\fmfdot{v5,v2,v3}
\end{fmfgraph}\end{center}} \hspace*{2mm}.
\eeq
Integrating this with the integration constant (\r{ITC}), we rederive
the three-loop result listed in Table I.
\subsubsection{Cross-Check}
The four graphical relations (\r{GRC}), (\r{CC1}), (\r{GRC2}) 
(\r{GRC4}) for the interaction negative free energy $W^{({\rm int})}$
are not independent from each other. Indeed, going back to the linear
functional differential equation (\r{CHECK1}) for $W$, we obtain together
with (\r{NR1}) and (\r{DEC}) the following topological identity for the number
of currents, lines, $3$- and $4$-vertices of each connected vacuum diagram:
\beq
\int_1 J_1 \frac{\delta W^{({\rm int})}}{\delta J_1} -2 \int_{12}
G_{12} \frac{\delta W^{({\rm int})}}{\delta G^{-1}_{12}} + 3
\int_{123} K_{123}\frac{\delta W^{({\rm int})}}{\delta K_{123}}
+ 4 \int_{1234} L_{1234}\frac{\delta W^{({\rm int})}}{\delta L_{1234}}
= 0 \, .
\eeq
Its graphical representation reads
\beq
\parbox{8mm}{\begin{center}
\begin{fmfgraph*}(5,5)
\setval
\fmfstraight
\fmfforce{0w,1/2h}{v1}
\fmfforce{1w,1/2h}{v2}
\fmf{plain}{v1,v2}
\fmfv{decor.shape=cross,decor.filled=shaded,decor.size=3thick}{v1}
\fmfv{decor.size=0, label=${\scs 1}$, l.dist=1mm, l.angle=0}{v2}
\end{fmfgraph*}
\end{center}}
\hspace*{2mm} \cdphi{W^{({\rm int})}}{1} \hspace*{1mm} -2 \hspace*{1mm}
\parbox{5.5mm}{\begin{center}
\begin{fmfgraph*}(2.5,5)
\setval
\fmfstraight
\fmfforce{1w,0h}{v1}
\fmfforce{1w,1h}{v2}
\fmf{plain,left=1}{v1,v2}
\fmfv{decor.size=0, label=${\scs 2}$, l.dist=1mm, l.angle=0}{v1}
\fmfv{decor.size=0, label=${\scs 1}$, l.dist=1mm, l.angle=0}{v2}
\end{fmfgraph*}
\end{center}}
\hspace*{0.3cm} \dphi{W^{({\rm int})}}{1}{2} \hspace*{1mm} +3 \hspace*{1mm}
\parbox{5mm}{\begin{center}
\begin{fmfgraph*}(2,4)
\setval
\fmfstraight
\fmfforce{0w,1/2h}{v1}
\fmfforce{1w,1/2h}{v2}
\fmfforce{1w,1.25h}{v3}
\fmfforce{1w,-0.25h}{v4}
\fmf{plain}{v1,v2}
\fmf{plain}{v1,v3}
\fmf{plain}{v1,v4}
\fmfv{decor.size=0, label=${\scs 2}$, l.dist=1mm, l.angle=0}{v2}
\fmfv{decor.size=0, label=${\scs 1}$, l.dist=1mm, l.angle=0}{v3}
\fmfv{decor.size=0, label=${\scs 3}$, l.dist=1mm, l.angle=0}{v4}
\fmfdot{v1}
\end{fmfgraph*}
\end{center}}
\hspace*{0.3cm} \dvertex{W^{({\rm int})}}{1}{2}{3}\hspace*{1mm} +4 
\hspace*{1mm}
\parbox{7mm}{\begin{center}
\begin{fmfgraph*}(3,3)
\setval
\fmfstraight
\fmfforce{0w,1/2h}{v1}
\fmfforce{1w,2h}{i1}
\fmfforce{1w,1h}{i2}
\fmfforce{1w,0h}{i3}
\fmfforce{1w,-1h}{i4}
\fmf{plain}{v1,i1}
\fmf{plain}{v1,i2}
\fmf{plain}{v1,i3}
\fmf{plain}{v1,i4}
\fmfdot{v1}
\fmfv{decor.size=0, label=${\scs 4}$, l.dist=1mm, l.angle=0}{i4}
\fmfv{decor.size=0, label=${\scs 3}$, l.dist=1mm, l.angle=0}{i3}
\fmfv{decor.size=0, label=${\scs 2}$, l.dist=1mm, l.angle=0}{i2}
\fmfv{decor.size=0, label=${\scs 1}$, l.dist=1mm, l.angle=0}{i1}
\end{fmfgraph*}
\end{center}}
\hspace*{2mm} \dvertexx{ W^{({\rm int})}}{2}{3}{4}{1}\hspace*{1mm} = 0
\, . \la{CHECK2}
\eeq
Inserting (\r{GRC}), (\r{CC1}), (\r{GRC2}), (\r{GRC4}) into (\r{CHECK2}),
we can check this relation term by term.
\end{fmffile}
\begin{fmffile}{fg7}
\section{Effective Energy}
In field theory one is often interested in the functional Legendre transform
of the negative free energy $W[J,G^{-1},K,L]$ with respect to the
current $J$ \cite{Verena,Amit,Zuber,Zinn}. 
To this end one introduces the new field
\beq
\la{LT1}
\Phi_1 [J,G^{-1},K,L] = \left. \frac{\delta W[J,G^{-1},K,L]}{\delta J_1}
\right|_{G^{-1},K,L} \, ,
\eeq
which implicitly defines $J$ as a functional of $\Phi$:
\beq
\la{LT2}
J_1 = J_1 [\Phi,G^{-1},K,L] \, .
\eeq
From Eq.~(\r{CR1}) we read off that $\Phi$ coincides with the $1$-point 
function of the theory, i.e. the field expectation value of the
fluctuations around the background field $\chi$ in Eq.~(\r{PH2})
in the presence of the current $J$. The functional 
Legendre transform of the negative free energy  $W[J,G^{-1},K,L]$
with respect to the current $J$
results in the effective energy
\beq
\Gamma [ \Phi , G^{-1}, K,L] & = & \int_1 J_1[ \Phi , G^{-1}, K,L] \,
\left. \frac{\delta  W\left[J[\Phi,G^{-1},K,L],G^{-1},K,L\right]}{\delta
J_1[ \Phi , G^{-1}, K,L]} \right|_{G^{-1},K,L} \no \\
& & - W\left[J[\Phi,G^{-1},K,L],G^{-1},K,L\right] \, ,
\eeq
which simplifies due to (\r{LT1}):
\beq
\la{EE}
\Gamma [ \Phi , G^{-1}, K,L] = \int_1 J_1[ \Phi , G^{-1}, K,L] \Phi_1
- W\left[J[\Phi,G^{-1},K,L],G^{-1},K,L\right] \, .
\eeq
Taking into account the functional chain rule, it leads to  
the equation of state 
\beq
\la{SE}
\left. \frac{\delta \Gamma [ \Phi , G^{-1}, K,L]}{\delta \Phi_1} 
\right|_{G^{-1}, K,L} = J_1 [\Phi,G^{-1},K,L] \, . 
\eeq
Performing a loop expansion, the respective contributions to the effective
energy (\r{EE}) may be displayed as one-particle irreducible vacuum diagrams
which are 
constructed according to the Feynman rules (\r{ES}), (\r{PRO}), (\r{3V}), 
(\r{4V}). In addition a dot with a wiggled line represents an integral over
the field expectation value
\beq
\parbox{10mm}{\begin{center}
\begin{fmfgraph*}(5,5)
\setval
\fmfforce{0w,1/2h}{v1}
\fmfforce{1w,1/2h}{v2}
\fmf{boson}{v1,v2}
\fmfdot{v1}
\end{fmfgraph*}
\end{center}}
\equiv \hspace*{0.1cm} \int_{1} \Phi_{1} \,\, . \la{FEV}
\eeq
For instance, if the cubic and the quartic interactions $K$ and $L$ vanish,
the zeroth order contribution to the
negative free energy (\r{W0}) leads  with (\r{LT1}) to the field expectation
value
\beq
\Phi_1^{(0)} [J,G^{-1},0,0] = \int_2 G_{12} J_2
\eeq
which is inverted as
\beq
J_1^{(0)} [\Phi,G^{-1},0,0] = \int_2 G_{12}^{-1} \Phi_2
\eeq
to result in the zeroth order contribution to the effective energy
\beq
\la{FEF}
\Gamma^{(0)} [ \Phi , G^{-1}, 0,0] = E [ 0 ] + 
\frac{1}{2} \mbox{Tr} \ln G^{-1} + \frac{1}{2} \int_{12} G_{12}^{-1}
\Phi_1 \Phi_2 \, .
\eeq
Its graphical representation reads by definition 
\beq
- \Gamma^{(0)} = 
\parbox{8mm}{\begin{center}
\begin{fmfgraph*}(5,5)
\setval
\fmfstraight
\fmfforce{1/2w,1/2h}{v1}
\fmfdot{v1}
\end{fmfgraph*}
\end{center}}
+ \frac{1}{2}\;
\parbox{8mm}{\centerline{
\begin{fmfgraph}(5,5)
\setval
\fmfi{plain}{reverse fullcircle scaled 1w shifted (0.5w,0.5h)}
\end{fmfgraph} }} 
+ \frac{1}{2} \hspace*{0.1cm}
\parbox{8mm}{\begin{center}
\begin{fmfgraph*}(5,5)
\setval
\fmfstraight
\fmfforce{0w,1/2h}{v1}
\fmfforce{1w,1/2h}{v2}
\fmf{boson}{v1,v2}
\fmfdot{v1,v2}
\end{fmfgraph*}
\end{center}} \, .
\eeq
In order to find the one-particle irreducible vacuum diagrams of the
effective energy together with their weights for  
non-vanishing cubic and quartic interactions $K$ and $L$, we proceed
as follows. We start in Subsection III.A with investigating the 
consequences of the functional Legendre transform with respect to the current
for functional derivatives and their compatibility relations.
With these result we derive in Subsection III.B a single nonlinear
functional differential equation for the effective energy which 
is converted into a recursion relation in Subsection III.C and graphically
solved in Subsection III.D.
In Subsection III.E we derive even simpler graphical recursion relations
for certain subsets of one-particle irreducible vacuum diagrams.
\subsection{Functional Legendre Transform}
In order to investigate in detail the field-theoretic consequences of the
functional Legendre transform, we
start with the effective energy $\Gamma [ \Phi , G^{-1}, K,L]$
and introduce the current $J$ via the equation of state (\r{SE}).
As this implicitly defines the field expectation value as a functional
of the current, i.e.
\beq
\Phi_1 = \Phi_1 [J,G^{-1},K,L] \, ,
\eeq
the negative free energy is recovered according to  
\beq
\la{FEE}
W [ J , G^{-1}, K,L] = \int_1 J_1 \Phi_1[ J , G^{-1}, K,L] 
- \Gamma \left[\Phi[J,G^{-1},K,L],G^{-1},K,L\right] \, .
\eeq
With this we derive useful relations between the functional
derivatives of the negative free energy $W$ and the effective energy
$\Gamma$, respectively.
\subsubsection{Functional Derivatives}
Taking into account the functional chain rule, the first
functional derivatives of the negative free energy $W$ read
(\r{LT1}) and
\beq
\left.\frac{\delta W[J,G^{-1},K,L]}{\delta G^{-1}_{12}}\right|_{J,K,L} &= & -
\left.
\frac{\delta \Gamma \left[\Phi[J,G^{-1},K,L],G^{-1},K,L\right]}{\delta
G^{-1}_{12}} \right|_{\Phi,K,L} \, , \la{D1}\\
\left.\frac{\delta W[J,G^{-1},K,L]}{\delta K_{123}}\right|_{J,G^{-1},L} &= & -
\left.
\frac{\delta \Gamma \left[\Phi[J,G^{-1},K,L],G^{-1},K,L\right]}{\delta
K_{123}} \right|_{\Phi,G^{-1},L} \, , \la{D2}\\
\left.\frac{\delta W[J,G^{-1},K,L]}{\delta L_{1234}}\right|_{J,G^{-1},K} 
&= & -
\left.
\frac{\delta \Gamma \left[\Phi[J,G^{-1},K,L],G^{-1},K,L\right]}{\delta
L_{1234}} \right|_{\Phi,G^{-1},K} \, . \la{D2b}
\eeq
To evaluate second functional derivatives of the 
negative free energy $W$ is 
more involved. At first we observe
\beq
\left.\frac{\delta^2 W[J,G^{-1},K,L]}{\delta J_2 \delta J_1} 
\right|_{G^{-1},K,L} = \left.\frac{\delta \Phi_1[J,G^{-1},K,L]}{\delta J_2} 
\right|_{G^{-1},K,L} \hspace*{3cm}
\nonumber \\ = \left( \left.
\frac{\delta J_2[\Phi[J,G^{-1},K,L], G^{-1},K,L]}{\delta \Phi_1[J,G^{-1},K,L]}
\right|_{G^{-1},K,L} \right)^{-1} 
= \left( \left.\frac{\delta^2 \Gamma 
\left[\Phi[J,G^{-1},K,L],G^{-1},K,L\right]}{\delta \Phi_1 [J,G^{-1},K,L]
\delta \Phi_2 [J,G^{-1},K,L]}
\right|_{G^{-1},K,L} \right)^{-1} \, ,
\la{SCH}
\eeq
where we used (\r{LT1}), (\r{SE}) and the fact that the derivative of
a functional equals the inverse of the derivative of the inverse functional.
To precise the meaning of relation (\r{SCH}), we rederive it from
another point of view. To this end we consider the functional identity
\beq
\left. \frac{\delta J_1 \left[ \Phi [ J , G^{-1}, 
K, L],G^{-1}, K, L\right]}{\delta J_2} \right|_{G^{-1}, K, L} = \delta_{12}
\eeq
and apply the functional chain rule  together with  (\r{LT1}), (\r{SE}),
so that we result in
\beq
\int_3 
\left.\frac{\delta^2 \Gamma 
\left[\Phi[J,G^{-1},K,L],G^{-1},K,L\right]}{\delta \Phi_1 [J,G^{-1},K,L]
\delta \Phi_3[J,G^{-1},K,L]}
\right|_{G^{-1},K,L} \, 
\left.\frac{\delta^2 W[J,G^{-1},K,L]}{\delta J_3 \delta J_2} 
\right|_{G^{-1},K,L} = \delta_{12} \, . \la{FID}
\eeq
Furthermore we obtain from (\r{D1}) by applying again the functional
chain rule and relation (\r{SCH})
\beq
&& \frac{\delta}{\delta J_3} \left( \left.
\frac{\delta W[J,G^{-1},K,L]}{\delta G^{-1}_{12}}\right|_{J,K,L} 
\right)_{G^{-1},K,L} 
= - \int_4 \left( \left.
\frac{\delta^2 \Gamma \left[\Phi[J,G^{-1},K,L],G^{-1},
K,L\right]}{\delta \Phi_3 [J,G^{-1},K,L]\delta \Phi_4[J,G^{-1},K,L]}
\right|_{G^{-1},K,L} \right)^{-1}\no\\
&& \hspace*{2cm}\times\,\, \frac{\delta}{\delta\Phi_4[J,G^{-1},K,L]}
\left( \left.
\frac{\delta \Gamma \left[\Phi[J,G^{-1},K,L],G^{-1},K,L
\right]]}{\delta G^{-1}_{12}} 
\right|_{\Phi,K,L} \right)_{G^{-1},K,L}
\la{D3}
\eeq
and, correspondingly,
\beq
\left.
\frac{\delta^2 W[J,G^{-1},K,L]}{\delta G^{-1}_{34}\delta G^{-1}_{12}}
\right|_{J,K,L} &= &- \left.
\frac{\delta^2 \Gamma
\left[\Phi[J,G^{-1},
K,L],G^{-1},K,L\right]}{\delta G^{-1}_{34} \delta G^{-1}_{12}} 
\right|_{\Phi,K,L}  \no \\
&&+ \int_{56}
\left( \left.
\frac{\delta^2 \Gamma \left[\Phi[J,G^{-1},K,L],G^{-1},
K,L\right]}{\delta \Phi_5[J,G^{-1},K,L]\delta \Phi_6[J,G^{-1},K,L]}
\right|_{G^{-1},K,L} \right)^{-1} \no \\
&&\times\,\, \frac{\delta}{\delta\Phi_5[J,G^{-1},K,L]}
\left( \left.
\frac{\delta \Gamma \left[\Phi[J,G^{-1},K,L],G^{-1},K,L\right]}{\delta
G^{-1}_{12}} 
\right|_{\Phi,K,L} \right)_{G^{-1},K,L}\no \\
&&\times\,\, \frac{\delta}{\delta\Phi_6[J,G^{-1},K,L]}
\left( \left.
\frac{\delta \Gamma \left[\Phi[J,G^{-1},K,L],G^{-1},K,L\right]}{\delta
G^{-1}_{34}} 
\right|_{\Phi,K,L} \right)_{G^{-1},K,L} \hspace*{3mm} . \la{D4}
\eeq
\subsubsection{Compatibility Relations}
Performing the functional Legendre transform with respect to the current, the
compatibility relation (\r{CP1}) between functional derivatives with
respect to the current $J$ and the kernel $G^{-1}$, we result in
\beq
\la{D5}
\left( \left.
\frac{\delta^2 \Gamma [\Phi,G^{-1},K,L]}{\delta \Phi_2 \delta \Phi_1}
\right|_{G^{-1},K,L} \right)^{-1} = 2 \, \left. \frac{\delta 
\Gamma [\Phi,G^{-1},K,L]}{\delta G^{-1}_{12}} 
\right|_{\Phi,K,L} - \Phi_1 \Phi_2
\eeq 
due to (\r{LT1}) (\r{D1}) and (\r{SCH}). 
The compatibility relation 
(\r{CP2}) between functional derivatives with
respect to the current $J$,  the kernel $G^{-1}$
and the $3$-vertex $K$ is converted
by using (\r{LT1}), (\r{D1}), (\r{D2}) and (\r{D3}) to 
\beq
&& \int_4 \frac{\delta}{\delta \Phi_4} \left( \left.
\frac{\delta\Gamma [\Phi,G^{-1},K,L]}{\delta G^{-1}_{12}} \right|_{\Phi,K,L}  
\right)_{G^{-1},K,L} 
\left( \left.
\frac{\delta^2 \Gamma [\Phi,G^{-1},K,L]}{\delta \Phi_3 \delta \Phi_4}
\right|_{G^{-1},K,L} \right)^{-1}
\no \\
&& \hspace*{2cm} = 3 \, \left.
\frac{\delta \Gamma [\Phi,G^{-1},K,L]}{\delta K_{123}}\right|_{\Phi,G^{-1},L} 
- \left.
\frac{\delta \Gamma [\Phi,G^{-1},K,L]}{\delta G^{-1}_{12}} 
\right|_{\Phi,K,L}\,\Phi_3 \, .
\la{D6}
\eeq
Furthermore we perform the functional Legendre transform of the
compatibility relation (\r{CP3}) between functional derivatives with
respect to the kernel $G^{-1}$ and the $4$-vertex $L$
which leads with (\r{D1}), (\r{D2b}) and (\r{D4}) to
\beq
&& \int_{56}
\frac{\delta}{\delta\Phi_5} \left( \left.
\frac{\delta \Gamma [\Phi,G^{-1},K,L]}{\delta G^{-1}_{12}} 
\right|_{\Phi,K,L} \right)_{G^{-1},K,L}
\, \frac{\delta}{\delta\Phi_6}
\left( \left. \frac{\delta \Gamma [\Phi,G^{-1},K,L}{\delta G^{-1}_{34}} 
\right|_{\Phi,K,L} \right)_{G^{-1},K,L} 
\, \left( \left.
\frac{\delta^2 \Gamma [\Phi,G^{-1},
K,L]}{\delta \Phi_5 \delta \Phi_6}
\right|_{G^{-1},K,L} \right)^{-1}  \no \\
&&  
= \left. 6 \,\frac{\delta \Gamma [\Phi,G^{-1},K,L]}{\delta L_{1234}}
\right|_{\Phi,G^{-1},K} 
+ \left. 
\frac{\delta^2 \Gamma [\Phi,G^{-1},K,L]}{\delta G^{-1}_{12} 
\delta G^{-1}_{34}} \right|_{\Phi,K,L} \left.
- \, 
\frac{\delta \Gamma [\Phi,G^{-1},K,L]}{\delta G^{-1}_{12}} \right|_{\Phi,K,L}
\, \left.
\frac{\delta \Gamma [\Phi,G^{-1},K,L]}{\delta G^{-1}_{34}} \right|_{\Phi,K,L}
\hspace*{3mm} . \la{D7}
\eeq
\subsubsection{Field-Theoretic Identities}
The functional Legendre transform has also consequences for the 
connected $n$-point functions (\r{CPF1})--(\r{CPF4}).
Taking into account (\r{LT1}), (\r{D1})--(\r{D2b}), we obtain
\beq
\Gb_1^{\rm c} &=& \Phi_1 \,, \la{I1} \\
\Gb_{12}^{\rm c} &=& 
2 \, \frac{\delta \Gamma}{\delta G^{-1}_{12}} - \Phi_1 \Phi_2
 \,, \la{I2} \\
\Gb_{123}^{\rm c} &=& 6 \, \frac{\delta \Gamma}{\delta K_{123}} 
- \Gb_{12}^{\rm c} \Phi_3
- \Gb_{13}^{\rm c} \Phi_2
- \Gb_{23}^{\rm c} \Phi_1 
- \Phi_1 \Phi_2 \Phi_3 \,, \la{I3}
\eeq
and, correspondingly,
\beq
\Gb_{1234}^{\rm c} & =& 
24 \,\frac{\delta \Gamma}{\delta L_{1234}} 
- \Gb_{12}^{\rm c}  \Gb_{34}^{\rm c} 
- \Gb_{13}^{\rm c}  \Gb_{24}^{\rm c} 
- \Gb_{14}^{\rm c}  \Gb_{23}^{\rm c} 
- \Gb_{123}^{\rm c}  \Phi_4 
- \Gb_{124}^{\rm c}  \Phi_3
- \Gb_{134}^{\rm c}  \Phi_2
- \Gb_{234}^{\rm c}  \Phi_1 
\no \\ && 
- \Gb_{12}^{\rm c} \Phi_3 \Phi_4
- \Gb_{13}^{\rm c} \Phi_2 \Phi_4
- \Gb_{14}^{\rm c} \Phi_2 \Phi_3
- \Gb_{23}^{\rm c} \Phi_1 \Phi_4
- \Gb_{24}^{\rm c} \Phi_1 \Phi_3
- \Gb_{34}^{\rm c} \Phi_1 \Phi_2
- \Phi_1 \Phi_2 \Phi_3 \Phi_4 \, .\la{I4}
\eeq
From (\r{I2}) we read off, for instance, that cutting a line of the
one-particle irreducible vacuum diagrams in all possible ways leads to the
diagrams contributing to the connected two-point function
$\Gb_{12}^{\rm c}$. Furthermore
the connected $n$-point functions 
$\Gb_{1 \cdots n}^{\rm c}$  are related to the one-particle
irreducible $n$-point functions $\Gamma_{1 \cdots n}$ 
which are defined as the $n$th functional derivative of the effective energy
$\Gamma$ with respect to the field expectation value $\Phi$:
\beq
\Gamma_{1 \cdots n}= \frac{\delta^n \Gamma}{\delta \Phi_1 \cdots
\delta \Phi_n} \, . 
\eeq
At first we get from (\r{SE}) 
\beq
\Gamma_1 &=& J_1 \,, \la{II1} 
\eeq
and from (\r{D5}), (\r{I2})
\beq
\Gamma_{12} &=& \Gb_{12}^{{\rm c}\,-1}   \,. \la{II2} 
\eeq
Separating the kernel $G^{-1}$ from the one-particle irreducible 
$2$-point function (\r{II2}) defines the self energy $\Sigma$ according to
\beq
\Gb_{12}^{{\rm c}\,-1} = G^{-1}_{12} - \Sigma_{12} \, ,\la{II3b} \, ,
\eeq
so that the connected two-point
function $\Gb_{12}^{\rm c}$ is reconstructed by the geometrical series
\beq
\Gb^{\rm c}_{12} = G_{12} + \int_{34} G_{13} \Sigma_{34} G_{42}
+ \int_{3456} G_{13} \Sigma_{34} G_{45} \Sigma_{56} G_{62} + 
\int_{345678} G_{13} \Sigma_{34} G_{45} \Sigma_{56} G_{67} \Sigma_{78}
G_{82} + \ldots \, .
\eeq
Furthermore we obtain by
applying one or two functional derivatives with respect to the current
$J$ to (\r{FID}) and by taking into account
the functional chain rule
\beq
\Gb_{123}^{\rm c} & = & - \int_{456} \Gb_{14}^{\rm c} \Gb_{25}^{\rm c}
\Gb_{36}^{\rm c} \Gamma_{456} \, , \la{II3} \\
\Gb_{1234}^{\rm c} & = & \int_{5678} \Gb_{15}^{\rm c} \Gb_{26}^{\rm c}
\Gb_{37}^{\rm c} \Gb_{48}^{\rm c} \left\{ - \Gamma_{5678} 
+ \int_{9\bar{1}} \Gb_{9\bar{1}}^{\rm c}
\left( \Gamma_{569} \Gamma_{78\bar{1}} +
\Gamma_{579} \Gamma_{68\bar{1}} +\Gamma_{589} \Gamma_{67\bar{1}} 
\right) \right\} \, , \la{II4} 
\eeq
where $\Gamma_{123}$ and $\Gamma_{1234}$
denote the one-particle irreducible three- and four-point
function, respectively.
\subsection{Functional Differential Equation for $\Gamma$}
Now we aim at deriving a functional differential equation for the
effective energy $\Gamma$. To this end we start with the first functional
differential equation (\r{NL1}) for $W$, which originates from the
identity (\r{ID1}), and perform the functional
Legendre transform with respect to
the current $J$. 
Inserting (\r{LT1}), (\r{SE}), (\r{D1}), (\r{D3}) and (\r{D4})
by taking into account the compatibility relation (\r{D5}) between
functional derivatives with respect to the field expectation value $\Phi$
and the kernel $G^{-1}$, we thus obtain
\beq
&&\delta_{12} + \Phi_1 \frac{\delta \Gamma}{\delta \Phi_2}  
- 2 \int_{3} G^{-1}_{13} \frac{\delta \Gamma}{\delta G^{-1}_{23}} = 
\int_{34} K_{134} \frac{\delta \Gamma}{\delta G^{-1}_{23}} \Phi_4
+ \int_{345} K_{134} 
\frac{\delta^2 \Gamma}{\delta G^{-1}_{23} \delta \Phi_5}
\left\{ 2 \, \frac{\delta \Gamma}{\delta G^{-1}_{45}} - \Phi_4 \Phi_5 \right\}
\no \\
&& \hspace*{0.8cm}+ \frac{2}{3} \int_{345} L_{1345} \left\{ -
\frac{\delta^2 \Gamma}{\delta G^{-1}_{23} \delta G^{-1}_{45}}
+ \frac{\delta \Gamma}{\delta G^{-1}_{23}}
\frac{\delta \Gamma}{\delta G^{-1}_{45}} \right\} 
+ \frac{2}{3} \int_{34567} L_{1345}  
\frac{\delta^2 \Gamma}{\delta G^{-1}_{23} \delta \Phi_6}
\frac{\delta^2 \Gamma}{\delta G^{-1}_{45} \delta \Phi_7}
\left\{ 2 \, \frac{\delta \Gamma}{\delta G^{-1}_{67}} - \Phi_6 \Phi_7 \right\}
\, ,
\la{EFL3}
\eeq
which corresponds to Eq.~(109) in Ref. \cite{Boris}. Then we reduce the
number of functional derivatives by inserting a combination of the two
compatibility relations (\r{D5}) and (\r{D6}), i.e.
\beq
\la{COMNE}
\int_4 \frac{\delta^2 \Gamma}{\delta G^{-1}_{12} \delta \Phi_4}
\left\{ 2 \, \frac{\delta \Gamma}{\delta G^{-1}_{34}} - \Phi_3 \Phi_4
\right\} = 3 \, \frac{\delta \Gamma}{\delta K_{123}}
- \frac{\delta \Gamma}{\delta G^{-1}_{12}} \Phi_3 \, ,
\eeq
in the last term of Eq.~(\r{EFL3}), so that we result in
\beq
&&\delta_{11} \int_1 + \int_1 \Phi_1 \frac{\delta \Gamma}{\delta \Phi_1} 
- 2 \int_{12} G^{-1}_{12} \frac{\delta \Gamma}{\delta G^{-1}_{12}} = 
\int_{123} K_{123} \Phi_3 \frac{\delta \Gamma}{\delta G^{-1}_{12}}
+ \int_{1234} K_{123} 
\frac{\delta^2 \Gamma}{\delta G^{-1}_{12} \delta \Phi_4}
\left\{ 2 \, \frac{\delta \Gamma}{\delta G^{-1}_{34}} - \Phi_3 \Phi_4 \right\}
\no \\
&& \hspace*{0.8cm}+ \frac{2}{3} \int_{1234} L_{1234} \left\{ -
\frac{\delta^2 \Gamma}{\delta G^{-1}_{12} \delta G^{-1}_{34}}
+ \frac{\delta \Gamma}{\delta G^{-1}_{12}}
\frac{\delta \Gamma}{\delta G^{-1}_{34}} \right\} 
+ \frac{2}{3} \int_{12345} L_{1234}  
\frac{\delta^2 \Gamma}{\delta G^{-1}_{12} \delta \Phi_5} \left\{
3 \, \frac{\delta \Gamma}{\delta K_{345}}
- \frac{\delta \Gamma}{\delta G^{-1}_{34}} \Phi_5 \right\} \, .
\la{EFLN3}
\eeq
In order to eliminate 
functional derivatives with respect to the field expectation value $\Phi$,
we consider the second functional differential equation (\r{CC}) for $W$,
which stems from the identity (\r{ID2}). Applying 
(\r{LT1}), (\r{SE}), (\r{D1}) and (\r{D2}), we obtain
\beq
\frac{\delta \Gamma}{\delta \Phi_1} = \int_2 G^{-1}_{12} \Phi_2
+ \int_{23} K_{123} \frac{\delta \Gamma}{\delta G^{-1}_{23}} + 
\int_{234} L_{1234} \frac{\delta \Gamma}{\delta K_{234}} \, ,
\la{EFL2}
\eeq
which  leads to
\beq
\frac{\delta^2 \Gamma}{\delta G^{-1}_{12} \Phi_3} = \frac{1}{2} \left\{
\delta_{13} \Phi_2 + \delta_{23} \Phi_1 \right\} + \int_{45} K_{345}
\frac{\delta^2 \Gamma}{\delta G^{-1}_{12} \delta G^{-1}_{45}} + 
\int_{456} L_{3456} 
\frac{\delta^2 \Gamma}{\delta G^{-1}_{12} \delta K_{456}}  \, .
\la{EFL4}
\eeq
Thus we can, indeed, eliminate functional derivatives with respect to the
field expectation value $\Phi$ on the right-hand side of Eq.~(\r{EFLN3}). 
In this way we end up with
a single nonlinear functional differential equation for the effective
energy $\Gamma$ which involves on the right-hand side 
functional derivatives with respect
to the kernel $G^{-1}$ and the cubic interaction $K$:
\beq
&&\delta_{11} \int_1 + \int_{1} \Phi_1 \frac{\delta \Gamma}{\delta \Phi_1} 
- 2 \int_{12} G^{-1}_{12} \frac{\delta \Gamma}{\delta G^{-1}_{12}}
= 
\int_{123456} K_{123} K_{456} 
\frac{\delta^2 \Gamma}{\delta G^{-1}_{12} \delta G^{-1}_{45}}
\left\{ 2 \frac{\delta \Gamma}{\delta 
G^{-1}_{36}} - \Phi_3 \Phi_6 \right\} \no \\
&& \hspace*{1cm}
+ 3 \int_{123} K_{123} \Phi_3 \frac{\delta \Gamma}{\delta G^{-1}_{12}} 
- \int_{123} K_{123} \Phi_1 \Phi_2 \Phi_3
+ \int_{1234567} K_{123} L_{4567} 
\frac{\delta^2 \Gamma}{\delta G^{-1}_{12} \delta K_{456}}
\left\{ 2 \frac{\delta \Gamma}{\delta 
G^{-1}_{37}} - \Phi_3 \Phi_7 \right\} \no \\
&&\hspace*{1cm} + \frac{2}{3} \int_{1234} L_{1234} \left\{ 
- \frac{\delta^2 \Gamma}{\delta G^{-1}_{12} \delta G^{-1}_{34}} 
+  \frac{\delta \Gamma}{\delta G^{-1}_{12}}
\frac{\delta \Gamma}{\delta G^{-1}_{34}}
\right\}
+ 2 \int_{1234} L_{1234} \Phi_4 \frac{\delta \Gamma}{\delta K_{123}}
- \frac{2}{3} \int_{1234} L_{1234} \Phi_3 \Phi_4 
\frac{\delta \Gamma}{\delta G^{-1}_{12}} \no \\
&&\hspace*{1cm}+ 2 \int_{1234567} K_{123} L_{4567} 
\frac{\delta^2 \Gamma}{\delta G^{-1}_{12} \delta G^{-1}_{45}}
\frac{\delta \Gamma}{\delta K_{367}}
- \frac{2}{3} \int_{1234567} K_{123} L_{4567} \Phi_3  
\frac{\delta^2 \Gamma}{\delta G^{-1}_{12} \delta G^{-1}_{45}}
\frac{\delta \Gamma}{\delta G^{-1}_{67}} \no \\
&&\hspace*{1cm}+ 2 \int_{12345678} L_{1234} L_{5678} 
\frac{\delta^2 \Gamma}{\delta G^{-1}_{12} \delta K_{567}}
\frac{\delta \Gamma}{\delta K_{348}}
- \frac{2}{3} \int_{12345678} L_{1234} L_{5678} \Phi_8
\frac{\delta^2 \Gamma}{\delta G^{-1}_{12} \delta K_{567}}
\frac{\delta \Gamma}{\delta G^{-1}_{34}} \, . \la{NONAME}
\eeq
Note that applying the compatibility relation (\r{D6}) between functional
derivatives with respect to the field expectation value $\Phi$, the
kernel $G^{-1}$ and the $3$-vertex $K$ to (\r{EFL2})
would lead to Eq.~(108) in Ref. \cite{Boris},
\beq
\frac{\delta \Gamma}{\delta \Phi_1} = \int_2 G^{-1}_{12} \Phi_2
+ \int_{23} K_{123} \frac{\delta \Gamma}{\delta G^{-1}_{23}}
+ \frac{1}{3} \int_{234} L_{1234} \Phi_4 
\frac{\delta \Gamma}{\delta G^{-1}_{23}} 
+ \frac{1}{3} \int_{2345} L_{1234} 
\frac{\delta^2 \Gamma}{\delta G^{-1}_{23} \delta \Phi_5} \left\{
2 \,\frac{\delta \Gamma}{\delta G^{-1}_{34}} - \Phi_3 \Phi_4 \right\} \, ,
\la{BEF}
\eeq
so that functional derivatives with respect to the field expectation
value $\Phi$ in (\r{EFL3}) could no longer be eliminated.
This procedure has been pursued in Ref.~\cite{Boris}, where the two coupled
nonlinear functional differential equations (\r{EFL3}) and (\r{BEF}) 
for the effective energy are investigated. Due to the last term in (\r{EFL3})
the highest nonlinearity within the approach of Ref.~\cite{Boris} is cubic,
whereas our  functional differential equation (\r{NONAME}) contains
at most only quadratic nonlinearities.
\subsection{Graphical Relation}
If the cubic and the quartic interactions $K$ and $L$ vanish,
Eq. (\r{NONAME}) is solved by the zeroth order contribution 
to the effective energy (\r{FEF}) which
has the functional derivatives
\beq
\frac{\delta \Gamma^{(0)}}{\delta \Phi_1} = \int_2 G^{-1}_{12} \Phi_2 \, ,
\hspace*{1cm}
\frac{\delta \Gamma^{(0)}}{\delta G^{-1}_{12}} = \frac{1}{2} \left\{
G_{12} + \Phi_1 \Phi_2 \right\} \, , \hspace*{1cm}
\frac{\delta^2 \Gamma^{(0)}}{\delta G^{-1}_{12} \delta G^{-1}_{34}}
= - \frac{1}{4} \left\{ G_{13} G_{24} + G_{14} G_{23} \right\} \, .
\la{EFNL4} 
\eeq
For non-vanishing cubic and quartic
interactions $K$ and $L$, the right-hand side in Eq.~(\r{NONAME}) produces
corrections to (\r{FEF}) which we shall denote with $\Gamma^{({\rm int})}$.
Thus the effective energy $\Gamma$ decomposes according to
\beq
\Gamma = \Gamma^{(0)} + \Gamma^{({\rm int})} \, . \la{EDEC}
\eeq
Inserting this into (\r{NONAME}) and using (\r{EFNL4}), we obtain 
together with (\r{NR1}) and (\r{NR2}) the 
following function differential equation for the interaction part of 
the effective energy $\Gamma^{({\rm int})}$:
\beq
&& \int_1 \Phi_1 \frac{\delta \Gamma^{({\rm int})}}{\delta \Phi_1} 
+ 2 \int_{12} G_{12} \frac{\delta \Gamma^{({\rm int})}}{\delta 
G_{12}} = \frac{1}{2} \int_{1234} L_{1234} G_{12} G_{34} 
- \frac{1}{2} \int_{123456} K_{123} K_{456} G_{14} G_{25} G_{36}
+ \frac{3}{2} \int_{123} K_{123} G_{12} \Phi_3 
\no \\ &&
+ \frac{1}{6} \int_{1234567} K_{123} L_{4567} G_{14} G_{25} G_{67} \Phi_3
+ \frac{1}{6} \int_{1234567} K_{123} L_{4567} G_{14} G_{25} \Phi_3
\Phi_6 \Phi_7
+ \frac{1}{2} \int_{123} K_{123} \Phi_1 \Phi_2 \Phi_3
\no \\ &&
- \frac{1}{6} \int_{1234} L_{1234} \Phi_1 \Phi_2 \Phi_3 \Phi_4
- 2 \int_{123456} L_{1234} G_{12} G_{35} G_{46} 
\frac{\delta \Gamma^{({\rm int})}}{\delta G_{56}}
- \frac{2}{3} \int_{12345678} L_{1234} G_{15} G_{26} G_{37} G_{48} 
\frac{\delta^2 \Gamma^{({\rm int})}}{\delta G_{56}\delta G_{78}}
\no\\&&
+ 3 \int_{12345678} K_{123} K_{456} G_{14} G_{25} G_{37} G_{68} 
\frac{\delta \Gamma^{({\rm int})}}{\delta G_{78}}
+\int_{123456789\bar{1}} K_{123} K_{456} G_{14} G_{27} G_{38} G_{59}
G_{6\bar{1}}
\frac{\delta^2 \Gamma^{({\rm int})}}{\delta G_{78}\delta G_{9\bar{1}}}
\no \\ && 
- \int_{1234567} K_{123}L_{4567} G_{16} G_{27}
\frac{\delta \Gamma^{({\rm int})}}{\delta K_{345}}
- \int_{123456789} K_{123} L_{4567} G_{34} G_{18} G_{29} 
\frac{\delta^2 \Gamma^{({\rm int})}}{\delta K_{567} \delta G_{89}}
\no \\ && 
-3 \int_{12345} K_{123} \Phi_3  G_{14} G_{25} 
\frac{\delta \Gamma}{\delta G_{45}}
+2 \int_{1234} L_{1234} \Phi_4 
\frac{\delta \Gamma^{({\rm int})}}{\delta K_{123}} 
- \frac{1}{3} \int_{123456789} K_{123} L_{4567} \Phi_3 G_{16} G_{27} G_{48}
G_{59} \frac{\delta \Gamma^{({\rm int})}}{\delta G_{89}} 
\no\\&&
- \frac{2}{3} \int_{123456789} K_{123} L_{4567} \Phi_3 G_{45}G_{16} G_{28} 
G_{79} \frac{\delta \Gamma^{({\rm int})}}{\delta G_{89}}
-  \frac{1}{3} \int_{123456789\bar{1}\bar{2}} K_{123} L_{4567} \Phi_3 
G_{45} G_{18} G_{29} G_{6\bar{1}} G_{7\bar{2}}
\frac{\delta^2 \Gamma^{({\rm int})}}{\delta G_{89}\delta G_{\bar{1}\bar{2}}}
\no\\&&
+  \frac{1}{3} \int_{123456789\bar{1}}L_{1234} L_{5678} \Phi_5 G_{34}
G_{19} G_{2\bar{1}} 
\frac{\delta^2 \Gamma^{({\rm int})}}{\delta K_{678}\delta G_{9\bar{1}}}
- \frac{2}{3} \int_{123456789} K_{123} L_{4567} \Phi_3 \Phi_4 \Phi_5
G_{16} G_{28} G_{79} 
\frac{\delta \Gamma^{({\rm int})}}{\delta G_{89}}
\no \\&&
- \frac{1}{3} \int_{123456789\bar{1}\bar{2}} K_{123} L_{4567}
\Phi_3 \Phi_4 \Phi_5 G_{18} G_{29} G_{6\bar{1}} G_{7\bar{2}}
\frac{\delta^2 \Gamma^{({\rm int})}}{\delta G_{89}\delta G_{\bar{1}\bar{2}}}
+ \frac{1}{3} \int_{123456789\bar{1}}L_{1234} L_{5678} \Phi_3 \Phi_4 \Phi_5
G_{19} G_{2\bar{1}}
\frac{\delta^2 \Gamma^{({\rm int})}}{\delta K_{678}\delta G_{9\bar{1}}}
\no \\&&
+\frac{2}{3} \int_{12345678} L_{1234} G_{15} G_{26} G_{37} G_{48} 
\frac{\delta \Gamma^{({\rm int})}}{\delta G_{56}}
\frac{\delta \Gamma^{({\rm int})}}{\delta G_{78}}
- 4 \int_{123456789\bar{1}} K_{123} K_{456} G_{15} G_{27} G_{68} G_{39}
G_{4\bar{1}}
\frac{\delta \Gamma^{({\rm int})}}{\delta G_{78}}
\frac{\delta \Gamma^{({\rm int})}}{\delta G_{9\bar{1}}}
\no \\&&
+2 \int_{123456789\bar{1}\bar{2}} K_{123} L_{4567} G_{18} G_{29} G_{3\bar{1}}
G_{4\bar{2}} 
\frac{\delta^2 \Gamma^{({\rm int})}}{\delta K_{567}\delta G_{89}}
\frac{\delta \Gamma^{({\rm int})}}{\delta G_{\bar{1}\bar{2}}}
+4 \int_{123456789} K_{123} L_{4567} G_{16} G_{28} G_{79}
\frac{\delta \Gamma^{({\rm int})}}{\delta K_{345}}
\frac{\delta \Gamma^{({\rm int})}}{\delta G_{89}} 
\no \\&&
+2 \int_{123456789\bar{1}\bar{2}} K_{123} L_{4567} G_{18} G_{29} G_{6\bar{1}}
G_{7\bar{2}} 
\frac{\delta \Gamma^{({\rm int})}}{\delta K_{345}}
\frac{\delta^2 \Gamma^{({\rm int})}}{\delta G_{89}\delta G_{\bar{1}\bar{2}}}
-2 \int_{123456789\bar{1}} L_{1234} L_{5678} G_{19} G_{2\bar{1}}
\frac{\delta \Gamma^{({\rm int})}}{\delta K_{345}}
\frac{\delta^2 \Gamma^{({\rm int})}}{\delta K_{678}\delta G_{9\bar{1}}}
\no \\&&
- 2 \int_{123456789\bar{1}\bar{2}\bar{3}} K_{123} K_{456} G_{17} G_{28} G_{59}
G_{6\bar{1}} G_{3\bar{2}} G_{4\bar{3}}
\frac{\delta^2 \Gamma^{({\rm int})}}{\delta G_{78}\delta G_{9\bar{1}}}
\frac{\delta \Gamma^{({\rm int})}}{\delta G_{\bar{2}\bar{3}}} 
\no \\&&
+ \frac{4}{3} \int_{123456789\bar{1}\bar{2}} K_{123} L_{4567} \Phi_3
G_{16} G_{28} G_{79} G_{4\bar{1}} G_{5\bar{2}}    
\frac{\delta \Gamma^{({\rm int})}}{\delta G_{89}}
\frac{\delta \Gamma^{({\rm int})}}{\delta G_{\bar{1}\bar{2}}}
\no \\&&
+ \frac{2}{3} \int_{123456789\bar{1}\bar{2}\bar{3}\bar{4}}
K_{123} L_{4567} \Phi_3 G_{18} G_{29} G_{6\bar{1}} G_{7\bar{2}} G_{4\bar{3}}
G_{5\bar{4}} 
\frac{\delta^2 \Gamma^{({\rm int})}}{\delta G_{89}\delta G_{\bar{1}\bar{2}}}
\frac{\delta \Gamma^{({\rm int})}}{\delta G_{\bar{3}\bar{4}}}
\no \\&&
- \frac{2}{3} \int_{123456789\bar{1}\bar{2}\bar{3}} L_{1234} L_{5678}
\Phi_5 G_{19} G_{2\bar{1}} G_{3\bar{2}} G_{4\bar{3}}
\frac{\delta^2 \Gamma^{({\rm int})}}{\delta K_{678}\delta G_{9\bar{1}}}
\frac{\delta \Gamma^{({\rm int})}}{\delta G_{\bar{2}\bar{3}}} \, .
\eeq
With the help of the graphical rules (\r{PRO}), (\r{3V}), (\r{4V}), 
(\r{FEV}), this functional differential equation can be written 
diagrammatically as follows:
\beq
&& 
\parbox{8mm}{\begin{center}
\begin{fmfgraph*}(5,5)
\setval
\fmfstraight
\fmfforce{0w,1/2h}{v1}
\fmfforce{1w,1/2h}{v2}
\fmf{boson}{v1,v2}
\fmfdot{v1}
\fmfv{decor.size=0, label=${\scs 1}$, l.dist=1mm, l.angle=0}{v2}
\end{fmfgraph*}
\end{center}}
\hspace*{2mm} \fdphi{\Gamma^{({\rm int})}}{1} \hspace*{1mm} + 2
\parbox{5.5mm}{\begin{center}
\begin{fmfgraph*}(2.5,5)
\setval
\fmfstraight
\fmfforce{1w,0h}{v1}
\fmfforce{1w,1h}{v2}
\fmf{plain,left=1}{v1,v2}
\fmfv{decor.size=0, label=${\scs 2}$, l.dist=1mm, l.angle=0}{v1}
\fmfv{decor.size=0, label=${\scs 1}$, l.dist=1mm, l.angle=0}{v2}
\end{fmfgraph*}
\end{center}}
\hspace*{0.3cm} \dphi{\Gamma^{({\rm int})}}{1}{2} = 
\hspace*{1mm}- \frac{1}{2}\hspace*{1mm}
\parbox{11mm}{\begin{center}
\begin{fmfgraph*}(8,4)
\setval
\fmfleft{i1}
\fmfright{o1}
\fmf{plain,left=1}{i1,v1,i1}
\fmf{plain,left=1}{o1,v1,o1}
\fmfdot{v1}
\end{fmfgraph*}\end{center}}
\hspace*{1mm}- \frac{1}{2}\hspace*{1mm}
\parbox{7mm}{\begin{center}
\begin{fmfgraph*}(4,4)
\setval
\fmfforce{0w,0.5h}{v1}
\fmfforce{1w,0.5h}{v2}
\fmf{plain,left=1}{v1,v2,v1}
\fmf{plain}{v1,v2}
\fmfdot{v1,v2}
\end{fmfgraph*}\end{center}}
\hspace*{1mm}- \frac{3}{2} \hspace*{1mm}
\parbox{11mm}{\begin{center}
\begin{fmfgraph*}(8,4)
\setval
\fmfforce{0w,1/2h}{v1}
\fmfforce{1/2w,1/2h}{v2}
\fmfforce{1w,1/2h}{v3}
\fmf{plain,left=1}{v2,v3,v2}
\fmf{boson}{v1,v2}
\fmfdot{v1,v2}
\end{fmfgraph*}\end{center}}
\hspace*{1mm}+ \frac{1}{6} \hspace*{1mm}
\parbox{15mm}{\begin{center}
\begin{fmfgraph*}(12,4)
\setval
\fmfforce{0w,1/2h}{v1}
\fmfforce{1/3w,1/2h}{v2}
\fmfforce{2/3w,1/2h}{v3}
\fmfforce{1w,1/2h}{v4}
\fmf{plain,left=1}{v2,v3,v2}
\fmf{plain,left=1}{v3,v4,v3}
\fmf{boson}{v1,v2}
\fmfdot{v1,v2,v3}
\end{fmfgraph*}\end{center}}
\hspace*{1mm} - \frac{1}{2}\hspace*{1mm}
\parbox{11mm}{\begin{center}
\begin{fmfgraph*}(6.928,12)
\setval
\fmfforce{1/2w,5/6h}{v1}
\fmfforce{1w,1/4h}{w1}
\fmfforce{0w,1/4h}{u1}
\fmfforce{1/2w,1/2h}{v2}
\fmf{boson}{v2,v1}
\fmf{boson}{v2,w1}
\fmf{boson}{v2,u1}
\fmfdot{v1,w1,u1,v2}
\end{fmfgraph*}\end{center}}
\hspace*{1mm} + \frac{1}{6}\hspace*{1mm}
\parbox{11mm}{\begin{center}
\begin{fmfgraph*}(8,8)
\setval
\fmfforce{0w,0h}{v1}
\fmfforce{1w,0h}{w1}
\fmfforce{1/2w,1h}{u1}
\fmfforce{1/2w,0h}{v2}
\fmfforce{1/2w,1/2h}{v3}
\fmf{plain,right=1}{v2,v3,v2}
\fmf{boson}{w1,v1}
\fmf{boson}{v3,u1}
\fmfdot{u1,v1,w1,v2,v3}
\end{fmfgraph*}\end{center}} 
\no \\ &&
\hspace*{1mm} + \frac{1}{6}\hspace*{1mm}
\parbox{11mm}{\begin{center}
\begin{fmfgraph*}(8,12)
\setval
\fmfforce{0w,1/2h}{v1}
\fmfforce{1w,1/2h}{w1}
\fmfforce{1/2w,1/6h}{u1}
\fmfforce{1/2w,5/6h}{x1}
\fmfforce{1/2w,1/2h}{v2}
\fmf{boson}{w1,v1}
\fmf{boson}{x1,u1}
\fmfdot{v1,u1,w1,x1,v2}
\end{fmfgraph*}\end{center}}
\hspace*{1mm}+2 \hspace*{1mm} 
\parbox{9mm}{\begin{center}
\begin{fmfgraph*}(6,4)
\setval
\fmfstraight
\fmfforce{0w,1/2h}{v1}
\fmfforce{4/6w,1/2h}{v2}
\fmfforce{1w,1h}{i2}
\fmfforce{1w,0h}{i1}
\fmf{plain}{i1,v2}
\fmf{plain}{v2,i2}
\fmf{plain,left}{v1,v2,v1}
\fmfdot{v2}
\fmfv{decor.size=0, label=${\scs 2}$, l.dist=1mm, l.angle=0}{i1}
\fmfv{decor.size=0, label=${\scs 1}$, l.dist=1mm, l.angle=0}{i2}
\end{fmfgraph*}
\end{center}}
\hspace*{3mm} \dphi{\Gamma^{({\rm int})}}{1}{2} 
\hspace*{1mm} + \frac{2}{3} \hspace*{1mm}
\parbox{7mm}{\begin{center}
\begin{fmfgraph*}(3,3)
\setval
\fmfstraight
\fmfforce{0w,1/2h}{v1}
\fmfforce{1w,2h}{i1}
\fmfforce{1w,1h}{i2}
\fmfforce{1w,0h}{i3}
\fmfforce{1w,-1h}{i4}
\fmf{plain}{v1,i1}
\fmf{plain}{v1,i2}
\fmf{plain}{v1,i3}
\fmf{plain}{v1,i4}
\fmfdot{v1}
\fmfv{decor.size=0, label=${\scs 4}$, l.dist=1mm, l.angle=0}{i4}
\fmfv{decor.size=0, label=${\scs 3}$, l.dist=1mm, l.angle=0}{i3}
\fmfv{decor.size=0, label=${\scs 2}$, l.dist=1mm, l.angle=0}{i2}
\fmfv{decor.size=0, label=${\scs 1}$, l.dist=1mm, l.angle=0}{i1}
\end{fmfgraph*}
\end{center}}
\hspace*{3mm} \ddphi{\Gamma^{({\rm int})}}{1}{2}{3}{4}
\hspace*{1mm} + 3 \hspace*{1mm}
\parbox{7mm}{\begin{center}
\begin{fmfgraph*}(6,4)
\setval
\fmfstraight
\fmfforce{1/3w,0h}{v1}
\fmfforce{1/3w,1h}{v2}
\fmfforce{1w,1h}{i2}
\fmfforce{1w,0h}{i1}
\fmf{plain}{i1,v1}
\fmf{plain}{v2,i2}
\fmf{plain,left}{v1,v2,v1}
\fmfdot{v2,v1}
\fmfv{decor.size=0, label=${\scs 2}$, l.dist=1mm, l.angle=0}{i1}
\fmfv{decor.size=0, label=${\scs 1}$, l.dist=1mm, l.angle=0}{i2}
\end{fmfgraph*}
\end{center}}
\hspace*{3mm} \dphi{\Gamma^{({\rm int})}}{1}{2}
\hspace*{1mm} + \hspace*{1mm}
\parbox{7mm}{\begin{center}
\begin{fmfgraph*}(3,9)
\setval
\fmfstraight
\fmfforce{1w,1h}{o1}
\fmfforce{1w,2/3h}{o2}
\fmfforce{0w,5/6h}{v1}
\fmfforce{0w,1/6h}{v2}
\fmfforce{1w,1/3h}{i1}
\fmfforce{1w,0h}{i2}
\fmf{plain}{v1,v2}
\fmf{plain}{v1,o1}
\fmf{plain}{v1,o2}
\fmf{plain}{v2,i1}
\fmf{plain}{v2,i2}
\fmfdot{v1,v2}
\fmfv{decor.size=0, label=${\scs 4}$, l.dist=1mm, l.angle=0}{i2}
\fmfv{decor.size=0, label=${\scs 3}$, l.dist=1mm, l.angle=0}{i1}
\fmfv{decor.size=0, label=${\scs 2}$, l.dist=1mm, l.angle=0}{o2}
\fmfv{decor.size=0, label=${\scs 1}$, l.dist=1mm, l.angle=0}{o1}
\end{fmfgraph*}
\end{center}}
\hspace*{3mm} \ddphi{\Gamma^{({\rm int})}}{1}{2}{3}{4} 
\no \\ &&
+ \hspace*{1mm} 
\parbox{10mm}{\begin{center}
\begin{fmfgraph*}(6,5.33333)
\setval
\fmfstraight
\fmfforce{2/6w,1/4h}{v1}
\fmfforce{2/6w,1h}{v2}
\fmfforce{1w,0h}{i1}
\fmfforce{1w,1/2h}{i2}
\fmfforce{1w,1h}{i3}
\fmf{plain,left=1}{v1,v2,v1}
\fmf{plain}{v1,i1}
\fmf{plain}{v1,i2}
\fmf{plain}{v2,i3}
\fmfdot{v1,v2}
\fmfv{decor.size=0, label=${\scs 2}$, l.dist=1mm, l.angle=0}{i2}
\fmfv{decor.size=0, label=${\scs 3}$, l.dist=1mm, l.angle=0}{i1}
\fmfv{decor.size=0, label=${\scs 1}$, l.dist=1mm, l.angle=0}{i3}
\end{fmfgraph*}
\end{center}}
\hspace*{3mm} \dvertex{\Gamma^{({\rm int})}}{1}{2}{3} 
\hspace*{1mm} + \hspace*{1mm}
\parbox{7mm}{\begin{center}
\begin{fmfgraph*}(3,12)
\setval
\fmfstraight
\fmfforce{1w,1h}{o0}
\fmfforce{1w,3/4h}{o1}
\fmfforce{1w,1/2h}{o2}
\fmfforce{0w,3/4h}{v1}
\fmfforce{0w,1/8h}{v2}
\fmfforce{1w,1/4h}{i1}
\fmfforce{1w,0h}{i2}
\fmf{plain}{v1,v2}
\fmf{plain}{v1,o0}
\fmf{plain}{v1,o1}
\fmf{plain}{v1,o2}
\fmf{plain}{v2,i1}
\fmf{plain}{v2,i2}
\fmfdot{v1,v2}
\fmfv{decor.size=0, label=${\scs 5}$, l.dist=1mm, l.angle=0}{i2}
\fmfv{decor.size=0, label=${\scs 4}$, l.dist=1mm, l.angle=0}{i1}
\fmfv{decor.size=0, label=${\scs 3}$, l.dist=1mm, l.angle=0}{o2}
\fmfv{decor.size=0, label=${\scs 2}$, l.dist=1mm, l.angle=0}{o1}
\fmfv{decor.size=0, label=${\scs 1}$, l.dist=1mm, l.angle=0}{o0}
\end{fmfgraph*}
\end{center}}
\hspace*{3mm} \ddvertex{\Gamma^{({\rm int})}}{1}{2}{3}{4}{5}
\hspace*{1mm} + 3 \hspace*{1mm}
\parbox{9mm}{\begin{center}
\begin{fmfgraph*}(6,4)
\setval
\fmfstraight
\fmfforce{0w,1/2h}{v1}
\fmfforce{4/6w,1/2h}{v2}
\fmfforce{1w,1h}{i1}
\fmfforce{1w,0h}{i2}
\fmf{boson}{v2,v1}
\fmf{plain}{v2,i1}
\fmf{plain}{v2,i2}
\fmfdot{v1,v2}
\fmfv{decor.size=0, label=${\scs 2}$, l.dist=1mm, l.angle=0}{i2}
\fmfv{decor.size=0, label=${\scs 1}$, l.dist=1mm, l.angle=0}{i1}
\end{fmfgraph*}
\end{center}}
\hspace*{2mm} \dphi{\Gamma^{({\rm int})}}{1}{2} 
\hspace*{1mm} + 2 \hspace*{1mm}
\parbox{9mm}{\begin{center}
\begin{fmfgraph*}(6,4)
\setval
\fmfstraight
\fmfforce{0w,1/2h}{v1}
\fmfforce{4/6w,1/2h}{v2}
\fmfforce{1w,1.25h}{i1}
\fmfforce{1w,1/2h}{i2}
\fmfforce{1w,-0.25h}{i3}
\fmf{boson}{v2,v1}
\fmf{plain}{v2,i1}
\fmf{plain}{v2,i2}
\fmf{plain}{v2,i3}
\fmfdot{v1,v2}
\fmfv{decor.size=0, label=${\scs 3}$, l.dist=1mm, l.angle=0}{i3}
\fmfv{decor.size=0, label=${\scs 2}$, l.dist=1mm, l.angle=0}{i2}
\fmfv{decor.size=0, label=${\scs 1}$, l.dist=1mm, l.angle=0}{i1}
\end{fmfgraph*}
\end{center}}
\hspace*{3mm} \dvertex{\Gamma^{({\rm int})}}{1}{2}{3} 
\hspace*{1mm} - \frac{1}{3} \hspace*{1mm}
\parbox{13mm}{\begin{center}
\begin{fmfgraph*}(10,4)
\setval
\fmfstraight
\fmfforce{0w,1/2h}{v1}
\fmfforce{4/10w,1/2h}{v2}
\fmfforce{8/10w,1/2h}{v3}
\fmfforce{1w,7/8h}{i1}
\fmfforce{1w,1/8h}{i2}
\fmf{boson}{v2,v1}
\fmf{plain,left=1}{v2,v3,v2}
\fmf{plain}{v3,i1}
\fmf{plain}{v3,i2}
\fmfdot{v1,v2,v3}
\fmfv{decor.size=0, label=${\scs 2}$, l.dist=1mm, l.angle=0}{i2}
\fmfv{decor.size=0, label=${\scs 1}$, l.dist=1mm, l.angle=0}{i1}
\end{fmfgraph*}
\end{center}}
\hspace*{3mm} \dphi{\Gamma^{({\rm int})}}{1}{2} 
\no \\ &&
\hspace*{1mm} - \frac{2}{3} \hspace*{1mm}
\parbox{9mm}{\begin{center}
\begin{fmfgraph*}(6,6)
\setval
\fmfstraight
\fmfforce{0w,0h}{v1}
\fmfforce{2/3w,0h}{v2}
\fmfforce{1w,0h}{v3}
\fmfforce{0w,2/3h}{v4}
\fmfforce{2/3w,2/3h}{v5}
\fmfforce{1w,2/3h}{v6}
\fmf{boson}{v2,v1}
\fmf{plain,left=1}{v4,v5,v4}
\fmf{plain}{v2,v3}
\fmf{plain}{v5,v6}
\fmf{plain}{v2,v5}
\fmfdot{v1,v2,v5}
\fmfv{decor.size=0, label=${\scs 2}$, l.dist=1mm, l.angle=0}{v3}
\fmfv{decor.size=0, label=${\scs 1}$, l.dist=1mm, l.angle=0}{v6}
\end{fmfgraph*}
\end{center}}
\hspace*{3mm} \dphi{\Gamma^{({\rm int})}}{1}{2} 
\hspace*{1mm} - \frac{1}{3} \hspace*{1mm}
\parbox{9mm}{\begin{center}
\begin{fmfgraph*}(6,4)
\setval
\fmfstraight
\fmfforce{0w,1/2h}{v1}
\fmfforce{4/6w,1/2h}{v2}
\fmfforce{1w,1h}{i1}
\fmfforce{1w,0h}{i2}
\fmf{boson}{v2,v1}
\fmf{plain}{v2,i1}
\fmf{plain}{v2,i2}
\fmfdot{v1,v2}
\fmfv{decor.size=0, label=${\scs 2}$, l.dist=1mm, l.angle=0}{i2}
\fmfv{decor.size=0, label=${\scs 1}$, l.dist=1mm, l.angle=0}{i1}
\end{fmfgraph*}
\end{center}}
\hspace*{2mm} \ddphi{\Gamma^{({\rm int})}}{1}{2}{3}{4} \hspace*{2mm} 
\parbox{9mm}{\begin{center}
\begin{fmfgraph*}(6,4)
\setval
\fmfstraight
\fmfforce{1w,1/2h}{v1}
\fmfforce{2/6w,1/2h}{v2}
\fmfforce{0w,1h}{i2}
\fmfforce{0w,0h}{i1}
\fmf{plain}{i1,v2}
\fmf{plain}{v2,i2}
\fmf{plain,left}{v1,v2,v1}
\fmfdot{v2}
\fmfv{decor.size=0, label=${\scs 4}$, l.dist=1mm, l.angle=-180}{i1}
\fmfv{decor.size=0, label=${\scs 3}$, l.dist=1mm, l.angle=-180}{i2}
\end{fmfgraph*}
\end{center}}
\hspace*{1mm} - \frac{1}{3} \hspace*{1mm}
\parbox{9mm}{\begin{center}
\begin{fmfgraph*}(6,4)
\setval
\fmfstraight
\fmfforce{0w,1/2h}{v1}
\fmfforce{4/6w,1/2h}{v2}
\fmfforce{1w,1.25h}{i1}
\fmfforce{1w,1/2h}{i2}
\fmfforce{1w,-0.25h}{i3}
\fmf{boson}{v2,v1}
\fmf{plain}{v2,i1}
\fmf{plain}{v2,i2}
\fmf{plain}{v2,i3}
\fmfdot{v1,v2}
\fmfv{decor.size=0, label=${\scs 3}$, l.dist=1mm, l.angle=0}{i3}
\fmfv{decor.size=0, label=${\scs 2}$, l.dist=1mm, l.angle=0}{i2}
\fmfv{decor.size=0, label=${\scs 1}$, l.dist=1mm, l.angle=0}{i1}
\end{fmfgraph*}
\end{center}}
\hspace*{2mm} \ddvertex{\Gamma^{({\rm int})}}{1}{2}{3}{4}{5}\hspace*{2mm} 
\parbox{9mm}{\begin{center}
\begin{fmfgraph*}(6,4)
\setval
\fmfstraight
\fmfforce{1w,1/2h}{v1}
\fmfforce{2/6w,1/2h}{v2}
\fmfforce{0w,1h}{i2}
\fmfforce{0w,0h}{i1}
\fmf{plain}{i1,v2}
\fmf{plain}{v2,i2}
\fmf{plain,left}{v1,v2,v1}
\fmfdot{v2}
\fmfv{decor.size=0, label=${\scs 5}$, l.dist=1mm, l.angle=-180}{i1}
\fmfv{decor.size=0, label=${\scs 4}$, l.dist=1mm, l.angle=-180}{i2}
\end{fmfgraph*}
\end{center}}
\hspace*{1mm} - \frac{2}{3} \hspace*{1mm}
\parbox{9mm}{\begin{center}
\begin{fmfgraph*}(6,7)
\setval
\fmfstraight
\fmfforce{0w,1h}{v1}
\fmfforce{2/3w,1h}{v2}
\fmfforce{1w,1h}{v3}
\fmfforce{2/3w,2/7h}{v4}
\fmfforce{0w,0h}{v5}
\fmfforce{0w,3.5/7h}{v6}
\fmfforce{1w,2/7h}{v7}
\fmf{boson}{v1,v2}
\fmf{plain}{v2,v3}
\fmf{plain}{v4,v7}
\fmf{boson}{v4,v5}
\fmf{boson}{v4,v6}
\fmf{plain}{v4,v2}
\fmfdot{v1,v2,v4,v5,v6}
\fmfv{decor.size=0, label=${\scs 1}$, l.dist=1mm, l.angle=0}{v3}
\fmfv{decor.size=0, label=${\scs 2}$, l.dist=1mm, l.angle=0}{v7}
\end{fmfgraph*}
\end{center}}
\hspace*{2mm} \dphi{\Gamma^{({\rm int})}}{1}{2} \hspace*{2mm} 
\no \\ &&
\hspace*{1mm} - \frac{1}{3} \hspace*{1mm}
\parbox{9mm}{\begin{center}
\begin{fmfgraph*}(6,4)
\setval
\fmfstraight
\fmfforce{0w,1/2h}{v1}
\fmfforce{4/6w,1/2h}{v2}
\fmfforce{1w,1h}{i1}
\fmfforce{1w,0h}{i2}
\fmf{boson}{v2,v1}
\fmf{plain}{v2,i1}
\fmf{plain}{v2,i2}
\fmfdot{v1,v2}
\fmfv{decor.size=0, label=${\scs 2}$, l.dist=1mm, l.angle=0}{i2}
\fmfv{decor.size=0, label=${\scs 1}$, l.dist=1mm, l.angle=0}{i1}
\end{fmfgraph*}
\end{center}}
\hspace*{2mm} \ddphi{\Gamma^{({\rm int})}}{1}{2}{3}{4} \hspace*{2mm} 
\parbox{8mm}{\begin{center}
\begin{fmfgraph*}(5,6)
\setval
\fmfstraight
\fmfforce{2/5w,1/2h}{v1}
\fmfforce{0w,1/6h}{v2}
\fmfforce{1w,0h}{v3}
\fmfforce{0w,5/6h}{v4}
\fmfforce{1w,1h}{v5}
\fmf{plain}{v1,v2}
\fmf{plain}{v1,v4}
\fmf{boson}{v1,v3}
\fmf{boson}{v1,v5}
\fmfdot{v1,v3,v5}
\fmfv{decor.size=0, label=${\scs 4}$, l.dist=1mm, l.angle=-180}{v2}
\fmfv{decor.size=0, label=${\scs 3}$, l.dist=1mm, l.angle=-180}{v4}
\end{fmfgraph*}
\end{center}}
\hspace*{2mm} \ddphi{\Gamma^{({\rm int})}}{1}{2}{3}{4} \hspace*{2mm} 
\hspace*{1mm} - \frac{1}{3} \hspace*{1mm}
\parbox{9mm}{\begin{center}
\begin{fmfgraph*}(6,4)
\setval
\fmfstraight
\fmfforce{0w,1/2h}{v1}
\fmfforce{4/6w,1/2h}{v2}
\fmfforce{1w,1.25h}{i1}
\fmfforce{1w,1/2h}{i2}
\fmfforce{1w,-0.25h}{i3}
\fmf{boson}{v2,v1}
\fmf{plain}{v2,i1}
\fmf{plain}{v2,i2}
\fmf{plain}{v2,i3}
\fmfdot{v1,v2}
\fmfv{decor.size=0, label=${\scs 3}$, l.dist=1mm, l.angle=0}{i3}
\fmfv{decor.size=0, label=${\scs 2}$, l.dist=1mm, l.angle=0}{i2}
\fmfv{decor.size=0, label=${\scs 1}$, l.dist=1mm, l.angle=0}{i1}
\end{fmfgraph*}
\end{center}}
\hspace*{2mm} \ddvertex{\Gamma^{({\rm int})}}{1}{2}{3}{4}{5}\hspace*{2mm} 
\parbox{8mm}{\begin{center}
\begin{fmfgraph*}(5,6)
\setval
\fmfstraight
\fmfforce{2/5w,1/2h}{v1}
\fmfforce{0w,1/6h}{v2}
\fmfforce{1w,0h}{v3}
\fmfforce{0w,5/6h}{v4}
\fmfforce{1w,1h}{v5}
\fmf{plain}{v1,v2}
\fmf{plain}{v1,v4}
\fmf{boson}{v1,v3}
\fmf{boson}{v1,v5}
\fmfdot{v1,v3,v5}
\fmfv{decor.size=0, label=${\scs 5}$, l.dist=1mm, l.angle=-180}{v2}
\fmfv{decor.size=0, label=${\scs 4}$, l.dist=1mm, l.angle=-180}{v4}
\end{fmfgraph*}
\end{center}}
\no \\ &&
\hspace*{1mm} - \frac{2}{3} \hspace*{1mm}
\hspace*{1mm} \dphi{\Gamma^{({\rm int})}}{1}{2}\hspace*{3mm}
\parbox{7mm}{\begin{center}
\begin{fmfgraph*}(4,4)
\setval
\fmfstraight
\fmfforce{0w,1h}{o1}
\fmfforce{0w,0h}{o2}
\fmfforce{1/2w,1/2h}{v1}
\fmfforce{1w,1h}{i1}
\fmfforce{1w,0h}{i2}
\fmf{plain}{v1,o1}
\fmf{plain}{v1,o2}
\fmf{plain}{v1,i1}
\fmf{plain}{v1,i2}
\fmfdot{v1}
\fmfv{decor.size=0, label=${\scs 4}$, l.dist=1mm, l.angle=0}{i2}
\fmfv{decor.size=0, label=${\scs 3}$, l.dist=1mm, l.angle=0}{i1}
\fmfv{decor.size=0, label=${\scs 2}$, l.dist=1mm, l.angle=-180}{o2}
\fmfv{decor.size=0, label=${\scs 1}$, l.dist=1mm, l.angle=-180}{o1}
\end{fmfgraph*}
\end{center}}
\hspace*{3mm} \dphi{\Gamma^{({\rm int})}}{3}{4} 
\hspace*{1mm} - 4 \hspace*{1mm}
\dphi{\Gamma^{({\rm int})}}{1}{2} \hspace*{2mm} 
\parbox{7mm}{\begin{center}
\begin{fmfgraph*}(4,4)
\setval
\fmfstraight
\fmfforce{0w,1h}{o1}
\fmfforce{0w,0h}{o2}
\fmfforce{1/2w,0h}{v1}
\fmfforce{1/2w,1h}{v2}
\fmfforce{1w,1h}{i1}
\fmfforce{1w,0h}{i2}
\fmf{plain}{v1,v2}
\fmf{plain}{v1,i2}
\fmf{plain}{v1,o2}
\fmf{plain}{v2,i1}
\fmf{plain}{v2,o1}
\fmfdot{v1,v2}
\fmfv{decor.size=0, label=${\scs 4}$, l.dist=1mm, l.angle=0}{i2}
\fmfv{decor.size=0, label=${\scs 3}$, l.dist=1mm, l.angle=0}{i1}
\fmfv{decor.size=0, label=${\scs 2}$, l.dist=1mm, l.angle=-180}{o2}
\fmfv{decor.size=0, label=${\scs 1}$, l.dist=1mm, l.angle=-180}{o1}
\end{fmfgraph*}
\end{center}}
\hspace*{2mm} \dphi{\Gamma^{({\rm int})}}{3}{4} 
- 2 \hspace*{1mm}
\dphi{\Gamma^{({\rm int})}}{1}{2} \hspace*{2mm} 
\parbox{9mm}{\begin{center}
\begin{fmfgraph*}(6,9)
\setval
\fmfstraight
\fmfforce{1w,1h}{o1}
\fmfforce{1w,2/3h}{o2}
\fmfforce{1/2w,5/6h}{v1}
\fmfforce{1/2w,1/6h}{v2}
\fmfforce{1w,1/3h}{i1}
\fmfforce{1w,0h}{i2}
\fmfforce{0w,5/6h}{z1}
\fmfforce{0w,1/6h}{z2}
\fmf{plain}{v1,o1}
\fmf{plain}{v1,o2}
\fmf{plain}{v2,i1}
\fmf{plain}{v2,i2}
\fmf{plain}{v1,z1}
\fmf{plain}{v2,z2}
\fmfdot{v1,v2}
\fmfv{decor.size=0, label=${\scs 2}$, l.dist=1mm, l.angle=-180}{z2}
\fmfv{decor.size=0, label=${\scs 1}$, l.dist=1mm, l.angle=-180}{z1}
\fmfv{decor.size=0, label=${\scs 6}$, l.dist=1mm, l.angle=0}{i2}
\fmfv{decor.size=0, label=${\scs 5}$, l.dist=1mm, l.angle=0}{i1}
\fmfv{decor.size=0, label=${\scs 4}$, l.dist=1mm, l.angle=0}{o2}
\fmfv{decor.size=0, label=${\scs 3}$, l.dist=1mm, l.angle=0}{o1}
\end{fmfgraph*}
\end{center}}
\hspace*{2mm} \ddphi{\Gamma^{({\rm int})}}{3}{4}{5}{6} \hspace*{2mm} 
\no \\ &&
\hspace*{1mm} - 2 \hspace*{1mm}
\dphi{\Gamma^{({\rm int})}}{1}{2} \hspace*{2mm} 
\parbox{9mm}{\begin{center}
\begin{fmfgraph*}(6,12)
\setval
\fmfstraight
\fmfforce{1w,1h}{o0}
\fmfforce{1w,3/4h}{o1}
\fmfforce{1w,1/2h}{o2}
\fmfforce{1/2w,3/4h}{v1}
\fmfforce{1/2w,1/8h}{v2}
\fmfforce{1w,1/4h}{i1}
\fmfforce{1w,0h}{i2}
\fmfforce{0w,3/4h}{z1}
\fmfforce{0w,1/8h}{z2}
\fmf{plain}{z1,v1}
\fmf{plain}{z2,v2}
\fmf{plain}{v1,o0}
\fmf{plain}{v1,o1}
\fmf{plain}{v1,o2}
\fmf{plain}{v2,i1}
\fmf{plain}{v2,i2}
\fmfdot{v1,v2}
\fmfv{decor.size=0, label=${\scs 7}$, l.dist=1mm, l.angle=0}{i2}
\fmfv{decor.size=0, label=${\scs 6}$, l.dist=1mm, l.angle=0}{i1}
\fmfv{decor.size=0, label=${\scs 5}$, l.dist=1mm, l.angle=0}{o2}
\fmfv{decor.size=0, label=${\scs 4}$, l.dist=1mm, l.angle=0}{o1}
\fmfv{decor.size=0, label=${\scs 3}$, l.dist=1mm, l.angle=0}{o0}
\fmfv{decor.size=0, label=${\scs 2}$, l.dist=1mm, l.angle=-180}{z2}
\fmfv{decor.size=0, label=${\scs 1}$, l.dist=1mm, l.angle=-180}{z1}
\end{fmfgraph*}
\end{center}}
\hspace*{2mm} \ddvertex{\Gamma^{({\rm int})}}{3}{4}{5}{6}{7} 
\hspace*{1mm} - 4 
\hspace*{2mm} \dphi{\Gamma^{({\rm int})}}{1}{2} \hspace*{2mm}
\parbox{10mm}{\begin{center}
\begin{fmfgraph*}(8,5.33333)
\setval
\fmfstraight
\fmfforce{4/8w,1/4h}{v1}
\fmfforce{4/8w,1h}{v2}
\fmfforce{1w,0h}{i1}
\fmfforce{1w,1/2h}{i2}
\fmfforce{1w,1h}{i3}
\fmfforce{0w,1/4h}{z1}
\fmfforce{0w,1h}{z2}
\fmf{plain}{v1,z1}
\fmf{plain}{z2,v2}
\fmf{plain}{v1,v2}
\fmf{plain}{v1,i1}
\fmf{plain}{v1,i2}
\fmf{plain}{v2,i3}
\fmfdot{v1,v2}
\fmfv{decor.size=0, label=${\scs 4}$, l.dist=1mm, l.angle=0}{i2}
\fmfv{decor.size=0, label=${\scs 5}$, l.dist=1mm, l.angle=0}{i1}
\fmfv{decor.size=0, label=${\scs 3}$, l.dist=1mm, l.angle=0}{i3}
\fmfv{decor.size=0, label=${\scs 2}$, l.dist=1mm, l.angle=-180}{z1}
\fmfv{decor.size=0, label=${\scs 1}$, l.dist=1mm, l.angle=-180}{z2}
\end{fmfgraph*}
\end{center}}
\hspace*{2mm} \dvertex{\Gamma^{({\rm int})}}{3}{4}{5} \hspace*{2mm} 
- 2 
\hspace*{1mm} \dvertex{\Gamma^{({\rm int})}}{1}{2}{3} \hspace*{2mm} 
\parbox{9mm}{\begin{center}
\begin{fmfgraph*}(6,9)
\setval
\fmfstraight
\fmfforce{1w,1h}{o1}
\fmfforce{1w,2/3h}{o2}
\fmfforce{1/2w,5/6h}{v1}
\fmfforce{1/2w,1/6h}{v2}
\fmfforce{1w,1/3h}{i1}
\fmfforce{1w,0h}{i2}
\fmfforce{0w,5/6h}{z1}
\fmfforce{0w,1/3h}{z2}
\fmfforce{0w,0h}{z3}
\fmf{plain}{v1,o1}
\fmf{plain}{v1,o2}
\fmf{plain}{v2,i1}
\fmf{plain}{v2,i2}
\fmf{plain}{v1,z1}
\fmf{plain}{v2,z2}
\fmf{plain}{v2,z3}
\fmfdot{v1,v2}
\fmfv{decor.size=0, label=${\scs 3}$, l.dist=1mm, l.angle=-180}{z3}
\fmfv{decor.size=0, label=${\scs 2}$, l.dist=1mm, l.angle=-180}{z2}
\fmfv{decor.size=0, label=${\scs 1}$, l.dist=1mm, l.angle=-180}{z1}
\fmfv{decor.size=0, label=${\scs 7}$, l.dist=1mm, l.angle=0}{i2}
\fmfv{decor.size=0, label=${\scs 6}$, l.dist=1mm, l.angle=0}{i1}
\fmfv{decor.size=0, label=${\scs 5}$, l.dist=1mm, l.angle=0}{o2}
\fmfv{decor.size=0, label=${\scs 4}$, l.dist=1mm, l.angle=0}{o1}
\end{fmfgraph*}
\end{center}}
\hspace*{2mm} \ddphi{\Gamma^{({\rm int})}}{4}{5}{6}{7} \hspace*{2mm} 
\no \\ &&
- 2 
\hspace*{1mm} \dvertex{\Gamma^{({\rm int})}}{1}{2}{3} \hspace*{2mm} 
\parbox{9mm}{\begin{center}
\begin{fmfgraph*}(6,12)
\setval
\fmfstraight
\fmfforce{1w,1h}{o0}
\fmfforce{1w,3/4h}{o1}
\fmfforce{1w,1/2h}{o2}
\fmfforce{1/2w,3/4h}{v1}
\fmfforce{1/2w,1/8h}{v2}
\fmfforce{1w,1/4h}{i1}
\fmfforce{1w,0h}{i2}
\fmfforce{0w,3/4h}{z1}
\fmfforce{0w,1/4h}{z2}
\fmfforce{0w,0h}{z3}
\fmf{plain}{z1,v1}
\fmf{plain}{z2,v2}
\fmf{plain}{z3,v2}
\fmf{plain}{v1,o0}
\fmf{plain}{v1,o1}
\fmf{plain}{v1,o2}
\fmf{plain}{v2,i1}
\fmf{plain}{v2,i2}
\fmfdot{v1,v2}
\fmfv{decor.size=0, label=${\scs 8}$, l.dist=1mm, l.angle=0}{i2}
\fmfv{decor.size=0, label=${\scs 7}$, l.dist=1mm, l.angle=0}{i1}
\fmfv{decor.size=0, label=${\scs 6}$, l.dist=1mm, l.angle=0}{o2}
\fmfv{decor.size=0, label=${\scs 5}$, l.dist=1mm, l.angle=0}{o1}
\fmfv{decor.size=0, label=${\scs 4}$, l.dist=1mm, l.angle=0}{o0}
\fmfv{decor.size=0, label=${\scs 2}$, l.dist=1mm, l.angle=-180}{z2}
\fmfv{decor.size=0, label=${\scs 1}$, l.dist=1mm, l.angle=-180}{z1}
\fmfv{decor.size=0, label=${\scs 3}$, l.dist=1mm, l.angle=-180}{z3}
\end{fmfgraph*}
\end{center}}
\hspace*{2mm} \ddvertex{\Gamma^{({\rm int})}}{4}{5}{6}{7}{8} \hspace*{2mm} 
+ \frac{2}{3} 
\parbox{9mm}{\begin{center}
\begin{fmfgraph*}(6,4)
\setval
\fmfstraight
\fmfforce{0w,1/2h}{v1}
\fmfforce{4/6w,1/2h}{v2}
\fmfforce{1w,1h}{i1}
\fmfforce{1w,0h}{i2}
\fmf{boson}{v2,v1}
\fmf{plain}{v2,i1}
\fmf{plain}{v2,i2}
\fmfdot{v1,v2}
\fmfv{decor.size=0, label=${\scs 2}$, l.dist=1mm, l.angle=0}{i2}
\fmfv{decor.size=0, label=${\scs 1}$, l.dist=1mm, l.angle=0}{i1}
\end{fmfgraph*}
\end{center}}
\hspace*{2mm} \ddphi{\Gamma^{({\rm int})}}{1}{2}{3}{4} \hspace*{2mm} 
\parbox{7mm}{\begin{center}
\begin{fmfgraph*}(4,4)
\setval
\fmfstraight
\fmfforce{0w,1h}{o1}
\fmfforce{0w,0h}{o2}
\fmfforce{1/2w,1/2h}{v1}
\fmfforce{1w,1h}{i1}
\fmfforce{1w,0h}{i2}
\fmf{plain}{v1,o1}
\fmf{plain}{v1,o2}
\fmf{plain}{v1,i1}
\fmf{plain}{v1,i2}
\fmfdot{v1}
\fmfforce{1w,1h}{i1}
\fmfforce{1w,0h}{i2}
\fmf{plain}{v1,o1}
\fmf{plain}{v1,o2}
\fmf{plain}{v1,i1}
\fmf{plain}{v1,i2}
\fmfdot{v1}
\fmfv{decor.size=0, label=${\scs 6}$, l.dist=1mm, l.angle=0}{i2}
\fmfv{decor.size=0, label=${\scs 5}$, l.dist=1mm, l.angle=0}{i1}
\fmfv{decor.size=0, label=${\scs 4}$, l.dist=1mm, l.angle=-180}{o2}
\fmfv{decor.size=0, label=${\scs 3}$, l.dist=1mm, l.angle=-180}{o1}
\end{fmfgraph*}
\end{center}}
\hspace*{2mm} \dphi{\Gamma^{({\rm int})}}{5}{6} \hspace*{2mm} 
\no \\ &&
+ \frac{4}{3} 
\hspace*{1mm} \dphi{\Gamma^{({\rm int})}}{1}{2} \hspace*{2mm} 
\parbox{9mm}{\begin{center}
\begin{fmfgraph*}(6,7)
\setval
\fmfstraight
\fmfforce{0w,1h}{v1}
\fmfforce{2/3w,1h}{v2}
\fmfforce{1w,1h}{v3}
\fmfforce{2/3w,2/7h}{v4}
\fmfforce{1/3w,0h}{v5}
\fmfforce{1/3w,4/7h}{v6}
\fmfforce{1w,2/7h}{v7}
\fmf{boson}{v1,v2}
\fmf{plain}{v2,v3}
\fmf{plain}{v4,v7}
\fmf{plain}{v4,v5}
\fmf{plain}{v4,v6}
\fmf{plain}{v4,v2}
\fmfdot{v1,v2,v4}
\fmfv{decor.size=0, label=${\scs 2}$, l.dist=1mm, l.angle=-180}{v5}
\fmfv{decor.size=0, label=${\scs 1}$, l.dist=1mm, l.angle=-180}{v6}
\fmfv{decor.size=0, label=${\scs 3}$, l.dist=1mm, l.angle=0}{v3}
\fmfv{decor.size=0, label=${\scs 4}$, l.dist=1mm, l.angle=0}{v7}
\end{fmfgraph*}
\end{center}}
\hspace*{2mm} \dphi{\Gamma^{({\rm int})}}{3}{4} \hspace*{2mm} 
- \frac{2}{3} 
\parbox{9mm}{\begin{center}
\begin{fmfgraph*}(6,4)
\setval
\fmfstraight
\fmfforce{0w,1/2h}{v1}
\fmfforce{4/6w,1/2h}{v2}
\fmfforce{1w,1.25h}{i1}
\fmfforce{1w,1/2h}{i2}
\fmfforce{1w,-0.25h}{i3}
\fmf{boson}{v2,v1}
\fmf{plain}{v2,i1}
\fmf{plain}{v2,i2}
\fmf{plain}{v2,i3}
\fmfdot{v1,v2}
\fmfv{decor.size=0, label=${\scs 3}$, l.dist=1mm, l.angle=0}{i3}
\fmfv{decor.size=0, label=${\scs 2}$, l.dist=1mm, l.angle=0}{i2}
\fmfv{decor.size=0, label=${\scs 1}$, l.dist=1mm, l.angle=0}{i1}
\end{fmfgraph*}
\end{center}}
\hspace*{2mm} \ddvertex{\Gamma^{({\rm int})}}{1}{2}{3}{4}{5} \hspace*{2mm} 
\parbox{7mm}{\begin{center}
\begin{fmfgraph*}(4,4)
\setval
\fmfstraight
\fmfforce{0w,1h}{o1}
\fmfforce{0w,0h}{o2}
\fmfforce{1/2w,1/2h}{v1}
\fmfforce{1w,1h}{i1}
\fmfforce{1w,0h}{i2}
\fmf{plain}{v1,o1}
\fmf{plain}{v1,o2}
\fmf{plain}{v1,i1}
\fmf{plain}{v1,i2}
\fmfdot{v1}
\fmfv{decor.size=0, label=${\scs 7}$, l.dist=1mm, l.angle=0}{i2}
\fmfv{decor.size=0, label=${\scs 6}$, l.dist=1mm, l.angle=0}{i1}
\fmfv{decor.size=0, label=${\scs 5}$, l.dist=1mm, l.angle=-180}{o2}
\fmfv{decor.size=0, label=${\scs 4}$, l.dist=1mm, l.angle=-180}{o1}
\end{fmfgraph*}
\end{center}}
\hspace*{2mm} \dphi{\Gamma^{({\rm int})}}{6}{7} \hspace*{2mm} \la{FFF} 
\hspace*{4mm} .
\eeq
The effect of the left-hand side is to count the number of field 
expectation values plus two times the number of lines of each 
one-particle irreducible vacuum diagram. The right-hand side contains
altogether 32 terms, 7 without $\Gamma^{({\rm int})}$, 15 linear in
$\Gamma^{({\rm int})}$ and 10 bilinear in $\Gamma^{({\rm int})}$.
\end{fmffile}
\begin{fmffile}{fg8}
\subsection{Loopwise Recursive Graphical Solution}
Now we show how (\r{FFF}) is graphically solved. To this end we expand
the interaction effective energy $\Gamma^{({\rm int})}$ with respect to
the number $n$ of field expection values $\Phi$  and the loop order $l$ 
\beq
\la{DECCE}
- \Gamma^{({\rm int})} = \sum_{n=0}^{\infty} \sum_{l=0}^{\infty} 
\Gamma^{(n,l)} \, .
\eeq
Here we can exclude the combinations $(n,l)\in \{ (0,0), (0,1), (1,0),
(2,0)\}$ as the corresponding expansion coefficients $\Gamma^{(n,l)}$ turn
out to be zero. With the help of (\r{DECCE}) we convert (\r{FFF})
into a graphical recursion relation for the expansion coefficients
$\Gamma^{(n,l)}$. As an example we consider the  graphical recursion relation
for the one-particle irreducible vacuum diagrams without field expectation
values $\Gamma^{(0,l)}$.  
For $n=0$ and $l=2$, Eq.~(\r{FFF}) reduces to
\beq
\parbox{8mm}{\begin{center}
\begin{fmfgraph*}(2.5,5)
\setval
\fmfstraight
\fmfforce{1w,0h}{v1}
\fmfforce{1w,1h}{v2}
\fmf{plain,left=1}{v1,v2}
\fmfv{decor.size=0, label=${\scs 2}$, l.dist=1mm, l.angle=0}{v1}
\fmfv{decor.size=0, label=${\scs 1}$, l.dist=1mm, l.angle=0}{v2}
\end{fmfgraph*}
\end{center}}
\hspace*{0.3cm} \dphi{\Gamma^{(0,2)}}{1}{2} = 
\frac{1}{4}
\parbox{11mm}{\begin{center}
\begin{fmfgraph*}(8,4)
\setval
\fmfleft{i1}
\fmfright{o1}
\fmf{plain,left=1}{i1,v1,i1}
\fmf{plain,left=1}{o1,v1,o1}
\fmfdot{v1}
\end{fmfgraph*}\end{center}}
+ \frac{1}{4}
\parbox{7mm}{\begin{center}
\begin{fmfgraph*}(4,4)
\setval
\fmfforce{0w,0.5h}{v1}
\fmfforce{1w,0.5h}{v2}
\fmf{plain,left=1}{v1,v2,v1}
\fmf{plain}{v1,v2}
\fmfdot{v1,v2}
\end{fmfgraph*}\end{center}} \, ,
\eeq
which is immediately solved by
\beq
\Gamma^{(0,2)} = 
\frac{1}{8}
\parbox{11mm}{\begin{center}
\begin{fmfgraph*}(8,4)
\setval
\fmfleft{i1}
\fmfright{o1}
\fmf{plain,left=1}{i1,v1,i1}
\fmf{plain,left=1}{o1,v1,o1}
\fmfdot{v1}
\end{fmfgraph*}\end{center}}
+ \frac{1}{12}
\parbox{7mm}{\begin{center}
\begin{fmfgraph*}(4,4)
\setval
\fmfforce{0w,0.5h}{v1}
\fmfforce{1w,0.5h}{v2}
\fmf{plain,left=1}{v1,v2,v1}
\fmf{plain}{v1,v2}
\fmfdot{v1,v2}
\end{fmfgraph*}\end{center}} \, ,
\la{EW2}
\eeq
as the first vacuum diagram contains 2 and the second vacuum diagram
3 lines. For $n=0$ and $l\ge3$ we obtain the graphical recursion relation
\beq
&& 
\parbox{5.5mm}{\begin{center}
\begin{fmfgraph*}(2.5,5)
\setval
\fmfstraight
\fmfforce{1w,0h}{v1}
\fmfforce{1w,1h}{v2}
\fmf{plain,left=1}{v1,v2}
\fmfv{decor.size=0, label=${\scs 2}$, l.dist=1mm, l.angle=0}{v1}
\fmfv{decor.size=0, label=${\scs 1}$, l.dist=1mm, l.angle=0}{v2}
\end{fmfgraph*}
\end{center}}
\hspace*{0.3cm} \dphi{\Gamma^{(0,l)}}{1}{2} = 
\hspace*{1mm} 
\parbox{9mm}{\begin{center}
\begin{fmfgraph*}(6,4)
\setval
\fmfstraight
\fmfforce{0w,1/2h}{v1}
\fmfforce{4/6w,1/2h}{v2}
\fmfforce{1w,1h}{i2}
\fmfforce{1w,0h}{i1}
\fmf{plain}{i1,v2}
\fmf{plain}{v2,i2}
\fmf{plain,left}{v1,v2,v1}
\fmfdot{v2}
\fmfv{decor.size=0, label=${\scs 2}$, l.dist=1mm, l.angle=0}{i1}
\fmfv{decor.size=0, label=${\scs 1}$, l.dist=1mm, l.angle=0}{i2}
\end{fmfgraph*}
\end{center}}
\hspace*{3mm} \dphi{\Gamma^{(0,l-1)}}{1}{2} 
\hspace*{1mm} + \frac{1}{3} \hspace*{1mm}
\parbox{7mm}{\begin{center}
\begin{fmfgraph*}(3,3)
\setval
\fmfstraight
\fmfforce{0w,1/2h}{v1}
\fmfforce{1w,2h}{i1}
\fmfforce{1w,1h}{i2}
\fmfforce{1w,0h}{i3}
\fmfforce{1w,-1h}{i4}
\fmf{plain}{v1,i1}
\fmf{plain}{v1,i2}
\fmf{plain}{v1,i3}
\fmf{plain}{v1,i4}
\fmfdot{v1}
\fmfv{decor.size=0, label=${\scs 4}$, l.dist=1mm, l.angle=0}{i4}
\fmfv{decor.size=0, label=${\scs 3}$, l.dist=1mm, l.angle=0}{i3}
\fmfv{decor.size=0, label=${\scs 2}$, l.dist=1mm, l.angle=0}{i2}
\fmfv{decor.size=0, label=${\scs 1}$, l.dist=1mm, l.angle=0}{i1}
\end{fmfgraph*}
\end{center}}
\hspace*{3mm} \ddphi{\Gamma^{(0,l-1)}}{1}{2}{3}{4}
\hspace*{1mm} + \frac{3}{2} \hspace*{1mm}
\parbox{7mm}{\begin{center}
\begin{fmfgraph*}(6,4)
\setval
\fmfstraight
\fmfforce{1/3w,0h}{v1}
\fmfforce{1/3w,1h}{v2}
\fmfforce{1w,1h}{i2}
\fmfforce{1w,0h}{i1}
\fmf{plain}{i1,v1}
\fmf{plain}{v2,i2}
\fmf{plain,left}{v1,v2,v1}
\fmfdot{v2,v1}
\fmfv{decor.size=0, label=${\scs 2}$, l.dist=1mm, l.angle=0}{i1}
\fmfv{decor.size=0, label=${\scs 1}$, l.dist=1mm, l.angle=0}{i2}
\end{fmfgraph*}
\end{center}}
\hspace*{3mm} \dphi{\Gamma^{(0,l-1)}}{1}{2}
\hspace*{1mm} + \frac{1}{2} \hspace*{1mm}
\parbox{7mm}{\begin{center}
\begin{fmfgraph*}(3,9)
\setval
\fmfstraight
\fmfforce{1w,1h}{o1}
\fmfforce{1w,2/3h}{o2}
\fmfforce{0w,5/6h}{v1}
\fmfforce{0w,1/6h}{v2}
\fmfforce{1w,1/3h}{i1}
\fmfforce{1w,0h}{i2}
\fmf{plain}{v1,v2}
\fmf{plain}{v1,o1}
\fmf{plain}{v1,o2}
\fmf{plain}{v2,i1}
\fmf{plain}{v2,i2}
\fmfdot{v1,v2}
\fmfv{decor.size=0, label=${\scs 4}$, l.dist=1mm, l.angle=0}{i2}
\fmfv{decor.size=0, label=${\scs 3}$, l.dist=1mm, l.angle=0}{i1}
\fmfv{decor.size=0, label=${\scs 2}$, l.dist=1mm, l.angle=0}{o2}
\fmfv{decor.size=0, label=${\scs 1}$, l.dist=1mm, l.angle=0}{o1}
\end{fmfgraph*}
\end{center}}
\hspace*{3mm} \ddphi{\Gamma^{(0,l-1)}}{1}{2}{3}{4} 
\no \\ &&
+ \frac{1}{2} \hspace*{1mm} 
\parbox{10mm}{\begin{center}
\begin{fmfgraph*}(6,5.33333)
\setval
\fmfstraight
\fmfforce{2/6w,1/4h}{v1}
\fmfforce{2/6w,1h}{v2}
\fmfforce{1w,0h}{i1}
\fmfforce{1w,1/2h}{i2}
\fmfforce{1w,1h}{i3}
\fmf{plain,left=1}{v1,v2,v1}
\fmf{plain}{v1,i1}
\fmf{plain}{v1,i2}
\fmf{plain}{v2,i3}
\fmfdot{v1,v2}
\fmfv{decor.size=0, label=${\scs 2}$, l.dist=1mm, l.angle=0}{i2}
\fmfv{decor.size=0, label=${\scs 3}$, l.dist=1mm, l.angle=0}{i1}
\fmfv{decor.size=0, label=${\scs 1}$, l.dist=1mm, l.angle=0}{i3}
\end{fmfgraph*}
\end{center}}
\hspace*{3mm} \dvertex{\Gamma^{(0,l-1)}}{1}{2}{3} 
\hspace*{1mm} + \frac{1}{2} \hspace*{1mm}
\parbox{7mm}{\begin{center}
\begin{fmfgraph*}(3,12)
\setval
\fmfstraight
\fmfforce{1w,1h}{o0}
\fmfforce{1w,3/4h}{o1}
\fmfforce{1w,1/2h}{o2}
\fmfforce{0w,3/4h}{v1}
\fmfforce{0w,1/8h}{v2}
\fmfforce{1w,1/4h}{i1}
\fmfforce{1w,0h}{i2}
\fmf{plain}{v1,v2}
\fmf{plain}{v1,o0}
\fmf{plain}{v1,o1}
\fmf{plain}{v1,o2}
\fmf{plain}{v2,i1}
\fmf{plain}{v2,i2}
\fmfdot{v1,v2}
\fmfv{decor.size=0, label=${\scs 5}$, l.dist=1mm, l.angle=0}{i2}
\fmfv{decor.size=0, label=${\scs 4}$, l.dist=1mm, l.angle=0}{i1}
\fmfv{decor.size=0, label=${\scs 3}$, l.dist=1mm, l.angle=0}{o2}
\fmfv{decor.size=0, label=${\scs 2}$, l.dist=1mm, l.angle=0}{o1}
\fmfv{decor.size=0, label=${\scs 1}$, l.dist=1mm, l.angle=0}{o0}
\end{fmfgraph*}
\end{center}}
\hspace*{3mm} \ddvertex{\Gamma^{(0,l-1)}}{1}{2}{3}{4}{5}
\hspace*{1mm} + \sum_{l'=2}^{l-2} \left\{  \frac{2}{3} \hspace*{1mm}
\hspace*{1mm} \dphi{\Gamma^{(0,l')}}{1}{2}\hspace*{3mm}
\parbox{7mm}{\begin{center}
\begin{fmfgraph*}(4,4)
\setval
\fmfstraight
\fmfforce{0w,1h}{o1}
\fmfforce{0w,0h}{o2}
\fmfforce{1/2w,1/2h}{v1}
\fmfforce{1w,1h}{i1}
\fmfforce{1w,0h}{i2}
\fmf{plain}{v1,o1}
\fmf{plain}{v1,o2}
\fmf{plain}{v1,i1}
\fmf{plain}{v1,i2}
\fmfdot{v1}
\fmfv{decor.size=0, label=${\scs 4}$, l.dist=1mm, l.angle=0}{i2}
\fmfv{decor.size=0, label=${\scs 3}$, l.dist=1mm, l.angle=0}{i1}
\fmfv{decor.size=0, label=${\scs 2}$, l.dist=1mm, l.angle=-180}{o2}
\fmfv{decor.size=0, label=${\scs 1}$, l.dist=1mm, l.angle=-180}{o1}
\end{fmfgraph*}
\end{center}}
\hspace*{3mm} \dphi{\Gamma^{(0,l-l'-1)}}{3}{4} 
\right. \no \\ &&
+ 2 \hspace*{1mm}
\dphi{\Gamma^{(0,l')}}{1}{2} \hspace*{2mm} 
\parbox{7mm}{\begin{center}
\begin{fmfgraph*}(4,4)
\setval
\fmfstraight
\fmfforce{0w,1h}{o1}
\fmfforce{0w,0h}{o2}
\fmfforce{1/2w,0h}{v1}
\fmfforce{1/2w,1h}{v2}
\fmfforce{1w,1h}{i1}
\fmfforce{1w,0h}{i2}
\fmf{plain}{v1,v2}
\fmf{plain}{v1,i2}
\fmf{plain}{v1,o2}
\fmf{plain}{v2,i1}
\fmf{plain}{v2,o1}
\fmfdot{v1,v2}
\fmfv{decor.size=0, label=${\scs 4}$, l.dist=1mm, l.angle=0}{i2}
\fmfv{decor.size=0, label=${\scs 3}$, l.dist=1mm, l.angle=0}{i1}
\fmfv{decor.size=0, label=${\scs 2}$, l.dist=1mm, l.angle=-180}{o2}
\fmfv{decor.size=0, label=${\scs 1}$, l.dist=1mm, l.angle=-180}{o1}
\end{fmfgraph*}
\end{center}}
\hspace*{2mm} \dphi{\Gamma^{(0,l-l'-1)}}{3}{4}  
+ \hspace*{1mm}
\dphi{\Gamma^{(0,l')}}{1}{2} \hspace*{2mm} 
\parbox{9mm}{\begin{center}
\begin{fmfgraph*}(6,9)
\setval
\fmfstraight
\fmfforce{1w,1h}{o1}
\fmfforce{1w,2/3h}{o2}
\fmfforce{1/2w,5/6h}{v1}
\fmfforce{1/2w,1/6h}{v2}
\fmfforce{1w,1/3h}{i1}
\fmfforce{1w,0h}{i2}
\fmfforce{0w,5/6h}{z1}
\fmfforce{0w,1/6h}{z2}
\fmf{plain}{v1,o1}
\fmf{plain}{v1,o2}
\fmf{plain}{v2,i1}
\fmf{plain}{v2,i2}
\fmf{plain}{v1,z1}
\fmf{plain}{v2,z2}
\fmfdot{v1,v2}
\fmfv{decor.size=0, label=${\scs 2}$, l.dist=1mm, l.angle=-180}{z2}
\fmfv{decor.size=0, label=${\scs 1}$, l.dist=1mm, l.angle=-180}{z1}
\fmfv{decor.size=0, label=${\scs 6}$, l.dist=1mm, l.angle=0}{i2}
\fmfv{decor.size=0, label=${\scs 5}$, l.dist=1mm, l.angle=0}{i1}
\fmfv{decor.size=0, label=${\scs 4}$, l.dist=1mm, l.angle=0}{o2}
\fmfv{decor.size=0, label=${\scs 3}$, l.dist=1mm, l.angle=0}{o1}
\end{fmfgraph*}
\end{center}}
\hspace*{2mm} \ddphi{\Gamma^{(0,l-l'-1)}}{3}{4}{5}{6} \hspace*{2mm} 
\hspace*{1mm} + \hspace*{1mm}
\dphi{\Gamma^{(0,l')}}{1}{2} \hspace*{2mm} 
\parbox{9mm}{\begin{center}
\begin{fmfgraph*}(6,12)
\setval
\fmfstraight
\fmfforce{1w,1h}{o0}
\fmfforce{1w,3/4h}{o1}
\fmfforce{1w,1/2h}{o2}
\fmfforce{1/2w,3/4h}{v1}
\fmfforce{1/2w,1/8h}{v2}
\fmfforce{1w,1/4h}{i1}
\fmfforce{1w,0h}{i2}
\fmfforce{0w,3/4h}{z1}
\fmfforce{0w,1/8h}{z2}
\fmf{plain}{z1,v1}
\fmf{plain}{z2,v2}
\fmf{plain}{v1,o0}
\fmf{plain}{v1,o1}
\fmf{plain}{v1,o2}
\fmf{plain}{v2,i1}
\fmf{plain}{v2,i2}
\fmfdot{v1,v2}
\fmfv{decor.size=0, label=${\scs 7}$, l.dist=1mm, l.angle=0}{i2}
\fmfv{decor.size=0, label=${\scs 6}$, l.dist=1mm, l.angle=0}{i1}
\fmfv{decor.size=0, label=${\scs 5}$, l.dist=1mm, l.angle=0}{o2}
\fmfv{decor.size=0, label=${\scs 4}$, l.dist=1mm, l.angle=0}{o1}
\fmfv{decor.size=0, label=${\scs 3}$, l.dist=1mm, l.angle=0}{o0}
\fmfv{decor.size=0, label=${\scs 2}$, l.dist=1mm, l.angle=-180}{z2}
\fmfv{decor.size=0, label=${\scs 1}$, l.dist=1mm, l.angle=-180}{z1}
\end{fmfgraph*}
\end{center}}
\hspace*{2mm} \ddvertex{\Gamma^{(0,l-l'-1)}}{3}{4}{5}{6}{7} 
\no \\ && \left.
+ 2
\hspace*{1mm} \dphi{\Gamma^{(0,l')}}{1}{2} \hspace*{1mm}
\parbox{10mm}{\begin{center}
\begin{fmfgraph*}(8,5.33333)
\setval
\fmfstraight
\fmfforce{4/8w,1/4h}{v1}
\fmfforce{4/8w,1h}{v2}
\fmfforce{1w,0h}{i1}
\fmfforce{1w,1/2h}{i2}
\fmfforce{1w,1h}{i3}
\fmfforce{0w,1/4h}{z1}
\fmfforce{0w,1h}{z2}
\fmf{plain}{v1,z1}
\fmf{plain}{z2,v2}
\fmf{plain}{v1,v2}
\fmf{plain}{v1,i1}
\fmf{plain}{v1,i2}
\fmf{plain}{v2,i3}
\fmfdot{v1,v2}
\fmfv{decor.size=0, label=${\scs 4}$, l.dist=1mm, l.angle=0}{i2}
\fmfv{decor.size=0, label=${\scs 5}$, l.dist=1mm, l.angle=0}{i1}
\fmfv{decor.size=0, label=${\scs 3}$, l.dist=1mm, l.angle=0}{i3}
\fmfv{decor.size=0, label=${\scs 2}$, l.dist=1mm, l.angle=-180}{z1}
\fmfv{decor.size=0, label=${\scs 1}$, l.dist=1mm, l.angle=-180}{z2}
\end{fmfgraph*}
\end{center}}
\hspace*{1mm} \dvertex{\Gamma^{(0,l-l'-1)}}{3}{4}{5} \hspace*{1mm} 
+  
\hspace*{1mm} \dvertex{\Gamma^{(0,l')}}{1}{2}{3} \hspace*{1mm} 
\parbox{9mm}{\begin{center}
\begin{fmfgraph*}(6,9)
\setval
\fmfstraight
\fmfforce{1w,1h}{o1}
\fmfforce{1w,2/3h}{o2}
\fmfforce{1/2w,5/6h}{v1}
\fmfforce{1/2w,1/6h}{v2}
\fmfforce{1w,1/3h}{i1}
\fmfforce{1w,0h}{i2}
\fmfforce{0w,5/6h}{z1}
\fmfforce{0w,1/3h}{z2}
\fmfforce{0w,0h}{z3}
\fmf{plain}{v1,o1}
\fmf{plain}{v1,o2}
\fmf{plain}{v2,i1}
\fmf{plain}{v2,i2}
\fmf{plain}{v1,z1}
\fmf{plain}{v2,z2}
\fmf{plain}{v2,z3}
\fmfdot{v1,v2}
\fmfv{decor.size=0, label=${\scs 3}$, l.dist=1mm, l.angle=-180}{z3}
\fmfv{decor.size=0, label=${\scs 2}$, l.dist=1mm, l.angle=-180}{z2}
\fmfv{decor.size=0, label=${\scs 1}$, l.dist=1mm, l.angle=-180}{z1}
\fmfv{decor.size=0, label=${\scs 7}$, l.dist=1mm, l.angle=0}{i2}
\fmfv{decor.size=0, label=${\scs 6}$, l.dist=1mm, l.angle=0}{i1}
\fmfv{decor.size=0, label=${\scs 5}$, l.dist=1mm, l.angle=0}{o2}
\fmfv{decor.size=0, label=${\scs 4}$, l.dist=1mm, l.angle=0}{o1}
\end{fmfgraph*}
\end{center}}
\hspace*{1mm} \ddphi{\Gamma^{(0,l-l'-1)}}{4}{5}{6}{7} \hspace*{1mm} 
+  
\hspace*{1mm} \dvertex{\Gamma^{(0,l')}}{1}{2}{3} \hspace*{1mm} 
\parbox{9mm}{\begin{center}
\begin{fmfgraph*}(6,12)
\setval
\fmfstraight
\fmfforce{1w,1h}{o0}
\fmfforce{1w,3/4h}{o1}
\fmfforce{1w,1/2h}{o2}
\fmfforce{1/2w,3/4h}{v1}
\fmfforce{1/2w,1/8h}{v2}
\fmfforce{1w,1/4h}{i1}
\fmfforce{1w,0h}{i2}
\fmfforce{0w,3/4h}{z1}
\fmfforce{0w,1/4h}{z2}
\fmfforce{0w,0h}{z3}
\fmf{plain}{z1,v1}
\fmf{plain}{z2,v2}
\fmf{plain}{z3,v2}
\fmf{plain}{v1,o0}
\fmf{plain}{v1,o1}
\fmf{plain}{v1,o2}
\fmf{plain}{v2,i1}
\fmf{plain}{v2,i2}
\fmfdot{v1,v2}
\fmfv{decor.size=0, label=${\scs 8}$, l.dist=1mm, l.angle=0}{i2}
\fmfv{decor.size=0, label=${\scs 7}$, l.dist=1mm, l.angle=0}{i1}
\fmfv{decor.size=0, label=${\scs 6}$, l.dist=1mm, l.angle=0}{o2}
\fmfv{decor.size=0, label=${\scs 5}$, l.dist=1mm, l.angle=0}{o1}
\fmfv{decor.size=0, label=${\scs 4}$, l.dist=1mm, l.angle=0}{o0}
\fmfv{decor.size=0, label=${\scs 2}$, l.dist=1mm, l.angle=-180}{z2}
\fmfv{decor.size=0, label=${\scs 1}$, l.dist=1mm, l.angle=-180}{z1}
\fmfv{decor.size=0, label=${\scs 3}$, l.dist=1mm, l.angle=-180}{z3}
\end{fmfgraph*}
\end{center}}
\hspace*{2mm} \ddvertex{\Gamma^{(0,l-l'-1)}}{4}{5}{6}{7}{8} \right\} \,.
\la{VACU}
\eeq
We observe that for either a vanishing cubic or quartic interaction this
graphical recursion relation only involves the graphical
operation of removing lines.
Proceeding to the loop order $l=3$, we have to evaluate from the vacuum 
diagrams (\r{EW2}) a one-line amputation
\beq
\la{ENNR1}
\dphi{\Gamma^{(0,2)}}{1}{2} = \frac{1}{4} 
\parbox{9mm}{\begin{center}
\begin{fmfgraph*}(6,4)
\setval
\fmfstraight
\fmfforce{0w,1/2h}{v1}
\fmfforce{4/6w,1/2h}{v2}
\fmfforce{1w,1h}{i2}
\fmfforce{1w,0h}{i1}
\fmf{plain}{i1,v2}
\fmf{plain}{v2,i2}
\fmf{plain,left}{v1,v2,v1}
\fmfdot{v2}
\fmfv{decor.size=0, label=${\scs 2}$, l.dist=1mm, l.angle=0}{i1}
\fmfv{decor.size=0, label=${\scs 1}$, l.dist=1mm, l.angle=0}{i2}
\end{fmfgraph*}
\end{center}}
\hspace*{1mm} + \frac{1}{4} \hspace*{1mm}
\parbox{7mm}{\begin{center}
\begin{fmfgraph*}(6,4)
\setval
\fmfstraight
\fmfforce{1/3w,0h}{v1}
\fmfforce{1/3w,1h}{v2}
\fmfforce{1w,1h}{i2}
\fmfforce{1w,0h}{i1}
\fmf{plain}{i1,v1}
\fmf{plain}{v2,i2}
\fmf{plain,left}{v1,v2,v1}
\fmfdot{v2,v1}
\fmfv{decor.size=0, label=${\scs 2}$, l.dist=1mm, l.angle=0}{i1}
\fmfv{decor.size=0, label=${\scs 1}$, l.dist=1mm, l.angle=0}{i2}
\end{fmfgraph*}
\end{center}}\hspace*{4mm} , 
\eeq
a two-line amputation
\beq
\la{ENNR2}
\ddphi{\Gamma^{(0,2)}}{1}{2}{3}{4} = \hspace*{1mm}\frac{1}{4} \hspace*{2mm}
\parbox{7mm}{\begin{center}
\begin{fmfgraph*}(4,4)
\setval
\fmfstraight
\fmfforce{0w,1h}{o1}
\fmfforce{0w,0h}{o2}
\fmfforce{1/2w,1/2h}{v1}
\fmfforce{1w,1h}{i1}
\fmfforce{1w,0h}{i2}
\fmf{plain}{v1,o1}
\fmf{plain}{v1,o2}
\fmf{plain}{v1,i1}
\fmf{plain}{v1,i2}
\fmfdot{v1}
\fmfv{decor.size=0, label=${\scs 4}$, l.dist=1mm, l.angle=0}{i2}
\fmfv{decor.size=0, label=${\scs 3}$, l.dist=1mm, l.angle=0}{i1}
\fmfv{decor.size=0, label=${\scs 2}$, l.dist=1mm, l.angle=-180}{o2}
\fmfv{decor.size=0, label=${\scs 1}$, l.dist=1mm, l.angle=-180}{o1}
\end{fmfgraph*}
\end{center}}
\hspace*{1mm} + \hspace*{2mm}\frac{1}{4} \hspace*{2mm}
\parbox{11mm}{\begin{center}
\begin{fmfgraph*}(8,4)
\setval
\fmfstraight
\fmfforce{0w,1h}{o1}
\fmfforce{0w,0h}{o2}
\fmfforce{1/4w,1/2h}{v1}
\fmfforce{3/4w,1/2h}{v2}
\fmfforce{1w,1h}{i1}
\fmfforce{1w,0h}{i2}
\fmf{plain}{v1,v2}
\fmf{plain}{v1,o1}
\fmf{plain}{v1,o2}
\fmf{plain}{v2,i1}
\fmf{plain}{v2,i2}
\fmfdot{v1,v2}
\fmfv{decor.size=0, label=${\scs 3}$, l.dist=1mm, l.angle=0}{i2}
\fmfv{decor.size=0, label=${\scs 2}$, l.dist=1mm, l.angle=0}{i1}
\fmfv{decor.size=0, label=${\scs 4}$, l.dist=1mm, l.angle=-180}{o2}
\fmfv{decor.size=0, label=${\scs 1}$, l.dist=1mm, l.angle=-180}{o1}
\end{fmfgraph*}
\end{center}} 
\hspace*{1mm}+ \hspace*{2mm}\frac{1}{4} \hspace*{2mm}
\parbox{11mm}{\begin{center}
\begin{fmfgraph*}(8,4)
\setval
\fmfstraight
\fmfforce{0w,1h}{o1}
\fmfforce{0w,0h}{o2}
\fmfforce{1/4w,1/2h}{v1}
\fmfforce{3/4w,1/2h}{v2}
\fmfforce{1w,1h}{i1}
\fmfforce{1w,0h}{i2}
\fmf{plain}{v1,v2}
\fmf{plain}{v1,o1}
\fmf{plain}{v1,o2}
\fmf{plain}{v2,i1}
\fmf{plain}{v2,i2}
\fmfdot{v1,v2}
\fmfv{decor.size=0, label=${\scs 4}$, l.dist=1mm, l.angle=0}{i2}
\fmfv{decor.size=0, label=${\scs 2}$, l.dist=1mm, l.angle=0}{i1}
\fmfv{decor.size=0, label=${\scs 3}$, l.dist=1mm, l.angle=-180}{o2}
\fmfv{decor.size=0, label=${\scs 1}$, l.dist=1mm, l.angle=-180}{o1}
\end{fmfgraph*}
\end{center}} \hspace*{4mm},
\eeq
a 3-vertex amputation
\beq
\la{ENNR3}
\dvertex{\Gamma^{(0,2)}}{1}{2}{3} = 
\frac{1}{6} \hspace*{2mm}
\parbox{8mm}{\centerline{
\begin{fmfgraph*}(5,4.33)
\setval
\fmfforce{1w,0h}{v1}
\fmfforce{0w,0h}{v2}
\fmfforce{0.5w,1h}{v3}
\fmfforce{0.5w,0.2886h}{vm}
\fmf{plain}{v1,vm,v2}
\fmf{plain}{v3,vm}
\fmfv{decor.size=0,label={\footnotesize 2},l.dist=0.5mm}{v1}
\fmfv{decor.size=0,label={\footnotesize 3},l.dist=0.5mm}{v2}
\fmfv{decor.size=0,label={\footnotesize 1},l.dist=0.5mm}{v3}
\fmfdot{vm}
\end{fmfgraph*}}}
\eeq
and the amputation of one line and one 3-vertex
\beq
\la{ENNR4}
\ddvertex{\Gamma^{(0,2)}}{1}{2}{3}{4}{5} & = &
\frac{1}{12}  \hspace*{2mm} 
{\displaystyle \left\{ \,
\parbox{8mm}{\centerline{
\begin{fmfgraph*}(5,18.33)
\setval
\fmfforce{1w,2/18.33h}{v1}
\fmfforce{0w,2/18.33h}{v2}
\fmfforce{0.5w,6.33/18.33h}{v3}
\fmfforce{0.5w,3.25/18.33h}{vm}
\fmfforce{0.5w,12.33/18.33h}{v4}
\fmfforce{0.5w,14.33/18.33h}{v5}
\fmfforce{0.5w,16.33/18.33h}{v6}
\fmf{plain}{v1,vm,v2}
\fmf{plain}{v3,vm}
\fmf{plain}{v4,v5}
\fmf{plain}{v6,v5}
\fmfv{decor.size=0,label={\footnotesize 2},l.dist=0.5mm, l.angle=-30}{v1}
\fmfv{decor.size=0,label={\footnotesize 3},l.dist=0.5mm, l.angle=-150}{v2}
\fmfv{decor.size=0,label={\footnotesize 5},l.dist=0.5mm, l.angle=90}{v3}
\fmfv{decor.size=0,label={\footnotesize 1},l.dist=0.5mm, l.angle=90}{v6}
\fmfv{decor.size=0,label={\footnotesize 4},l.dist=0.5mm, l.angle=-90}{v4}
\fmfv{decor.shape=circle,decor.filled=empty,decor.size=0.6mm}{v5}
\fmfdot{vm}
\end{fmfgraph*}}}
+ \,5\, \mbox{perm.} \right\} } \, .
\eeq
With this we obtain from Eq.~(\r{VACU}) for $l=3$
\beq
\parbox{5.5mm}{\begin{center}
\begin{fmfgraph*}(2.5,5)
\setval
\fmfstraight
\fmfforce{1w,0h}{v1}
\fmfforce{1w,1h}{v2}
\fmf{plain,left=1}{v1,v2}
\fmfv{decor.size=0, label=${\scs 2}$, l.dist=1mm, l.angle=0}{v1}
\fmfv{decor.size=0, label=${\scs 1}$, l.dist=1mm, l.angle=0}{v2}
\end{fmfgraph*}
\end{center}}
\hspace*{0.3cm} \dphi{\Gamma^{(0,3)}}{1}{2}  &=& 
\hspace*{1mm} \frac{1}{4}
\parbox{9mm}{\begin{center}
\begin{fmfgraph}(4,4)
\setval
\fmfforce{0w,0h}{v1}
\fmfforce{1w,0h}{v2}
\fmfforce{1w,1h}{v3}
\fmfforce{0w,1h}{v4}
\fmf{plain,right=1}{v1,v3,v1}
\fmf{plain}{v1,v3}
\fmf{plain}{v2,v4}
\fmfdot{v1,v2,v3,v4}
\end{fmfgraph}\end{center}} 
\hspace*{1mm} + \frac{3}{8} \hspace*{1mm}
\parbox{9mm}{\begin{center}
\begin{fmfgraph}(4,4)
\setval
\fmfforce{0w,0h}{v1}
\fmfforce{1w,0h}{v2}
\fmfforce{1w,1h}{v3}
\fmfforce{0w,1h}{v4}
\fmf{plain,right=1}{v1,v3,v1}
\fmf{plain,right=0.4}{v1,v4}
\fmf{plain,left=0.4}{v2,v3}
\fmfdot{v1,v2,v3,v4}
\end{fmfgraph}\end{center}} 
\hspace*{1mm} + \frac{5}{8} \hspace*{1mm} 
\parbox{9mm}{\begin{center}
\begin{fmfgraph}(6,6)
\setval
\fmfforce{0w,1/2h}{v1}
\fmfforce{1w,1/2h}{v2}
\fmfforce{1/2w,1h}{v3}
\fmf{plain,left=1}{v1,v2,v1}
\fmf{plain,left=0.4}{v3,v1}
\fmf{plain,left=0.4}{v2,v3}
\fmfdot{v1,v2,v3}
\end{fmfgraph}\end{center}}
\hspace*{1mm} + \frac{5}{8} \hspace*{1mm} 
\parbox{11mm}{\begin{center}
\begin{fmfgraph}(4,8)
\setval
\fmfforce{0w,1/4h}{v1}
\fmfforce{1w,1/4h}{v2}
\fmfforce{1/2w,1/2h}{v3}
\fmfforce{1/2w,1h}{v4}
\fmf{plain,left=1}{v1,v2,v1}
\fmf{plain,left=1}{v3,v4,v3}
\fmf{plain}{v1,v2}
\fmfdot{v2,v3,v1}
\end{fmfgraph}\end{center}} 
\hspace*{1mm}+ \frac{1}{12} \hspace*{1mm}
\parbox{9mm}{\begin{center}
\begin{fmfgraph}(6,4)
\setval
\fmfforce{0w,0.5h}{v1}
\fmfforce{1w,0.5h}{v2}
\fmf{plain,left=1}{v1,v2,v1}
\fmf{plain,left=0.4}{v1,v2,v1}
\fmfdot{v1,v2}
\end{fmfgraph}\end{center}} 
\hspace*{1mm}+ \frac{1}{4} \hspace*{1mm}
\parbox{15mm}{\begin{center}
\begin{fmfgraph}(12,4)
\setval
\fmfleft{i1}
\fmfright{o1}
\fmf{plain,left=1}{i1,v1,i1}
\fmf{plain,left=1}{v1,v2,v1}
\fmf{plain,left=1}{o1,v2,o1}
\fmfdot{v1,v2}
\end{fmfgraph}\end{center}} \hspace*{2mm},
\eeq
which leads to the one-particle irreducible 
vacuum diagrams listed in Table III together
with the subsequent loop order $l=4$. In a similar way, the graphical
relation (\r{FFF}) is iterated to construct the 
one-particle irreducible vacuum
diagrams which involve field expectation values. Table IV 
depicts the resulting
diagrams for the respective first two loop orders with $n=1,2,3,4$.
\subsection{Simpler Recursion Relations}
The graphical relation (\r{FFF}) allows us in principle to 
construct all one-particle
irreducible vacuum diagrams contributing to the interaction
effective energy $\Gamma^{({\rm int})}$. However, the iteration of (\r{FFF})
is in practice a tedious task, 
as quite often different terms lead to the same topological diagram so
that the corresponding contributions pile up to its proper weight.
Thus it would be advantageous to obtain simpler graphical 
relations for certain subsets of one-particle irreducible 
vacuum diagrams. 
In the following we aim at deriving graphical recursion relations which
rely on counting graphical elements of diagrams such as the
field expectation value, the 3- and the 4-vertices.
\subsubsection{Counting Field Expectation Values}
Applying (\r{NR1}) and (\r{EDEC}) to
Eq.~(\r{EFL2}), we immediately obtain a simple linear functional differential
equation similar to Eq.~(\r{CWW})
\beq
\int_1 \Phi_1 \frac{\delta \Gamma^{({\rm int})}}{\delta \Phi_1} &= &
\frac{1}{2} \int_{123} K_{123} G_{12} \Phi_3
+ \frac{1}{2} \int_{123} K_{123} \Phi_1 \Phi_2 \Phi_3
\nonumber \\
& & - \int_{12345} K_{123} G_{14} G_{25} \Phi_3 \frac{\delta
\Gamma^{({\rm int})}}{\delta G_{45}} 
+ \int_{1234} L_{1234} \Phi_1 \frac{\delta \Gamma^{({\rm int})}}{\delta
K_{234}} \hspace*{0.4cm} ,
\eeq
where the term on the left-hand side counts the number of field expectation
values in each
one-particle irreducible 
vacuum diagram. It can be diagramatically written as follows
\beq
\parbox{8mm}{\begin{center}
\begin{fmfgraph*}(5,5)
\setval
\fmfstraight
\fmfforce{0w,1/2h}{v1}
\fmfforce{1w,1/2h}{v2}
\fmf{boson}{v1,v2}
\fmfdot{v1}
\fmfv{decor.size=0, label=${\scs 1}$, l.dist=1mm, l.angle=0}{v2}
\end{fmfgraph*}
\end{center}}
\hspace*{2mm} \fdphi{\Gamma^{({\rm int})}}{1} =  - \frac{1}{2}
\parbox{11mm}{\begin{center}
\begin{fmfgraph*}(8,4)
\setval
\fmfforce{0w,1/2h}{v1}
\fmfforce{1/2w,1/2h}{v2}
\fmfforce{1w,1/2h}{v3}
\fmf{plain,left=1}{v2,v3,v2}
\fmf{boson}{v1,v2}
\fmfdot{v2,v1}
\end{fmfgraph*}\end{center}}
- \frac{1}{2} 
\parbox{11mm}{\begin{center}
\begin{fmfgraph*}(6.928,12)
\setval
\fmfforce{1/2w,5/6h}{v1}
\fmfforce{1w,1/4h}{w1}
\fmfforce{0w,1/4h}{u1}
\fmfforce{1/2w,1/2h}{v2}
\fmf{boson}{v2,v1}
\fmf{boson}{v2,w1}
\fmf{boson}{v2,u1}
\fmfdot{v2,v1,w1,u1}
\end{fmfgraph*}\end{center}}
+ 
\parbox{9mm}{\begin{center}
\begin{fmfgraph*}(6,4)
\setval
\fmfstraight
\fmfforce{0w,1/2h}{v1}
\fmfforce{4/6w,1/2h}{v2}
\fmfforce{1w,1h}{i1}
\fmfforce{1w,0h}{i2}
\fmf{boson}{v2,v1}
\fmf{plain}{v2,i1}
\fmf{plain}{v2,i2}
\fmfdot{v2,v1}
\fmfv{decor.size=0, label=${\scs 2}$, l.dist=1mm, l.angle=0}{i2}
\fmfv{decor.size=0, label=${\scs 1}$, l.dist=1mm, l.angle=0}{i1}
\end{fmfgraph*}
\end{center}}
\hspace*{3mm} \dphi{\Gamma^{({\rm int})}}{1}{2}
+
\parbox{9mm}{\begin{center}
\begin{fmfgraph*}(6,4)
\setval
\fmfstraight
\fmfforce{0w,1/2h}{v1}
\fmfforce{4/6w,1/2h}{v2}
\fmfforce{1w,1.25h}{i1}
\fmfforce{1w,1/2h}{i2}
\fmfforce{1w,-0.25h}{i3}
\fmf{boson}{v2,v1}
\fmf{plain}{v2,i1}
\fmf{plain}{v2,i2}
\fmf{plain}{v2,i3}
\fmfdot{v2,v1}
\fmfv{decor.size=0, label=${\scs 3}$, l.dist=1mm, l.angle=0}{i3}
\fmfv{decor.size=0, label=${\scs 2}$, l.dist=1mm, l.angle=0}{i2}
\fmfv{decor.size=0, label=${\scs 1}$, l.dist=1mm, l.angle=0}{i1}
\end{fmfgraph*}
\end{center}}
\hspace*{3mm} \dvertex{\Gamma^{({\rm int})}}{1}{2}{3} \hspace*{4mm} ,
\la{CA1}
\eeq
where the right-hand side contains only 4 terms, 2 without 
$\Gamma^{({\rm int})}$ and 2 linear in $\Gamma^{({\rm int})}$. 
With the decomposition (\r{DECCE}), the graphical relation which 
is derived from (\r{CA1}) reads for $n > 0$
\beq
\Gamma^{(n,l)} = \frac{1}{n} {\displaystyle \left\{ 
\parbox{9mm}{\begin{center}
\begin{fmfgraph*}(6,4)
\setval
\fmfstraight
\fmfforce{0w,1/2h}{v1}
\fmfforce{4/6w,1/2h}{v2}
\fmfforce{1w,1h}{i1}
\fmfforce{1w,0h}{i2}
\fmf{boson}{v2,v1}
\fmf{plain}{v2,i1}
\fmf{plain}{v2,i2}
\fmfdot{v1,v2}
\fmfv{decor.size=0, label=${\scs 2}$, l.dist=1mm, l.angle=0}{i2}
\fmfv{decor.size=0, label=${\scs 1}$, l.dist=1mm, l.angle=0}{i1}
\end{fmfgraph*}
\end{center}}
\hspace*{3mm} \dphi{\Gamma^{(n-1,l)}}{1}{2}
+
\parbox{9mm}{\begin{center}
\begin{fmfgraph*}(6,4)
\setval
\fmfstraight
\fmfforce{0w,1/2h}{v1}
\fmfforce{4/6w,1/2h}{v2}
\fmfforce{1w,1.25h}{i1}
\fmfforce{1w,1/2h}{i2}
\fmfforce{1w,-0.25h}{i3}
\fmf{boson}{v2,v1}
\fmf{plain}{v2,i1}
\fmf{plain}{v2,i2}
\fmf{plain}{v2,i3}
\fmfdot{v2,v1}
\fmfv{decor.size=0, label=${\scs 3}$, l.dist=1mm, l.angle=0}{i3}
\fmfv{decor.size=0, label=${\scs 2}$, l.dist=1mm, l.angle=0}{i2}
\fmfv{decor.size=0, label=${\scs 1}$, l.dist=1mm, l.angle=0}{i1}
\end{fmfgraph*}
\end{center}}
\hspace*{3mm} \dvertex{\Gamma^{(n-1,l)}}{1}{2}{3}
\right\} }
\eeq
which is iterated starting from
\beq
\Gamma^{(1,1)} = \frac{1}{2}
\parbox{11mm}{\begin{center}
\begin{fmfgraph*}(8,4)
\setval
\fmfforce{0w,1/2h}{v1}
\fmfforce{1/2w,1/2h}{v2}
\fmfforce{1w,1/2h}{v3}
\fmf{plain,left=1}{v2,v3,v2}
\fmf{boson}{v1,v2}
\fmfdot{v2,v1}
\end{fmfgraph*}\end{center}}
\, , \hspace*{1cm} \Gamma^{(3,0)} = \frac{1}{6}
\parbox{11mm}{\begin{center}
\begin{fmfgraph*}(6.928,12)
\setval
\fmfforce{1/2w,5/6h}{v1}
\fmfforce{1w,1/4h}{w1}
\fmfforce{0w,1/4h}{u1}
\fmfforce{1/2w,1/2h}{v2}
\fmf{boson}{v2,v1}
\fmf{boson}{v2,w1}
\fmf{boson}{v2,u1}
\fmfdot{v2,v1,w1,u1}
\end{fmfgraph*}\end{center}}
\eeq
and the one-particle irreducible 
vacuum diagrams without field expectation values $\Gamma^{(0,l)}$. As a first
example, we consider the case $n=1$ and $l=2$, insert (\r{ENNR1}), 
(\r{ENNR3}) into 
\beq
\Gamma^{(1,2)} =  
\parbox{9mm}{\begin{center}
\begin{fmfgraph*}(6,4)
\setval
\fmfstraight
\fmfforce{0w,1/2h}{v1}
\fmfforce{4/6w,1/2h}{v2}
\fmfforce{1w,1h}{i1}
\fmfforce{1w,0h}{i2}
\fmf{boson}{v2,v1}
\fmf{plain}{v2,i1}
\fmf{plain}{v2,i2}
\fmfdot{v1,v2}
\fmfv{decor.size=0, label=${\scs 2}$, l.dist=1mm, l.angle=0}{i2}
\fmfv{decor.size=0, label=${\scs 1}$, l.dist=1mm, l.angle=0}{i1}
\end{fmfgraph*}
\end{center}}
\hspace*{3mm} \dphi{\Gamma^{(0,2)}}{1}{2}
+
\parbox{9mm}{\begin{center}
\begin{fmfgraph*}(6,4)
\setval
\fmfstraight
\fmfforce{0w,1/2h}{v1}
\fmfforce{4/6w,1/2h}{v2}
\fmfforce{1w,1.25h}{i1}
\fmfforce{1w,1/2h}{i2}
\fmfforce{1w,-0.25h}{i3}
\fmf{boson}{v2,v1}
\fmf{plain}{v2,i1}
\fmf{plain}{v2,i2}
\fmf{plain}{v2,i3}
\fmfdot{v2,v1}
\fmfv{decor.size=0, label=${\scs 3}$, l.dist=1mm, l.angle=0}{i3}
\fmfv{decor.size=0, label=${\scs 2}$, l.dist=1mm, l.angle=0}{i2}
\fmfv{decor.size=0, label=${\scs 1}$, l.dist=1mm, l.angle=0}{i1}
\end{fmfgraph*}
\end{center}}
\hspace*{3mm} \dvertex{\Gamma^{(0,2)}}{1}{2}{3}
\eeq
and thus obtain
\beq
\Gamma^{(1,2)} = \frac{1}{4} \hspace*{1mm}
\parbox{11mm}{\begin{center}
\begin{fmfgraph*}(8,4)
\setval
\fmfforce{0w,1/2h}{v1}
\fmfforce{1/2w,1/2h}{v2}
\fmfforce{1w,1/2h}{v3}
\fmfforce{3/4w,0h}{v4}
\fmfforce{3/4w,1h}{v5}
\fmf{plain,left=1}{v2,v3,v2}
\fmf{boson}{v1,v2}
\fmf{plain}{v4,v5}
\fmfdot{v1,v2,v4,v5}
\end{fmfgraph*}\end{center}}
\hspace*{1mm}+ \frac{1}{6} \hspace*{1mm}
\parbox{11mm}{\begin{center}
\begin{fmfgraph*}(8,4)
\setval
\fmfforce{0w,1/2h}{v1}
\fmfforce{1/2w,1/2h}{v2}
\fmfforce{1w,1/2h}{v3}
\fmf{plain,left=1}{v2,v3,v2}
\fmf{boson}{v1,v2}
\fmf{plain}{v2,v3}
\fmfdot{v1,v2,v3}
\end{fmfgraph*}\end{center}}
\hspace*{1mm}+ \frac{1}{4} \hspace*{1mm}
\parbox{15mm}{\begin{center}
\begin{fmfgraph*}(12,4)
\setval
\fmfforce{0w,1/2h}{v1}
\fmfforce{1/3w,1/2h}{v2}
\fmfforce{2/3w,1/2h}{v3}
\fmfforce{1w,1/2h}{v4}
\fmf{plain,left=1}{v2,v3,v2}
\fmf{plain,left=1}{v3,v4,v3}
\fmf{boson}{v1,v2}
\fmfdot{v1,v2,v3}
\end{fmfgraph*}\end{center}}
\hspace*{4mm} .
\eeq
In the second example we set $n=2$ and $l=1$, evaluate from $\Gamma^{(1,1)}$  
a line as well as a 3-vertex amputation
\beq
\dphi{\Gamma^{(1,1)}}{1}{2} = \frac{1}{2} \hspace*{1mm}
\parbox{9mm}{\begin{center}
\begin{fmfgraph*}(6,4)
\setval
\fmfstraight
\fmfforce{0w,1/2h}{v1}
\fmfforce{4/6w,1/2h}{v2}
\fmfforce{1w,1h}{i1}
\fmfforce{1w,0h}{i2}
\fmf{boson}{v2,v1}
\fmf{plain}{v2,i1}
\fmf{plain}{v2,i2}
\fmfdot{v2,v1}
\fmfv{decor.size=0, label=${\scs 2}$, l.dist=1mm, l.angle=0}{i2}
\fmfv{decor.size=0, label=${\scs 1}$, l.dist=1mm, l.angle=0}{i1}
\end{fmfgraph*}
\end{center}}
\hspace*{3mm} \, , \hspace*{5mm}
\dvertex{\Gamma^{(1,1)}}{1}{2}{3} = \frac{1}{6} {\displaystyle \left\{
\parbox{8mm}{\centerline{
\begin{fmfgraph*}(4,4)
\setval
\fmfforce{0w,1h}{v1}
\fmfforce{1w,1h}{v2}
\fmfforce{0w,0h}{v3}
\fmfforce{1w,0h}{v4}
\fmf{plain}{v2,v1}
\fmf{boson}{v4,v3}
\fmfdot{v3}
\fmfv{decor.size=0,label={\footnotesize 1},l.dist=0.5mm, l.angle=-180}{v1}
\fmfv{decor.size=0,label={\footnotesize 2},l.dist=0.5mm, l.angle=0}{v2}
\fmfv{decor.size=0,label={\footnotesize 3},l.dist=0.5mm, l.angle=0}{v4}
\end{fmfgraph*}}}
\hspace*{2mm} + \, 2 \, \mbox{perm.} \right\} }
\eeq
and insert this into
\beq
\Gamma^{(2,1)} = \frac{1}{2} {\displaystyle \left\{ 
\parbox{9mm}{\begin{center}
\begin{fmfgraph*}(6,4)
\setval
\fmfstraight
\fmfforce{0w,1/2h}{v1}
\fmfforce{4/6w,1/2h}{v2}
\fmfforce{1w,1h}{i1}
\fmfforce{1w,0h}{i2}
\fmf{boson}{v2,v1}
\fmf{plain}{v2,i1}
\fmf{plain}{v2,i2}
\fmfdot{v1,v2}
\fmfv{decor.size=0, label=${\scs 2}$, l.dist=1mm, l.angle=0}{i2}
\fmfv{decor.size=0, label=${\scs 1}$, l.dist=1mm, l.angle=0}{i1}
\end{fmfgraph*}
\end{center}}
\hspace*{3mm} \dphi{\Gamma^{(1,1)}}{1}{2}
+
\parbox{9mm}{\begin{center}
\begin{fmfgraph*}(6,4)
\setval
\fmfstraight
\fmfforce{0w,1/2h}{v1}
\fmfforce{4/6w,1/2h}{v2}
\fmfforce{1w,1.25h}{i1}
\fmfforce{1w,1/2h}{i2}
\fmfforce{1w,-0.25h}{i3}
\fmf{boson}{v2,v1}
\fmf{plain}{v2,i1}
\fmf{plain}{v2,i2}
\fmf{plain}{v2,i3}
\fmfdot{v1,v2}
\fmfv{decor.size=0, label=${\scs 3}$, l.dist=1mm, l.angle=0}{i3}
\fmfv{decor.size=0, label=${\scs 2}$, l.dist=1mm, l.angle=0}{i2}
\fmfv{decor.size=0, label=${\scs 1}$, l.dist=1mm, l.angle=0}{i1}
\end{fmfgraph*}
\end{center}}
\hspace*{3mm} \dvertex{\Gamma^{(1,1)}}{1}{2}{3}
\right\} } \hspace*{4mm} ,
\eeq
so that we result in
\beq
\Gamma^{(2,1)} = \frac{1}{4} \hspace*{1mm}
\parbox{11mm}{\begin{center}
\begin{fmfgraph*}(8,4)
\setval
\fmfforce{0w,0h}{v1}
\fmfforce{1/2w,0h}{v2}
\fmfforce{1/2w,1h}{v3}
\fmfforce{1w,0h}{w1}
\fmf{plain,left=1}{v2,v3,v2}
\fmf{boson}{v1,w1}
\fmfdot{v1,w1,v2}
\end{fmfgraph*}\end{center}}
\hspace*{1mm} + \frac{1}{4} \hspace*{1mm}
\parbox{15mm}{\begin{center}
\begin{fmfgraph*}(12,4)
\setval
\fmfforce{0w,1/2h}{v1}
\fmfforce{1/3w,1/2h}{v2}
\fmfforce{2/3w,1/2h}{v3}
\fmfforce{1w,1/2h}{w1}
\fmf{plain,left=1}{v2,v3,v2}
\fmf{boson}{v1,v2}
\fmf{boson}{v3,w1}
\fmfdot{v1,w1,v2,v3}
\end{fmfgraph*}\end{center}}
\hspace*{0.5cm} \, .
\eeq
\end{fmffile}
\begin{fmffile}{fg9}
\subsubsection{Counting $3$-Vertices}
The combination (\r{COMNE}) of the two
compatibility relations (\r{D5}) and (\r{D6})  
leads together with (\r{EFL4}) to 
\beq
\int_{123} K_{123} \frac{\delta \Gamma}{\delta K_{123}} &= &- \frac{1}{3} 
\int_{123} K_{123} \Phi_1 \Phi_2 \Phi_3 + \int_{123} K_{123} \Phi_1
\frac{\delta \Gamma}{\delta G^{-1}_{23}} 
+ \frac{1}{3} \int_{123456} K_{123} K_{456} 
\frac{\delta^2 \Gamma}{\delta G^{-1}_{12} \delta G^{-1}_{56}}
\left\{ 2 \,\frac{\delta \Gamma}{\delta G^{-1}_{34}} - \Phi_3 \Phi_4
\right\} 
\no \\ && 
+ \frac{1}{3} \int_{1234567} K_{123} L_{4567} 
\frac{\delta^2 \Gamma}{\delta G^{-1}_{12} \delta K_{567}}
\left\{ 2 \,\frac{\delta \Gamma}{\delta G^{-1}_{34}} - \Phi_3 \Phi_4
\right\} \, .
\eeq
Applying (\r{NR1}), (\r{NR2}) and the decomposition (\r{EDEC}), we obtain
a functional differential equation which is based on counting $3$-vertices:
\beq
&&\int_{123} K_{123} \frac{\delta \Gamma^{({\rm int})}}{\delta K_{123}} 
= - \frac{1}{6} \int_{123456} K_{123} K_{456} G_{14} G_{25} G_{36} 
+ \frac{1}{2} \int_{123} K_{123} G_{12} \Phi_3 
+ \frac{1}{6} \int_{123} K_{123} \Phi_1 \Phi_2 \Phi_3  
\no \\ && \hspace*{0.4cm}
- \int_{12345} K_{123} G_{14} G_{25} \Phi_3 
\frac{\delta \Gamma^{({\rm int})}}{\delta G_{45}}
+ \int_{12345678} K_{123} K_{456} G_{14} G_{25} G_{37} G_{68} 
\frac{\delta \Gamma^{({\rm int})}}{\delta G_{78}}
\no \\ &&  \hspace*{0.4cm}
- \frac{4}{3} \int_{123456789\bar{1}} K_{123} K_{456} G_{15} G_{27} G_{68} 
G_{39} G_{4\bar{1}} 
\frac{\delta \Gamma^{({\rm int})}}{\delta G_{78}}
\frac{\delta \Gamma^{({\rm int})}}{\delta G_{9\bar{1}}}
+ \frac{1}{3} \int_{123456789\bar{1}} K_{123} K_{456} G_{14} G_{27} G_{38} 
G_{59} G_{6\bar{1}} 
\frac{\delta^2 \Gamma^{({\rm int})}}{\delta G_{78} \delta G_{9\bar{1}}}
\no \\ && \hspace*{0.4cm}
- \frac{2}{3} \int_{123456789\bar{1}\bar{2}\bar{3}} K_{123} K_{456} G_{17} 
G_{28} G_{59} G_{6\bar{1}} G_{3\bar{2}} G_{4\bar{3}} 
\frac{\delta^2 \Gamma^{({\rm int})}}{\delta G_{78}\delta G_{9\bar{1}} }
\frac{\delta \Gamma^{({\rm int})}}{\delta G_{\bar{2}\bar{3}}}
\no \\ && \hspace*{0.4cm}
- \frac{1}{3} \int_{123456789} K_{123} L_{4567} G_{14} G_{28} G_{39} 
\frac{\delta^2 \Gamma^{({\rm int})}}{\delta K_{567} \delta G_{89}}
+ \frac{2}{3} \int_{123456789\bar{1}\bar{2}} K_{123} L_{4567} G_{18} 
G_{29} G_{3\bar{1}} G_{4\bar{2}} 
\frac{\delta^2 \Gamma^{({\rm int})}}{\delta K_{567} \delta G_{89}}
\frac{\delta \Gamma^{({\rm int})}}{\delta G_{\bar{1}\bar{2}}} 
\, .
\eeq
The corresponding graphical representation reads
\beq
&&
\parbox{5mm}{\begin{center}
\begin{fmfgraph*}(2,4)
\setval
\fmfstraight
\fmfforce{0w,1/2h}{v1}
\fmfforce{1w,1/2h}{v2}
\fmfforce{1w,1.25h}{v3}
\fmfforce{1w,-0.25h}{v4}
\fmf{plain}{v1,v2}
\fmf{plain}{v1,v3}
\fmf{plain}{v1,v4}
\fmfv{decor.size=0, label=${\scs 2}$, l.dist=1mm, l.angle=0}{v2}
\fmfv{decor.size=0, label=${\scs 1}$, l.dist=1mm, l.angle=0}{v3}
\fmfv{decor.size=0, label=${\scs 3}$, l.dist=1mm, l.angle=0}{v4}
\fmfdot{v1}
\end{fmfgraph*}
\end{center}}
\hspace*{0.3cm} \dvertex{\Gamma^{({\rm int})}}{1}{2}{3} = 
\hspace*{1mm} - \frac{1}{6}\hspace*{1mm}
\parbox{7mm}{\begin{center}
\begin{fmfgraph*}(4,4)
\setval
\fmfforce{0w,0.5h}{v1}
\fmfforce{1w,0.5h}{v2}
\fmf{plain,left=1}{v1,v2,v1}
\fmf{plain}{v1,v2}
\fmfdot{v1,v2}
\end{fmfgraph*}\end{center}}
\hspace*{1mm} - \frac{1}{2} \hspace*{1mm}
\parbox{11mm}{\begin{center}
\begin{fmfgraph*}(8,4)
\setval
\fmfforce{0w,1/2h}{v1}
\fmfforce{1/2w,1/2h}{v2}
\fmfforce{1w,1/2h}{v3}
\fmf{plain,left=1}{v2,v3,v2}
\fmf{boson}{v1,v2}
\fmfdot{v1,v2}
\end{fmfgraph*}\end{center}}
\hspace*{1mm} - \frac{1}{6}\hspace*{1mm}
\parbox{11mm}{\begin{center}
\begin{fmfgraph*}(6.928,12)
\setval
\fmfforce{1/2w,5/6h}{v1}
\fmfforce{1w,1/4h}{w1}
\fmfforce{0w,1/4h}{u1}
\fmfforce{1/2w,1/2h}{v2}
\fmf{boson}{v2,v1}
\fmf{boson}{v2,w1}
\fmf{boson}{v2,u1}
\fmfdot{v1,u1,w1,v2}
\end{fmfgraph*}\end{center}}
\hspace*{1mm} + \hspace*{1mm}
\parbox{7mm}{\begin{center}
\begin{fmfgraph*}(6,4)
\setval
\fmfstraight
\fmfforce{1/3w,0h}{v1}
\fmfforce{1/3w,1h}{v2}
\fmfforce{1w,1h}{i2}
\fmfforce{1w,0h}{i1}
\fmf{plain}{i1,v1}
\fmf{plain}{v2,i2}
\fmf{plain,left}{v1,v2,v1}
\fmfdot{v2,v1}
\fmfv{decor.size=0, label=${\scs 2}$, l.dist=1mm, l.angle=0}{i1}
\fmfv{decor.size=0, label=${\scs 1}$, l.dist=1mm, l.angle=0}{i2}
\end{fmfgraph*}
\end{center}}
\hspace*{3mm} \dphi{\Gamma^{({\rm int})}}{1}{2}
\hspace*{1mm} + \hspace*{1mm}
\parbox{9mm}{\begin{center}
\begin{fmfgraph*}(6,4)
\setval
\fmfstraight
\fmfforce{0w,1/2h}{v1}
\fmfforce{4/6w,1/2h}{v2}
\fmfforce{1w,1h}{i1}
\fmfforce{1w,0h}{i2}
\fmf{boson}{v2,v1}
\fmf{plain}{v2,i1}
\fmf{plain}{v2,i2}
\fmfdot{v1,v2}
\fmfv{decor.size=0, label=${\scs 2}$, l.dist=1mm, l.angle=0}{i2}
\fmfv{decor.size=0, label=${\scs 1}$, l.dist=1mm, l.angle=0}{i1}
\end{fmfgraph*}
\end{center}}
\hspace*{3mm} \dphi{\Gamma^{({\rm int})}}{1}{2}
\no \\ &&
\hspace*{10mm} + \frac{1}{3} \hspace*{1mm}
\parbox{7mm}{\begin{center}
\begin{fmfgraph*}(3,9)
\setval
\fmfstraight
\fmfforce{1w,1h}{o1}
\fmfforce{1w,2/3h}{o2}
\fmfforce{0w,5/6h}{v1}
\fmfforce{0w,1/6h}{v2}
\fmfforce{1w,1/3h}{i1}
\fmfforce{1w,0h}{i2}
\fmf{plain}{v1,v2}
\fmf{plain}{v1,o1}
\fmf{plain}{v1,o2}
\fmf{plain}{v2,i1}
\fmf{plain}{v2,i2}
\fmfdot{v1,v2}
\fmfv{decor.size=0, label=${\scs 4}$, l.dist=1mm, l.angle=0}{i2}
\fmfv{decor.size=0, label=${\scs 3}$, l.dist=1mm, l.angle=0}{i1}
\fmfv{decor.size=0, label=${\scs 2}$, l.dist=1mm, l.angle=0}{o2}
\fmfv{decor.size=0, label=${\scs 1}$, l.dist=1mm, l.angle=0}{o1}
\end{fmfgraph*}
\end{center}}
\hspace*{3mm} \ddphi{\Gamma^{({\rm int})}}{1}{2}{3}{4} 
\hspace*{1mm} + \frac{1}{3} \hspace*{1mm}
\parbox{7mm}{\begin{center}
\begin{fmfgraph*}(3,12)
\setval
\fmfstraight
\fmfforce{1w,1h}{o0}
\fmfforce{1w,3/4h}{o1}
\fmfforce{1w,1/2h}{o2}
\fmfforce{0w,3/4h}{v1}
\fmfforce{0w,1/8h}{v2}
\fmfforce{1w,1/4h}{i1}
\fmfforce{1w,0h}{i2}
\fmf{plain}{v1,v2}
\fmf{plain}{v1,o0}
\fmf{plain}{v1,o1}
\fmf{plain}{v1,o2}
\fmf{plain}{v2,i1}
\fmf{plain}{v2,i2}
\fmfdot{v1,v2}
\fmfv{decor.size=0, label=${\scs 5}$, l.dist=1mm, l.angle=0}{i2}
\fmfv{decor.size=0, label=${\scs 4}$, l.dist=1mm, l.angle=0}{i1}
\fmfv{decor.size=0, label=${\scs 3}$, l.dist=1mm, l.angle=0}{o2}
\fmfv{decor.size=0, label=${\scs 2}$, l.dist=1mm, l.angle=0}{o1}
\fmfv{decor.size=0, label=${\scs 1}$, l.dist=1mm, l.angle=0}{o0}
\end{fmfgraph*}
\end{center}}
\hspace*{3mm} \ddvertex{\Gamma^{({\rm int})}}{1}{2}{3}{4}{5}
\hspace*{1mm} - \frac{4}{3} 
\hspace*{1mm} \dphi{\Gamma^{({\rm int})}}{1}{2}\hspace*{3mm}
\parbox{9mm}{\begin{center}
\begin{fmfgraph*}(6,4)
\setval
\fmfstraight
\fmfforce{1w,1h}{o1}
\fmfforce{1w,0h}{o2}
\fmfforce{1/2w,1h}{v1}
\fmfforce{1/2w,0h}{v2}
\fmfforce{0w,1h}{i1}
\fmfforce{0w,0h}{i2}
\fmf{plain}{v1,v2}
\fmf{plain}{o1,i1}
\fmf{plain}{i2,o2}
\fmfdot{v1,v2}
\fmfv{decor.size=0, label=${\scs 2}$, l.dist=1mm, l.angle=-180}{i2}
\fmfv{decor.size=0, label=${\scs 1}$, l.dist=1mm, l.angle=-180}{i1}
\fmfv{decor.size=0, label=${\scs 4}$, l.dist=1mm, l.angle=0}{o2}
\fmfv{decor.size=0, label=${\scs 3}$, l.dist=1mm, l.angle=0}{o1}
\end{fmfgraph*}
\end{center}}
\hspace*{3mm} \dphi{\Gamma^{({\rm int})}}{3}{4}\hspace*{1mm}
\no \\ & &
\hspace*{10mm} - \frac{2}{3} 
\hspace*{1mm} \dphi{\Gamma^{({\rm int})}}{1}{2}\hspace*{3mm}
\parbox{9mm}{\begin{center}
\begin{fmfgraph*}(6,9)
\setval
\fmfstraight
\fmfforce{1w,1h}{o1}
\fmfforce{1w,2/3h}{o2}
\fmfforce{1/2w,5/6h}{v1}
\fmfforce{1/2w,1/6h}{v2}
\fmfforce{1w,1/3h}{i1}
\fmfforce{1w,0h}{i2}
\fmfforce{0w,5/6h}{k1}
\fmfforce{0w,1/6h}{k2}
\fmf{plain}{v1,k1}
\fmf{plain}{v2,k2}
\fmf{plain}{v1,o1}
\fmf{plain}{v1,o2}
\fmf{plain}{v2,i1}
\fmf{plain}{v2,i2}
\fmfdot{v1,v2}
\fmfv{decor.size=0, label=${\scs 6}$, l.dist=1mm, l.angle=0}{i2}
\fmfv{decor.size=0, label=${\scs 5}$, l.dist=1mm, l.angle=0}{i1}
\fmfv{decor.size=0, label=${\scs 4}$, l.dist=1mm, l.angle=0}{o2}
\fmfv{decor.size=0, label=${\scs 3}$, l.dist=1mm, l.angle=0}{o1}
\fmfv{decor.size=0, label=${\scs 2}$, l.dist=1mm, l.angle=-180}{k2}
\fmfv{decor.size=0, label=${\scs 1}$, l.dist=1mm, l.angle=-180}{k1}
\end{fmfgraph*}
\end{center}}
\hspace*{3mm} \ddphi{\Gamma^{({\rm int})}}{3}{4}{5}{6}\hspace*{1mm}
\hspace*{1mm} - \frac{2}{3} \hspace*{1mm}
\hspace*{1mm} \dphi{\Gamma^{({\rm int})}}{1}{2}\hspace*{3mm}
\parbox{9mm}{\begin{center}
\begin{fmfgraph*}(6,12)
\setval
\fmfstraight
\fmfforce{1w,1h}{o0}
\fmfforce{1w,3/4h}{o1}
\fmfforce{1w,1/2h}{o2}
\fmfforce{1/2w,3/4h}{v1}
\fmfforce{1/2w,1/8h}{v2}
\fmfforce{1w,1/4h}{i1}
\fmfforce{1w,0h}{i2}
\fmfforce{0w,3/4h}{k1}
\fmfforce{0w,1/8h}{k2}
\fmf{plain}{v1,k1}
\fmf{plain}{v2,k2}
\fmf{plain}{v1,o0}
\fmf{plain}{v1,o1}
\fmf{plain}{v1,o2}
\fmf{plain}{v2,i1}
\fmf{plain}{v2,i2}
\fmfdot{v1,v2}
\fmfv{decor.size=0, label=${\scs 7}$, l.dist=1mm, l.angle=0}{i2}
\fmfv{decor.size=0, label=${\scs 6}$, l.dist=1mm, l.angle=0}{i1}
\fmfv{decor.size=0, label=${\scs 5}$, l.dist=1mm, l.angle=0}{o2}
\fmfv{decor.size=0, label=${\scs 4}$, l.dist=1mm, l.angle=0}{o1}
\fmfv{decor.size=0, label=${\scs 3}$, l.dist=1mm, l.angle=0}{o0}
\fmfv{decor.size=0, label=${\scs 2}$, l.dist=1mm, l.angle=-180}{k2}
\fmfv{decor.size=0, label=${\scs 1}$, l.dist=1mm, l.angle=-180}{k1}
\end{fmfgraph*}
\end{center}}
\hspace*{3mm} \ddvertex{\Gamma^{({\rm int})}}{3}{4}{5}{6}{7}
\hspace*{4mm}. \la{GRR}
\eeq
The right-hand side consists of only 10 out of 32 terms from Eq.~(\r{FFF}),
3 without $\Gamma^{({\rm int})}$, 4 linear in $\Gamma^{({\rm int})}$ and 3
bilinear in  $\Gamma^{({\rm int})}$. Therefore the iteration of the graphical
relation (\r{GRR}) is simpler than (\r{FFF}). To demonstrate 
this, we perform the decomposition (\r{DECCE}) and restrict ourselves again
to the one-particle irreducible 
vacuum diagrams without field expectation values
$\Gamma^{(0,l)}$. For $n=0$ and $l=2$ Eq.~(\r{GRR}) reduces to 
\beq
\parbox{5mm}{\begin{center}
\begin{fmfgraph*}(2,4)
\setval
\fmfstraight
\fmfforce{0w,1/2h}{v1}
\fmfforce{1w,1/2h}{v2}
\fmfforce{1w,1.25h}{v3}
\fmfforce{1w,-0.25h}{v4}
\fmf{plain}{v1,v2}
\fmf{plain}{v1,v3}
\fmf{plain}{v1,v4}
\fmfv{decor.size=0, label=${\scs 2}$, l.dist=1mm, l.angle=0}{v2}
\fmfv{decor.size=0, label=${\scs 1}$, l.dist=1mm, l.angle=0}{v3}
\fmfv{decor.size=0, label=${\scs 3}$, l.dist=1mm, l.angle=0}{v4}
\fmfdot{v1}
\end{fmfgraph*}
\end{center}}
\hspace*{0.3cm} \dvertex{\Gamma^{(0,2)}}{1}{2}{3} = 
\hspace*{1mm} \frac{1}{6}\hspace*{1mm}
\parbox{7mm}{\begin{center}
\begin{fmfgraph*}(4,4)
\setval
\fmfforce{0w,0.5h}{v1}
\fmfforce{1w,0.5h}{v2}
\fmf{plain,left=1}{v1,v2,v1}
\fmf{plain}{v1,v2}
\fmfdot{v1,v2}
\end{fmfgraph*}\end{center}}
\hspace*{4mm} ,
\la{GRR1}
\eeq
whereas for $n=0$ and $l\ge3$ we obtain
\end{fmffile}
\begin{fmffile}{fg10}
\beq
\parbox{5mm}{\begin{center}
\begin{fmfgraph*}(2,4)
\setval
\fmfstraight
\fmfforce{0w,1/2h}{v1}
\fmfforce{1w,1/2h}{v2}
\fmfforce{1w,1.25h}{v3}
\fmfforce{1w,-0.25h}{v4}
\fmf{plain}{v1,v2}
\fmf{plain}{v1,v3}
\fmf{plain}{v1,v4}
\fmfv{decor.size=0, label=${\scs 2}$, l.dist=1mm, l.angle=0}{v2}
\fmfv{decor.size=0, label=${\scs 1}$, l.dist=1mm, l.angle=0}{v3}
\fmfv{decor.size=0, label=${\scs 3}$, l.dist=1mm, l.angle=0}{v4}
\fmfdot{v1}
\end{fmfgraph*}
\end{center}}
\hspace*{0.3cm} \dvertex{\Gamma^{(0,l)}}{1}{2}{3}& =& 
\parbox{7mm}{\begin{center}
\begin{fmfgraph*}(6,4)
\setval
\fmfstraight
\fmfforce{1/3w,0h}{v1}
\fmfforce{1/3w,1h}{v2}
\fmfforce{1w,1h}{i2}
\fmfforce{1w,0h}{i1}
\fmf{plain}{i1,v1}
\fmf{plain}{v2,i2}
\fmf{plain,left}{v1,v2,v1}
\fmfdot{v2,v1}
\fmfv{decor.size=0, label=${\scs 2}$, l.dist=1mm, l.angle=0}{i1}
\fmfv{decor.size=0, label=${\scs 1}$, l.dist=1mm, l.angle=0}{i2}
\end{fmfgraph*}
\end{center}}
\hspace*{3mm} \dphi{\Gamma^{(0,l-1)}}{1}{2}
\hspace*{1mm} + \frac{1}{3} \hspace*{1mm}
\parbox{7mm}{\begin{center}
\begin{fmfgraph*}(3,9)
\setval
\fmfstraight
\fmfforce{1w,1h}{o1}
\fmfforce{1w,2/3h}{o2}
\fmfforce{0w,5/6h}{v1}
\fmfforce{0w,1/6h}{v2}
\fmfforce{1w,1/3h}{i1}
\fmfforce{1w,0h}{i2}
\fmf{plain}{v1,v2}
\fmf{plain}{v1,o1}
\fmf{plain}{v1,o2}
\fmf{plain}{v2,i1}
\fmf{plain}{v2,i2}
\fmfdot{v1,v2}
\fmfv{decor.size=0, label=${\scs 4}$, l.dist=1mm, l.angle=0}{i2}
\fmfv{decor.size=0, label=${\scs 3}$, l.dist=1mm, l.angle=0}{i1}
\fmfv{decor.size=0, label=${\scs 2}$, l.dist=1mm, l.angle=0}{o2}
\fmfv{decor.size=0, label=${\scs 1}$, l.dist=1mm, l.angle=0}{o1}
\end{fmfgraph*}
\end{center}}
\hspace*{3mm} \ddphi{\Gamma^{(0,l-1)}}{1}{2}{3}{4} 
\hspace*{1mm} + \frac{1}{3} \hspace*{1mm}
\parbox{7mm}{\begin{center}
\begin{fmfgraph*}(3,12)
\setval
\fmfstraight
\fmfforce{1w,1h}{o0}
\fmfforce{1w,3/4h}{o1}
\fmfforce{1w,1/2h}{o2}
\fmfforce{0w,3/4h}{v1}
\fmfforce{0w,1/8h}{v2}
\fmfforce{1w,1/4h}{i1}
\fmfforce{1w,0h}{i2}
\fmf{plain}{v1,v2}
\fmf{plain}{v1,o0}
\fmf{plain}{v1,o1}
\fmf{plain}{v1,o2}
\fmf{plain}{v2,i1}
\fmf{plain}{v2,i2}
\fmfdot{v1,v2}
\fmfv{decor.size=0, label=${\scs 5}$, l.dist=1mm, l.angle=0}{i2}
\fmfv{decor.size=0, label=${\scs 4}$, l.dist=1mm, l.angle=0}{i1}
\fmfv{decor.size=0, label=${\scs 3}$, l.dist=1mm, l.angle=0}{o2}
\fmfv{decor.size=0, label=${\scs 2}$, l.dist=1mm, l.angle=0}{o1}
\fmfv{decor.size=0, label=${\scs 1}$, l.dist=1mm, l.angle=0}{o0}
\end{fmfgraph*}
\end{center}}
\hspace*{3mm} \ddvertex{\Gamma^{(0,l-1)}}{1}{2}{3}{4}{5} 
\no \\ && 
+ \sum_{l'=2}^{l-2} \left\{ \frac{4}{3} 
\hspace*{1mm} \dphi{\Gamma^{(0,l')}}{1}{2}\hspace*{3mm}
\parbox{9mm}{\begin{center}
\begin{fmfgraph*}(6,4)
\setval
\fmfstraight
\fmfforce{1w,1h}{o1}
\fmfforce{1w,0h}{o2}
\fmfforce{1/2w,1h}{v1}
\fmfforce{1/2w,0h}{v2}
\fmfforce{0w,1h}{i1}
\fmfforce{0w,0h}{i2}
\fmf{plain}{v1,v2}
\fmf{plain}{o1,i1}
\fmf{plain}{i2,o2}
\fmfdot{v1,v2}
\fmfv{decor.size=0, label=${\scs 2}$, l.dist=1mm, l.angle=-180}{i2}
\fmfv{decor.size=0, label=${\scs 1}$, l.dist=1mm, l.angle=-180}{i1}
\fmfv{decor.size=0, label=${\scs 4}$, l.dist=1mm, l.angle=0}{o2}
\fmfv{decor.size=0, label=${\scs 3}$, l.dist=1mm, l.angle=0}{o1}
\end{fmfgraph*}
\end{center}}
\hspace*{3mm} \dphi{\Gamma^{(0,l-l'-1)}}{3}{4}
\hspace*{1mm} + \frac{2}{3} 
\hspace*{1mm} \dphi{\Gamma^{(0,l)}}{1}{2}\hspace*{3mm}
\parbox{9mm}{\begin{center}
\begin{fmfgraph*}(6,9)
\setval
\fmfstraight
\fmfforce{1w,1h}{o1}
\fmfforce{1w,2/3h}{o2}
\fmfforce{1/2w,5/6h}{v1}
\fmfforce{1/2w,1/6h}{v2}
\fmfforce{1w,1/3h}{i1}
\fmfforce{1w,0h}{i2}
\fmfforce{0w,5/6h}{k1}
\fmfforce{0w,1/6h}{k2}
\fmf{plain}{v1,k1}
\fmf{plain}{v2,k2}
\fmf{plain}{v1,o1}
\fmf{plain}{v1,o2}
\fmf{plain}{v2,i1}
\fmf{plain}{v2,i2}
\fmfdot{v1,v2}
\fmfv{decor.size=0, label=${\scs 6}$, l.dist=1mm, l.angle=0}{i2}
\fmfv{decor.size=0, label=${\scs 5}$, l.dist=1mm, l.angle=0}{i1}
\fmfv{decor.size=0, label=${\scs 4}$, l.dist=1mm, l.angle=0}{o2}
\fmfv{decor.size=0, label=${\scs 3}$, l.dist=1mm, l.angle=0}{o1}
\fmfv{decor.size=0, label=${\scs 2}$, l.dist=1mm, l.angle=-180}{k2}
\fmfv{decor.size=0, label=${\scs 1}$, l.dist=1mm, l.angle=-180}{k1}
\end{fmfgraph*}
\end{center}}
\hspace*{3mm} \ddphi{\Gamma^{(0,l-l'-1)}}{3}{4}{5}{6}\hspace*{1mm} \right.
\no \\ &&
+ \frac{2}{3} \hspace*{1mm} \left.
\hspace*{1mm} \dphi{\Gamma^{(0,l')}}{1}{2}\hspace*{3mm}
\parbox{9mm}{\begin{center}
\begin{fmfgraph*}(6,12)
\setval
\fmfstraight
\fmfforce{1w,1h}{o0}
\fmfforce{1w,3/4h}{o1}
\fmfforce{1w,1/2h}{o2}
\fmfforce{1/2w,3/4h}{v1}
\fmfforce{1/2w,1/8h}{v2}
\fmfforce{1w,1/4h}{i1}
\fmfforce{1w,0h}{i2}
\fmfforce{0w,3/4h}{k1}
\fmfforce{0w,1/8h}{k2}
\fmf{plain}{v1,k1}
\fmf{plain}{v2,k2}
\fmf{plain}{v1,o0}
\fmf{plain}{v1,o1}
\fmf{plain}{v1,o2}
\fmf{plain}{v2,i1}
\fmf{plain}{v2,i2}
\fmfdot{v1,v2}
\fmfv{decor.size=0, label=${\scs 7}$, l.dist=1mm, l.angle=0}{i2}
\fmfv{decor.size=0, label=${\scs 6}$, l.dist=1mm, l.angle=0}{i1}
\fmfv{decor.size=0, label=${\scs 5}$, l.dist=1mm, l.angle=0}{o2}
\fmfv{decor.size=0, label=${\scs 4}$, l.dist=1mm, l.angle=0}{o1}
\fmfv{decor.size=0, label=${\scs 3}$, l.dist=1mm, l.angle=0}{o0}
\fmfv{decor.size=0, label=${\scs 2}$, l.dist=1mm, l.angle=-180}{k2}
\fmfv{decor.size=0, label=${\scs 1}$, l.dist=1mm, l.angle=-180}{k1}
\end{fmfgraph*}
\end{center}}
\hspace*{3mm} \ddvertex{\Gamma^{(0,l-l'-1)}}{3}{4}{5}{6}{7} \right\} \, .
\la{GRR2}
\eeq
Integrating (\r{GRR1}) and (\r{GRR2}), we have to take into account as a
yet undetermined integration constant all those one-particle irreducible
vacuum diagrams
$\tilde{\Gamma}^{(0,l)}$ which only consist of $4$-vertices. A comparison
with (\r{FFF}) shows that those vacuum diagrams follow from the graphical
recursion relation 
\beq
\tilde{\Gamma}^{(0,l)} = \frac{1}{2(l-1)} \left\{ 
\hspace*{1mm} 
\parbox{9mm}{\begin{center}
\begin{fmfgraph*}(6,4)
\setval
\fmfstraight
\fmfforce{0w,1/2h}{v1}
\fmfforce{4/6w,1/2h}{v2}
\fmfforce{1w,1h}{i2}
\fmfforce{1w,0h}{i1}
\fmf{plain}{i1,v2}
\fmf{plain}{v2,i2}
\fmf{plain,left}{v1,v2,v1}
\fmfdot{v2}
\fmfv{decor.size=0, label=${\scs 2}$, l.dist=1mm, l.angle=0}{i1}
\fmfv{decor.size=0, label=${\scs 1}$, l.dist=1mm, l.angle=0}{i2}
\end{fmfgraph*}
\end{center}}
\hspace*{3mm} \dphi{\tilde{\Gamma}^{(0,l-1)}}{1}{2} 
\hspace*{1mm} + \frac{1}{3} \hspace*{1mm}
\parbox{7mm}{\begin{center}
\begin{fmfgraph*}(3,3)
\setval
\fmfstraight
\fmfforce{0w,1/2h}{v1}
\fmfforce{1w,2h}{i1}
\fmfforce{1w,1h}{i2}
\fmfforce{1w,0h}{i3}
\fmfforce{1w,-1h}{i4}
\fmf{plain}{v1,i1}
\fmf{plain}{v1,i2}
\fmf{plain}{v1,i3}
\fmf{plain}{v1,i4}
\fmfdot{v1}
\fmfv{decor.size=0, label=${\scs 4}$, l.dist=1mm, l.angle=0}{i4}
\fmfv{decor.size=0, label=${\scs 3}$, l.dist=1mm, l.angle=0}{i3}
\fmfv{decor.size=0, label=${\scs 2}$, l.dist=1mm, l.angle=0}{i2}
\fmfv{decor.size=0, label=${\scs 1}$, l.dist=1mm, l.angle=0}{i1}
\end{fmfgraph*}
\end{center}}
\hspace*{3mm} \ddphi{\tilde{\Gamma}^{(0,l-1)}}{1}{2}{3}{4}
\hspace*{1mm} +  \sum_{l'=2}^{l-2} \hspace*{1mm} \frac{1}{3} 
\hspace*{1mm} \dphi{\tilde{\Gamma}^{(0,l')}}{1}{2}\hspace*{3mm}
\parbox{7mm}{\begin{center}
\begin{fmfgraph*}(4,4)
\setval
\fmfstraight
\fmfforce{0w,1h}{o1}
\fmfforce{0w,0h}{o2}
\fmfforce{1/2w,1/2h}{v1}
\fmfforce{1w,1h}{i1}
\fmfforce{1w,0h}{i2}
\fmf{plain}{v1,o1}
\fmf{plain}{v1,o2}
\fmf{plain}{v1,i1}
\fmf{plain}{v1,i2}
\fmfdot{v1}
\fmfv{decor.size=0, label=${\scs 4}$, l.dist=1mm, l.angle=0}{i2}
\fmfv{decor.size=0, label=${\scs 3}$, l.dist=1mm, l.angle=0}{i1}
\fmfv{decor.size=0, label=${\scs 2}$, l.dist=1mm, l.angle=-180}{o2}
\fmfv{decor.size=0, label=${\scs 1}$, l.dist=1mm, l.angle=-180}{o1}
\end{fmfgraph*}
\end{center}}
\hspace*{3mm} \dphi{\Gamma^{(0,l-l'-1)}}{3}{4} \right\} \hspace*{4mm} ,
\la{TIT}
\eeq
which is to be iterated starting from
\beq
\la{TIST}
\tilde{\Gamma}^{(0,2)} = 
\frac{1}{8}
\parbox{11mm}{\begin{center}
\begin{fmfgraph*}(8,4)
\setval
\fmfleft{i1}
\fmfright{o1}
\fmf{plain,left=1}{i1,v1,i1}
\fmf{plain,left=1}{o1,v1,o1}
\fmfdot{v1}
\end{fmfgraph*}\end{center}} \hspace*{4mm} .
\eeq
Thus the vacuum diagrams $\tilde{\Gamma}^{(0,l)}$ coincide with
$\tilde{W}^{(0,l)}$ which are defined by (\r{TI}), (\r{TIS}) and which have
been already discussed in \cite{PHI4}. Therefore we assume
from now on that the one-particle irreducible
vacuum diagrams $\tilde{\Gamma}^{(0,l)}$ above
the critical point are known and construct with them the one-particle
irreducible vacuum
diagrams $\Gamma^{(0,l)}$ below the critical point. Integrating (\r{GRR1}) 
with the integration constant (\r{TIST}) leads, indeed, to the correct 
two-loop result (\r{EW2}). Subsequently we insert (\r{ENNR1}), 
(\r{ENNR2}), (\r{ENNR4}) 
into (\r{TIT}) with $l=3$ and obtain
\beq
\parbox{5mm}{\begin{center}
\begin{fmfgraph*}(2,4)
\setval
\fmfstraight
\fmfforce{0w,1/2h}{v1}
\fmfforce{1w,1/2h}{v2}
\fmfforce{1w,1.25h}{v3}
\fmfforce{1w,-0.25h}{v4}
\fmf{plain}{v1,v2}
\fmf{plain}{v1,v3}
\fmf{plain}{v1,v4}
\fmfv{decor.size=0, label=${\scs 2}$, l.dist=1mm, l.angle=0}{v2}
\fmfv{decor.size=0, label=${\scs 1}$, l.dist=1mm, l.angle=0}{v3}
\fmfv{decor.size=0, label=${\scs 3}$, l.dist=1mm, l.angle=0}{v4}
\fmfdot{v1}
\end{fmfgraph*}
\end{center}}
\hspace*{0.3cm} \dvertex{\Gamma^{(0,3)}}{1}{2}{3} &=& 
\hspace*{1mm} \frac{1}{6}
\parbox{9mm}{\begin{center}
\begin{fmfgraph}(4,4)
\setval
\fmfforce{0w,0h}{v1}
\fmfforce{1w,0h}{v2}
\fmfforce{1w,1h}{v3}
\fmfforce{0w,1h}{v4}
\fmf{plain,right=1}{v1,v3,v1}
\fmf{plain}{v1,v3}
\fmf{plain}{v2,v4}
\fmfdot{v1,v2,v3,v4}
\end{fmfgraph}\end{center}} 
\hspace*{1mm} + \frac{1}{4} \hspace*{1mm}
\parbox{9mm}{\begin{center}
\begin{fmfgraph}(4,4)
\setval
\fmfforce{0w,0h}{v1}
\fmfforce{1w,0h}{v2}
\fmfforce{1w,1h}{v3}
\fmfforce{0w,1h}{v4}
\fmf{plain,right=1}{v1,v3,v1}
\fmf{plain,right=0.4}{v1,v4}
\fmf{plain,left=0.4}{v2,v3}
\fmfdot{v1,v2,v3,v4}
\end{fmfgraph}\end{center}} 
\hspace*{1mm} + \frac{1}{4} \hspace*{1mm} 
\parbox{9mm}{\begin{center}
\begin{fmfgraph}(6,6)
\setval
\fmfforce{0w,1/2h}{v1}
\fmfforce{1w,1/2h}{v2}
\fmfforce{1/2w,1h}{v3}
\fmf{plain,left=1}{v1,v2,v1}
\fmf{plain,left=0.4}{v3,v1}
\fmf{plain,left=0.4}{v2,v3}
\fmfdot{v1,v2,v3}
\end{fmfgraph}\end{center}}
+ \frac{1}{4} \hspace*{1mm} 
\parbox{11mm}{\begin{center}
\begin{fmfgraph}(4,8)
\setval
\fmfforce{0w,1/4h}{v1}
\fmfforce{1w,1/4h}{v2}
\fmfforce{1/2w,1/2h}{v3}
\fmfforce{1/2w,1h}{v4}
\fmf{plain,left=1}{v1,v2,v1}
\fmf{plain,left=1}{v3,v4,v3}
\fmf{plain}{v1,v2}
\fmfdot{v2,v3,v1}
\end{fmfgraph}\end{center}} 
\hspace*{2mm}.
\eeq
Integrating this with the integration constant $\tilde{\Gamma}^{(0,3)}$
which coincides with (\r{TWOL}), we rederive
the three-loop result listed in Table III.
\end{fmffile}
\begin{fmffile}{fg11}
\subsubsection{Counting $4$-Vertices}
Combining the two compatibility relations (\r{D6}) and (\r{D7})
leads together with (\r{EFL4}) to the functional differential equation
\beq
&& \int_{1234} L_{1234} \frac{\delta \Gamma}{\delta L_{1234}}
= 
\frac{1}{6} \int_{1234} L_{1234} \left\{ 
- \frac{\delta^2 \Gamma}{\delta G^{-1}_{12} \delta G^{-1}_{34}} 
+  \frac{\delta \Gamma}{\delta G^{-1}_{12}}
\frac{\delta \Gamma}{\delta G^{-1}_{34}}
\right\}
+ \frac{1}{6} \int_{1234} L_{1234} \Phi_4 
\left\{ 3\,\frac{\delta \Gamma}{\delta K_{123}}
- \frac{\delta \Gamma}{\delta G^{-1}_{12}} \Phi_3 \right\} \no \\
&& \hspace*{1cm} + \frac{1}{6} \int_{1234567} K_{123} L_{4567} 
\frac{\delta^2 \Gamma}{\delta G^{-1}_{12} \delta G^{-1}_{67}} \left\{
3\, \frac{\delta \Gamma}{\delta K_{345}}
- \frac{\delta \Gamma}{\delta G^{-1}_{34}} \Phi_5 \right\} \no \\
&& \hspace*{1cm}+ \frac{1}{6} \int_{12345678} L_{1234} L_{5678} 
\frac{\delta^2 \Gamma}{\delta G^{-1}_{12} \delta K_{678}} \left\{
3\,\frac{\delta \Gamma}{\delta K_{345}}
- \frac{\delta \Gamma}{\delta G^{-1}_{34}} \Phi_5 \right\}  \, . \la{NONAMEN}
\eeq
Applying (\r{NR1}), (\r{NR2}) and (\r{DEC}), we obtain
a functional differential equation which is based on counting $4$-vertices
\beq
&& \int_{1234} L_{1234} \frac{\delta \Gamma^{({\rm int})}}{\delta L_{1234}}
= \frac{1}{8} \int_{1234} L_{1234} G_{12} G_{34} 
+ \frac{1}{24} \int_{1234567} K_{123} L_{4567} G_{14} G_{25} G_{67} \Phi_3
\no \\ &&
+ \frac{1}{24} \int_{1234567} K_{123} L_{4567} G_{14} G_{25} \Phi_3
\Phi_6 \Phi_7
- \frac{1}{24} \int_{1234} L_{1234} \Phi_1 \Phi_2 \Phi_3 \Phi_4
- \frac{1}{2} \int_{123456} L_{1234} G_{12} G_{35} G_{46} 
\frac{\delta \Gamma^{({\rm int})}}{\delta G_{56}}
\no \\ &&
- \frac{1}{6} \int_{12345678} L_{1234} G_{15} G_{26} G_{37} G_{48} 
\frac{\delta^2 \Gamma^{({\rm int})}}{\delta G_{56}\delta G_{78}}
- \frac{1}{4} \int_{1234567} K_{123}L_{4567} G_{16} G_{27}
\frac{\delta \Gamma^{({\rm int})}}{\delta K_{345}}
+ \frac{1}{2} \int_{1234} L_{1234} \Phi_4 
\frac{\delta \Gamma^{({\rm int})}}{\delta K_{123}} 
\no\\&&
- \frac{1}{12} \int_{123456789} K_{123} L_{4567} \Phi_3 G_{16} G_{27} G_{48}
G_{59} \frac{\delta \Gamma^{({\rm int})}}{\delta G_{89}} 
- \frac{1}{6} \int_{123456789} K_{123} L_{4567} \Phi_3 G_{45}G_{16} G_{28} 
G_{79} \frac{\delta \Gamma^{({\rm int})}}{\delta G_{89}}
\no\\&&
-  \frac{1}{12} \int_{123456789\bar{1}\bar{2}} K_{123} L_{4567} \Phi_3 
G_{45} G_{18} G_{29} G_{6\bar{1}} G_{7\bar{2}}
\frac{\delta^2 \Gamma^{({\rm int})}}{\delta G_{89}\delta G_{\bar{1}\bar{2}}}
+  \frac{1}{12} \int_{123456789\bar{1}}L_{1234} L_{5678} \Phi_5 G_{34}
G_{19} G_{2\bar{1}} 
\frac{\delta^2 \Gamma^{({\rm int})}}{\delta K_{678}\delta G_{9\bar{1}}}
\no\\&&
- \frac{1}{6} \int_{123456789} K_{123} L_{4567} \Phi_3 \Phi_4 \Phi_5
G_{16} G_{28} G_{79} 
\frac{\delta \Gamma^{({\rm int})}}{\delta G_{89}}
- \frac{1}{12} \int_{123456789\bar{1}\bar{2}} K_{123} L_{4567}
\Phi_3 \Phi_4 \Phi_5 G_{18} G_{29} G_{6\bar{1}} G_{7\bar{2}}
\frac{\delta^2 \Gamma^{({\rm int})}}{\delta G_{89}\delta G_{\bar{1}\bar{2}}}
\no \\&&
+ \frac{1}{12} \int_{123456789\bar{1}}L_{1234} L_{5678} \Phi_3 \Phi_4 \Phi_5
G_{19} G_{2\bar{1}}
\frac{\delta^2 \Gamma^{({\rm int})}}{\delta K_{678}\delta G_{9\bar{1}}}
+\frac{1}{6} \int_{12345678} L_{1234} G_{15} G_{26} G_{37} G_{48} 
\frac{\delta \Gamma^{({\rm int})}}{\delta G_{56}}
\frac{\delta \Gamma^{({\rm int})}}{\delta G_{78}}
\no \\&&
+ \frac{1}{2} 
\int_{123456789\bar{1}\bar{2}} K_{123} L_{4567} G_{18} G_{29} G_{3\bar{1}}
G_{4\bar{2}} 
\frac{\delta^2 \Gamma^{({\rm int})}}{\delta K_{567}\delta G_{89}}
\frac{\delta \Gamma^{({\rm int})}}{\delta G_{\bar{1}\bar{2}}}
+ \int_{123456789} K_{123} L_{4567} G_{16} G_{28} G_{79}
\frac{\delta \Gamma^{({\rm int})}}{\delta K_{345}}
\frac{\delta \Gamma^{({\rm int})}}{\delta G_{89}} 
\no \\&&
- \frac{1}{2}
\int_{123456789\bar{1}} L_{1234} L_{5678} G_{19} G_{2\bar{1}}
\frac{\delta \Gamma^{({\rm int})}}{\delta K_{345}}
\frac{\delta^2 \Gamma^{({\rm int})}}{\delta K_{678}\delta G_{9\bar{1}}}
+ \frac{1}{3} \int_{123456789\bar{1}\bar{2}} K_{123} L_{4567} \Phi_3
G_{16} G_{28} G_{79} G_{4\bar{1}} G_{5\bar{2}}    
\frac{\delta \Gamma^{({\rm int})}}{\delta G_{89}}
\frac{\delta \Gamma^{({\rm int})}}{\delta G_{\bar{1}\bar{2}}}
\no \\&&
+ \frac{1}{6} \int_{123456789\bar{1}\bar{2}\bar{3}\bar{4}}
K_{123} L_{4567} \Phi_3 G_{18} G_{29} G_{6\bar{1}} G_{7\bar{2}} G_{4\bar{3}}
G_{5\bar{4}} 
\frac{\delta^2 \Gamma^{({\rm int})}}{\delta G_{89}\delta G_{\bar{1}\bar{2}}}
\frac{\delta \Gamma^{({\rm int})}}{\delta G_{\bar{3}\bar{4}}}
\no \\&&
- \frac{1}{6} \int_{123456789\bar{1}\bar{2}\bar{3}} L_{1234} L_{5678}
\Phi_5 G_{19} G_{2\bar{1}} G_{3\bar{2}} G_{4\bar{3}}
\frac{\delta^2 \Gamma^{({\rm int})}}{\delta K_{678}\delta G_{9\bar{1}}}
\frac{\delta \Gamma^{({\rm int})}}{\delta G_{\bar{2}\bar{3}}} \, .
\eeq
The corresponding graphical representation reads
\beq
&& 
\parbox{7mm}{\begin{center}
\begin{fmfgraph*}(3,3)
\setval
\fmfstraight
\fmfforce{0w,1/2h}{v1}
\fmfforce{1w,2h}{i1}
\fmfforce{1w,1h}{i2}
\fmfforce{1w,0h}{i3}
\fmfforce{1w,-1h}{i4}
\fmf{plain}{v1,i1}
\fmf{plain}{v1,i2}
\fmf{plain}{v1,i3}
\fmf{plain}{v1,i4}
\fmfdot{v1}
\fmfv{decor.size=0, label=${\scs 4}$, l.dist=1mm, l.angle=0}{i4}
\fmfv{decor.size=0, label=${\scs 3}$, l.dist=1mm, l.angle=0}{i3}
\fmfv{decor.size=0, label=${\scs 2}$, l.dist=1mm, l.angle=0}{i2}
\fmfv{decor.size=0, label=${\scs 1}$, l.dist=1mm, l.angle=0}{i1}
\end{fmfgraph*}
\end{center}}
\hspace*{2mm} \dvertexx{\Gamma^{({\rm int})}}{2}{3}{4}{1} = 
\hspace*{1mm}- \frac{1}{8}\hspace*{1mm}
\parbox{11mm}{\begin{center}
\begin{fmfgraph*}(8,4)
\setval
\fmfleft{i1}
\fmfright{o1}
\fmf{plain,left=1}{i1,v1,i1}
\fmf{plain,left=1}{o1,v1,o1}
\fmfdot{v1}
\end{fmfgraph*}\end{center}}
\hspace*{1mm}+ \frac{1}{24} \hspace*{1mm}
\parbox{15mm}{\begin{center}
\begin{fmfgraph*}(12,4)
\setval
\fmfforce{0w,1/2h}{v1}
\fmfforce{1/3w,1/2h}{v2}
\fmfforce{2/3w,1/2h}{v3}
\fmfforce{1w,1/2h}{v4}
\fmf{plain,left=1}{v2,v3,v2}
\fmf{plain,left=1}{v3,v4,v3}
\fmf{boson}{v1,v2}
\fmfdot{v1,v2,v3}
\end{fmfgraph*}\end{center}}
\hspace*{1mm} + \frac{1}{24}\hspace*{1mm}
\parbox{11mm}{\begin{center}
\begin{fmfgraph*}(8,8)
\setval
\fmfforce{0w,0h}{v1}
\fmfforce{1w,0h}{w1}
\fmfforce{1/2w,1h}{u1}
\fmfforce{1/2w,0h}{v2}
\fmfforce{1/2w,1/2h}{v3}
\fmf{plain,right=1}{v2,v3,v2}
\fmf{boson}{w1,v1}
\fmf{boson}{v3,u1}
\fmfdot{u1,v1,w1,v2,v3}
\end{fmfgraph*}\end{center}} 
\hspace*{1mm} + \frac{1}{24}\hspace*{1mm}
\parbox{11mm}{\begin{center}
\begin{fmfgraph*}(8,12)
\setval
\fmfforce{0w,1/2h}{v1}
\fmfforce{1w,1/2h}{w1}
\fmfforce{1/2w,1/6h}{u1}
\fmfforce{1/2w,5/6h}{x1}
\fmfforce{1/2w,1/2h}{v2}
\fmf{boson}{w1,v1}
\fmf{boson}{x1,u1}
\fmfdot{v1,u1,w1,x1,v2}
\end{fmfgraph*}\end{center}}
\hspace*{1mm}+ \frac{1}{2} \hspace*{1mm} 
\parbox{9mm}{\begin{center}
\begin{fmfgraph*}(6,4)
\setval
\fmfstraight
\fmfforce{0w,1/2h}{v1}
\fmfforce{4/6w,1/2h}{v2}
\fmfforce{1w,1h}{i2}
\fmfforce{1w,0h}{i1}
\fmf{plain}{i1,v2}
\fmf{plain}{v2,i2}
\fmf{plain,left}{v1,v2,v1}
\fmfdot{v2}
\fmfv{decor.size=0, label=${\scs 2}$, l.dist=1mm, l.angle=0}{i1}
\fmfv{decor.size=0, label=${\scs 1}$, l.dist=1mm, l.angle=0}{i2}
\end{fmfgraph*}
\end{center}}
\hspace*{3mm} \dphi{\Gamma^{({\rm int})}}{1}{2} 
\no \\ && \hspace*{0.5cm}
\hspace*{1mm} + \frac{1}{6} \hspace*{1mm}
\parbox{7mm}{\begin{center}
\begin{fmfgraph*}(3,3)
\setval
\fmfstraight
\fmfforce{0w,1/2h}{v1}
\fmfforce{1w,2h}{i1}
\fmfforce{1w,1h}{i2}
\fmfforce{1w,0h}{i3}
\fmfforce{1w,-1h}{i4}
\fmf{plain}{v1,i1}
\fmf{plain}{v1,i2}
\fmf{plain}{v1,i3}
\fmf{plain}{v1,i4}
\fmfdot{v1}
\fmfv{decor.size=0, label=${\scs 4}$, l.dist=1mm, l.angle=0}{i4}
\fmfv{decor.size=0, label=${\scs 3}$, l.dist=1mm, l.angle=0}{i3}
\fmfv{decor.size=0, label=${\scs 2}$, l.dist=1mm, l.angle=0}{i2}
\fmfv{decor.size=0, label=${\scs 1}$, l.dist=1mm, l.angle=0}{i1}
\end{fmfgraph*}
\end{center}}
\hspace*{3mm} \ddphi{\Gamma^{({\rm int})}}{1}{2}{3}{4}
+ \frac{1}{4} \hspace*{1mm} 
\parbox{10mm}{\begin{center}
\begin{fmfgraph*}(6,5.33333)
\setval
\fmfstraight
\fmfforce{2/6w,1/4h}{v1}
\fmfforce{2/6w,1h}{v2}
\fmfforce{1w,0h}{i1}
\fmfforce{1w,1/2h}{i2}
\fmfforce{1w,1h}{i3}
\fmf{plain,left=1}{v1,v2,v1}
\fmf{plain}{v1,i1}
\fmf{plain}{v1,i2}
\fmf{plain}{v2,i3}
\fmfdot{v1,v2}
\fmfv{decor.size=0, label=${\scs 2}$, l.dist=1mm, l.angle=0}{i2}
\fmfv{decor.size=0, label=${\scs 3}$, l.dist=1mm, l.angle=0}{i1}
\fmfv{decor.size=0, label=${\scs 1}$, l.dist=1mm, l.angle=0}{i3}
\end{fmfgraph*}
\end{center}}
\hspace*{3mm} \dvertex{\Gamma^{({\rm int})}}{1}{2}{3} 
\hspace*{1mm} + \frac{1}{2} \hspace*{1mm}
\parbox{9mm}{\begin{center}
\begin{fmfgraph*}(6,4)
\setval
\fmfstraight
\fmfforce{0w,1/2h}{v1}
\fmfforce{4/6w,1/2h}{v2}
\fmfforce{1w,1.25h}{i1}
\fmfforce{1w,1/2h}{i2}
\fmfforce{1w,-0.25h}{i3}
\fmf{boson}{v2,v1}
\fmf{plain}{v2,i1}
\fmf{plain}{v2,i2}
\fmf{plain}{v2,i3}
\fmfdot{v1,v2}
\fmfv{decor.size=0, label=${\scs 3}$, l.dist=1mm, l.angle=0}{i3}
\fmfv{decor.size=0, label=${\scs 2}$, l.dist=1mm, l.angle=0}{i2}
\fmfv{decor.size=0, label=${\scs 1}$, l.dist=1mm, l.angle=0}{i1}
\end{fmfgraph*}
\end{center}}
\hspace*{3mm} \dvertex{\Gamma^{({\rm int})}}{1}{2}{3} 
\hspace*{1mm} - \frac{1}{12} \hspace*{1mm}
\parbox{13mm}{\begin{center}
\begin{fmfgraph*}(10,4)
\setval
\fmfstraight
\fmfforce{0w,1/2h}{v1}
\fmfforce{4/10w,1/2h}{v2}
\fmfforce{8/10w,1/2h}{v3}
\fmfforce{1w,7/8h}{i1}
\fmfforce{1w,1/8h}{i2}
\fmf{boson}{v2,v1}
\fmf{plain,left=1}{v2,v3,v2}
\fmf{plain}{v3,i1}
\fmf{plain}{v3,i2}
\fmfdot{v1,v2,v3}
\fmfv{decor.size=0, label=${\scs 2}$, l.dist=1mm, l.angle=0}{i2}
\fmfv{decor.size=0, label=${\scs 1}$, l.dist=1mm, l.angle=0}{i1}
\end{fmfgraph*}
\end{center}}
\hspace*{3mm} \dphi{\Gamma^{({\rm int})}}{1}{2} 
\no \\ && \hspace*{0.5cm}
\hspace*{1mm} - \frac{1}{6} \hspace*{1mm}
\parbox{9mm}{\begin{center}
\begin{fmfgraph*}(6,6)
\setval
\fmfstraight
\fmfforce{0w,0h}{v1}
\fmfforce{2/3w,0h}{v2}
\fmfforce{1w,0h}{v3}
\fmfforce{0w,2/3h}{v4}
\fmfforce{2/3w,2/3h}{v5}
\fmfforce{1w,2/3h}{v6}
\fmf{boson}{v2,v1}
\fmf{plain,left=1}{v4,v5,v4}
\fmf{plain}{v2,v3}
\fmf{plain}{v5,v6}
\fmf{plain}{v2,v5}
\fmfdot{v1,v2,v5}
\fmfv{decor.size=0, label=${\scs 2}$, l.dist=1mm, l.angle=0}{v3}
\fmfv{decor.size=0, label=${\scs 1}$, l.dist=1mm, l.angle=0}{v6}
\end{fmfgraph*}
\end{center}}
\hspace*{3mm} \dphi{\Gamma^{({\rm int})}}{1}{2} 
\hspace*{1mm} - \frac{1}{12} \hspace*{1mm}
\parbox{9mm}{\begin{center}
\begin{fmfgraph*}(6,4)
\setval
\fmfstraight
\fmfforce{0w,1/2h}{v1}
\fmfforce{4/6w,1/2h}{v2}
\fmfforce{1w,1h}{i1}
\fmfforce{1w,0h}{i2}
\fmf{boson}{v2,v1}
\fmf{plain}{v2,i1}
\fmf{plain}{v2,i2}
\fmfdot{v1,v2}
\fmfv{decor.size=0, label=${\scs 2}$, l.dist=1mm, l.angle=0}{i2}
\fmfv{decor.size=0, label=${\scs 1}$, l.dist=1mm, l.angle=0}{i1}
\end{fmfgraph*}
\end{center}}
\hspace*{2mm} \ddphi{\Gamma^{({\rm int})}}{1}{2}{3}{4} \hspace*{2mm} 
\parbox{9mm}{\begin{center}
\begin{fmfgraph*}(6,4)
\setval
\fmfstraight
\fmfforce{1w,1/2h}{v1}
\fmfforce{2/6w,1/2h}{v2}
\fmfforce{0w,1h}{i2}
\fmfforce{0w,0h}{i1}
\fmf{plain}{i1,v2}
\fmf{plain}{v2,i2}
\fmf{plain,left}{v1,v2,v1}
\fmfdot{v2}
\fmfv{decor.size=0, label=${\scs 4}$, l.dist=1mm, l.angle=-180}{i1}
\fmfv{decor.size=0, label=${\scs 3}$, l.dist=1mm, l.angle=-180}{i2}
\end{fmfgraph*}
\end{center}}
\hspace*{1mm} - \frac{1}{12} \hspace*{1mm}
\parbox{9mm}{\begin{center}
\begin{fmfgraph*}(6,4)
\setval
\fmfstraight
\fmfforce{0w,1/2h}{v1}
\fmfforce{4/6w,1/2h}{v2}
\fmfforce{1w,1.25h}{i1}
\fmfforce{1w,1/2h}{i2}
\fmfforce{1w,-0.25h}{i3}
\fmf{boson}{v2,v1}
\fmf{plain}{v2,i1}
\fmf{plain}{v2,i2}
\fmf{plain}{v2,i3}
\fmfdot{v1,v2}
\fmfv{decor.size=0, label=${\scs 3}$, l.dist=1mm, l.angle=0}{i3}
\fmfv{decor.size=0, label=${\scs 2}$, l.dist=1mm, l.angle=0}{i2}
\fmfv{decor.size=0, label=${\scs 1}$, l.dist=1mm, l.angle=0}{i1}
\end{fmfgraph*}
\end{center}}
\hspace*{2mm} \ddvertex{\Gamma^{({\rm int})}}{1}{2}{3}{4}{5}\hspace*{2mm} 
\parbox{9mm}{\begin{center}
\begin{fmfgraph*}(6,4)
\setval
\fmfstraight
\fmfforce{1w,1/2h}{v1}
\fmfforce{2/6w,1/2h}{v2}
\fmfforce{0w,1h}{i2}
\fmfforce{0w,0h}{i1}
\fmf{plain}{i1,v2}
\fmf{plain}{v2,i2}
\fmf{plain,left}{v1,v2,v1}
\fmfdot{v2}
\fmfv{decor.size=0, label=${\scs 5}$, l.dist=1mm, l.angle=-180}{i1}
\fmfv{decor.size=0, label=${\scs 4}$, l.dist=1mm, l.angle=-180}{i2}
\end{fmfgraph*}
\end{center}}
\no \\ && \hspace*{0.5cm}
\hspace*{1mm} - \frac{1}{6} \hspace*{1mm}
\parbox{9mm}{\begin{center}
\begin{fmfgraph*}(6,7)
\setval
\fmfstraight
\fmfforce{0w,1h}{v1}
\fmfforce{2/3w,1h}{v2}
\fmfforce{1w,1h}{v3}
\fmfforce{2/3w,2/7h}{v4}
\fmfforce{0w,0h}{v5}
\fmfforce{0w,3.5/7h}{v6}
\fmfforce{1w,2/7h}{v7}
\fmf{boson}{v1,v2}
\fmf{plain}{v2,v3}
\fmf{plain}{v4,v7}
\fmf{boson}{v4,v5}
\fmf{boson}{v4,v6}
\fmf{plain}{v4,v2}
\fmfdot{v1,v2,v4,v5,v6}
\fmfv{decor.size=0, label=${\scs 1}$, l.dist=1mm, l.angle=0}{v3}
\fmfv{decor.size=0, label=${\scs 2}$, l.dist=1mm, l.angle=0}{v7}
\end{fmfgraph*}
\end{center}}
\hspace*{2mm} \dphi{\Gamma^{({\rm int})}}{1}{2} \hspace*{2mm} 
\hspace*{1mm} - \frac{1}{12} \hspace*{1mm}
\parbox{9mm}{\begin{center}
\begin{fmfgraph*}(6,4)
\setval
\fmfstraight
\fmfforce{0w,1/2h}{v1}
\fmfforce{4/6w,1/2h}{v2}
\fmfforce{1w,1h}{i1}
\fmfforce{1w,0h}{i2}
\fmf{boson}{v2,v1}
\fmf{plain}{v2,i1}
\fmf{plain}{v2,i2}
\fmfdot{v1,v2}
\fmfv{decor.size=0, label=${\scs 2}$, l.dist=1mm, l.angle=0}{i2}
\fmfv{decor.size=0, label=${\scs 1}$, l.dist=1mm, l.angle=0}{i1}
\end{fmfgraph*}
\end{center}}
\hspace*{2mm} \ddphi{\Gamma^{({\rm int})}}{1}{2}{3}{4} \hspace*{2mm} 
\parbox{8mm}{\begin{center}
\begin{fmfgraph*}(5,6)
\setval
\fmfstraight
\fmfforce{2/5w,1/2h}{v1}
\fmfforce{0w,1/6h}{v2}
\fmfforce{1w,0h}{v3}
\fmfforce{0w,5/6h}{v4}
\fmfforce{1w,1h}{v5}
\fmf{plain}{v1,v2}
\fmf{plain}{v1,v4}
\fmf{boson}{v1,v3}
\fmf{boson}{v1,v5}
\fmfdot{v1,v3,v5}
\fmfv{decor.size=0, label=${\scs 4}$, l.dist=1mm, l.angle=-180}{v2}
\fmfv{decor.size=0, label=${\scs 3}$, l.dist=1mm, l.angle=-180}{v4}
\end{fmfgraph*}
\end{center}}
\hspace*{1mm} - \frac{1}{12} \hspace*{1mm}
\parbox{9mm}{\begin{center}
\begin{fmfgraph*}(6,4)
\setval
\fmfstraight
\fmfforce{0w,1/2h}{v1}
\fmfforce{4/6w,1/2h}{v2}
\fmfforce{1w,1.25h}{i1}
\fmfforce{1w,1/2h}{i2}
\fmfforce{1w,-0.25h}{i3}
\fmf{boson}{v2,v1}
\fmf{plain}{v2,i1}
\fmf{plain}{v2,i2}
\fmf{plain}{v2,i3}
\fmfdot{v1,v2}
\fmfv{decor.size=0, label=${\scs 3}$, l.dist=1mm, l.angle=0}{i3}
\fmfv{decor.size=0, label=${\scs 2}$, l.dist=1mm, l.angle=0}{i2}
\fmfv{decor.size=0, label=${\scs 1}$, l.dist=1mm, l.angle=0}{i1}
\end{fmfgraph*}
\end{center}}
\hspace*{2mm} \ddvertex{\Gamma^{({\rm int})}}{1}{2}{3}{4}{5}\hspace*{2mm} 
\parbox{8mm}{\begin{center}
\begin{fmfgraph*}(5,6)
\setval
\fmfstraight
\fmfforce{2/5w,1/2h}{v1}
\fmfforce{0w,1/6h}{v2}
\fmfforce{1w,0h}{v3}
\fmfforce{0w,5/6h}{v4}
\fmfforce{1w,1h}{v5}
\fmf{plain}{v1,v2}
\fmf{plain}{v1,v4}
\fmf{boson}{v1,v3}
\fmf{boson}{v1,v5}
\fmfdot{v1,v3,v5}
\fmfv{decor.size=0, label=${\scs 5}$, l.dist=1mm, l.angle=-180}{v2}
\fmfv{decor.size=0, label=${\scs 4}$, l.dist=1mm, l.angle=-180}{v4}
\end{fmfgraph*}
\end{center}}
\no \\ && \hspace*{0.5cm}
\hspace*{1mm} - \frac{1}{6} \hspace*{1mm}
\hspace*{1mm} \dphi{\Gamma^{({\rm int})}}{1}{2}\hspace*{3mm}
\parbox{7mm}{\begin{center}
\begin{fmfgraph*}(4,4)
\setval
\fmfstraight
\fmfforce{0w,1h}{o1}
\fmfforce{0w,0h}{o2}
\fmfforce{1/2w,1/2h}{v1}
\fmfforce{1w,1h}{i1}
\fmfforce{1w,0h}{i2}
\fmf{plain}{v1,o1}
\fmf{plain}{v1,o2}
\fmf{plain}{v1,i1}
\fmf{plain}{v1,i2}
\fmfdot{v1}
\fmfv{decor.size=0, label=${\scs 4}$, l.dist=1mm, l.angle=0}{i2}
\fmfv{decor.size=0, label=${\scs 3}$, l.dist=1mm, l.angle=0}{i1}
\fmfv{decor.size=0, label=${\scs 2}$, l.dist=1mm, l.angle=-180}{o2}
\fmfv{decor.size=0, label=${\scs 1}$, l.dist=1mm, l.angle=-180}{o1}
\end{fmfgraph*}
\end{center}}
\hspace*{3mm} \dphi{\Gamma^{({\rm int})}}{3}{4} 
- 
\hspace*{2mm} \dphi{\Gamma^{({\rm int})}}{1}{2} \hspace*{2mm}
\parbox{10mm}{\begin{center}
\begin{fmfgraph*}(8,5.33333)
\setval
\fmfstraight
\fmfforce{4/8w,1/4h}{v1}
\fmfforce{4/8w,1h}{v2}
\fmfforce{1w,0h}{i1}
\fmfforce{1w,1/2h}{i2}
\fmfforce{1w,1h}{i3}
\fmfforce{0w,1/4h}{z1}
\fmfforce{0w,1h}{z2}
\fmf{plain}{v1,z1}
\fmf{plain}{z2,v2}
\fmf{plain}{v1,v2}
\fmf{plain}{v1,i1}
\fmf{plain}{v1,i2}
\fmf{plain}{v2,i3}
\fmfdot{v1,v2}
\fmfv{decor.size=0, label=${\scs 4}$, l.dist=1mm, l.angle=0}{i2}
\fmfv{decor.size=0, label=${\scs 5}$, l.dist=1mm, l.angle=0}{i1}
\fmfv{decor.size=0, label=${\scs 3}$, l.dist=1mm, l.angle=0}{i3}
\fmfv{decor.size=0, label=${\scs 2}$, l.dist=1mm, l.angle=-180}{z1}
\fmfv{decor.size=0, label=${\scs 1}$, l.dist=1mm, l.angle=-180}{z2}
\end{fmfgraph*}
\end{center}}
\hspace*{2mm} \dvertex{\Gamma^{({\rm int})}}{3}{4}{5} \hspace*{2mm} 
- \frac{1}{2} 
\hspace*{1mm} \dvertex{\Gamma^{({\rm int})}}{1}{2}{3} \hspace*{2mm} 
\parbox{9mm}{\begin{center}
\begin{fmfgraph*}(6,9)
\setval
\fmfstraight
\fmfforce{1w,1h}{o1}
\fmfforce{1w,2/3h}{o2}
\fmfforce{1/2w,5/6h}{v1}
\fmfforce{1/2w,1/6h}{v2}
\fmfforce{1w,1/3h}{i1}
\fmfforce{1w,0h}{i2}
\fmfforce{0w,5/6h}{z1}
\fmfforce{0w,1/3h}{z2}
\fmfforce{0w,0h}{z3}
\fmf{plain}{v1,o1}
\fmf{plain}{v1,o2}
\fmf{plain}{v2,i1}
\fmf{plain}{v2,i2}
\fmf{plain}{v1,z1}
\fmf{plain}{v2,z2}
\fmf{plain}{v2,z3}
\fmfdot{v1,v2}
\fmfv{decor.size=0, label=${\scs 3}$, l.dist=1mm, l.angle=-180}{z3}
\fmfv{decor.size=0, label=${\scs 2}$, l.dist=1mm, l.angle=-180}{z2}
\fmfv{decor.size=0, label=${\scs 1}$, l.dist=1mm, l.angle=-180}{z1}
\fmfv{decor.size=0, label=${\scs 7}$, l.dist=1mm, l.angle=0}{i2}
\fmfv{decor.size=0, label=${\scs 6}$, l.dist=1mm, l.angle=0}{i1}
\fmfv{decor.size=0, label=${\scs 5}$, l.dist=1mm, l.angle=0}{o2}
\fmfv{decor.size=0, label=${\scs 4}$, l.dist=1mm, l.angle=0}{o1}
\end{fmfgraph*}
\end{center}}
\hspace*{2mm} \ddphi{\Gamma^{({\rm int})}}{4}{5}{6}{7} \hspace*{2mm} 
\no \\ && \hspace*{0.5cm}
- \frac{1}{2} 
\hspace*{1mm} \dvertex{\Gamma^{({\rm int})}}{1}{2}{3} \hspace*{2mm} 
\parbox{9mm}{\begin{center}
\begin{fmfgraph*}(6,12)
\setval
\fmfstraight
\fmfforce{1w,1h}{o0}
\fmfforce{1w,3/4h}{o1}
\fmfforce{1w,1/2h}{o2}
\fmfforce{1/2w,3/4h}{v1}
\fmfforce{1/2w,1/8h}{v2}
\fmfforce{1w,1/4h}{i1}
\fmfforce{1w,0h}{i2}
\fmfforce{0w,3/4h}{z1}
\fmfforce{0w,1/4h}{z2}
\fmfforce{0w,0h}{z3}
\fmf{plain}{z1,v1}
\fmf{plain}{z2,v2}
\fmf{plain}{z3,v2}
\fmf{plain}{v1,o0}
\fmf{plain}{v1,o1}
\fmf{plain}{v1,o2}
\fmf{plain}{v2,i1}
\fmf{plain}{v2,i2}
\fmfdot{v1,v2}
\fmfv{decor.size=0, label=${\scs 8}$, l.dist=1mm, l.angle=0}{i2}
\fmfv{decor.size=0, label=${\scs 7}$, l.dist=1mm, l.angle=0}{i1}
\fmfv{decor.size=0, label=${\scs 6}$, l.dist=1mm, l.angle=0}{o2}
\fmfv{decor.size=0, label=${\scs 5}$, l.dist=1mm, l.angle=0}{o1}
\fmfv{decor.size=0, label=${\scs 4}$, l.dist=1mm, l.angle=0}{o0}
\fmfv{decor.size=0, label=${\scs 2}$, l.dist=1mm, l.angle=-180}{z2}
\fmfv{decor.size=0, label=${\scs 1}$, l.dist=1mm, l.angle=-180}{z1}
\fmfv{decor.size=0, label=${\scs 3}$, l.dist=1mm, l.angle=-180}{z3}
\end{fmfgraph*}
\end{center}}
\hspace*{2mm} \ddvertex{\Gamma^{({\rm int})}}{4}{5}{6}{7}{8} \hspace*{2mm} 
+ \frac{1}{3} 
\hspace*{1mm} \dphi{\Gamma^{({\rm int})}}{1}{2} \hspace*{2mm} 
\parbox{9mm}{\begin{center}
\begin{fmfgraph*}(6,7)
\setval
\fmfstraight
\fmfforce{0w,1h}{v1}
\fmfforce{2/3w,1h}{v2}
\fmfforce{1w,1h}{v3}
\fmfforce{2/3w,2/7h}{v4}
\fmfforce{1/3w,0h}{v5}
\fmfforce{1/3w,4/7h}{v6}
\fmfforce{1w,2/7h}{v7}
\fmf{boson}{v1,v2}
\fmf{plain}{v2,v3}
\fmf{plain}{v4,v7}
\fmf{plain}{v4,v5}
\fmf{plain}{v4,v6}
\fmf{plain}{v4,v2}
\fmfdot{v1,v2,v4}
\fmfv{decor.size=0, label=${\scs 2}$, l.dist=1mm, l.angle=-180}{v5}
\fmfv{decor.size=0, label=${\scs 1}$, l.dist=1mm, l.angle=-180}{v6}
\fmfv{decor.size=0, label=${\scs 3}$, l.dist=1mm, l.angle=0}{v3}
\fmfv{decor.size=0, label=${\scs 4}$, l.dist=1mm, l.angle=0}{v7}
\end{fmfgraph*}
\end{center}}
\hspace*{2mm} \dphi{\Gamma^{({\rm int})}}{3}{4} \hspace*{2mm} 
\no \\ && \hspace*{0.5cm}
+ \frac{1}{6} 
\parbox{9mm}{\begin{center}
\begin{fmfgraph*}(6,4)
\setval
\fmfstraight
\fmfforce{0w,1/2h}{v1}
\fmfforce{4/6w,1/2h}{v2}
\fmfforce{1w,1h}{i1}
\fmfforce{1w,0h}{i2}
\fmf{boson}{v2,v1}
\fmf{plain}{v2,i1}
\fmf{plain}{v2,i2}
\fmfdot{v1,v2}
\fmfv{decor.size=0, label=${\scs 2}$, l.dist=1mm, l.angle=0}{i2}
\fmfv{decor.size=0, label=${\scs 1}$, l.dist=1mm, l.angle=0}{i1}
\end{fmfgraph*}
\end{center}}
\hspace*{2mm} \ddphi{\Gamma^{({\rm int})}}{1}{2}{3}{4} \hspace*{2mm} 
\parbox{7mm}{\begin{center}
\begin{fmfgraph*}(4,4)
\setval
\fmfstraight
\fmfforce{0w,1h}{o1}
\fmfforce{0w,0h}{o2}
\fmfforce{1/2w,1/2h}{v1}
\fmfforce{1w,1h}{i1}
\fmfforce{1w,0h}{i2}
\fmf{plain}{v1,o1}
\fmf{plain}{v1,o2}
\fmf{plain}{v1,i1}
\fmf{plain}{v1,i2}
\fmfdot{v1}
\fmfforce{1w,1h}{i1}
\fmfforce{1w,0h}{i2}
\fmf{plain}{v1,o1}
\fmf{plain}{v1,o2}
\fmf{plain}{v1,i1}
\fmf{plain}{v1,i2}
\fmfdot{v1}
\fmfv{decor.size=0, label=${\scs 6}$, l.dist=1mm, l.angle=0}{i2}
\fmfv{decor.size=0, label=${\scs 5}$, l.dist=1mm, l.angle=0}{i1}
\fmfv{decor.size=0, label=${\scs 4}$, l.dist=1mm, l.angle=-180}{o2}
\fmfv{decor.size=0, label=${\scs 3}$, l.dist=1mm, l.angle=-180}{o1}
\end{fmfgraph*}
\end{center}}
\hspace*{2mm} \dphi{\Gamma^{({\rm int})}}{5}{6} \hspace*{2mm} 
- \frac{1}{6} 
\parbox{9mm}{\begin{center}
\begin{fmfgraph*}(6,4)
\setval
\fmfstraight
\fmfforce{0w,1/2h}{v1}
\fmfforce{4/6w,1/2h}{v2}
\fmfforce{1w,1.25h}{i1}
\fmfforce{1w,1/2h}{i2}
\fmfforce{1w,-0.25h}{i3}
\fmf{boson}{v2,v1}
\fmf{plain}{v2,i1}
\fmf{plain}{v2,i2}
\fmf{plain}{v2,i3}
\fmfdot{v1,v2}
\fmfv{decor.size=0, label=${\scs 3}$, l.dist=1mm, l.angle=0}{i3}
\fmfv{decor.size=0, label=${\scs 2}$, l.dist=1mm, l.angle=0}{i2}
\fmfv{decor.size=0, label=${\scs 1}$, l.dist=1mm, l.angle=0}{i1}
\end{fmfgraph*}
\end{center}}
\hspace*{2mm} \ddvertex{\Gamma^{({\rm int})}}{1}{2}{3}{4}{5} \hspace*{2mm} 
\parbox{7mm}{\begin{center}
\begin{fmfgraph*}(4,4)
\setval
\fmfstraight
\fmfforce{0w,1h}{o1}
\fmfforce{0w,0h}{o2}
\fmfforce{1/2w,1/2h}{v1}
\fmfforce{1w,1h}{i1}
\fmfforce{1w,0h}{i2}
\fmf{plain}{v1,o1}
\fmf{plain}{v1,o2}
\fmf{plain}{v1,i1}
\fmf{plain}{v1,i2}
\fmfdot{v1}
\fmfv{decor.size=0, label=${\scs 7}$, l.dist=1mm, l.angle=0}{i2}
\fmfv{decor.size=0, label=${\scs 6}$, l.dist=1mm, l.angle=0}{i1}
\fmfv{decor.size=0, label=${\scs 5}$, l.dist=1mm, l.angle=-180}{o2}
\fmfv{decor.size=0, label=${\scs 4}$, l.dist=1mm, l.angle=-180}{o1}
\end{fmfgraph*}
\end{center}}
\hspace*{2mm} \dphi{\Gamma^{({\rm int})}}{6}{7} \hspace*{2mm}
\hspace*{4mm} . \la{4VRE}
\eeq
The right-hand side consists of only 22 out of 32 terms from Eq.~(\r{FFF}),
4 without $\Gamma^{({\rm int})}$, 11 linear in $\Gamma^{({\rm int})}$ and 7
bilinear in  $\Gamma^{({\rm int})}$. 
We demonstrate the iteration of the 
graphical relation (\r{4VRE}) for the one-particle irreducible
vacuum diagrams without field expectation values $\Gamma^{(0,l)}$.
For $n=0$ and $l=2$ Eq.~(\ref{4VRE}) reduces to
\beq
\parbox{7mm}{\begin{center}
\begin{fmfgraph*}(3,3)
\setval
\fmfstraight
\fmfforce{0w,1/2h}{v1}
\fmfforce{1w,2h}{i1}
\fmfforce{1w,1h}{i2}
\fmfforce{1w,0h}{i3}
\fmfforce{1w,-1h}{i4}
\fmf{plain}{v1,i1}
\fmf{plain}{v1,i2}
\fmf{plain}{v1,i3}
\fmf{plain}{v1,i4}
\fmfdot{v1}
\fmfv{decor.size=0, label=${\scs 4}$, l.dist=1mm, l.angle=0}{i4}
\fmfv{decor.size=0, label=${\scs 3}$, l.dist=1mm, l.angle=0}{i3}
\fmfv{decor.size=0, label=${\scs 2}$, l.dist=1mm, l.angle=0}{i2}
\fmfv{decor.size=0, label=${\scs 1}$, l.dist=1mm, l.angle=0}{i1}
\end{fmfgraph*}
\end{center}}
\hspace*{2mm} \dvertexx{ \Gamma
^{(0,2)}}{2}{3}{4}{1}\hspace*{1mm} = 
\hspace*{1mm} \frac{1}{8}\hspace*{1mm}
\parbox{11mm}{\begin{center}
\begin{fmfgraph*}(8,4)
\setval
\fmfleft{i1}
\fmfright{o1}
\fmf{plain,left=1}{i1,v1,i1}
\fmf{plain,left=1}{o1,v1,o1}
\fmfdot{v1}
\end{fmfgraph*}\end{center}} \la{GVE1} \, ,
\eeq
whereas for $n=0$ and $l \ge 3$ we obtain the graphical recursion relation
\beq
&&
\parbox{7mm}{\begin{center}
\begin{fmfgraph*}(3,3)
\setval
\fmfstraight
\fmfforce{0w,1/2h}{v1}
\fmfforce{1w,2h}{i1}
\fmfforce{1w,1h}{i2}
\fmfforce{1w,0h}{i3}
\fmfforce{1w,-1h}{i4}
\fmf{plain}{v1,i1}
\fmf{plain}{v1,i2}
\fmf{plain}{v1,i3}
\fmf{plain}{v1,i4}
\fmfdot{v1}
\fmfv{decor.size=0, label=${\scs 4}$, l.dist=1mm, l.angle=0}{i4}
\fmfv{decor.size=0, label=${\scs 3}$, l.dist=1mm, l.angle=0}{i3}
\fmfv{decor.size=0, label=${\scs 2}$, l.dist=1mm, l.angle=0}{i2}
\fmfv{decor.size=0, label=${\scs 1}$, l.dist=1mm, l.angle=0}{i1}
\end{fmfgraph*}
\end{center}}
\hspace*{2mm} \dvertexx{ \Gamma^{(0,l)}}{2}{3}{4}{1}\hspace*{1mm} 
= \hspace*{1mm}
\frac{1}{2} \hspace*{1mm} 
\parbox{9mm}{\begin{center}
\begin{fmfgraph*}(6,4)
\setval
\fmfstraight
\fmfforce{0w,1/2h}{v1}
\fmfforce{4/6w,1/2h}{v2}
\fmfforce{1w,1h}{i2}
\fmfforce{1w,0h}{i1}
\fmf{plain}{i1,v2}
\fmf{plain}{v2,i2}
\fmf{plain,left}{v1,v2,v1}
\fmfdot{v2}
\fmfv{decor.size=0, label=${\scs 2}$, l.dist=1mm, l.angle=0}{i1}
\fmfv{decor.size=0, label=${\scs 1}$, l.dist=1mm, l.angle=0}{i2}
\end{fmfgraph*}
\end{center}}
\hspace*{3mm} \dphi{\Gamma^{(0,l-1)}}{1}{2} 
+ \frac{1}{4} \hspace*{1mm} 
\parbox{10mm}{\begin{center}
\begin{fmfgraph*}(6,5.33333)
\setval
\fmfstraight
\fmfforce{2/6w,1/4h}{v1}
\fmfforce{2/6w,1h}{v2}
\fmfforce{1w,0h}{i1}
\fmfforce{1w,1/2h}{i2}
\fmfforce{1w,1h}{i3}
\fmf{plain,left=1}{v1,v2,v1}
\fmf{plain}{v1,i1}
\fmf{plain}{v1,i2}
\fmf{plain}{v2,i3}
\fmfdot{v1,v2}
\fmfv{decor.size=0, label=${\scs 2}$, l.dist=1mm, l.angle=0}{i2}
\fmfv{decor.size=0, label=${\scs 3}$, l.dist=1mm, l.angle=0}{i1}
\fmfv{decor.size=0, label=${\scs 1}$, l.dist=1mm, l.angle=0}{i3}
\end{fmfgraph*}
\end{center}}
\hspace*{3mm} \dvertex{\Gamma^{(0,l-1)}}{1}{2}{3} 
\hspace*{1mm}  + \frac{1}{6} \hspace*{1mm}
\parbox{7mm}{\begin{center}
\begin{fmfgraph*}(3,3)
\setval
\fmfstraight
\fmfforce{0w,1/2h}{v1}
\fmfforce{1w,2h}{i1}
\fmfforce{1w,1h}{i2}
\fmfforce{1w,0h}{i3}
\fmfforce{1w,-1h}{i4}
\fmf{plain}{v1,i1}
\fmf{plain}{v1,i2}
\fmf{plain}{v1,i3}
\fmf{plain}{v1,i4}
\fmfdot{v1}
\fmfv{decor.size=0, label=${\scs 4}$, l.dist=1mm, l.angle=0}{i4}
\fmfv{decor.size=0, label=${\scs 3}$, l.dist=1mm, l.angle=0}{i3}
\fmfv{decor.size=0, label=${\scs 2}$, l.dist=1mm, l.angle=0}{i2}
\fmfv{decor.size=0, label=${\scs 1}$, l.dist=1mm, l.angle=0}{i1}
\end{fmfgraph*}
\end{center}}
\hspace*{3mm} \ddphi{\Gamma^{(0,l-1)}}{1}{2}{3}{4}
\no \vspace*{0.3mm} \\ && \hspace*{1cm}
\hspace*{1mm} + \sum_{l'=2}^{l-2} \left\{ \frac{1}{6} \hspace*{1mm}
\hspace*{1mm} \dphi{\Gamma^{(0,l')}}{1}{2}\hspace*{3mm}
\parbox{7mm}{\begin{center}
\begin{fmfgraph*}(4,4)
\setval
\fmfstraight
\fmfforce{0w,1h}{o1}
\fmfforce{0w,0h}{o2}
\fmfforce{1/2w,1/2h}{v1}
\fmfforce{1w,1h}{i1}
\fmfforce{1w,0h}{i2}
\fmf{plain}{v1,o1}
\fmf{plain}{v1,o2}
\fmf{plain}{v1,i1}
\fmf{plain}{v1,i2}
\fmfdot{v1}
\fmfv{decor.size=0, label=${\scs 4}$, l.dist=1mm, l.angle=0}{i2}
\fmfv{decor.size=0, label=${\scs 3}$, l.dist=1mm, l.angle=0}{i1}
\fmfv{decor.size=0, label=${\scs 2}$, l.dist=1mm, l.angle=-180}{o2}
\fmfv{decor.size=0, label=${\scs 1}$, l.dist=1mm, l.angle=-180}{o1}
\end{fmfgraph*}
\end{center}}
\hspace*{3mm} \dphi{\Gamma^{(0,l-l'-1)}}{3}{4} 
+ 
\hspace*{2mm} \dphi{\Gamma^{(0,l')}}{1}{2} \hspace*{2mm}
\parbox{10mm}{\begin{center}
\begin{fmfgraph*}(8,5.33333)
\setval
\fmfstraight
\fmfforce{4/8w,1/4h}{v1}
\fmfforce{4/8w,1h}{v2}
\fmfforce{1w,0h}{i1}
\fmfforce{1w,1/2h}{i2}
\fmfforce{1w,1h}{i3}
\fmfforce{0w,1/4h}{z1}
\fmfforce{0w,1h}{z2}
\fmf{plain}{v1,z1}
\fmf{plain}{z2,v2}
\fmf{plain}{v1,v2}
\fmf{plain}{v1,i1}
\fmf{plain}{v1,i2}
\fmf{plain}{v2,i3}
\fmfdot{v1,v2}
\fmfv{decor.size=0, label=${\scs 4}$, l.dist=1mm, l.angle=0}{i2}
\fmfv{decor.size=0, label=${\scs 5}$, l.dist=1mm, l.angle=0}{i1}
\fmfv{decor.size=0, label=${\scs 3}$, l.dist=1mm, l.angle=0}{i3}
\fmfv{decor.size=0, label=${\scs 2}$, l.dist=1mm, l.angle=-180}{z1}
\fmfv{decor.size=0, label=${\scs 1}$, l.dist=1mm, l.angle=-180}{z2}
\end{fmfgraph*}
\end{center}}
\hspace*{2mm} \dvertex{\Gamma^{(l-l'-1)}}{3}{4}{5} \hspace*{2mm} 
\right.
\no \\ && \hspace*{1cm} \left.
+ \frac{1}{2} 
\hspace*{1mm} \dvertex{\Gamma^{(0,l')}}{1}{2}{3} \hspace*{2mm} 
\parbox{9mm}{\begin{center}
\begin{fmfgraph*}(6,9)
\setval
\fmfstraight
\fmfforce{1w,1h}{o1}
\fmfforce{1w,2/3h}{o2}
\fmfforce{1/2w,5/6h}{v1}
\fmfforce{1/2w,1/6h}{v2}
\fmfforce{1w,1/3h}{i1}
\fmfforce{1w,0h}{i2}
\fmfforce{0w,5/6h}{z1}
\fmfforce{0w,1/3h}{z2}
\fmfforce{0w,0h}{z3}
\fmf{plain}{v1,o1}
\fmf{plain}{v1,o2}
\fmf{plain}{v2,i1}
\fmf{plain}{v2,i2}
\fmf{plain}{v1,z1}
\fmf{plain}{v2,z2}
\fmf{plain}{v2,z3}
\fmfdot{v1,v2}
\fmfv{decor.size=0, label=${\scs 3}$, l.dist=1mm, l.angle=-180}{z3}
\fmfv{decor.size=0, label=${\scs 2}$, l.dist=1mm, l.angle=-180}{z2}
\fmfv{decor.size=0, label=${\scs 1}$, l.dist=1mm, l.angle=-180}{z1}
\fmfv{decor.size=0, label=${\scs 7}$, l.dist=1mm, l.angle=0}{i2}
\fmfv{decor.size=0, label=${\scs 6}$, l.dist=1mm, l.angle=0}{i1}
\fmfv{decor.size=0, label=${\scs 5}$, l.dist=1mm, l.angle=0}{o2}
\fmfv{decor.size=0, label=${\scs 4}$, l.dist=1mm, l.angle=0}{o1}
\end{fmfgraph*}
\end{center}}
\hspace*{2mm} \ddphi{\Gamma^{(l-l'-1)}}{4}{5}{6}{7} \hspace*{2mm} 
+ \frac{1}{2} 
\hspace*{1mm} \dvertex{\Gamma^{(0,l')}}{1}{2}{3} \hspace*{2mm} 
\parbox{9mm}{\begin{center}
\begin{fmfgraph*}(6,12)
\setval
\fmfstraight
\fmfforce{1w,1h}{o0}
\fmfforce{1w,3/4h}{o1}
\fmfforce{1w,1/2h}{o2}
\fmfforce{1/2w,3/4h}{v1}
\fmfforce{1/2w,1/8h}{v2}
\fmfforce{1w,1/4h}{i1}
\fmfforce{1w,0h}{i2}
\fmfforce{0w,3/4h}{z1}
\fmfforce{0w,1/4h}{z2}
\fmfforce{0w,0h}{z3}
\fmf{plain}{z1,v1}
\fmf{plain}{z2,v2}
\fmf{plain}{z3,v2}
\fmf{plain}{v1,o0}
\fmf{plain}{v1,o1}
\fmf{plain}{v1,o2}
\fmf{plain}{v2,i1}
\fmf{plain}{v2,i2}
\fmfdot{v1,v2}
\fmfv{decor.size=0, label=${\scs 8}$, l.dist=1mm, l.angle=0}{i2}
\fmfv{decor.size=0, label=${\scs 7}$, l.dist=1mm, l.angle=0}{i1}
\fmfv{decor.size=0, label=${\scs 6}$, l.dist=1mm, l.angle=0}{o2}
\fmfv{decor.size=0, label=${\scs 5}$, l.dist=1mm, l.angle=0}{o1}
\fmfv{decor.size=0, label=${\scs 4}$, l.dist=1mm, l.angle=0}{o0}
\fmfv{decor.size=0, label=${\scs 2}$, l.dist=1mm, l.angle=-180}{z2}
\fmfv{decor.size=0, label=${\scs 1}$, l.dist=1mm, l.angle=-180}{z1}
\fmfv{decor.size=0, label=${\scs 3}$, l.dist=1mm, l.angle=-180}{z3}
\end{fmfgraph*}
\end{center}}
\hspace*{2mm} \ddvertex{\Gamma^{(0,l-l'-1)}}{4}{5}{6}{7}{8} \right\}
\hspace*{2mm}  \, . \la{GVE2}
\eeq
Integrating (\r{GVE1}) and (\r{GVE2}), we have to take into account
as a yet undetermined integration constant all those one-particle
irreducible vacuum
diagrams $\tilde{\Gamma}^{(0,l)}$ 
which only consist of $3$-vertices. A comparison
with (\r{FFF}) shows that those diagrams follow from the graphical 
recursion relation
\beq
\tilde{\Gamma}^{(0,l)}\hspace*{1mm}& =&\hspace*{1mm} 
\frac{1}{3(l-1)} \, \left\{ 
\frac{3}{2}
\parbox{7mm}{\begin{center}
\begin{fmfgraph*}(6,4)
\setval
\fmfstraight
\fmfforce{1/3w,0h}{v1}
\fmfforce{1/3w,1h}{v2}
\fmfforce{1w,1h}{i2}
\fmfforce{1w,0h}{i1}
\fmf{plain}{i1,v1}
\fmf{plain}{v2,i2}
\fmf{plain,left}{v1,v2,v1}
\fmfdot{v2,v1}
\fmfv{decor.size=0, label=${\scs 2}$, l.dist=1mm, l.angle=0}{i1}
\fmfv{decor.size=0, label=${\scs 1}$, l.dist=1mm, l.angle=0}{i2}
\end{fmfgraph*}
\end{center}}
\hspace*{3mm} \dphi{\tilde{\Gamma}^{(0,l-1)}}{1}{2}
+ \frac{1}{2} \hspace*{1mm}
\parbox{7mm}{\begin{center}
\begin{fmfgraph*}(3,9)
\setval
\fmfstraight
\fmfforce{1w,1h}{o1}
\fmfforce{1w,2/3h}{o2}
\fmfforce{0w,5/6h}{v1}
\fmfforce{0w,1/6h}{v2}
\fmfforce{1w,1/3h}{i1}
\fmfforce{1w,0h}{i2}
\fmf{plain}{v1,v2}
\fmf{plain}{v1,o1}
\fmf{plain}{v1,o2}
\fmf{plain}{v2,i1}
\fmf{plain}{v2,i2}
\fmfdot{v1,v2}
\fmfv{decor.size=0, label=${\scs 4}$, l.dist=1mm, l.angle=0}{i2}
\fmfv{decor.size=0, label=${\scs 3}$, l.dist=1mm, l.angle=0}{i1}
\fmfv{decor.size=0, label=${\scs 2}$, l.dist=1mm, l.angle=0}{o2}
\fmfv{decor.size=0, label=${\scs 1}$, l.dist=1mm, l.angle=0}{o1}
\end{fmfgraph*}
\end{center}}
\hspace*{3mm} \ddphi{\tilde{\Gamma}^{(0,l-1)}}{1}{2}{3}{4} 
\right. \no \\ & & \left.  
\hspace*{1cm} + \sum_{l'=2}^{l-2} \left(
2 \hspace*{1mm}
\dphi{\tilde{\Gamma}^{(0,l')}}{1}{2} \hspace*{2mm} 
\parbox{7mm}{\begin{center}
\begin{fmfgraph*}(4,4)
\setval
\fmfstraight
\fmfforce{0w,1h}{o1}
\fmfforce{0w,0h}{o2}
\fmfforce{1/2w,0h}{v1}
\fmfforce{1/2w,1h}{v2}
\fmfforce{1w,1h}{i1}
\fmfforce{1w,0h}{i2}
\fmf{plain}{v1,v2}
\fmf{plain}{v1,i2}
\fmf{plain}{v1,o2}
\fmf{plain}{v2,i1}
\fmf{plain}{v2,o1}
\fmfdot{v1,v2}
\fmfv{decor.size=0, label=${\scs 4}$, l.dist=1mm, l.angle=0}{i2}
\fmfv{decor.size=0, label=${\scs 3}$, l.dist=1mm, l.angle=0}{i1}
\fmfv{decor.size=0, label=${\scs 2}$, l.dist=1mm, l.angle=-180}{o2}
\fmfv{decor.size=0, label=${\scs 1}$, l.dist=1mm, l.angle=-180}{o1}
\end{fmfgraph*}
\end{center}}
\hspace*{2mm} \dphi{\tilde{\Gamma}^{(0,l-l'-1)}}{3}{4}  
+ \hspace*{1mm}
\dphi{\tilde{\Gamma}^{(0,l')}}{1}{2} \hspace*{2mm} 
\parbox{9mm}{\begin{center}
\begin{fmfgraph*}(6,9)
\setval
\fmfstraight
\fmfforce{1w,1h}{o1}
\fmfforce{1w,2/3h}{o2}
\fmfforce{1/2w,5/6h}{v1}
\fmfforce{1/2w,1/6h}{v2}
\fmfforce{1w,1/3h}{i1}
\fmfforce{1w,0h}{i2}
\fmfforce{0w,5/6h}{z1}
\fmfforce{0w,1/6h}{z2}
\fmf{plain}{v1,o1}
\fmf{plain}{v1,o2}
\fmf{plain}{v2,i1}
\fmf{plain}{v2,i2}
\fmf{plain}{v1,z1}
\fmf{plain}{v2,z2}
\fmfdot{v1,v2}
\fmfv{decor.size=0, label=${\scs 2}$, l.dist=1mm, l.angle=-180}{z2}
\fmfv{decor.size=0, label=${\scs 1}$, l.dist=1mm, l.angle=-180}{z1}
\fmfv{decor.size=0, label=${\scs 6}$, l.dist=1mm, l.angle=0}{i2}
\fmfv{decor.size=0, label=${\scs 5}$, l.dist=1mm, l.angle=0}{i1}
\fmfv{decor.size=0, label=${\scs 4}$, l.dist=1mm, l.angle=0}{o2}
\fmfv{decor.size=0, label=${\scs 3}$, l.dist=1mm, l.angle=0}{o1}
\end{fmfgraph*}
\end{center}}
\hspace*{2mm} \ddphi{\tilde{\Gamma}^{(0,l-l'-1)}}{3}{4}{5}{6} \right) \right\}
\hspace*{2mm}, \la{YTYE}
\eeq
which is to be iterated starting from
\beq
\tilde{\Gamma}^{(0,2)}\hspace*{1mm} =  \hspace*{1mm}\frac{1}{12}\hspace*{1mm}
\parbox{7mm}{\begin{center}
\begin{fmfgraph*}(4,4)
\setval
\fmfforce{0w,0.5h}{v1}
\fmfforce{1w,0.5h}{v2}
\fmf{plain,left=1}{v1,v2,v1}
\fmf{plain}{v1,v2}
\fmfdot{v1,v2}
\end{fmfgraph*}\end{center}}
\hspace*{4mm} . \la{SEE}
\eeq
In the first iteration step we have to evaluate the amputation of one
or two lines from (\r{SEE})
\beq
\la{ENNNR1}
\dphi{\tilde{\Gamma}^{(0,2)}}{1}{2} &=&  \frac{1}{4} \hspace*{1mm}
\parbox{9mm}{\begin{center}
\begin{fmfgraph*}(6,4)
\setval
\fmfstraight
\fmfforce{1/3w,0h}{v1}
\fmfforce{1/3w,1h}{v2}
\fmfforce{1w,1h}{i2}
\fmfforce{1w,0h}{i1}
\fmf{plain}{i1,v1}
\fmf{plain}{v2,i2}
\fmf{plain,left}{v1,v2,v1}
\fmfdot{v2,v1}
\fmfv{decor.size=0, label=${\scs 2}$, l.dist=1mm, l.angle=0}{i1}
\fmfv{decor.size=0, label=${\scs 1}$, l.dist=1mm, l.angle=0}{i2}
\end{fmfgraph*}
\end{center}}
\hspace*{4mm}  \no \\
\la{ENNNR2}
\ddphi{\tilde{\Gamma}^{(0,2)}}{1}{2}{3}{4} &=&  \frac{1}{4} \hspace*{2mm}
\parbox{11mm}{\begin{center}
\begin{fmfgraph*}(8,4)
\setval
\fmfstraight
\fmfforce{0w,1h}{o1}
\fmfforce{0w,0h}{o2}
\fmfforce{1/4w,1/2h}{v1}
\fmfforce{3/4w,1/2h}{v2}
\fmfforce{1w,1h}{i1}
\fmfforce{1w,0h}{i2}
\fmf{plain}{v1,v2}
\fmf{plain}{v1,o1}
\fmf{plain}{v1,o2}
\fmf{plain}{v2,i1}
\fmf{plain}{v2,i2}
\fmfdot{v1,v2}
\fmfv{decor.size=0, label=${\scs 4}$, l.dist=1mm, l.angle=0}{i2}
\fmfv{decor.size=0, label=${\scs 2}$, l.dist=1mm, l.angle=0}{i1}
\fmfv{decor.size=0, label=${\scs 3}$, l.dist=1mm, l.angle=-180}{o2}
\fmfv{decor.size=0, label=${\scs 1}$, l.dist=1mm, l.angle=-180}{o1}
\end{fmfgraph*}
\end{center}}
\hspace*{2mm}+ \hspace*{2mm}\frac{1}{4} \hspace*{2mm}
\parbox{11mm}{\begin{center}
\begin{fmfgraph*}(8,4)
\setval
\fmfstraight
\fmfforce{0w,1h}{o1}
\fmfforce{0w,0h}{o2}
\fmfforce{1/4w,1/2h}{v1}
\fmfforce{3/4w,1/2h}{v2}
\fmfforce{1w,1h}{i1}
\fmfforce{1w,0h}{i2}
\fmf{plain}{v1,v2}
\fmf{plain}{v1,o1}
\fmf{plain}{v1,o2}
\fmf{plain}{v2,i1}
\fmf{plain}{v2,i2}
\fmfdot{v1,v2}
\fmfv{decor.size=0, label=${\scs 3}$, l.dist=1mm, l.angle=0}{i2}
\fmfv{decor.size=0, label=${\scs 2}$, l.dist=1mm, l.angle=0}{i1}
\fmfv{decor.size=0, label=${\scs 4}$, l.dist=1mm, l.angle=-180}{o2}
\fmfv{decor.size=0, label=${\scs 1}$, l.dist=1mm, l.angle=-180}{o1}
\end{fmfgraph*}
\end{center}}
\hspace*{4mm} , 
\eeq
and insert this into (\r{YTYE}) for $l=3$
\beq
\tilde{\Gamma}^{(0,3)}\hspace*{1mm} =\hspace*{1mm} \frac{1}{6} \, 
{\displaystyle \left\{ \frac{3}{2}
\parbox{7mm}{\begin{center}
\begin{fmfgraph*}(6,4)
\setval
\fmfstraight
\fmfforce{1/3w,0h}{v1}
\fmfforce{1/3w,1h}{v2}
\fmfforce{1w,1h}{i2}
\fmfforce{1w,0h}{i1}
\fmf{plain}{i1,v1}
\fmf{plain}{v2,i2}
\fmf{plain,left}{v1,v2,v1}
\fmfdot{v2,v1}
\fmfv{decor.size=0, label=${\scs 2}$, l.dist=1mm, l.angle=0}{i1}
\fmfv{decor.size=0, label=${\scs 1}$, l.dist=1mm, l.angle=0}{i2}
\end{fmfgraph*}
\end{center}}
\hspace*{3mm} \dphi{\tilde{\Gamma}^{(0,2)}}{1}{2}
+ \frac{1}{2} \hspace*{1mm}
\parbox{7mm}{\begin{center}
\begin{fmfgraph*}(3,9)
\setval
\fmfstraight
\fmfforce{1w,1h}{o1}
\fmfforce{1w,2/3h}{o2}
\fmfforce{0w,5/6h}{v1}
\fmfforce{0w,1/6h}{v2}
\fmfforce{1w,1/3h}{i1}
\fmfforce{1w,0h}{i2}
\fmf{plain}{v1,v2}
\fmf{plain}{v1,o1}
\fmf{plain}{v1,o2}
\fmf{plain}{v2,i1}
\fmf{plain}{v2,i2}
\fmfdot{v1,v2}
\fmfv{decor.size=0, label=${\scs 4}$, l.dist=1mm, l.angle=0}{i2}
\fmfv{decor.size=0, label=${\scs 3}$, l.dist=1mm, l.angle=0}{i1}
\fmfv{decor.size=0, label=${\scs 2}$, l.dist=1mm, l.angle=0}{o2}
\fmfv{decor.size=0, label=${\scs 1}$, l.dist=1mm, l.angle=0}{o1}
\end{fmfgraph*}
\end{center}}
\hspace*{3mm} \ddphi{\tilde{\Gamma}^{(0,2)}}{1}{2}{3}{4} 
\right\} }\hspace*{2mm},  
\la{YTE}
\eeq
to obtain
\beq
\tilde{\Gamma}^{(0,3)} = \frac{1}{24}
\parbox{9mm}{\begin{center}
\begin{fmfgraph}(4,4)
\setval
\fmfforce{0w,0h}{v1}
\fmfforce{1w,0h}{v2}
\fmfforce{1w,1h}{v3}
\fmfforce{0w,1h}{v4}
\fmf{plain,right=1}{v1,v3,v1}
\fmf{plain}{v1,v3}
\fmf{plain}{v2,v4}
\fmfdot{v1,v2,v3,v4}
\end{fmfgraph}\end{center}} 
\hspace*{1mm} + \frac{1}{16} \hspace*{1mm}
\parbox{9mm}{\begin{center}
\begin{fmfgraph}(4,4)
\setval
\fmfforce{0w,0h}{v1}
\fmfforce{1w,0h}{v2}
\fmfforce{1w,1h}{v3}
\fmfforce{0w,1h}{v4}
\fmf{plain,right=1}{v1,v3,v1}
\fmf{plain,right=0.4}{v1,v4}
\fmf{plain,left=0.4}{v2,v3}
\fmfdot{v1,v2,v3,v4}
\end{fmfgraph}\end{center}} 
\hspace*{4mm} .\la{ITCE}
\eeq
Integrating (\r{GVE1}) with the integration constant (\r{SEE}) leads,
indeed, to the correct two-loop result (\r{EW2}). Subsequently we  
insert (\r{ENNR1})--(\r{ENNR4}) into (\r{GVE2}) with $l=3$ and obtain
\beq
\parbox{7mm}{\begin{center}
\begin{fmfgraph*}(3,3)
\setval
\fmfstraight
\fmfforce{0w,1/2h}{v1}
\fmfforce{1w,2h}{i1}
\fmfforce{1w,1h}{i2}
\fmfforce{1w,0h}{i3}
\fmfforce{1w,-1h}{i4}
\fmf{plain}{v1,i1}
\fmf{plain}{v1,i2}
\fmf{plain}{v1,i3}
\fmf{plain}{v1,i4}
\fmfdot{v1}
\fmfv{decor.size=0, label=${\scs 4}$, l.dist=1mm, l.angle=0}{i4}
\fmfv{decor.size=0, label=${\scs 3}$, l.dist=1mm, l.angle=0}{i3}
\fmfv{decor.size=0, label=${\scs 2}$, l.dist=1mm, l.angle=0}{i2}
\fmfv{decor.size=0, label=${\scs 1}$, l.dist=1mm, l.angle=0}{i1}
\end{fmfgraph*}
\end{center}}
\hspace*{2mm} \dvertexx{ \Gamma^{(0,3)}}{2}{3}{4}{1} = 
\frac{1}{24}\hspace*{1mm}
\parbox{9mm}{\begin{center}
\begin{fmfgraph}(6,4)
\setval
\fmfforce{0w,0.5h}{v1}
\fmfforce{1w,0.5h}{v2}
\fmf{plain,left=1}{v1,v2,v1}
\fmf{plain,left=0.4}{v1,v2,v1}
\fmfdot{v1,v2}
\end{fmfgraph}\end{center}} 
\hspace*{1mm}+ \frac{1}{8}\hspace*{1mm}
\parbox{15mm}{\begin{center}
\begin{fmfgraph}(12,4)
\setval
\fmfleft{i1}
\fmfright{o1}
\fmf{plain,left=1}{i1,v1,i1}
\fmf{plain,left=1}{v1,v2,v1}
\fmf{plain,left=1}{o1,v2,o1}
\fmfdot{v1,v2}
\end{fmfgraph}\end{center}} 
+\frac{1}{8} \hspace*{1mm} 
\parbox{9mm}{\begin{center}
\begin{fmfgraph}(6,6)
\setval
\fmfforce{0w,1/2h}{v1}
\fmfforce{1w,1/2h}{v2}
\fmfforce{1/2w,1h}{v3}
\fmf{plain,left=1}{v1,v2,v1}
\fmf{plain,left=0.4}{v3,v1}
\fmf{plain,left=0.4}{v2,v3}
\fmfdot{v1,v2,v3}
\end{fmfgraph}\end{center}}
\hspace*{1mm}+ \frac{1}{8} \hspace*{1mm} 
\parbox{11mm}{\begin{center}
\begin{fmfgraph}(4,8)
\setval
\fmfforce{0w,1/4h}{v1}
\fmfforce{1w,1/4h}{v2}
\fmfforce{1/2w,1/2h}{v3}
\fmfforce{1/2w,1h}{v4}
\fmf{plain,left=1}{v1,v2,v1}
\fmf{plain,left=1}{v3,v4,v3}
\fmf{plain}{v1,v2}
\fmfdot{v2,v3,v1}
\end{fmfgraph}\end{center}} \hspace*{2mm}.
\eeq
Integrating this with the integration constant (\r{ITCE}), we rederive
the three-loop result listed in Table III.
\subsubsection{Cross-Check}
The graphical relations (\r{FFF}), (\r{GRR}), 
(\r{4VRE}) for the interaction negative effective energy $\Gamma^{({\rm int})}$
are not independent from each other. Indeed, going back to the linear
functional differential equation (\r{CHECK1}) for $W$, the functional
Legendre transform with respect to the current leads to
\beq
\delta_{11} \int_1 + \int_1 \Phi_1 \frac{\delta \Gamma}{\delta \Phi_1}
-2 \int_{12} G^{-1}_{12} \frac{\delta \Gamma}{\delta G^{-1}_{12}}
-3 \int_{123} K_{123} \frac{\delta \Gamma}{\delta K_{123}}
-4 \int_{1234} L_{1234} \frac{\delta \Gamma}{\delta L_{1234}}
= 0 \, , \la{CHDI}
\eeq
so that
we obtain together
with (\r{NR1}) and (\r{EDEC}) the following topological identity for the 
number of field expectation values, lines, $3$- and $4$-vertices of each 
one-particle irreducible vacuum diagram:
\beq
\int_1 \Phi_1 \frac{\delta \Gamma^{({\rm int})}}{\delta \Phi_1} +2 \int_{12}
G_{12} \frac{\delta \Gamma^{({\rm int})}}{\delta G^{-1}_{12}} - 3
\int_{123} K_{123}\frac{\delta \Gamma^{({\rm int})}}{\delta K_{123}}
- 4 \int_{1234} L_{1234}\frac{\delta \Gamma^{({\rm int})}}{\delta L_{1234}}
= 0 \, .
\eeq
Its graphical representation is
\beq
\parbox{8mm}{\begin{center}
\begin{fmfgraph*}(5,5)
\setval
\fmfstraight
\fmfforce{0w,1/2h}{v1}
\fmfforce{1w,1/2h}{v2}
\fmf{boson}{v1,v2}
\fmfdot{v1}
\fmfv{decor.size=0, label=${\scs 1}$, l.dist=1mm, l.angle=0}{v2}
\end{fmfgraph*}
\end{center}}
\hspace*{2mm} \fdphi{\Gamma^{({\rm int})}}{1} \hspace*{1mm} +2 \hspace*{1mm}
\parbox{5.5mm}{\begin{center}
\begin{fmfgraph*}(2.5,5)
\setval
\fmfstraight
\fmfforce{1w,0h}{v1}
\fmfforce{1w,1h}{v2}
\fmf{plain,left=1}{v1,v2}
\fmfv{decor.size=0, label=${\scs 2}$, l.dist=1mm, l.angle=0}{v1}
\fmfv{decor.size=0, label=${\scs 1}$, l.dist=1mm, l.angle=0}{v2}
\end{fmfgraph*}
\end{center}}
\hspace*{0.3cm} \dphi{\Gamma^{({\rm int})}}{1}{2} 
\hspace*{1mm} - 3 \hspace*{1mm}
\parbox{5mm}{\begin{center}
\begin{fmfgraph*}(2,4)
\setval
\fmfstraight
\fmfforce{0w,1/2h}{v1}
\fmfforce{1w,1/2h}{v2}
\fmfforce{1w,1.25h}{v3}
\fmfforce{1w,-0.25h}{v4}
\fmf{plain}{v1,v2}
\fmf{plain}{v1,v3}
\fmf{plain}{v1,v4}
\fmfv{decor.size=0, label=${\scs 2}$, l.dist=1mm, l.angle=0}{v2}
\fmfv{decor.size=0, label=${\scs 1}$, l.dist=1mm, l.angle=0}{v3}
\fmfv{decor.size=0, label=${\scs 3}$, l.dist=1mm, l.angle=0}{v4}
\fmfdot{v1}
\end{fmfgraph*}
\end{center}}
\hspace*{0.3cm} \dvertex{\Gamma^{({\rm int})}}{1}{2}{3}\hspace*{1mm} - 4 
\hspace*{1mm}
\parbox{7mm}{\begin{center}
\begin{fmfgraph*}(3,3)
\setval
\fmfstraight
\fmfforce{0w,1/2h}{v1}
\fmfforce{1w,2h}{i1}
\fmfforce{1w,1h}{i2}
\fmfforce{1w,0h}{i3}
\fmfforce{1w,-1h}{i4}
\fmf{plain}{v1,i1}
\fmf{plain}{v1,i2}
\fmf{plain}{v1,i3}
\fmf{plain}{v1,i4}
\fmfdot{v1}
\fmfv{decor.size=0, label=${\scs 4}$, l.dist=1mm, l.angle=0}{i4}
\fmfv{decor.size=0, label=${\scs 3}$, l.dist=1mm, l.angle=0}{i3}
\fmfv{decor.size=0, label=${\scs 2}$, l.dist=1mm, l.angle=0}{i2}
\fmfv{decor.size=0, label=${\scs 1}$, l.dist=1mm, l.angle=0}{i1}
\end{fmfgraph*}
\end{center}}
\hspace*{2mm} \dvertexx{\Gamma^{({\rm int})}}{2}{3}{4}{1}\hspace*{1mm} = 0
\, . \la{CHECK3}
\eeq
Inserting (\r{FFF}), (\r{GRR}), (\r{4VRE}), we can check this relation term
by term.
\end{fmffile}
\section{Summary}
In this work we have presented a method
for determining the connected
and one-particle irreducible vacuum diagrams together with their proper
weights in the ordered phase of the euclidean multicomponent scalar
$\phi^4$-theory. Whereas in the disordered, symmetric phase it is sufficient
to deal with even field powers in the energy functional by using functional
derivatives with respect to free correlation function \cite{PHI4}, the
situation is more complicated in the ordered, broken-symmetry phase. Due to
the non-zero field expectation value both odd and even field powers appear 
in the energy functional, so it is necessary to extend the symmetric
treatment by introducing a second type of functional derivative.\\

We have based the construction on
functional derivatives with respect to both the free correlation function
and the $3$-vertex
in contrast to a previous solution of the same problem in Ref.~\cite{Boris},
where functional
derivatives with respect to both the free correlation function and the external
current was used. 
Our approach has turned out to be conceptually
easier, more transparent, and possibly more efficient than the one used
in Ref.~\cite{Boris} for the following reasons.
Whereas we obtain {\em one} nonlinear
graphical recursion relation for the connected and the one-particle 
irreducible vacuum diagrams, Ref.~\cite{Boris} had to solve {\em two} coupled
nonlinear graphical recursion relations. In particular, the determination of
the loop contributions to the interacting part of the free energy 
$W^{({\rm int})}$ and the effective energy
$\Gamma^{({\rm int})}$ necessitates the
construction of the diagrams of one-point functions in an intermediate step,
whereas our recursive approach only involves the desired vacuum diagrams.
Another advantage of our method is that it only involves
{\em quadratic} nonlinearities for the recursive graphical
construction of one-particle irreducible vacuum diagrams, whereas there 
appears a {\em cubic} nonlinearity in the 
corresponding procedure of Ref.~\cite{Boris}.\\

The second major aspect of our paper is to provide a systematic comparison
between different graphical recursion relations which are based on counting
the number of currents, field expectation values, lines, $3$- and $4$-vertices
of connected and one-particle vacuum diagrams. By doing so, we have obtained
a result which is relevant for the calculation of universal amplitude
ratios \cite{Muenster,Dohm} by an additive renormalization of the 
vacuum energy above \cite{Kastening1,Kastening2} and below the 
critical point. According to Eq.~(5.35a) in Ref.~\cite{Kleinert3},
this calculation can be performed by evaluating the one-particle irreducible
vacuum diagrams of the effective energy $\Gamma[\Phi]$ if the background
field $\chi$ in (\r{PH2}) is identified with the field expectation value
and if the expectation value $\Phi$ of the fluctuations around this 
background field $\chi$ is set equal to zero.
The simplest and most efficient
approach to construct the relevant vacuum diagrams proceeds in two steps.
Having determined the vacuum diagrams above the critical point
by solving (\r{TIT}), the
remaining one-particle irreducible vacuum diagrams below the critical point
are obtained by solving (\r{GRR2}).
\section{Acknowledgement}
We are grateful to Dr. Boris Kastening for sending his manuscript \cite{Boris}
prior to publication.

\newpage
\begin{fmffile}{fg12}
\begin{table}[t]
\begin{center}
\begin{tabular}{|cc|c|}
\,\,\,$l$\,\,\,
& \,\,\,$p$\,\,\, &
$W^{(0,l,p)}$
\\
\hline
$2$ & $0$ &
\hspace{-10pt}
\rule[-10pt]{0pt}{26pt}
\begin{tabular}{@{}c}
$\mbox{}$ \\
$1/12$ \\ ${\scs ( 0, 0, 1 , 0 ; 2 )}$\\
$\mbox{}$ 
\end{tabular}
\parbox{7mm}{\begin{center}
\begin{fmfgraph}(4,4)
\setval
\fmfforce{0w,0.5h}{v1}
\fmfforce{1w,0.5h}{v2}
\fmf{plain,left=1}{v1,v2,v1}
\fmf{plain}{v1,v2}
\fmfdot{v1,v2}
\end{fmfgraph}\end{center}}
\hspace*{1mm}
\begin{tabular}{@{}c}
$1/8$ \\ ${\scs ( 2, 0, 0 , 0 ; 2 )}$
\end{tabular}
\parbox{15mm}{\begin{center}
\begin{fmfgraph}(12,4)
\setval
\fmfforce{0w,0.5h}{v1}
\fmfforce{1/3w,0.5h}{v2}
\fmfforce{2/3w,0.5h}{v3}
\fmfforce{1w,0.5h}{v4}
\fmf{plain,left=1}{v1,v2,v1}
\fmf{plain,left=1}{v3,v4,v3}
\fmf{plain}{v2,v3}
\fmfdot{v2,v3}
\end{fmfgraph}\end{center}}
\\ 
$2$ & $1$ &
\hspace{-10pt}
\rule[-10pt]{0pt}{26pt}
\begin{tabular}{@{}c}
$\mbox{}$ \\
$1/8$ \\ ${\scs ( 2, 1, 0 , 0 ; 1 )}$\\
$\mbox{}$ 
\end{tabular}
\parbox{11mm}{\begin{center}
\begin{fmfgraph}(8,4)
\setval
\fmfleft{i1}
\fmfright{o1}
\fmf{plain,left=1}{i1,v1,i1}
\fmf{plain,left=1}{o1,v1,o1}
\fmfdot{v1}
\end{fmfgraph}\end{center}}
\\
\hline
\hspace{-10pt}
\rule[-10pt]{0pt}{26pt}
$3$ & $0$  & 
\begin{tabular}{@{}c}
$\mbox{}$\\
$1/24$ \\ ${\scs ( 0, 0, 0 , 0 ; 24 )}$\\
$\mbox{}$
\end{tabular}
\parbox{9mm}{\begin{center}
\begin{fmfgraph}(4,4)
\setval
\fmfforce{0w,0h}{v1}
\fmfforce{1w,0h}{v2}
\fmfforce{1w,1h}{v3}
\fmfforce{0w,1h}{v4}
\fmf{plain,right=1}{v1,v3,v1}
\fmf{plain}{v1,v3}
\fmf{plain}{v2,v4}
\fmfdot{v1,v2,v3,v4}
\end{fmfgraph}\end{center}} 
\hspace*{1mm}
\begin{tabular}{@{}c}
$1/16$ \\ ${\scs ( 0, 2, 0 , 0 ; 4 )}$
\end{tabular}
\parbox{9mm}{\begin{center}
\begin{fmfgraph}(4,4)
\setval
\fmfforce{0w,0h}{v1}
\fmfforce{1w,0h}{v2}
\fmfforce{1w,1h}{v3}
\fmfforce{0w,1h}{v4}
\fmf{plain,right=1}{v1,v3,v1}
\fmf{plain,right=0.4}{v1,v4}
\fmf{plain,left=0.4}{v2,v3}
\fmfdot{v1,v2,v3,v4}
\end{fmfgraph}\end{center}} 
\hspace*{1mm}
\begin{tabular}{@{}c}
$1/8$ \\ ${\scs ( 1, 1, 0 , 0 ; 2 )}$
\end{tabular}
\parbox{15mm}{\begin{center}
\begin{fmfgraph}(12,4)
\setval
\fmfforce{0w,1/2h}{v1}
\fmfforce{1/3w,1/2h}{v2}
\fmfforce{2/3w,1/2h}{v3}
\fmfforce{5/6w,0h}{v4}
\fmfforce{5/6w,1h}{v5}
\fmfforce{1w,1/2h}{v6}
\fmf{plain,left=1}{v1,v2,v1}
\fmf{plain,left=1}{v3,v6,v3}
\fmf{plain}{v2,v3}
\fmf{plain}{v4,v5}
\fmfdot{v2,v3,v4,v5}
\end{fmfgraph}\end{center}} 
\hspace*{1mm}
\begin{tabular}{@{}c}
$1/16$ \\ ${\scs ( 2, 1, 0 , 0 ; 2 )}$
\end{tabular}
\parbox{23mm}{\begin{center}
\begin{fmfgraph}(20,4)
\setval
\fmfforce{0w,1/2h}{v1}
\fmfforce{1/5w,1/2h}{v2}
\fmfforce{2/5w,1/2h}{v3}
\fmfforce{3/5w,1/2h}{v4}
\fmfforce{4/5w,1/2h}{v5}
\fmfforce{1w,1/2h}{v6}
\fmf{plain,left=1}{v1,v2,v1}
\fmf{plain,left=1}{v3,v4,v3}
\fmf{plain,left=1}{v5,v6,v5}
\fmf{plain}{v2,v3}
\fmf{plain}{v4,v5}
\fmfdot{v2,v3,v4,v5}
\end{fmfgraph}\end{center}} 
\hspace*{1mm}
\begin{tabular}{@{}c}
$1/48$ \\ ${\scs ( 3, 0, 0 , 0 ; 6 )}$
\end{tabular}
\parbox{17mm}{\begin{center}
\begin{fmfgraph}(13.856,12)
\setval
\fmfforce{0w,0h}{v1}
\fmfforce{1/4w,1/6h}{v2}
\fmfforce{1/2w,1/3h}{v3}
\fmfforce{3/4w,1/6h}{v4}
\fmfforce{1w,0h}{v5}
\fmfforce{1/2w,2/3h}{v6}
\fmfforce{1/2w,1h}{v7}
\fmf{plain,left=1}{v1,v2,v1}
\fmf{plain,left=1}{v4,v5,v4}
\fmf{plain,left=1}{v6,v7,v6}
\fmf{plain}{v2,v3}
\fmf{plain}{v4,v3}
\fmf{plain}{v3,v6}
\fmfdot{v2,v3,v4,v6}
\end{fmfgraph}\end{center}} 
\\ 
$3$ & $1$ & 
\begin{tabular}{@{}c}
$\mbox{}$ \\
$1/8$ \\ ${\scs ( 0, 2, 0 , 0 ; 2 )}$\\
$\mbox{}$
\end{tabular}
\parbox{9mm}{\begin{center}
\begin{fmfgraph}(6,6)
\setval
\fmfforce{0w,1/2h}{v1}
\fmfforce{1w,1/2h}{v2}
\fmfforce{1/2w,1h}{v3}
\fmf{plain,left=1}{v1,v2,v1}
\fmf{plain,left=0.4}{v3,v1}
\fmf{plain,left=0.4}{v2,v3}
\fmfdot{v1,v2,v3}
\end{fmfgraph}\end{center}}
\hspace*{1mm}
\begin{tabular}{@{}c}
$1/8$ \\ ${\scs ( 1, 1, 0 , 0 ; 2 )}$
\end{tabular}
\parbox{11mm}{\begin{center}
\begin{fmfgraph}(4,8)
\setval
\fmfforce{0w,1/4h}{v1}
\fmfforce{1w,1/4h}{v2}
\fmfforce{1/2w,1/2h}{v3}
\fmfforce{1/2w,1h}{v4}
\fmf{plain,left=1}{v1,v2,v1}
\fmf{plain,left=1}{v3,v4,v3}
\fmf{plain}{v1,v2}
\fmfdot{v2,v3,v1}
\end{fmfgraph}\end{center}} 
\hspace*{1mm}
\begin{tabular}{@{}c}
$1/12$ \\ ${\scs ( 1, 0, 1 , 0 ; 1 )}$
\end{tabular}
\parbox{15mm}{\begin{center}
\begin{fmfgraph}(12,4)
\setval
\fmfforce{0w,1/2h}{v1}
\fmfforce{1/3w,1/2h}{v2}
\fmfforce{2/3w,1/2h}{v3}
\fmfforce{1w,1/2h}{v4}
\fmf{plain,left=1}{v1,v2,v1}
\fmf{plain}{v2,v4}
\fmf{plain,left=1}{v3,v4,v3}
\fmfdot{v4,v2,v3}
\end{fmfgraph}\end{center}} 
\hspace*{1mm}
\begin{tabular}{@{}c}
$1/8$ \\ ${\scs ( 2 , 1 , 0 , 0 ; 1 )}$
\end{tabular}
\parbox{19mm}{\begin{center}
\begin{fmfgraph}(16,4)
\setval
\fmfforce{0w,1/2h}{v1}
\fmfforce{1/4w,1/2h}{v2}
\fmfforce{1/2w,1/2h}{v3}
\fmfforce{3/4w,1/2h}{v4}
\fmfforce{1w,1/2h}{v5}
\fmf{plain,left=1}{v1,v2,v1}
\fmf{plain,left=1}{v2,v3,v2}
\fmf{plain}{v3,v4}
\fmf{plain,left=1}{v4,v5,v4}
\fmfdot{v4,v2,v3}
\end{fmfgraph}\end{center}} 
\hspace*{1mm}
\begin{tabular}{@{}c}
$1/16$ \\ ${\scs ( 3, 0, 0 , 0 ; 2 )}$
\end{tabular}
\parbox{19mm}{\begin{center}
\begin{fmfgraph}(16,6)
\setval
\fmfforce{0w,1/3h}{v1}
\fmfforce{1/4w,1/3h}{v2}
\fmfforce{1/2w,1/3h}{v3}
\fmfforce{1/2w,1h}{v4}
\fmfforce{1w,1/3h}{v6}
\fmfforce{3/4w,1/3h}{v5}
\fmf{plain,left=1}{v1,v2,v1}
\fmf{plain}{v2,v5}
\fmf{plain,left=1}{v3,v4,v3}
\fmf{plain,left=1}{v5,v6,v5}
\fmfdot{v5,v2,v3}
\end{fmfgraph}\end{center}} 
\\
$3$ &  $2$ &
\begin{tabular}{@{}c}
$\mbox{}$\\
$1/48$ \\ ${\scs ( 0, 0, 0 , 1 ; 2 )}$\\
$\mbox{}$
\end{tabular}
\parbox{9mm}{\begin{center}
\begin{fmfgraph}(6,4)
\setval
\fmfforce{0w,0.5h}{v1}
\fmfforce{1w,0.5h}{v2}
\fmf{plain,left=1}{v1,v2,v1}
\fmf{plain,left=0.4}{v1,v2,v1}
\fmfdot{v1,v2}
\end{fmfgraph}\end{center}} 
\hspace*{1mm}
\begin{tabular}{@{}c}
$1/16$ \\ ${\scs ( 2, 1, 0 , 0 ; 2 )}$
\end{tabular}
\parbox{15mm}{\begin{center}
\begin{fmfgraph}(12,4)
\setval
\fmfleft{i1}
\fmfright{o1}
\fmf{plain,left=1}{i1,v1,i1}
\fmf{plain,left=1}{v1,v2,v1}
\fmf{plain,left=1}{o1,v2,o1}
\fmfdot{v1,v2}
\end{fmfgraph}\end{center}}
\\ \hline 
\hspace{-10pt}
\rule[-10pt]{0pt}{26pt}
$4$ & $0$ &
\begin{tabular}{@{}c}
$\mbox{}$\\
$1/72$ \\ ${\scs ( 0, 0, 0 , 0 ; 72 )}$\\
$\mbox{}$
\end{tabular}
\parbox{9mm}{\begin{center}
\begin{fmfgraph}(6,6)
\setval
\fmfforce{0w,0.5h}{v1}
\fmfforce{0.25w,0.933h}{v2}
\fmfforce{0.75w,0.933h}{v3}
\fmfforce{1w,0.5h}{v4}
\fmfforce{0.75w,0.067h}{v5}
\fmfforce{0.25w,0.067h}{v6}
\fmf{plain,right=1}{v1,v4,v1}
\fmf{plain}{v1,v4}
\fmf{plain}{v2,v5}
\fmf{plain}{v3,v6}
\fmfdot{v1,v2,v3,v4,v5,v6}
\end{fmfgraph}
\end{center}}
\hspace*{2mm}
\begin{tabular}{@{}c}
$1/12$ \\ ${\scs ( 0, 0, 0 , 0 ; 12 )}$
\end{tabular}
\parbox{9mm}{\begin{center}
\begin{fmfgraph}(6,6)
\setval
\fmfforce{0w,0.5h}{v1}
\fmfforce{0.25w,0.933h}{v2}
\fmfforce{0.75w,0.933h}{v3}
\fmfforce{1w,0.5h}{v4}
\fmfforce{0.75w,0.067h}{v5}
\fmfforce{0.25w,0.067h}{v6}
\fmf{plain,right=1}{v1,v4,v1}
\fmf{plain}{v1,v4}
\fmf{plain}{v2,v6}
\fmf{plain}{v3,v5}
\fmfdot{v1,v2,v3,v4,v5,v6}
\end{fmfgraph}
\end{center}}
\hspace*{2mm}
\begin{tabular}{@{}c}
$1/48$ \\ ${\scs ( 0, 3, 0 , 0 ; 6 )}$
\end{tabular}
\parbox{9mm}{\begin{center}
\begin{fmfgraph}(6,6)
\setval
\fmfforce{0w,0.5h}{v1}
\fmfforce{0.25w,0.933h}{v2}
\fmfforce{0.75w,0.933h}{v3}
\fmfforce{1w,0.5h}{v4}
\fmfforce{0.75w,0.067h}{v5}
\fmfforce{0.25w,0.067h}{v6}
\fmf{plain,right=1}{v1,v4,v1}
\fmf{plain,right=0.7}{v2,v3}
\fmf{plain,right=0.7}{v4,v5}
\fmf{plain,right=0.7}{v6,v1}
\fmfdot{v1,v2,v3,v4,v5,v6}
\end{fmfgraph}
\end{center}} 
\hspace*{2mm}
\begin{tabular}{@{}c}
$1/16$ \\ ${\scs ( 0, 2, 0 , 0 ; 4 )}$
\end{tabular}
\parbox{9mm}{\begin{center}
\begin{fmfgraph}(6,6)
\setval
\fmfforce{0w,0.5h}{v1}
\fmfforce{0.25w,0.933h}{v2}
\fmfforce{0.75w,0.933h}{v3}
\fmfforce{1w,0.5h}{v4}
\fmfforce{0.75w,0.067h}{v5}
\fmfforce{0.25w,0.067h}{v6}
\fmf{plain,right=1}{v1,v4,v1}
\fmf{plain,right=0.7}{v2,v3}
\fmf{plain}{v1,v4}
\fmf{plain,right=0.7}{v5,v6}
\fmfdot{v1,v2,v3,v4,v5,v6}
\end{fmfgraph}
\end{center}}
\hspace*{2mm}
\begin{tabular}{@{}c}
$1/8$ \\ ${\scs ( 0, 1, 0 , 0 ; 4 )}$
\end{tabular}
\parbox{9mm}{\begin{center}
\begin{fmfgraph}(6,6)
\setval
\fmfforce{0w,0.5h}{v1}
\fmfforce{0.25w,0.933h}{v2}
\fmfforce{0.75w,0.933h}{v3}
\fmfforce{1w,0.5h}{v4}
\fmfforce{0.75w,0.067h}{v5}
\fmfforce{0.25w,0.067h}{v6}
\fmf{plain,right=1}{v1,v4,v1}
\fmf{plain,right=0.7}{v2,v3}
\fmf{plain,right=0.2}{v4,v6}
\fmf{plain,right=0.2}{v5,v1}
\fmfdot{v1,v2,v3,v4,v5,v6}
\end{fmfgraph}
\end{center}}
\hspace*{1mm}
\begin{tabular}{@{}c}
$1/32$ \\ ${\scs ( 0, 2, 0 , 0 ; 8 )}$
\end{tabular}
\parbox{7mm}{\begin{center}
\begin{fmfgraph}(4,12)
\setval
\fmfforce{0w,2/12h}{v1}
\fmfforce{1w,2/12h}{v2}
\fmfforce{1/2w,4/12h}{v3}
\fmfforce{1/2w,8/12h}{v4}
\fmfforce{0w,10/12h}{v5}
\fmfforce{1w,10/12h}{v6}
\fmf{plain,right=1}{v1,v2,v1}
\fmf{plain}{v1,v2}
\fmf{plain}{v3,v4}
\fmf{plain}{v5,v6}
\fmf{plain,right=1}{v6,v5,v6}
\fmfdot{v1,v2,v3,v4,v5,v6}
\end{fmfgraph}
\end{center}}
\\ & & 
\begin{tabular}{@{}c}
$\mbox{}$\\
$1/16$ \\ ${\scs ( 1 , 2 , 0 , 0 ; 2 )}$\\
$\mbox{}$
\end{tabular}
\parbox{9mm}{\begin{center}
\begin{fmfgraph}(6,14)
\setval
\fmfforce{1/2w,0h}{v1}
\fmfforce{1/2w,6/14h}{v2}
\fmfforce{1/2w,10/14h}{v3}
\fmfforce{1/2w,1h}{v4}
\fmfforce{0.155w,0.36h}{v5}
\fmfforce{0.155w,0.068h}{v6}
\fmfforce{0.845w,0.36h}{v7}
\fmfforce{0.845w,0.068h}{v8}
\fmf{plain}{v2,v3}
\fmf{plain,right=1}{v1,v2,v1}
\fmf{plain,right=1}{v4,v3,v4}
\fmf{plain,right=0.4}{v6,v5}
\fmf{plain,right=0.4}{v7,v8}
\fmfdot{v2,v3,v5,v6,v7,v8}
\end{fmfgraph}
\end{center}}
\hspace*{1mm}
\begin{tabular}{@{}c}
$1/8$ \\ ${\scs ( 1 , 0 , 0 , 0 ; 4 )}$
\end{tabular}
\parbox{9mm}{\begin{center}
\begin{fmfgraph}(6,14)
\setval
\fmfforce{1/2w,0h}{v1}
\fmfforce{1/2w,6/14h}{v2}
\fmfforce{1/2w,10/14h}{v3}
\fmfforce{1/2w,1h}{v4}
\fmfforce{0.155w,0.36h}{v5}
\fmfforce{0.155w,0.068h}{v6}
\fmfforce{0.845w,0.36h}{v7}
\fmfforce{0.845w,0.068h}{v8}
\fmf{plain}{v2,v3}
\fmf{plain,right=1}{v1,v2,v1}
\fmf{plain,right=1}{v4,v3,v4}
\fmf{plain}{v6,v7}
\fmf{plain}{v5,v8}
\fmfdot{v2,v3,v5,v6,v7,v8}
\end{fmfgraph}
\end{center}}
\hspace*{1mm}
\begin{tabular}{@{}c}
$1/8$ \\ ${\scs ( 1 , 1 , 0 , 0 ; 2 )}$
\end{tabular}
\parbox{9mm}{\begin{center}
\begin{fmfgraph}(6,14)
\setval
\fmfforce{1/2w,0h}{v1}
\fmfforce{1/2w,6/14h}{v2}
\fmfforce{1/2w,10/14h}{v3}
\fmfforce{1/2w,1h}{v4}
\fmfforce{0.155w,0.36h}{v5}
\fmfforce{0.155w,0.068h}{v6}
\fmfforce{0.845w,0.36h}{v7}
\fmfforce{0.845w,0.068h}{v8}
\fmf{plain}{v2,v3}
\fmf{plain,right=1}{v1,v2,v1}
\fmf{plain,right=1}{v4,v3,v4}
\fmf{plain,right=0.4}{v5,v7}
\fmf{plain,right=0.4}{v8,v6}
\fmfdot{v2,v3,v5,v6,v7,v8}
\end{fmfgraph}
\end{center}}
\hspace*{1mm}
\begin{tabular}{@{}c}
$1/16$ \\ ${\scs ( 2 , 1 , 0 , 0 ; 2 )}$
\end{tabular}
\parbox{9mm}{\begin{center}
\begin{fmfgraph}(6,14)
\setval
\fmfforce{0w,3/14h}{v1}
\fmfforce{1w,3/14h}{v2}
\fmfforce{0.9/6w,5/14h}{v5}
\fmfforce{0.9/6w,1/14h}{v6}
\fmfforce{5.1/6w,5/14h}{v7}
\fmfforce{5.1/6w,1/14h}{v8}
\fmfforce{5.1/6w,9/14h}{v3}
\fmfforce{5.1/6w,13/14h}{v4}
\fmfforce{0.9/6w,9/14h}{v9}
\fmfforce{0.9/6w,13/14h}{v10}
\fmf{plain,left=1}{v1,v2,v1}
\fmf{plain}{v3,v7}
\fmf{plain}{v5,v9}
\fmf{plain,left=1}{v9,v10,v9}
\fmf{plain,left=1}{v3,v4,v3}
\fmf{plain,left=0.5}{v6,v8}
\fmfdot{v3,v5,v6,v7,v8,v9}
\end{fmfgraph}\end{center}} 
\hspace*{1mm}
\begin{tabular}{@{}c}
$1/16$ \\ ${\scs ( 2 , 0 , 0 , 0 ; 4 )}$
\end{tabular}
\parbox{23mm}{\begin{center}
\begin{fmfgraph}(20,4)
\setval
\fmfforce{0w,1/2h}{v1}
\fmfforce{1/5w,1/2h}{v2}
\fmfforce{2/5w,1/2h}{v3}
\fmfforce{3/5w,1/2h}{v4}
\fmfforce{4/5w,1/2h}{v5}
\fmfforce{1w,1/2h}{v6}
\fmfforce{1/2w,0h}{v7}
\fmfforce{1/2w,1h}{v8}
\fmf{plain,right=1}{v1,v2,v1}
\fmf{plain}{v3,v2}
\fmf{plain,right=1}{v7,v8,v7}
\fmf{plain}{v7,v8}
\fmf{plain}{v4,v5}
\fmf{plain,right=1}{v6,v5,v6}
\fmfdot{v2,v3,v4,v5,v8,v7}
\end{fmfgraph}\end{center}} 
\hspace*{1mm}
\begin{tabular}{@{}c}
$1/16$ \\ ${\scs ( 1 , 2 , 0 , 0 ; 2 )}$
\end{tabular}
\parbox{7mm}{\begin{center}
\begin{fmfgraph}(4,20)
\setval
\fmfforce{0w,1/10h}{v1}
\fmfforce{1w,1/10h}{v2}
\fmfforce{1/2w,1/5h}{v3}
\fmfforce{1/2w,2/5h}{v4}
\fmfforce{1/2w,3/5h}{v5}
\fmfforce{1/2w,4/5h}{v6}
\fmfforce{1/2w,1h}{v7}
\fmf{plain,right=1}{v1,v2,v1}
\fmf{plain}{v1,v2}
\fmf{plain}{v3,v4}
\fmf{plain,right=1}{v4,v5,v4}
\fmf{plain}{v6,v5}
\fmf{plain,right=1}{v6,v7,v6}
\fmfdot{v1,v2,v3,v4,v5,v6}
\end{fmfgraph}\end{center}} 
\\ & & 
\hspace*{1mm}
\begin{tabular}{@{}c}
$\mbox{}$ \\
$1/32$ \\ ${\scs ( 2 , 1 , 0 , 0 ; 4 )}$\\
$\mbox{}$ 
\end{tabular}
\parbox{11mm}{\begin{center}
\begin{fmfgraph}(8,14)
\setval
\fmfforce{2/8w,2/14h}{v1}
\fmfforce{6/8w,2/14h}{v2}
\fmfforce{1/2w,4/14h}{v3}
\fmfforce{1/2w,8/14h}{v4}
\fmfforce{6.828/8w,10.828/14h}{v5}
\fmfforce{1.171/8w,10.828/14h}{v6}
\fmfforce{9.657/8w,13.657/14h}{v7}
\fmfforce{-1.657/8w,13.657/14h}{v8}
\fmf{plain,right=1}{v1,v2,v1}
\fmf{plain}{v1,v2}
\fmf{plain}{v3,v4}
\fmf{plain}{v4,v5}
\fmf{plain}{v4,v6}
\fmf{plain,right=1}{v7,v5,v7}
\fmf{plain,right=1}{v6,v8,v6}
\fmfdot{v1,v2,v3,v4,v5,v6}
\end{fmfgraph}\end{center}} 
\begin{tabular}{@{}c}
$1/32$ \\ ${\scs ( 2 , 2 , 0 , 0 ; 2 )}$
\end{tabular}
\parbox{31mm}{\begin{center}
\begin{fmfgraph}(28,4)
\setval
\fmfforce{0w,1/2h}{v1}
\fmfforce{1/7w,1/2h}{v2}
\fmfforce{2/7w,1/2h}{v3}
\fmfforce{3/7w,1/2h}{v4}
\fmfforce{4/7w,1/2h}{v5}
\fmfforce{5/7w,1/2h}{v6}
\fmfforce{6/7w,1/2h}{v7}
\fmfforce{1w,1/2h}{v8}
\fmf{plain,right=1}{v1,v2,v1}
\fmf{plain}{v2,v3}
\fmf{plain,right=1}{v4,v3,v4}
\fmf{plain}{v4,v5}
\fmf{plain,right=1}{v6,v5,v6}
\fmf{plain}{v6,v7}
\fmf{plain,right=1}{v8,v7,v8}
\fmfdot{v2,v3,v4,v5,v6,v7}
\end{fmfgraph}\end{center}} 
\hspace*{1mm}
\begin{tabular}{@{}c}
$1/48$ \\ ${\scs ( 3 , 0 , 0 , 0 ; 6 )}$
\end{tabular}
\parbox{21mm}{\begin{center}
\begin{fmfgraph}(20,20)
\setval
\fmfforce{1/2w,12/20h}{v1}
\fmfforce{8.268/20w,9/20h}{v2}
\fmfforce{11.732/20w,9/20h}{v3}
\fmfforce{1/2w,8/20h}{v4}
\fmfforce{1/2w,16/20h}{v5}
\fmfforce{1/2w,1h}{v6}
\fmfforce{4.804/20w,7/20h}{v7}
\fmfforce{1.34/20w,5/20h}{v8}
\fmfforce{15.196/20w,7/20h}{v9}
\fmfforce{18.66/20w,5/20h}{v10}
\fmf{plain,right=1}{v1,v4,v1}
\fmf{plain,right=1}{v6,v5,v6}
\fmf{plain,right=1}{v8,v7,v8}
\fmf{plain,right=1}{v10,v9,v10}
\fmf{plain}{v1,v5}
\fmf{plain}{v2,v7}
\fmf{plain}{v3,v9}
\fmfdot{v1,v2,v3,v5,v7,v9}
\end{fmfgraph}\end{center}} 
\begin{tabular}{@{}c}
$1/32$ \\ ${\scs ( 3, 1, 0 , 0 ; 2 )}$
\end{tabular}
\parbox{14mm}{\begin{center}
\begin{fmfgraph}(12,22)
\setval
\fmfforce{0.343/12w,0.343/22h}{v1}
\fmfforce{11.657/12w,0.343/22h}{v2}
\fmfforce{3.172/12w,3.17/22h}{v3}
\fmfforce{8.828/12w,3.17/22h}{v4}
\fmfforce{1/2w,6/22h}{v5}
\fmfforce{1/2w,10/22h}{v6}
\fmfforce{1/2w,14/22h}{v7}
\fmfforce{1/2w,18/22h}{v8}
\fmfforce{1/2w,1h}{v9}
\fmf{plain,right=1}{v1,v3,v1}
\fmf{plain}{v5,v4}
\fmf{plain}{v5,v3}
\fmf{plain,right=1}{v2,v4,v2}
\fmf{plain}{v5,v6}
\fmf{plain,right=1}{v7,v6,v7}
\fmf{plain}{v7,v8}
\fmf{plain,right=1}{v9,v8,v9}
\fmfdot{v3,v4,v5,v6,v7,v8}
\end{fmfgraph}\end{center}} 
\begin{tabular}{@{}c}
$1/128$ \\ ${\scs ( 4, 0, 0 , 0 ; 8 )}$
\end{tabular}
\parbox{13mm}{\begin{center}
\begin{fmfgraph}(10,16)
\setval
\fmfforce{2/10w,0h}{v1}
\fmfforce{2/10w,1/4h}{v2}
\fmfforce{2/10w,2/4h}{v3}
\fmfforce{2/10w,3/4h}{v4}
\fmfforce{2/10w,1h}{v5}
\fmfforce{8/10w,0h}{v6}
\fmfforce{8/10w,1/4h}{v7}
\fmfforce{8/10w,2/4h}{v8}
\fmfforce{8/10w,3/4h}{v9}
\fmfforce{8/10w,1h}{v10}
\fmf{plain,right=1}{v1,v2,v1}
\fmf{plain}{v2,v4}
\fmf{plain,right=1}{v5,v4,v5}
\fmf{plain,right=1}{v7,v6,v7}
\fmf{plain}{v7,v9}
\fmf{plain,right=1}{v10,v9,v10}
\fmf{plain}{v3,v8}
\fmfdot{v2,v3,v4,v7,v8,v9}
\end{fmfgraph}\end{center}} 
\\ $4$ & $1$ &
\begin{tabular}{@{}c}
$\mbox{}$ \\
$1/8$ \\ ${\scs ( 0, 0, 0 , 0 ; 8 )}$\\
$\mbox{}$
\end{tabular}
\parbox{9mm}{\begin{center}
\begin{fmfgraph}(4,4)
\setval
\fmfforce{0w,0h}{v1}
\fmfforce{1w,0h}{v2}
\fmfforce{1w,1h}{v3}
\fmfforce{0w,1h}{v4}
\fmfforce{1/2w,1/2h}{v5}
\fmf{plain,right=1}{v1,v3,v1}
\fmf{plain}{v1,v3}
\fmf{plain}{v2,v4}
\fmfdot{v1,v2,v3,v4,v5}
\end{fmfgraph}\end{center}} 
\hspace*{1mm}
\begin{tabular}{@{}c}
$1/4$ \\ ${\scs ( 0, 2, 0 , 0 ; 2 )}$
\end{tabular}
\parbox{13mm}{\begin{center}
\begin{fmfgraph}(8,8)
\setval
\fmfforce{1/2w,1h}{v1}
\fmfforce{1/2w,0h}{v2}
\fmfforce{0.85w,0.85h}{v3}
\fmfforce{0.15w,0.85h}{v4}
\fmfforce{1/2w,0.7h}{v5}
\fmfforce{1/2w,0.3h}{v6}
\fmf{plain,left=1}{v1,v2,v1}
\fmf{plain,left=1}{v5,v6,v5}
\fmf{plain}{v6,v2}
\fmf{plain,right=0.4}{v5,v4}
\fmf{plain,right=0.4}{v3,v5}
\fmfdot{v2,v3,v4,v5,v6}
\end{fmfgraph}\end{center}} 
\hspace*{1mm}
\begin{tabular}{@{}c}
$1/16$ \\ ${\scs ( 0, 3, 0 , 0 ; 2 )}$
\end{tabular}
\parbox{9mm}{\begin{center}
\begin{fmfgraph}(6,6)
\setval
\fmfforce{1/2w,1h}{v1}
\fmfforce{1/2w,0h}{v2}
\fmfforce{0.25w,0.933h}{v3}
\fmfforce{0.75w,0.933h}{v4}
\fmfforce{0w,0.5h}{v5}
\fmfforce{1w,0.5h}{v6}
\fmf{plain,right=1}{v1,v2,v1}
\fmf{plain,right=0.7}{v3,v4}
\fmf{plain,right=0.4}{v2,v5}
\fmf{plain,right=0.4}{v6,v2}
\fmfdot{v2,v3,v4,v5,v6}
\end{fmfgraph}
\end{center}}
\hspace*{1mm}
\begin{tabular}{@{}c}
$1/32$ \\ ${\scs ( 0, 2, 0 , 0 ; 8 )}$
\end{tabular}
\parbox{9mm}{\begin{center}
\begin{fmfgraph}(4,8)
\setval
\fmfforce{1/2w,1h}{v1}
\fmfforce{1/2w,1/2h}{v2}
\fmfforce{1/2w,0h}{v3}
\fmfforce{1w,3/4h}{v4}
\fmfforce{0w,3/4h}{v5}
\fmfforce{1w,1/4h}{v6}
\fmfforce{0w,1/4h}{v7}
\fmf{plain,right=1}{v1,v2,v1}
\fmf{plain,right=1}{v3,v2,v3}
\fmf{plain}{v4,v5}
\fmf{plain}{v6,v7}
\fmfdot{v2,v4,v5,v6,v7}
\end{fmfgraph}
\end{center}}
\hspace*{1mm}
\begin{tabular}{@{}c}
$1/4$ \\ ${\scs ( 0, 2, 0 , 0 ; 1 )}$
\end{tabular}
\parbox{9mm}{\begin{center}
\begin{fmfgraph}(6,6)
\setval
\fmfforce{0w,1/2h}{v1}
\fmfforce{1/2w,1h}{v2}
\fmfforce{1/2w,0h}{v3}
\fmfforce{0.9w,0.75h}{v4}
\fmfforce{0.9w,0.25h}{v5}
\fmf{plain,right=1}{v2,v3,v2}
\fmf{plain}{v2,v3}
\fmf{plain,right=0.4}{v1,v2}
\fmf{plain,right=0.4}{v4,v5}
\fmfdot{v1,v3,v2,v4,v5}
\end{fmfgraph}
\end{center}}
\hspace*{1mm}
\begin{tabular}{@{}c}
$1/16$\\ 
${\scs ( 1, 2, 0 , 0 ; 2 )}$
\end{tabular}
\parbox{9mm}{\begin{center}
\begin{fmfgraph}(6,10)
\setval
\fmfforce{0w,0.3h}{v1}
\fmfforce{1w,0.3h}{v2}
\fmfforce{0.5w,0.6h}{v3}
\fmfforce{0.5w,1h}{v4}
\fmfforce{0.15w,0.5h}{v5}
\fmfforce{0.15w,0.1h}{v6}
\fmfforce{0.85w,0.5h}{v7}
\fmfforce{0.85w,0.1h}{v8}
\fmf{plain,left=1}{v1,v2,v1}
\fmf{plain,left=0.5}{v5,v6}
\fmf{plain,left=0.5}{v8,v7}
\fmf{plain,left=1}{v3,v4,v3}
\fmfdot{v3,v5,v6,v7,v8}
\end{fmfgraph}\end{center}} 
\\ & &
\begin{tabular}{@{}c}
$\mbox{}$\\
$1/8$\\ 
${\scs ( 1, 1, 0 , 0 ; 2 )}$\\
$\mbox{}$
\end{tabular}
\parbox{9mm}{\begin{center}
\begin{fmfgraph}(6,10)
\setval
\fmfforce{0w,0.3h}{v1}
\fmfforce{1w,0.3h}{v2}
\fmfforce{0.5w,0.6h}{v3}
\fmfforce{0.5w,1h}{v4}
\fmfforce{0.15w,0.5h}{v5}
\fmfforce{0.15w,0.1h}{v6}
\fmfforce{0.85w,0.5h}{v7}
\fmfforce{0.85w,0.1h}{v8}
\fmf{plain,left=1}{v1,v2,v1}
\fmf{plain,right=0.5}{v5,v7}
\fmf{plain,right=0.5}{v8,v6}
\fmf{plain,left=1}{v3,v4,v3}
\fmfdot{v3,v5,v6,v7,v8}
\end{fmfgraph}\end{center}} 
\hspace*{1mm}
\begin{tabular}{@{}c}
$1/8$\\ 
${\scs ( 1, 0, 0 , 0 ; 4 )}$
\end{tabular}
\parbox{9mm}{\begin{center}
\begin{fmfgraph}(6,10)
\setval
\fmfforce{0w,0.3h}{v1}
\fmfforce{1w,0.3h}{v2}
\fmfforce{0.5w,0.6h}{v3}
\fmfforce{0.5w,1h}{v4}
\fmfforce{0.15w,0.5h}{v5}
\fmfforce{0.15w,0.1h}{v6}
\fmfforce{0.85w,0.5h}{v7}
\fmfforce{0.85w,0.1h}{v8}
\fmf{plain,left=1}{v1,v2,v1}
\fmf{plain}{v5,v8}
\fmf{plain}{v7,v6}
\fmf{plain,left=1}{v3,v4,v3}
\fmfdot{v3,v5,v6,v7,v8}
\end{fmfgraph}\end{center}} 
\hspace*{1mm}
\begin{tabular}{@{}c}
$1/8$\\ 
${\scs ( 1, 2, 0 , 0 ; 1 )}$
\end{tabular}
\parbox{9mm}{\begin{center}
\begin{fmfgraph}(6,14)
\setval
\fmfforce{0w,3/14h}{v1}
\fmfforce{1w,3/14h}{v2}
\fmfforce{0.15w,5/14h}{v5}
\fmfforce{0.15w,1/14h}{v6}
\fmfforce{0.85w,5/14h}{v7}
\fmfforce{0.85w,1/14h}{v8}
\fmfforce{0.85w,9/14h}{v3}
\fmfforce{0.85w,13/14h}{v4}
\fmf{plain,left=1}{v1,v2,v1}
\fmf{plain}{v3,v7}
\fmf{plain,left=1}{v3,v4,v3}
\fmf{plain,right=0.5}{v5,v7}
\fmf{plain,right=0.5}{v8,v6}
\fmfdot{v3,v5,v6,v7,v8}
\end{fmfgraph}\end{center}} 
\hspace*{1mm}
\begin{tabular}{@{}c}
$1/12$\\ 
${\scs ( 1, 0, 0 , 0 ; 6 )}$
\end{tabular}
\parbox{9mm}{\begin{center}
\begin{fmfgraph}(6,14)
\setval
\fmfforce{0w,3/14h}{v1}
\fmfforce{1w,3/14h}{v2}
\fmfforce{1/2w,0h}{v3}
\fmfforce{1/2w,6/14h}{v4}
\fmfforce{1/2w,10/14h}{v5}
\fmfforce{1/2w,1h}{v6}
\fmf{plain,left=1}{v1,v2,v1}
\fmf{plain}{v1,v2}
\fmf{plain}{v3,v5}
\fmf{plain,left=1}{v5,v6,v5}
\fmfdot{v1,v2,v3,v4,v5}
\end{fmfgraph}\end{center}} 
\hspace*{1mm}
\begin{tabular}{@{}c}
$1/24$ \\ ${\scs ( 0, 1, 1 , 0 ; 2 )}$
\end{tabular}
\parbox{7mm}{\begin{center}
\begin{fmfgraph}(4,12)
\setval
\fmfforce{0w,1/6h}{v1}
\fmfforce{1w,1/6h}{v2}
\fmfforce{1/2w,1/3h}{v3}
\fmfforce{1/2w,2/3h}{v4}
\fmfforce{1/2w,1h}{v5}
\fmf{plain,left=1}{v1,v2,v1}
\fmf{plain}{v1,v2}
\fmf{plain}{v5,v3}
\fmf{plain,left=1}{v4,v5,v4}
\fmfdot{v1,v2,v3,v4,v5}
\end{fmfgraph}\end{center}}
\hspace*{1mm}
\begin{tabular}{@{}c}
$1/8$ \\ ${\scs ( 2, 0, 0 , 0 ; 2 )}$
\end{tabular}
\parbox{7mm}{\begin{center}
\begin{fmfgraph}(4,16)
\setval
\fmfforce{1/2w,0h}{v1}
\fmfforce{1/2w,1/4h}{v2}
\fmfforce{0w,3/8h}{v3}
\fmfforce{1w,3/8h}{v4}
\fmfforce{1/2w,2/4h}{v5}
\fmfforce{1/2w,3/4h}{v6}
\fmfforce{1/2w,1h}{v7}
\fmf{plain,left=1}{v1,v2,v1}
\fmf{plain,left=1}{v2,v5,v2}
\fmf{plain}{v4,v3}
\fmf{plain}{v6,v5}
\fmf{plain,left=1}{v6,v7,v6}
\fmfdot{v2,v3,v4,v5,v6}
\end{fmfgraph}\end{center}}
\\ & & 
\begin{tabular}{@{}c}
$\mbox{}$\\
$1/16$ \\ ${\scs ( 1, 2, 0 , 0 ; 2 )}$\\
$\mbox{}$
\end{tabular}
\parbox{7mm}{\begin{center}
\begin{fmfgraph}(4,16)
\setval
\fmfforce{0w,1/8h}{v1}
\fmfforce{1w,1/8h}{v2}
\fmfforce{1/2w,1/4h}{v3}
\fmfforce{1/2w,2/4h}{v4}
\fmfforce{1/2w,3/4h}{v5}
\fmfforce{1/2w,4/4h}{v6}
\fmf{plain,left=1}{v1,v2,v1}
\fmf{plain}{v1,v2}
\fmf{plain,left=1}{v3,v4,v3}
\fmf{plain}{v4,v5}
\fmf{plain,left=1}{v6,v5,v6}
\fmfdot{v1,v2,v3,v4,v5}
\end{fmfgraph}\end{center}}
\hspace*{1mm} 
\begin{tabular}{@{}c}
$1/16$ \\ ${\scs ( 1, 2, 0 , 0 ; 2 )}$
\end{tabular}
\parbox{7mm}{\begin{center}
\begin{fmfgraph}(4,16)
\setval
\fmfforce{0w,1/8h}{v1}
\fmfforce{1w,1/8h}{v2}
\fmfforce{1/2w,1/4h}{v3}
\fmfforce{1/2w,2/4h}{v4}
\fmfforce{1/2w,3/4h}{v5}
\fmfforce{1/2w,4/4h}{v6}
\fmf{plain,left=1}{v1,v2,v1}
\fmf{plain}{v1,v2}
\fmf{plain,left=1}{v5,v4,v5}
\fmf{plain}{v4,v3}
\fmf{plain,left=1}{v6,v5,v6}
\fmfdot{v1,v2,v3,v4,v5}
\end{fmfgraph}\end{center}}
\hspace*{1mm}
\begin{tabular}{@{}c}
$1/4$ \\ ${\scs ( 1, 1, 0 , 0 ; 1 )}$
\end{tabular}
\parbox{9mm}{\begin{center}
\begin{fmfgraph}(6,14)
\setval
\fmfforce{0w,3/14h}{v1}
\fmfforce{1/2w,6/14h}{v2}
\fmfforce{1w,3/14h}{v4}
\fmfforce{1/2w,0h}{v5}
\fmfforce{1/2w,10/14h}{v6}
\fmfforce{1/2w,1h}{v7}
\fmf{plain,right=1}{v2,v5,v2}
\fmf{plain,right=1}{v7,v6,v7}
\fmf{plain,right=0.4}{v4,v5,v4}
\fmf{plain}{v1,v4}
\fmf{plain}{v2,v6}
\fmfdot{v1,v2,v4,v5,v6}
\end{fmfgraph}
\end{center}}
\hspace*{1mm}
\begin{tabular}{@{}c}
$1/16$\\ 
${\scs ( 1, 2, 0 , 0 ; 2 )}$
\end{tabular}
\parbox{9mm}{\begin{center}
\begin{fmfgraph}(6,14)
\setval
\fmfforce{0w,3/14h}{v1}
\fmfforce{1w,3/14h}{v2}
\fmfforce{0.5w,6/14h}{v3}
\fmfforce{0.5w,10/14h}{v4}
\fmfforce{0.5w,0h}{v5}
\fmfforce{0.5w,1h}{v6}
\fmf{plain,left=1}{v1,v2,v1}
\fmf{plain,left=0.4}{v1,v5}
\fmf{plain,left=0.4}{v5,v2}
\fmf{plain,left=1}{v4,v6,v4}
\fmf{plain}{v3,v4}
\fmfdot{v1,v2,v3,v4,v5}
\end{fmfgraph}\end{center}} 
\hspace*{1mm}
\begin{tabular}{@{}c}
$1/8$ \\ ${\scs ( 2, 1, 0 , 0 ; 1 )}$
\end{tabular}
\parbox{9mm}{\begin{center}
\begin{fmfgraph}(6,14)
\setval
\fmfforce{0w,3.9/14h}{v1}
\fmfforce{1w,3.9/14h}{v2}
\fmfforce{0.2w,6.24/14h}{v3}
\fmfforce{0.8w,6.24/14h}{v4}
\fmfforce{0.2w,10.24/14h}{v5}
\fmfforce{0.2w,14.24/14h}{v6}
\fmfforce{1.2w,9.1/14h}{v7}
\fmf{plain,left=1}{v1,v2,v1}
\fmf{plain,left=1}{v6,v5,v6}
\fmf{plain,left=1}{v7,v4,v7}
\fmf{plain}{v3,v5}
\fmf{plain}{v1,v2}
\fmfdot{v1,v2,v3,v4,v5}
\end{fmfgraph}\end{center}}
\hspace*{1mm}
\begin{tabular}{@{}c}
$1/16$ \\ ${\scs ( 2, 1, 0 , 0 ; 2 )}$
\end{tabular}
\parbox{9mm}{\begin{center}
\begin{fmfgraph}(6,16)
\setval
\fmfforce{0w,1/8h}{v1}
\fmfforce{2/3w,1/8h}{v2}
\fmfforce{1/3w,1/4h}{v3}
\fmfforce{1/3w,2/4h}{v4}
\fmfforce{1/3w,3/4h}{v5}
\fmfforce{1/3w,4/4h}{v6}
\fmfforce{1w,2/4h}{v7}
\fmf{plain,left=1}{v1,v2,v1}
\fmf{plain}{v1,v2}
\fmf{plain}{v5,v3}
\fmf{plain,left=1}{v6,v5,v6}
\fmf{plain,left=1}{v7,v4,v7}
\fmfdot{v1,v2,v3,v4,v5}
\end{fmfgraph}\end{center}}
\\ && 
\begin{tabular}{@{}c}
$\mbox{}$\\
$1/32$ \\ ${\scs ( 2, 1, 0 , 0 ; 4 )}$\\
$\mbox{}$
\end{tabular}
\parbox{19mm}{\begin{center}
\begin{fmfgraph}(16,6)
\setval
\fmfforce{0w,2/3h}{v1}
\fmfforce{1/4w,2/3h}{v2}
\fmfforce{2/4w,2/3h}{v3}
\fmfforce{3/4w,2/3h}{v4}
\fmfforce{4/4w,2/3h}{v5}
\fmfforce{6/16w,1/3h}{v6}
\fmfforce{10/16w,1/3h}{v7}
\fmf{plain,left=1}{v1,v2,v1}
\fmf{plain}{v4,v2}
\fmf{plain,left=1}{v5,v4,v5}
\fmf{plain}{v6,v7}
\fmf{plain,left=1}{v6,v7,v6}
\fmfdot{v2,v3,v4,v6,v7}
\end{fmfgraph}\end{center}}
\hspace*{1mm}
\begin{tabular}{@{}c}
$1/16$ \\ ${\scs ( 2, 2, 0 , 0 ; 1 )}$
\end{tabular}
\parbox{27mm}{\begin{center}
\begin{fmfgraph}(24,4)
\setval
\fmfforce{0w,1/2h}{v1}
\fmfforce{1/6w,1/2h}{v2}
\fmfforce{2/6w,1/2h}{v3}
\fmfforce{3/6w,1/2h}{v4}
\fmfforce{4/6w,1/2h}{v5}
\fmfforce{5/6w,1/2h}{v6}
\fmfforce{1w,1/2h}{v7}
\fmf{plain,left=1}{v1,v2,v1}
\fmf{plain}{v2,v3}
\fmf{plain,left=1}{v3,v4,v3}
\fmf{plain}{v4,v5}
\fmf{plain,left=1}{v5,v6,v5}
\fmf{plain,left=1}{v6,v7,v6}
\fmfdot{v2,v3,v4,v5,v6}
\end{fmfgraph}\end{center}}
\hspace*{1mm}
\begin{tabular}{@{}c}
$1/32$ \\ ${\scs ( 2, 2, 0 , 0 ; 2 )}$
\end{tabular}
\parbox{27mm}{\begin{center}
\begin{fmfgraph}(24,4)
\setval
\fmfforce{0w,1/2h}{v1}
\fmfforce{1/6w,1/2h}{v2}
\fmfforce{2/6w,1/2h}{v3}
\fmfforce{3/6w,1/2h}{v4}
\fmfforce{4/6w,1/2h}{v5}
\fmfforce{5/6w,1/2h}{v6}
\fmfforce{1w,1/2h}{v7}
\fmf{plain,left=1}{v1,v2,v1}
\fmf{plain}{v2,v3}
\fmf{plain,left=1}{v3,v4,v3}
\fmf{plain,left=1}{v5,v4,v5}
\fmf{plain}{v5,v6}
\fmf{plain,left=1}{v7,v6,v7}
\fmfdot{v2,v3,v4,v5,v6}
\end{fmfgraph}\end{center}}
\hspace*{1mm}
\begin{tabular}{@{}c}
$1/16$ \\ ${\scs ( 3, 0, 0 , 0 ; 2 )}$
\end{tabular}
\parbox{24mm}{\begin{center}
\begin{fmfgraph}(20,8)
\setval
\fmfforce{0w,1/4h}{v1}
\fmfforce{1/5w,1/4h}{v2}
\fmfforce{2/5w,1/4h}{v3}
\fmfforce{3/5w,1/4h}{v4}
\fmfforce{4/5w,1/4h}{v5}
\fmfforce{1w,1/4h}{v6}
\fmfforce{1/2w,1/2h}{v7}
\fmfforce{1/2w,1h}{v8}
\fmf{plain,left=1}{v1,v2,v1}
\fmf{plain}{v2,v3}
\fmf{plain,left=1}{v3,v4,v3}
\fmf{plain}{v5,v4}
\fmf{plain,left=1}{v5,v6,v5}
\fmf{plain,left=1}{v7,v8,v7}
\fmfdot{v2,v3,v4,v5,v7}
\end{fmfgraph}\end{center}}
\\ && 
\begin{tabular}{@{}c}
$\mbox{}$\\
$1/24$ \\ ${\scs ( 1, 1, 1 , 0 ; 1 )}$\\
$\mbox{}$
\end{tabular}
\parbox{24mm}{\begin{center}
\begin{fmfgraph}(20,4)
\setval
\fmfforce{0w,1/2h}{v1}
\fmfforce{1/5w,1/2h}{v2}
\fmfforce{2/5w,1/2h}{v3}
\fmfforce{3/5w,1/2h}{v4}
\fmfforce{4/5w,1/2h}{v5}
\fmfforce{1w,1/2h}{v6}
\fmf{plain,left=1}{v1,v2,v1}
\fmf{plain}{v2,v3}
\fmf{plain,left=1}{v3,v4,v3}
\fmf{plain}{v6,v4}
\fmf{plain,left=1}{v5,v6,v5}
\fmfdot{v2,v3,v4,v5,v6}
\end{fmfgraph}\end{center}}
\hspace*{1mm} 
\begin{tabular}{@{}c}
$1/8$ \\ ${\scs ( 2, 1, 0 , 0 ; 1 )}$
\end{tabular}
\parbox{16mm}{\begin{center}
\begin{fmfgraph}(12,12)
\setval
\fmfforce{0w,1/6h}{v1}
\fmfforce{1/3w,1/6h}{v2}
\fmfforce{2/3w,1/6h}{v3}
\fmfforce{3/3w,1/6h}{v4}
\fmfforce{1/6w,1/3h}{v5}
\fmfforce{1/6w,2/3h}{v6}
\fmfforce{1/6w,3/3h}{v7}
\fmf{plain,left=1}{v1,v2,v1}
\fmf{plain}{v1,v3}
\fmf{plain,left=1}{v3,v4,v3}
\fmf{plain}{v6,v5}
\fmf{plain,left=1}{v7,v6,v7}
\fmfdot{v1,v2,v3,v5,v6}
\end{fmfgraph}\end{center}}
\hspace*{1mm}
\begin{tabular}{@{}c}
$1/48$ \\ ${\scs ( 2, 0, 1 , 0 ; 2 )}$
\end{tabular}
\parbox{20mm}{\begin{center}
\begin{fmfgraph}(16,10)
\setval
\fmfforce{0w,2/10h}{v1}
\fmfforce{1/4w,2/10h}{v2}
\fmfforce{2/4w,2/10h}{v3}
\fmfforce{3/4w,2/10h}{v4}
\fmfforce{4/4w,2/10h}{v5}
\fmfforce{1/2w,6/10h}{v6}
\fmfforce{1/2w,10/10h}{v7}
\fmf{plain,left=1}{v1,v2,v1}
\fmf{plain}{v2,v4}
\fmf{plain,left=1}{v5,v4,v5}
\fmf{plain}{v3,v7}
\fmf{plain,left=1}{v7,v6,v7}
\fmfdot{v2,v3,v4,v6,v7}
\end{fmfgraph}\end{center}}
\hspace*{1mm}
\begin{tabular}{@{}c}
$1/16$ \\ ${\scs ( 3, 1, 0 , 0 ; 1 )}$
\end{tabular}
\parbox{27mm}{\begin{center}
\begin{fmfgraph}(24,6)
\setval
\fmfforce{0w,1/3h}{v1}
\fmfforce{1/6w,1/3h}{v2}
\fmfforce{2/6w,1/3h}{v3}
\fmfforce{3/6w,1/3h}{v4}
\fmfforce{4/6w,1/3h}{v5}
\fmfforce{5/6w,1/3h}{v6}
\fmfforce{1w,1/3h}{v7}
\fmfforce{4/6w,1h}{v8}
\fmf{plain,left=1}{v1,v2,v1}
\fmf{plain}{v2,v3}
\fmf{plain,left=1}{v3,v4,v3}
\fmf{plain}{v6,v4}
\fmf{plain,left=1}{v7,v6,v7}
\fmf{plain,left=1}{v8,v5,v8}
\fmfdot{v2,v3,v4,v5,v6}
\end{fmfgraph}\end{center}}
\\ &&
\begin{tabular}{@{}c}
$\mbox{}$\\
$1/32$ \\ ${\scs ( 3, 1, 0 , 0 ; 2 )}$\\
$\mbox{}$
\end{tabular}
\parbox{19mm}{\begin{center}
\begin{fmfgraph}(16,14)
\setval
\fmfforce{0w,2/14h}{v1}
\fmfforce{1/4w,2/14h}{v2}
\fmfforce{2/4w,2/14h}{v3}
\fmfforce{3/4w,2/14h}{v4}
\fmfforce{1w,2/14h}{v5}
\fmfforce{1/2w,6/14h}{v6}
\fmfforce{1/2w,10/14h}{v7}
\fmfforce{1/2w,1h}{v8}
\fmf{plain,left=1}{v1,v2,v1}
\fmf{plain}{v2,v4}
\fmf{plain,left=1}{v5,v4,v5}
\fmf{plain,left=1}{v3,v6,v3}
\fmf{plain}{v7,v6}
\fmf{plain,left=1}{v8,v7,v8}
\fmfdot{v2,v3,v4,v7,v6}
\end{fmfgraph}\end{center}}
\hspace*{1mm}
\begin{tabular}{@{}c}
$1/32$ \\ ${\scs ( 3, 1, 0 , 0 ; 2 )}$
\end{tabular}
\parbox{19mm}{\begin{center}
\begin{fmfgraph}(16,14)
\setval
\fmfforce{0w,2/14h}{v1}
\fmfforce{1/4w,2/14h}{v2}
\fmfforce{2/4w,2/14h}{v3}
\fmfforce{3/4w,2/14h}{v4}
\fmfforce{1w,2/14h}{v5}
\fmfforce{1/2w,6/14h}{v6}
\fmfforce{1/2w,10/14h}{v7}
\fmfforce{1/2w,1h}{v8}
\fmf{plain,left=1}{v1,v2,v1}
\fmf{plain}{v2,v4}
\fmf{plain,left=1}{v5,v4,v5}
\fmf{plain}{v3,v6}
\fmf{plain,left=1}{v7,v6,v7}
\fmf{plain,left=1}{v8,v7,v8}
\fmfdot{v2,v3,v4,v7,v6}
\end{fmfgraph}\end{center}}
\hspace*{1mm}
\begin{tabular}{@{}c}
$1/32$ \\ ${\scs ( 4, 0, 0 , 0 ; 2 )}$
\end{tabular}
\parbox{19mm}{\begin{center}
\begin{fmfgraph}(16,14)
\setval
\fmfforce{0w,2/14h}{v1}
\fmfforce{1/4w,2/14h}{v2}
\fmfforce{2/4w,2/14h}{v3}
\fmfforce{3/4w,2/14h}{v4}
\fmfforce{1w,2/14h}{v5}
\fmfforce{1/2w,6/14h}{v6}
\fmfforce{1/2w,10/14h}{v7}
\fmfforce{1/2w,1h}{v8}
\fmfforce{12/16w,6/14h}{v9}
\fmf{plain,left=1}{v1,v2,v1}
\fmf{plain}{v2,v4}
\fmf{plain,left=1}{v5,v4,v5}
\fmf{plain}{v3,v7}
\fmf{plain,left=1}{v8,v7,v8}
\fmf{plain,left=1}{v9,v6,v9}
\fmfdot{v2,v3,v4,v7,v6}
\end{fmfgraph}\end{center}}
\hspace*{1mm}
\begin{tabular}{@{}c}
$1/384$ \\ ${\scs ( 4, 0, 0 , 0 ; 24 )}$
\end{tabular}
\parbox{19mm}{\begin{center}
\begin{fmfgraph}(16,16)
\setval
\fmfforce{1/2w,0h}{v1}
\fmfforce{1/2w,1/4h}{v2}
\fmfforce{0w,1/2h}{v3}
\fmfforce{1/4w,1/2h}{v4}
\fmfforce{2/4w,1/2h}{v5}
\fmfforce{3/4w,1/2h}{v6}
\fmfforce{1w,1/2h}{v7}
\fmfforce{1/2w,3/4h}{v8}
\fmfforce{1/2w,1h}{v9}
\fmf{plain,left=1}{v1,v2,v1}
\fmf{plain}{v2,v8}
\fmf{plain,left=1}{v3,v4,v3}
\fmf{plain}{v4,v6}
\fmf{plain,left=1}{v6,v7,v6}
\fmf{plain,left=1}{v9,v8,v9}
\fmfdot{v2,v4,v5,v6,v8}
\end{fmfgraph}\end{center}}
\\ $4$ & $2$ &
\hspace*{1mm}
\begin{tabular}{@{}c}
$\mbox{}$ \\
$1/8$ \\ ${\scs ( 0, 1, 0 , 0 ; 4 )}$\\
$\mbox{}$
\end{tabular}
\parbox{9mm}{\begin{center}
\begin{fmfgraph}(6,6)
\setval
\fmfforce{0w,0.5h}{v1}
\fmfforce{1w,0.5h}{v2}
\fmfforce{1/2w,0.7h}{v3}
\fmfforce{1/2w,0.3h}{v4}
\fmf{plain,left=1}{v1,v2,v1}
\fmf{plain,left=0.4}{v1,v2,v1}
\fmf{plain}{v4,v3}
\fmfdot{v1,v2,v3,v4}
\end{fmfgraph}\end{center}} 
\hspace*{1mm}
\begin{tabular}{@{}c}
$1/16$ \\ ${\scs ( 0, 3, 0 , 0 ; 2 )}$
\end{tabular}
\parbox{9mm}{\begin{center}
\begin{fmfgraph}(4.5,4.5)
\setval
\fmfforce{0w,0h}{v1}
\fmfforce{1w,0h}{v2}
\fmfforce{1w,1h}{v3}
\fmfforce{0w,1h}{v4}
\fmf{plain,left=1}{v1,v3,v1}
\fmf{plain,right=0.4}{v3,v2}
\fmf{plain,right=0.4}{v4,v3}
\fmf{plain,right=0.4}{v1,v4}
\fmfdot{v1,v2,v3,v4}
\end{fmfgraph}\end{center}} 
\hspace*{1mm}
\begin{tabular}{@{}c}
$1/8$ \\ ${\scs ( 0, 2, 0 , 0 ; 2 )}$
\end{tabular}
\parbox{9mm}{\begin{center}
\begin{fmfgraph}(4.5,4.5)
\setval
\fmfforce{0w,0h}{v1}
\fmfforce{1w,0h}{v2}
\fmfforce{1w,1h}{v3}
\fmfforce{0w,1h}{v4}
\fmf{plain,left=1}{v1,v3,v1}
\fmf{plain,right=0.4}{v3,v2}
\fmf{plain}{v1,v3}
\fmf{plain,right=0.4}{v1,v4}
\fmfdot{v1,v2,v3,v4}
\end{fmfgraph}\end{center}} 
\hspace*{1mm}
\begin{tabular}{@{}c}
$1/24$ \\ ${\scs ( 0, 1, 1 , 0 ; 2 )}$
\end{tabular}
\parbox{9mm}{\begin{center}
\begin{fmfgraph}(4,10)
\setval
\fmfforce{0w,1/5h}{v1}
\fmfforce{1w,1/5h}{v2}
\fmfforce{0w,4/5h}{v3}
\fmfforce{1w,4/5h}{v4}
\fmf{plain,left=1}{v4,v3,v4}
\fmf{plain,left=1}{v1,v2,v1}
\fmf{plain}{v4,v3}
\fmf{plain,right=0.3}{v3,v1}
\fmf{plain,right=0.3}{v2,v4}
\fmfdot{v1,v2,v3,v4}
\end{fmfgraph}\end{center}} 
\hspace*{1mm}
\begin{tabular}{@{}c}
$1/4$ \\ ${\scs ( 1, 1, 0 , 0 ; 1 )}$
\end{tabular}
\parbox{9mm}{\begin{center}
\begin{fmfgraph}(6,10)
\setval
\fmfforce{0w,3/10h}{v1}
\fmfforce{1/2w,6/10h}{v2}
\fmfforce{1w,3/10h}{v4}
\fmfforce{1/2w,0h}{v5}
\fmfforce{1/2w,1h}{v6}
\fmf{plain,right=1}{v2,v5,v2}
\fmf{plain,right=1}{v2,v6,v2}
\fmf{plain,right=0.4}{v4,v5,v4}
\fmf{plain}{v1,v4}
\fmfdot{v1,v2,v4,v5}
\end{fmfgraph}
\end{center}}
\hspace*{1mm}
\begin{tabular}{@{}c}
$1/16$\\ 
${\scs ( 1, 2, 0 , 0 ; 2 )}$
\end{tabular}
\parbox{9mm}{\begin{center}
\begin{fmfgraph}(6,10)
\setval
\fmfforce{0w,0.3h}{v1}
\fmfforce{1w,0.3h}{v2}
\fmfforce{0.5w,0.6h}{v3}
\fmfforce{0.5w,1h}{v4}
\fmfforce{0.5w,0h}{v5}
\fmf{plain,left=1}{v1,v2,v1}
\fmf{plain,left=0.4}{v1,v5}
\fmf{plain,left=0.4}{v5,v2}
\fmf{plain,left=1}{v3,v4,v3}
\fmfdot{v1,v2,v3,v5}
\end{fmfgraph}\end{center}} 
\hspace*{1mm}
\\ &&
\begin{tabular}{@{}c}
$\mbox{}$\\
$1/16$ \\ ${\scs ( 1, 2, 0 , 0 ; 2 )}$\\
$\mbox{}$
\end{tabular}
\parbox{7mm}{\begin{center}
\begin{fmfgraph}(4,12)
\setval
\fmfforce{0w,1/6h}{v1}
\fmfforce{1w,1/6h}{v2}
\fmfforce{1/2w,1/3h}{v3}
\fmfforce{1/2w,2/3h}{v4}
\fmfforce{1/2w,1h}{v5}
\fmf{plain,left=1}{v1,v2,v1}
\fmf{plain,left=1}{v3,v4,v3}
\fmf{plain,left=1}{v4,v5,v4}
\fmf{plain}{v1,v2}
\fmfdot{v1,v2,v3,v4}
\end{fmfgraph}\end{center}} 
\hspace*{1mm}
\begin{tabular}{@{}c}
$1/16$ \\ ${\scs ( 2, 0, 0 , 0 ; 4 )}$
\end{tabular}
\parbox{15mm}{\begin{center}
\begin{fmfgraph}(12,4)
\setval
\fmfforce{0w,1/2h}{v1}
\fmfforce{1/3w,1/2h}{v2}
\fmfforce{1/2w,0h}{v3}
\fmfforce{1/2w,1h}{v4}
\fmfforce{2/3w,1/2h}{v5}
\fmfforce{1w,1/2h}{v6}
\fmf{plain,left=1}{v1,v2,v1}
\fmf{plain,left=1}{v2,v5,v2}
\fmf{plain,left=1}{v5,v6,v5}
\fmf{plain}{v3,v4}
\fmfdot{v2,v3,v4,v5}
\end{fmfgraph}\end{center}} 
\hspace*{1mm}
\begin{tabular}{@{}c}
$1/16$ \\ ${\scs ( 2, 1, 0 , 0 ; 2 )}$
\end{tabular}
\parbox{15mm}{\begin{center}
\begin{fmfgraph}(6,9.75)
\setval
\fmfforce{0w,0.3h}{v1}
\fmfforce{1w,0.3h}{v2}
\fmfforce{0.2w,0.55h}{v3}
\fmfforce{0.8w,0.55h}{v4}
\fmf{plain,left=1}{v1,v2,v1}
\fmf{plain}{v1,v2}
\fmfi{plain}{reverse fullcircle scaled 0.7w shifted (1.02w,0.72h)}
\fmfi{plain}{reverse fullcircle scaled 0.7w shifted (-0.02w,0.72h)}
\fmfdot{v1,v2,v3,v4}
\end{fmfgraph}\end{center}}
\hspace*{1mm}
\begin{tabular}{@{}c}
$1/16$ \\ ${\scs ( 2, 2, 0 , 0 ; 1 )}$
\end{tabular}
\parbox{23mm}{\begin{center}
\begin{fmfgraph}(20,8)
\setval
\fmfforce{0w,1/2h}{v1}
\fmfforce{1/5w,1/2h}{v2}
\fmfforce{2/5w,1/2h}{v3}
\fmfforce{3/5w,1/2h}{v4}
\fmfforce{4/5w,1/2h}{v5}
\fmfforce{1w,1/2h}{v6}
\fmf{plain,left=1}{v1,v2,v1}
\fmf{plain}{v2,v3}
\fmf{plain,left=1}{v3,v4,v3}
\fmf{plain,left=1}{v4,v5,v4}
\fmf{plain,left=1}{v5,v6,v5}
\fmfdot{v2,v3,v4,v5}
\end{fmfgraph}\end{center}} 
\hspace*{1mm}
\begin{tabular}{@{}c}
$1/16$ \\ ${\scs ( 3, 0, 0 , 0 ; 2 )}$
\end{tabular}
\parbox{15mm}{\begin{center}
\begin{fmfgraph}(12,16)
\setval
\fmfforce{1/2w,1/4h}{v1}
\fmfforce{1/2w,2/4h}{v2}
\fmfforce{1/2w,3/4h}{v3}
\fmfforce{0.355662432w,5/16h}{v4}
\fmfforce{0.64433568w,5/16h}{v5}
\fmfforce{0.067w,3/16h}{v6}
\fmfforce{0.933w,3/16h}{v7}
\fmfforce{1/2w,1h}{v8}
\fmf{plain,left=1}{v1,v2,v1}
\fmf{plain}{v2,v3}
\fmf{plain,left=1}{v8,v3,v8}
\fmf{plain,left=1}{v4,v6,v4}
\fmf{plain,left=1}{v5,v7,v5}
\fmfdot{v2,v3,v4,v5}
\end{fmfgraph}\end{center}} 
\hspace*{1mm}
\\ & &
\begin{tabular}{@{}c}
$\mbox{}$\\
$1/24$ \\ 
${\scs ( 1, 0, 1 , 0 ; 2 )}$\\
$\mbox{}$
\end{tabular}
\parbox{9mm}{\begin{center}
\begin{fmfgraph}(6,14)
\setval
\fmfforce{0w,3/14h}{v1}
\fmfforce{1w,3/14h}{v2}
\fmfforce{0.5w,6/14h}{v3}
\fmfforce{0.5w,10/14h}{v4}
\fmfforce{0.5w,1h}{v5}
\fmf{plain,left=1}{v1,v2,v1}
\fmf{plain,left=0.4}{v1,v2,v1}
\fmf{plain}{v4,v3}
\fmf{plain,left=1}{v5,v4,v5}
\fmfdot{v1,v2,v3,v4}
\end{fmfgraph}\end{center}} 
\hspace*{1mm}
\begin{tabular}{@{}c}
$1/8$ \\ 
${\scs ( 2, 1, 0 , 0 ; 1 )}$
\end{tabular}
\parbox{15mm}{\begin{center}
\begin{fmfgraph}(12,8)
\setval
\fmfforce{0w,1/4h}{v1}
\fmfforce{1/3w,1/4h}{v2}
\fmfforce{2/3w,1/4h}{v3}
\fmfforce{1w,1/4h}{v4}
\fmfforce{5/6w,1/2h}{v5}
\fmfforce{5/6w,1h}{v6}
\fmf{plain,left=1}{v1,v2,v1}
\fmf{plain,left=1}{v3,v4,v3}
\fmf{plain,left=1}{v5,v6,v5}
\fmf{plain}{v4,v2}
\fmfdot{v2,v3,v4,v5}
\end{fmfgraph}\end{center}} 
\hspace*{1mm}
\begin{tabular}{@{}c}
$1/72$ \\ 
${\scs ( 0, 0, 2 , 0 ; 2 )}$
\end{tabular}
\parbox{15mm}{\begin{center}
\begin{fmfgraph}(12,4)
\setval
\fmfforce{0w,1/2h}{v1}
\fmfforce{1/3w,1/2h}{v2}
\fmfforce{2/3w,1/2h}{v3}
\fmfforce{1w,1/2h}{v4}
\fmf{plain,left=1}{v1,v2,v1}
\fmf{plain,left=1}{v3,v4,v3}
\fmf{plain}{v1,v4}
\fmfdot{v1,v2,v3,v4}
\end{fmfgraph}\end{center}} 
\hspace*{1mm}
\begin{tabular}{@{}c}
$1/24$ \\ ${\scs ( 1, 1, 1 , 0 ; 1 )}$
\end{tabular}
\parbox{23mm}{\begin{center}
\begin{fmfgraph}(16,4)
\setval
\fmfforce{0w,1/2h}{v1}
\fmfforce{1/4w,1/2h}{v2}
\fmfforce{2/4w,1/2h}{v3}
\fmfforce{3/4w,1/2h}{v4}
\fmfforce{1w,1/2h}{v5}
\fmf{plain,left=1}{v1,v2,v1}
\fmf{plain,left=1}{v3,v2,v3}
\fmf{plain,left=1}{v4,v5,v4}
\fmf{plain}{v3,v5}
\fmfdot{v2,v3,v4,v5}
\end{fmfgraph}\end{center}} 
\hspace*{1mm}
\begin{tabular}{@{}c}
$1/8$ \\ ${\scs ( 1, 2, 0 , 0 ; 1 )}$
\end{tabular}
\parbox{17mm}{\begin{center}
\begin{fmfgraph}(14,6)
\setval
\fmfforce{8/14w,1/2h}{v1}
\fmfforce{1w,1/2h}{v2}
\fmfforce{11/14w,1h}{v3}
\fmfforce{0w,1/2h}{v4}
\fmfforce{4/14w,1/2h}{v5}
\fmf{plain,left=1}{v1,v2,v1}
\fmf{plain,left=1}{v4,v5,v4}
\fmf{plain,left=0.4}{v3,v1}
\fmf{plain,left=0.4}{v2,v3}
\fmf{plain}{v5,v1}
\fmfdot{v1,v2,v3,v5}
\end{fmfgraph}\end{center}}
\hspace*{1mm} $\mbox{}$
\\ &&
\hspace*{1mm}
\begin{tabular}{@{}c}
$\mbox{}$\\
$1/32$ \\ 
${\scs ( 2, 2, 0 , 0 ; 2 )}$\\
$\mbox{}$
\end{tabular}
\parbox{23mm}{\begin{center}
\begin{fmfgraph}(20,4)
\setval
\fmfforce{0w,1/2h}{v1}
\fmfforce{1/5w,1/2h}{v2}
\fmfforce{2/5w,1/2h}{v3}
\fmfforce{3/5w,1/2h}{v4}
\fmfforce{4/5w,1/2h}{v5}
\fmfforce{1w,1/2h}{v6}
\fmf{plain,left=1}{v1,v2,v1}
\fmf{plain,left=1}{v2,v3,v2}
\fmf{plain}{v4,v3}
\fmf{plain,left=1}{v5,v4,v5}
\fmf{plain,left=1}{v5,v6,v5}
\fmf{plain}{v4,v3}
\fmfdot{v2,v3,v4,v5}
\end{fmfgraph}\end{center}} 
\begin{tabular}{@{}c}
$1/16$ \\ 
${\scs ( 3, 1, 0 , 0 ; 1 )}$
\end{tabular}
\parbox{23mm}{\begin{center}
\begin{fmfgraph}(20,6)
\setval
\fmfforce{0w,1/3h}{v1}
\fmfforce{1/5w,1/3h}{v2}
\fmfforce{2/5w,1/3h}{v3}
\fmfforce{3/5w,1/3h}{v4}
\fmfforce{4/5w,1/3h}{v5}
\fmfforce{1w,1/3h}{v6}
\fmfforce{2/5w,1h}{v7}
\fmf{plain,left=1}{v1,v2,v1}
\fmf{plain,left=1}{v6,v5,v6}
\fmf{plain,left=1}{v5,v4,v5}
\fmf{plain,left=1}{v3,v7,v3}
\fmf{plain}{v4,v2}
\fmfdot{v2,v3,v4,v5}
\end{fmfgraph}\end{center}} 
\hspace*{1mm}
\begin{tabular}{@{}c}
$1/32$ \\ 
${\scs ( 3, 1, 0 , 0 ; 2 )}$
\end{tabular}
\parbox{19mm}{\begin{center}
\begin{fmfgraph}(16,10)
\setval
\fmfforce{0w,1/5h}{v1}
\fmfforce{1/4w,1/5h}{v2}
\fmfforce{2/4w,1/5h}{v3}
\fmfforce{3/4w,1/5h}{v4}
\fmfforce{1w,1/5h}{v5}
\fmfforce{1/2w,3/5h}{v6}
\fmfforce{1/2w,1h}{v7}
\fmf{plain,left=1}{v1,v2,v1}
\fmf{plain,left=1}{v4,v5,v4}
\fmf{plain,left=1}{v3,v6,v3}
\fmf{plain,left=1}{v6,v7,v6}
\fmf{plain}{v4,v2}
\fmfdot{v2,v3,v4,v6}
\end{fmfgraph}\end{center}} 
\hspace*{1mm}
\begin{tabular}{@{}c}
$1/24$ \\ 
${\scs ( 2, 0, 1 , 0 ; 1 )}$
\end{tabular}
\parbox{19mm}{\begin{center}
\begin{fmfgraph}(16,6)
\setval
\fmfforce{0w,1/3h}{v1}
\fmfforce{1/4w,1/3h}{v2}
\fmfforce{2/4w,1/3h}{v3}
\fmfforce{3/4w,1/3h}{v4}
\fmfforce{1w,1/3h}{v5}
\fmfforce{1/2w,1h}{v6}
\fmf{plain,left=1}{v1,v2,v1}
\fmf{plain,left=1}{v4,v5,v4}
\fmf{plain,left=1}{v3,v6,v3}
\fmf{plain}{v5,v2}
\fmfdot{v2,v3,v4,v5}
\end{fmfgraph}\end{center}} 
\\ &&
\begin{tabular}{@{}c}
$\mbox{}$\\
$1/48$ \\ 
${\scs ( 2, 0, 1 , 0 ; 2 )}$\\
$\mbox{}$
\end{tabular}
\parbox{23mm}{\begin{center}
\begin{fmfgraph}(20,4)
\setval
\fmfforce{0w,1/2h}{v1}
\fmfforce{1/5w,1/2h}{v2}
\fmfforce{2/5w,1/2h}{v3}
\fmfforce{3/5w,1/2h}{v4}
\fmfforce{4/5w,1/2h}{v5}
\fmfforce{1w,1/2h}{v6}
\fmf{plain,left=1}{v1,v2,v1}
\fmf{plain,left=1}{v4,v3,v4}
\fmf{plain,left=1}{v5,v6,v5}
\fmf{plain}{v5,v2}
\fmfdot{v2,v3,v4,v5}
\end{fmfgraph}\end{center}} 
\hspace*{1mm}
\begin{tabular}{@{}c}
$1/32$\\ 
${\scs ( 4, 0, 0 , 0 ; 2 )}$
\end{tabular}
\parbox{23mm}{\begin{center}
\begin{fmfgraph}(22,6)
\setval
\fmfforce{0w,1/3h}{v1}
\fmfforce{4/22w,1/3h}{v2}
\fmfforce{8/22w,1/3h}{v3}
\fmfforce{14/22w,1/3h}{v4}
\fmfforce{18/22w,1/3h}{v5}
\fmfforce{1w,1/3h}{v6}
\fmfforce{8/22w,1h}{v7}
\fmfforce{14/22w,1h}{v8}
\fmf{plain,left=1}{v1,v2,v1}
\fmf{plain,left=1}{v6,v5,v6}
\fmf{plain,left=1}{v8,v4,v8}
\fmf{plain,left=1}{v3,v7,v3}
\fmf{plain}{v5,v2}
\fmfdot{v2,v3,v4,v5}
\end{fmfgraph}\end{center}} 
\\
$4$& $3$ &
\begin{tabular}{@{}c}
$\mbox{}$ \\
$1/48$ \\ 
${\scs ( 0, 3, 0 , 0 ; 6 )}$ \\
$\mbox{}$
\end{tabular}
\parbox{9mm}{\begin{center}
\begin{fmfgraph}(6,6)
\setval
\fmfforce{0.5w,0h}{v1}
\fmfforce{0.5w,1h}{v2}
\fmfforce{0.066987w,0.25h}{v3}
\fmfforce{0.93301w,0.25h}{v4}
\fmf{plain,left=1}{v1,v2,v1}
\fmf{plain}{v2,v3}
\fmf{plain}{v3,v4}
\fmf{plain}{v2,v4}
\fmfdot{v2,v3,v4}
\end{fmfgraph}
\end{center}} 
\hspace*{1mm}
\begin{tabular}{@{}c}
$1/24$\\ 
${\scs ( 1, 0, 1 , 0 ; 2 )}$
\end{tabular}
\parbox{9mm}{\begin{center}
\begin{fmfgraph}(6,10)
\setval
\fmfforce{0w,0.3h}{v1}
\fmfforce{1w,0.3h}{v2}
\fmfforce{0.5w,0.6h}{v3}
\fmfforce{0.5w,1h}{v4}
\fmf{plain,left=1}{v1,v2,v1}
\fmf{plain,left=0.4}{v1,v2,v1}
\fmf{plain,left=1}{v3,v4,v3}
\fmfdot{v1,v2,v3}
\end{fmfgraph}\end{center}} 
\hspace*{1mm}
\begin{tabular}{@{}c}
$1/48$\\ 
${\scs ( 3, 0, 0 , 0 ; 6 )}$
\end{tabular}
\parbox{15mm}{\begin{center}
\begin{fmfgraph}(12,12)
\setval
\fmfforce{1/2w,1/3h}{v1}
\fmfforce{1/2w,2/3h}{v2}
\fmfforce{1/2w,1h}{v3}
\fmfforce{0.355662432w,0.416666666h}{v4}
\fmfforce{0.64433568w,0.416666666h}{v5}
\fmfforce{0.067w,1/4h}{v6}
\fmfforce{0.933w,1/4h}{v7}
\fmf{plain,left=1}{v1,v2,v1}
\fmf{plain,left=1}{v2,v3,v2}
\fmf{plain,left=1}{v4,v6,v4}
\fmf{plain,left=1}{v5,v7,v5}
\fmfdot{v2,v4,v5}
\end{fmfgraph}\end{center}} 
\hspace*{1mm}
\begin{tabular}{@{}c}
$1/32$\\ 
${\scs ( 2, 2, 0 , 0 ; 2 )}$
\end{tabular}
\parbox{19mm}{\begin{center}
\begin{fmfgraph}(16,4)
\setval
\fmfleft{i1}
\fmfright{o1}
\fmf{plain,left=1}{i1,v1,i1}
\fmf{plain,left=1}{v1,v2,v1}
\fmf{plain,left=1}{v2,v3,v2}
\fmf{plain,left=1}{o1,v3,o1}
\fmfdot{v1,v2,v3}
\end{fmfgraph}\end{center}}
\hspace*{1mm}
\end{tabular}
\end{center}
\caption{Connected vacuum diagrams and their 
weights of the $\phi^3$-$\phi^4$-theory without currents
up to four loops. Withing each loop order $l$ the diagrams are
distinguished with respect to the number $p$ of $4$-vertices.
Each diagram is characterized by the
vector $(S,D,T,F;N$) whose components specify the number of self-, double,
triple, fourfold connections, and of the identical vertex permutations,
respectively.}
\end{table}
\end{fmffile}
\newpage
\begin{fmffile}{fg13}

\begin{table}[t]
\begin{center}
\begin{tabular}{|ccc|c|}
\,\,\,$n$\,\,\,
&\,\,\,$l$\,\,\,
& \,\,\,$p$\,\,\, &
$W^{(n ,l,p)}$
\\
\hline
$1$ & $1$ & $0$ &
\hspace{-10pt}
\rule[-10pt]{0pt}{26pt}
\begin{tabular}{@{}c}
$1/2$ \\ ${\scs ( 1, 0, 0 , 0 , 1 )}$
\end{tabular}
\parbox{11mm}{\begin{center}
\begin{fmfgraph*}(8,4)
\setval
\fmfforce{0w,1/2h}{v1}
\fmfforce{1/2w,1/2h}{v2}
\fmfforce{1w,1/2h}{v3}
\fmf{plain,left=1}{v2,v3,v2}
\fmf{plain}{v1,v2}
\fmfdot{v2}
\fmfv{decor.shape=cross,decor.filled=shaded,decor.size=3thick}{v1}
\end{fmfgraph*}\end{center}}
\\ \hline 
$1$ &$2$ & $1$ &
\hspace{-10pt}
\rule[-10pt]{0pt}{26pt}
\begin{tabular}{@{}c}
$1/6$ \\ ${\scs ( 0, 0, 1 , 0 ; 1 )}$
\end{tabular}
\parbox{11mm}{\begin{center}
\begin{fmfgraph*}(8,4)
\setval
\fmfforce{0w,1/2h}{v1}
\fmfforce{1/2w,1/2h}{v2}
\fmfforce{1w,1/2h}{v3}
\fmf{plain,left=1}{v2,v3,v2}
\fmf{plain}{v1,v3}
\fmfdot{v2,v3}
\fmfv{decor.shape=cross,decor.filled=shaded,decor.size=3thick}{v1}
\end{fmfgraph*}\end{center}}
\hspace*{1mm}
\begin{tabular}{@{}c}
$1/4$ \\ ${\scs ( 1, 1, 0 , 0 ; 1 )}$
\end{tabular}
\parbox{15mm}{\begin{center}
\begin{fmfgraph*}(12,4)
\setval
\fmfforce{0w,1/2h}{v1}
\fmfforce{1/3w,1/2h}{v2}
\fmfforce{2/3w,1/2h}{v3}
\fmfforce{1w,1/2h}{v4}
\fmf{plain,left=1}{v2,v3,v2}
\fmf{plain,left=1}{v3,v4,v3}
\fmf{plain}{v1,v2}
\fmfdot{v2,v3}
\fmfv{decor.shape=cross,decor.filled=shaded,decor.size=3thick}{v1}
\end{fmfgraph*}\end{center}}
\hspace*{1mm}
\begin{tabular}{@{}c}
$1/4$ \\ ${\scs ( 2, 0, 0 , 0 ; 1 )}$
\end{tabular}
\parbox{15mm}{\begin{center}
\begin{fmfgraph*}(12,6)
\setval
\fmfforce{0w,1/3h}{v1}
\fmfforce{1/3w,1/3h}{v2}
\fmfforce{2/3w,1/3h}{v3}
\fmfforce{1w,1/3h}{v4}
\fmfforce{1/3w,1h}{v5}
\fmf{plain,left=1}{v2,v5,v2}
\fmf{plain,left=1}{v3,v4,v3}
\fmf{plain}{v1,v3}
\fmfdot{v2,v3}
\fmfv{decor.shape=cross,decor.filled=shaded,decor.size=3thick}{v1}
\end{fmfgraph*}\end{center}}
\\ $1$ & $2$ & $0$ &
\begin{tabular}{@{}c}
$1/4$ \\ ${\scs ( 0, 1, 0 , 0 ; 2 )}$
\end{tabular}
\parbox{11mm}{\begin{center}
\begin{fmfgraph*}(8,4)
\setval
\fmfforce{0w,1/2h}{v1}
\fmfforce{1/2w,1/2h}{v2}
\fmfforce{1w,1/2h}{v3}
\fmfforce{3/4w,0h}{v4}
\fmfforce{3/4w,1h}{v5}
\fmf{plain,left=1}{v2,v3,v2}
\fmf{plain}{v1,v2}
\fmf{plain}{v4,v5}
\fmfdot{v2,v4,v5}
\fmfv{decor.shape=cross,decor.filled=shaded,decor.size=3thick}{v1}
\end{fmfgraph*}\end{center}}
\hspace*{1mm}
\begin{tabular}{@{}c}
$1/4$ \\ ${\scs ( 1, 1, 0 , 0 ; 2 )}$
\end{tabular}
\parbox{19mm}{\begin{center}
\begin{fmfgraph*}(16,4)
\setval
\fmfforce{0w,1/2h}{v1}
\fmfforce{1/4w,1/2h}{v2}
\fmfforce{1/2w,1/2h}{v3}
\fmfforce{3/4w,1/2h}{v4}
\fmfforce{1w,1/2h}{v5}
\fmf{plain,left=1}{v2,v3,v2}
\fmf{plain,left=1}{v4,v5,v4}
\fmf{plain}{v1,v2}
\fmf{plain}{v3,v4}
\fmfdot{v2,v4,v3}
\fmfv{decor.shape=cross,decor.filled=shaded,decor.size=3thick}{v1}
\end{fmfgraph*}\end{center}}
\hspace*{1mm}
\begin{tabular}{@{}c}
$1/8$ \\ ${\scs ( 2, 0, 0 , 0 ; 2 )}$
\end{tabular}
\parbox{11mm}{\begin{center}
\begin{fmfgraph*}(8,13.856)
\setval
\fmfforce{0w,1/2h}{v1}
\fmfforce{1/2w,1/2h}{v2}
\fmfforce{3/4w,1/4h}{v3}
\fmfforce{3/4w,3/4h}{v4}
\fmfforce{1w,0h}{v5}
\fmfforce{1w,1h}{v6}
\fmf{plain,left=1}{v3,v5,v3}
\fmf{plain,left=1}{v4,v6,v4}
\fmf{plain}{v1,v2}
\fmf{plain}{v2,v3}
\fmf{plain}{v2,v4}
\fmfdot{v2,v4,v3}
\fmfv{decor.shape=cross,decor.filled=shaded,decor.size=3thick}{v1}
\end{fmfgraph*}\end{center}} \\ \hline\hline
$2$ & $1$ & $1$ &
\hspace{-10pt}
\rule[-10pt]{0pt}{26pt}
\begin{tabular}{@{}c}
$1/4$ \\ ${\scs ( 1, 0, 0 , 0 ; 2 )}$
\end{tabular}
\parbox{11mm}{\begin{center}
\begin{fmfgraph*}(8,4)
\setval
\fmfforce{0w,0h}{v1}
\fmfforce{1/2w,0h}{v2}
\fmfforce{1/2w,1h}{v3}
\fmfforce{1w,0h}{w1}
\fmf{plain,left=1}{v2,v3,v2}
\fmf{plain}{v1,w1}
\fmfdot{v2}
\fmfv{decor.shape=cross,decor.filled=shaded,decor.size=3thick}{v1}
\fmfv{decor.shape=cross,decor.filled=shaded,decor.size=3thick}{w1}
\end{fmfgraph*}\end{center}}
\\ $2$ &$1$ & $0$ &
\begin{tabular}{@{}c}
$1/4$ \\ ${\scs ( 0, 1, 0 , 0 ; 2 )}$
\end{tabular}
\parbox{15mm}{\begin{center}
\begin{fmfgraph*}(12,4)
\setval
\fmfforce{0w,1/2h}{v1}
\fmfforce{1/3w,1/2h}{v2}
\fmfforce{2/3w,1/2h}{v3}
\fmfforce{1w,1/2h}{w1}
\fmf{plain,left=1}{v2,v3,v2}
\fmf{plain}{v1,v2}
\fmf{plain}{v3,w1}
\fmfdot{v2,v3}
\fmfv{decor.shape=cross,decor.filled=shaded,decor.size=3thick}{v1}
\fmfv{decor.shape=cross,decor.filled=shaded,decor.size=3thick}{w1}
\end{fmfgraph*}\end{center}}
\hspace*{1mm}
\begin{tabular}{@{}c}
$1/4$ \\ ${\scs ( 1, 0, 0 , 0 ; 2 )}$
\end{tabular}
\parbox{11mm}{\begin{center}
\begin{fmfgraph*}(8,8)
\setval
\fmfforce{0w,0h}{v1}
\fmfforce{1/2w,0h}{v2}
\fmfforce{1/2w,1/2h}{v3}
\fmfforce{1/2w,1h}{v4}
\fmfforce{1w,0h}{w1}
\fmf{plain,left=1}{v3,v4,v3}
\fmf{plain}{v1,w1}
\fmf{plain}{v2,v3}
\fmfdot{v2,v3}
\fmfv{decor.shape=cross,decor.filled=shaded,decor.size=3thick}{v1}
\fmfv{decor.shape=cross,decor.filled=shaded,decor.size=3thick}{w1}
\end{fmfgraph*}\end{center}}
\\ \hline $2$ &$2$ & $2$ &
\hspace{-10pt}
\rule[-10pt]{0pt}{26pt}
\begin{tabular}{@{}c}
$1/12$ \\ ${\scs ( 0, 0, 1 , 0 ; 2 )}$
\end{tabular}
\parbox{15mm}{\begin{center}
\begin{fmfgraph*}(12,4)
\setval
\fmfforce{0w,1/2h}{v1}
\fmfforce{1/3w,1/2h}{v2}
\fmfforce{2/3w,1/2h}{v3}
\fmfforce{1w,1/2h}{w1}
\fmf{plain,left=1}{v2,v3,v2}
\fmf{plain}{v1,w1}
\fmfdot{v2,v3}
\fmfv{decor.shape=cross,decor.filled=shaded,decor.size=3thick}{v1}
\fmfv{decor.shape=cross,decor.filled=shaded,decor.size=3thick}{w1}
\end{fmfgraph*}\end{center}}
\hspace*{1mm}
\begin{tabular}{@{}c}
$1/8$ \\ ${\scs ( 1, 1, 0 , 0 ; 2 )}$
\end{tabular}
\parbox{11mm}{\begin{center}
\begin{fmfgraph*}(8,8)
\setval
\fmfforce{0w,0h}{v1}
\fmfforce{1/2w,0h}{v2}
\fmfforce{1/2w,1/2h}{v3}
\fmfforce{1/2w,1h}{v4}
\fmfforce{1w,0h}{w1}
\fmf{plain,left=1}{v2,v3,v2}
\fmf{plain,left=1}{v3,v4,v3}
\fmf{plain}{v1,w1}
\fmfdot{v2,v3}
\fmfv{decor.shape=cross,decor.filled=shaded,decor.size=3thick}{v1}
\fmfv{decor.shape=cross,decor.filled=shaded,decor.size=3thick}{w1}
\end{fmfgraph*}\end{center}}
\hspace*{1mm}
\begin{tabular}{@{}c}
$1/8$ \\ ${\scs ( 2, 0, 0 , 0 ; 2 )}$
\end{tabular}
\parbox{17mm}{\begin{center}
\begin{fmfgraph*}(14,4)
\setval
\fmfforce{0w,0h}{v1}
\fmfforce{2/7w,0h}{v2}
\fmfforce{5/7w,0h}{v3}
\fmfforce{2/7w,1h}{v4}
\fmfforce{5/7w,1h}{v5}
\fmfforce{1w,0h}{w1}
\fmf{plain,left=1}{v2,v4,v2}
\fmf{plain,left=1}{v3,v5,v3}
\fmf{plain}{v1,w1}
\fmfdot{v2,v3}
\fmfv{decor.shape=cross,decor.filled=shaded,decor.size=3thick}{v1}
\fmfv{decor.shape=cross,decor.filled=shaded,decor.size=3thick}{w1}
\end{fmfgraph*}\end{center}}
\\ $2$ &$2$ & $1$ &
\hspace{-10pt}
\rule[-10pt]{0pt}{26pt}
\begin{tabular}{@{}c}
$1/8$ \\ ${\scs ( 0, 2, 0 , 0 ; 2 )}$
\end{tabular}
\parbox{19mm}{\begin{center}
\begin{fmfgraph*}(16,4)
\setval
\fmfforce{0w,1/2h}{v1}
\fmfforce{1/4w,1/2h}{v2}
\fmfforce{1/2w,1/2h}{v3}
\fmfforce{3/4w,1/2h}{v4}
\fmfforce{1w,1/2h}{w1}
\fmf{plain,left=1}{v2,v3,v2}
\fmf{plain,left=1}{v3,v4,v3}
\fmf{plain}{v1,v2}
\fmf{plain}{v4,w1}
\fmfdot{v2,v3,v4}
\fmfv{decor.shape=cross,decor.filled=shaded,decor.size=3thick}{v1}
\fmfv{decor.shape=cross,decor.filled=shaded,decor.size=3thick}{w1}
\end{fmfgraph*}\end{center}}
\hspace*{1mm}
\begin{tabular}{@{}c}
$1/8$ \\ ${\scs ( 0, 1, 0 , 0 ; 4 )}$
\end{tabular}
\parbox{15mm}{\begin{center}
\begin{fmfgraph*}(8,4)
\setval
\fmfforce{0w,0h}{v1}
\fmfforce{1/2w,0h}{v2}
\fmfforce{1w,0h}{w1}
\fmfforce{1/2w,1h}{v3}
\fmfforce{1/4w,1/2h}{v4}
\fmfforce{3/4w,1/2h}{v5}
\fmf{plain,left=1}{v2,v3,v2}
\fmf{plain}{v1,w1}
\fmf{plain}{v4,v5}
\fmfdot{v2,v4,v5}
\fmfv{decor.shape=cross,decor.filled=shaded,decor.size=3thick}{v1}
\fmfv{decor.shape=cross,decor.filled=shaded,decor.size=3thick}{w1}
\end{fmfgraph*}\end{center}}
\hspace*{1mm}
\begin{tabular}{@{}c}
$1/2$ \\ ${\scs ( 0, 1, 0 , 0 ; 1 )}$
\end{tabular}
\parbox{14.5mm}{\begin{center}
\begin{fmfgraph*}(11.464,3)
\setval
\fmfforce{0w,0h}{v1}
\fmfforce{4/11.464w,0h}{v2}
\fmfforce{7.464/11.464w,0h}{v3}
\fmfforce{1/2w,1h}{v4}
\fmfforce{1/2w,-1/3h}{v7}
\fmfforce{1w,0h}{w1}
\fmf{plain,right=1}{v4,v7,v4}
\fmf{plain}{v1,v2}
\fmf{plain}{v4,v2}
\fmf{plain}{v3,w1}
\fmfdot{v2,v3,v4}
\fmfv{decor.shape=cross,decor.filled=shaded,decor.size=3thick}{v1}
\fmfv{decor.shape=cross,decor.filled=shaded,decor.size=3thick}{w1}
\end{fmfgraph*}\end{center}}
\hspace*{1mm}
\begin{tabular}{@{}c}
$1/4$ \\ ${\scs ( 1, 0, 0 , 0 ; 2 )}$
\end{tabular}
\parbox{14.5mm}{\begin{center}
\begin{fmfgraph*}(11.464,7)
\setval
\fmfforce{0w,0h}{v1}
\fmfforce{4/11.464w,0h}{v2}
\fmfforce{7.464/11.464w,0h}{v3}
\fmfforce{1/2w,3/7h}{v4}
\fmfforce{1/2w,1h}{v5}
\fmfforce{1/2w,-1/7h}{v7}
\fmfforce{1w,0h}{w1}
\fmf{plain,right=1}{v4,v7,v4}
\fmf{plain,left=1}{v4,v5,v4}
\fmf{plain}{v1,v2}
\fmf{plain}{v3,w1}
\fmfdot{v2,v3,v4}
\fmfv{decor.shape=cross,decor.filled=shaded,decor.size=3thick}{v1}
\fmfv{decor.shape=cross,decor.filled=shaded,decor.size=3thick}{w1}
\end{fmfgraph*}\end{center}}
\hspace*{1mm}
\begin{tabular}{@{}c}
$1/12$ \\ ${\scs ( 0, 0, 1 , 0 ;2 )}$
\end{tabular}
\parbox{11mm}{\begin{center}
\begin{fmfgraph*}(8,8)
\setval
\fmfforce{0w,0h}{v1}
\fmfforce{1/2w,0h}{v2}
\fmfforce{1/2w,1/2h}{v3}
\fmfforce{1/2w,1h}{v4}
\fmfforce{1w,0h}{w1}
\fmf{plain,right=1}{v4,v3,v4}
\fmf{plain}{v1,w1}
\fmf{plain}{v4,v2}
\fmfdot{v2,v3,v4}
\fmfv{decor.shape=cross,decor.filled=shaded,decor.size=3thick}{v1}
\fmfv{decor.shape=cross,decor.filled=shaded,decor.size=3thick}{w1}
\end{fmfgraph*}\end{center}}
\\ &&&
\begin{tabular}{@{}c}
$1/4$ \\ ${\scs ( 1, 1, 0 , 0 ; 1 )}$
\end{tabular}
\parbox{19mm}{\begin{center}
\begin{fmfgraph*}(16,6)
\setval
\fmfforce{0w,1/3h}{v1}
\fmfforce{1/4w,1/3h}{v2}
\fmfforce{1/2w,1/3h}{v3}
\fmfforce{3/4w,1/3h}{v4}
\fmfforce{3/4w,1h}{v5}
\fmfforce{1w,1/3h}{w1}
\fmf{plain,right=1}{v2,v3,v2}
\fmf{plain,left=1}{v4,v5,v4}
\fmf{plain}{v1,v2}
\fmf{plain}{w1,v3}
\fmfdot{v2,v3,v4}
\fmfv{decor.shape=cross,decor.filled=shaded,decor.size=3thick}{v1}
\fmfv{decor.shape=cross,decor.filled=shaded,decor.size=3thick}{w1}
\end{fmfgraph*}\end{center}}
\hspace*{1mm}
\begin{tabular}{@{}c}
$1/8$ \\ ${\scs ( 1, 1, 0 , 0 ; 2 )}$
\end{tabular}
\parbox{11mm}{\begin{center}
\begin{fmfgraph*}(8,12)
\setval
\fmfforce{0w,0h}{v1}
\fmfforce{1/2w,0h}{v2}
\fmfforce{1/2w,1/3h}{v3}
\fmfforce{1/2w,2/3h}{v4}
\fmfforce{1/2w,1h}{v5}
\fmfforce{1w,0h}{w1}
\fmf{plain,right=1}{v3,v4,v3}
\fmf{plain,left=1}{v4,v5,v4}
\fmf{plain}{v1,w1}
\fmf{plain}{v2,v3}
\fmfdot{v2,v3,v4}
\fmfv{decor.shape=cross,decor.filled=shaded,decor.size=3thick}{v1}
\fmfv{decor.shape=cross,decor.filled=shaded,decor.size=3thick}{w1}
\end{fmfgraph*}\end{center}}
\hspace*{1mm}
\begin{tabular}{@{}c}
$1/8$ \\ ${\scs ( 1, 1, 0 , 0 ; 2 )}$
\end{tabular}
\parbox{11mm}{\begin{center}
\begin{fmfgraph*}(8,12)
\setval
\fmfforce{0w,0h}{v1}
\fmfforce{1/2w,0h}{v2}
\fmfforce{1w,0h}{w1}
\fmfforce{1/2w,1/3h}{v3}
\fmfforce{1/2w,2/3h}{v4}
\fmfforce{1/2w,1h}{v5}
\fmf{plain,left=1}{v2,v3,v2}
\fmf{plain,left=1}{v4,v5,v4}
\fmf{plain}{v3,v4}
\fmf{plain}{v1,w1}
\fmfdot{v2,v3,v4}
\fmfv{decor.shape=cross,decor.filled=shaded,decor.size=3thick}{v1}
\fmfv{decor.shape=cross,decor.filled=shaded,decor.size=3thick}{w1}
\end{fmfgraph*}\end{center}}
\hspace*{1mm}
\begin{tabular}{@{}c}
$1/4$ \\ ${\scs ( 1, 1, 0 , 0 ; 1 )}$
\end{tabular}
\parbox{15mm}{\begin{center}
\begin{fmfgraph*}(12,10)
\setval
\fmfforce{0w,1/5h}{v1}
\fmfforce{1/3w,1/5h}{v2}
\fmfforce{2/3w,1/5h}{v3}
\fmfforce{2/3w,3/5h}{v4}
\fmfforce{2/3w,1h}{v5}
\fmfforce{1w,1/5h}{w1}
\fmf{plain,left=1}{v2,v3,v2}
\fmf{plain,left=1}{v4,v5,v4}
\fmf{plain}{w1,v3}
\fmf{plain}{v3,v4}
\fmf{plain}{v1,v2}
\fmfdot{v2,v3,v4}
\fmfv{decor.shape=cross,decor.filled=shaded,decor.size=3thick}{v1}
\fmfv{decor.shape=cross,decor.filled=shaded,decor.size=3thick}{w1}
\end{fmfgraph*}\end{center}}
\hspace*{1mm}
\begin{tabular}{@{}c}
$1/8$ \\ ${\scs ( 2, 0, 0 , 0 ; 2 )}$
\end{tabular}
\parbox{11mm}{\begin{center}
\begin{fmfgraph*}(8,12)
\setval
\fmfforce{0w,0h}{v1}
\fmfforce{1/2w,0h}{v2}
\fmfforce{1w,0h}{w1}
\fmfforce{1/2w,1/3h}{v3}
\fmfforce{1/2w,2/3h}{v4}
\fmfforce{1/2w,1h}{v5}
\fmfforce{0w,1/3h}{v6}
\fmf{plain,left=1}{v6,v3,v6}
\fmf{plain,left=1}{v4,v5,v4}
\fmf{plain}{v2,v4}
\fmf{plain}{v1,w1}
\fmfdot{v2,v3,v4}
\fmfv{decor.shape=cross,decor.filled=shaded,decor.size=3thick}{v1}
\fmfv{decor.shape=cross,decor.filled=shaded,decor.size=3thick}{w1}
\end{fmfgraph*}\end{center}}
\\ &&&
\begin{tabular}{@{}c}
$1/4$ \\ ${\scs ( 2, 0, 0 , 0 ; 1 )}$
\end{tabular}
\parbox{15mm}{\begin{center}
\begin{fmfgraph*}(12,8)
\setval
\fmfforce{0w,0h}{v1}
\fmfforce{1/3w,0h}{v2}
\fmfforce{1/3w,1/2h}{v3}
\fmfforce{1/3w,1h}{v4}
\fmfforce{2/3w,0h}{v5}
\fmfforce{2/3w,1/2h}{v6}
\fmfforce{1w,0h}{w1}
\fmf{plain,left=1}{v4,v3,v4}
\fmf{plain,left=1}{v6,v5,v6}
\fmf{plain}{v2,v3}
\fmf{plain}{v1,w1}
\fmfdot{v2,v3,v5}
\fmfv{decor.shape=cross,decor.filled=shaded,decor.size=3thick}{v1}
\fmfv{decor.shape=cross,decor.filled=shaded,decor.size=3thick}{w1}
\end{fmfgraph*}\end{center}}
\hspace*{1mm}
\begin{tabular}{@{}c}
$1/16$ \\ ${\scs ( 2, 0, 0 , 0 ; 4 )}$
\end{tabular}
\parbox{11mm}{\begin{center}
\begin{fmfgraph*}(8,16)
\setval
\fmfforce{0w,1/2h}{v1}
\fmfforce{1/2w,1/2h}{v2}
\fmfforce{1/2w,3/4h}{v3}
\fmfforce{1/2w,1h}{v4}
\fmfforce{1/2w,1/4h}{v5}
\fmfforce{1/2w,0h}{v6}
\fmfforce{1w,1/2h}{w1}
\fmf{plain,left=1}{v4,v3,v4}
\fmf{plain,left=1}{v6,v5,v6}
\fmf{plain}{v5,v3}
\fmf{plain}{v1,w1}
\fmfdot{v2,v3,v5}
\fmfv{decor.shape=cross,decor.filled=shaded,decor.size=3thick}{v1}
\fmfv{decor.shape=cross,decor.filled=shaded,decor.size=3thick}{w1}
\end{fmfgraph*}\end{center}}
\\ $2$ &$2$ & $0$ &
\begin{tabular}{@{}c}
$1/4$ \\ ${\scs ( 0, 0, 0 , 0 ; 4 )}$
\end{tabular}
\parbox{15mm}{\begin{center}
\begin{fmfgraph*}(12,4)
\setval
\fmfforce{0w,1/2h}{v1}
\fmfforce{1/3w,1/2h}{v2}
\fmfforce{2/3w,1/2h}{v3}
\fmfforce{1w,1/2h}{w1}
\fmfforce{1/2w,0h}{v4}
\fmfforce{1/2w,1h}{v5}
\fmf{plain,left=1}{v2,v3,v2}
\fmf{plain}{v1,v2}
\fmf{plain}{v3,w1}
\fmf{plain}{v4,v5}
\fmfdot{v2,v3,v4,v5}
\fmfv{decor.shape=cross,decor.filled=shaded,decor.size=3thick}{v1}
\fmfv{decor.shape=cross,decor.filled=shaded,decor.size=3thick}{w1}
\end{fmfgraph*}\end{center}}
\hspace*{1mm}
\begin{tabular}{@{}c}
$1/4$ \\ ${\scs ( 0, 1, 0 , 0 ; 2 )}$
\end{tabular}
\parbox{15mm}{\begin{center}
\begin{fmfgraph}(12,4)
\setval
\fmfforce{1/3w,0h}{v1}
\fmfforce{2/3w,0h}{v2}
\fmfforce{2/3w,1h}{v3}
\fmfforce{1/3w,1h}{v4}
\fmfforce{0w,0h}{v5}
\fmfforce{1w,0h}{v6}
\fmf{plain,right=1}{v1,v3,v1}
\fmf{plain}{v1,v5}
\fmf{plain}{v2,v6}
\fmf{plain,right=0.5}{v4,v3}
\fmfdot{v1,v2,v3,v4}
\fmfv{decor.shape=cross,decor.filled=shaded,decor.size=3thick}{v5}
\fmfv{decor.shape=cross,decor.filled=shaded,decor.size=3thick}{v6}
\end{fmfgraph}\end{center}} 
\hspace*{1mm}
\begin{tabular}{@{}c}
$1/8$ \\ ${\scs ( 0, 1, 0 , 0 ; 4 )}$
\end{tabular}
\parbox{11mm}{\begin{center}
\begin{fmfgraph*}(8,8)
\setval
\fmfforce{0w,0h}{v1}
\fmfforce{1/2w,0h}{v2}
\fmfforce{1w,0h}{w1}
\fmfforce{1/4w,3/4h}{v3}
\fmfforce{3/4w,3/4h}{v4}
\fmfforce{1/2w,1/2h}{v5}
\fmf{plain,left=1}{v3,v4,v3}
\fmf{plain}{v1,w1}
\fmf{plain}{v3,v4}
\fmf{plain}{v5,v2}
\fmfdot{v2,v3,v4,v5}
\fmfv{decor.shape=cross,decor.filled=shaded,decor.size=3thick}{v1}
\fmfv{decor.shape=cross,decor.filled=shaded,decor.size=3thick}{w1}
\end{fmfgraph*}\end{center}}
\hspace*{1mm}
\begin{tabular}{@{}c}
$1/8$ \\ ${\scs ( 0, 2, 0 , 0 ; 2 )}$
\end{tabular}
\parbox{23mm}{\begin{center}
\begin{fmfgraph*}(20,4)
\setval
\fmfforce{0w,1/2h}{v1}
\fmfforce{1/5w,1/2h}{v2}
\fmfforce{2/5w,1/2h}{v3}
\fmfforce{3/5w,1/2h}{v4}
\fmfforce{4/5w,1/2h}{v5}
\fmfforce{1w,1/2h}{w1}
\fmf{plain,left=1}{v2,v3,v2}
\fmf{plain,left=1}{v4,v5,v4}
\fmf{plain}{v1,v2}
\fmf{plain}{v3,v4}
\fmf{plain}{v5,w1}
\fmfdot{v2,v3,v4,v5}
\fmfv{decor.shape=cross,decor.filled=shaded,decor.size=3thick}{v1}
\fmfv{decor.shape=cross,decor.filled=shaded,decor.size=3thick}{w1}
\end{fmfgraph*}\end{center}}
\hspace*{1mm}
\begin{tabular}{@{}c}
$1/8$ \\ ${\scs ( 1, 1, 0 , 0 ; 2 )}$
\end{tabular}
\parbox{11mm}{\begin{center}
\begin{fmfgraph*}(8,16)
\setval
\fmfforce{0w,0h}{v1}
\fmfforce{1/2w,0h}{v2}
\fmfforce{1/2w,1/4h}{v3}
\fmfforce{1/2w,1/2h}{v4}
\fmfforce{1/2w,3/4h}{v5}
\fmfforce{1/2w,1h}{v6}
\fmfforce{1w,0h}{w1}
\fmf{plain,left=1}{v3,v4,v3}
\fmf{plain,left=1}{v5,v6,v5}
\fmf{plain}{v2,v3}
\fmf{plain}{v4,v5}
\fmf{plain}{v1,w1}
\fmfdot{v2,v3,v4,v5}
\fmfv{decor.shape=cross,decor.filled=shaded,decor.size=3thick}{v1}
\fmfv{decor.shape=cross,decor.filled=shaded,decor.size=3thick}{w1}
\end{fmfgraph*}\end{center}}
\\ & && 
\begin{tabular}{@{}c}
$1/4$ \\ ${\scs ( 1, 1, 0 , 0 ; 1 )}$
\end{tabular}
\parbox{19mm}{\begin{center}
\begin{fmfgraph*}(16,10)
\setval
\fmfforce{0w,1/5h}{v1}
\fmfforce{1/4w,1/5h}{v2}
\fmfforce{1/4w,3/5h}{v3}
\fmfforce{1/4w,1h}{v4}
\fmfforce{1/2w,1/5h}{v5}
\fmfforce{3/4w,1/5h}{v6}
\fmfforce{1w,1/5h}{w1}
\fmf{plain,left=1}{v3,v4,v3}
\fmf{plain,left=1}{v5,v6,v5}
\fmf{plain}{v6,w1}
\fmf{plain}{v2,v3}
\fmf{plain}{v1,v5}
\fmfdot{v2,v3,v5,v6}
\fmfv{decor.shape=cross,decor.filled=shaded,decor.size=3thick}{v1}
\fmfv{decor.shape=cross,decor.filled=shaded,decor.size=3thick}{w1}
\end{fmfgraph*}\end{center}}
\hspace*{1mm}
\begin{tabular}{@{}c}
$1/4$ \\ ${\scs ( 1, 0, 0 , 0 ; 2 )}$
\end{tabular}
\parbox{14.5mm}{\begin{center}
\begin{fmfgraph*}(11.464,11)
\setval
\fmfforce{0w,0h}{v1}
\fmfforce{4/11.464w,0h}{v2}
\fmfforce{7.464/11.464w,0h}{v3}
\fmfforce{1/2w,3/11h}{v4}
\fmfforce{1/2w,7/11h}{v5}
\fmfforce{1/2w,1h}{v6}
\fmfforce{1/2w,-1/11h}{v7}
\fmfforce{1w,0h}{w1}
\fmf{plain,right=1}{v4,v7,v4}
\fmf{plain,left=1}{v5,v6,v5}
\fmf{plain}{v1,v2}
\fmf{plain}{v3,w1}
\fmf{plain}{v4,v5}
\fmfdot{v2,v3,v4,v5}
\fmfv{decor.shape=cross,decor.filled=shaded,decor.size=3thick}{v1}
\fmfv{decor.shape=cross,decor.filled=shaded,decor.size=3thick}{w1}
\end{fmfgraph*}\end{center}}
\hspace*{1mm}
\begin{tabular}{@{}c}
$1/8$ \\ ${\scs ( 2, 0, 0 , 0 ; 2 )}$
\end{tabular}
\parbox{17mm}{\begin{center}
\begin{fmfgraph*}(14,8)
\setval
\fmfforce{0w,0h}{v1}
\fmfforce{4/14w,0h}{v2}
\fmfforce{4/14w,1/2h}{v3}
\fmfforce{4/14w,1h}{v4}
\fmfforce{10/14w,0h}{v5}
\fmfforce{10/14w,1/2h}{v6}
\fmfforce{10/14w,1h}{v7}
\fmfforce{1w,0h}{w1}
\fmf{plain,right=1}{v3,v4,v3}
\fmf{plain,left=1}{v6,v7,v6}
\fmf{plain}{v1,w1}
\fmf{plain}{v2,v3}
\fmf{plain}{v6,v5}
\fmfdot{v2,v3,v6,v5}
\fmfv{decor.shape=cross,decor.filled=shaded,decor.size=3thick}{v1}
\fmfv{decor.shape=cross,decor.filled=shaded,decor.size=3thick}{w1}
\end{fmfgraph*}\end{center}}
\hspace*{1mm}
\begin{tabular}{@{}c}
$1/16$ \\ ${\scs ( 2, 0, 0 , 0 ; 4 )}$
\end{tabular}
\hspace*{3mm}
\parbox{11mm}{\begin{center}
\begin{fmfgraph*}(8,8)
\setval
\fmfforce{0w,0h}{v1}
\fmfforce{1/2w,0h}{v2}
\fmfforce{1/2w,1/2h}{v3}
\fmfforce{0.067w,3/4h}{v4}
\fmfforce{0.933w,3/4h}{v5}
\fmfforce{-0.366w,1h}{v6}
\fmfforce{1.366w,1h}{v7}
\fmfforce{1w,0h}{w1}
\fmf{plain,right=1}{v4,v6,v4}
\fmf{plain,left=1}{v5,v7,v5}
\fmf{plain}{v1,w1}
\fmf{plain}{v2,v3}
\fmf{plain}{v3,v4}
\fmf{plain}{v3,v5}
\fmfdot{v2,v3,v4,v5}
\fmfv{decor.shape=cross,decor.filled=shaded,decor.size=3thick}{v1}
\fmfv{decor.shape=cross,decor.filled=shaded,decor.size=3thick}{w1}
\end{fmfgraph*}\end{center}} \\ \hline\hline
$3$ &$0$ & $0$ &
\hspace{-10pt}
\rule[-10pt]{0pt}{26pt}
\begin{tabular}{@{}c}
$1/6$ \\ ${\scs ( 0, 0, 0 , 0 ; 6 )}$
\end{tabular}
\parbox{11mm}{\begin{center}
\begin{fmfgraph*}(6.928,12)
\setval
\fmfforce{1/2w,5/6h}{v1}
\fmfforce{1w,1/4h}{w1}
\fmfforce{0w,1/4h}{u1}
\fmfforce{1/2w,1/2h}{v2}
\fmf{plain}{v2,v1}
\fmf{plain}{v2,w1}
\fmf{plain}{v2,u1}
\fmfdot{v2}
\fmfv{decor.shape=cross,decor.filled=shaded,decor.size=3thick}{v1}
\fmfv{decor.shape=cross,decor.filled=shaded,decor.size=3thick}{w1}
\fmfv{decor.shape=cross,decor.filled=shaded,decor.size=3thick}{u1}
\end{fmfgraph*}\end{center}}
\\ \hline $3$ &$1$ & $1$ &
\hspace{-10pt}
\rule[-10pt]{0pt}{26pt}
\begin{tabular}{@{}c}
$1/6$ \\ ${\scs ( 0, 0, 0 , 0 ; 6 )}$
\end{tabular}
\parbox{15mm}{\begin{center}
\begin{fmfgraph}(12,12)
\setval
\fmfforce{1/2w,1/3h}{v1}
\fmfforce{1/2w,2/3h}{v2}
\fmfforce{1/2w,1h}{v3}
\fmfforce{0.355662432w,0.416666666h}{v4}
\fmfforce{0.64433568w,0.416666666h}{v5}
\fmfforce{0.067w,1/4h}{v6}
\fmfforce{0.933w,1/4h}{v7}
\fmf{plain,left=1}{v1,v2,v1}
\fmf{plain}{v2,v3}
\fmf{plain}{v4,v6}
\fmf{plain}{v5,v7}
\fmfdot{v2,v4,v5}
\fmfv{decor.shape=cross,decor.filled=shaded,decor.size=3thick}{v3}
\fmfv{decor.shape=cross,decor.filled=shaded,decor.size=3thick}{v6}
\fmfv{decor.shape=cross,decor.filled=shaded,decor.size=3thick}{v7}
\end{fmfgraph}\end{center}} 
\hspace*{1mm}
\begin{tabular}{@{}c}
$1/4$ \\ ${\scs ( 0, 1, 0 , 0 ; 2 )}$
\end{tabular}
\parbox{11mm}{\begin{center}
\begin{fmfgraph*}(8,12)
\setval
\fmfforce{0w,0h}{v1}
\fmfforce{1w,0h}{w1}
\fmfforce{1/2w,1h}{u1}
\fmfforce{1/2w,0h}{v2}
\fmfforce{1/2w,1/3h}{v3}
\fmfforce{1/2w,2/3h}{v4}
\fmf{plain,right=1}{v4,v3,v4}
\fmf{plain}{w1,v1}
\fmf{plain}{v2,v3}
\fmf{plain}{v4,u1}
\fmfdot{v2,v3,v4}
\fmfv{decor.shape=cross,decor.filled=shaded,decor.size=3thick}{v1}
\fmfv{decor.shape=cross,decor.filled=shaded,decor.size=3thick}{w1}
\fmfv{decor.shape=cross,decor.filled=shaded,decor.size=3thick}{u1}
\end{fmfgraph*}\end{center}}
\hspace*{1mm}
\begin{tabular}{@{}c}
$1/4$ \\ ${\scs ( 1, 0, 0 , 0 ; 2 )}$
\end{tabular}
\parbox{17mm}{\begin{center}
\begin{fmfgraph*}(14,8)
\setval
\fmfforce{0w,0h}{v1}
\fmfforce{4/14w,0h}{v2}
\fmfforce{4/14w,1/2h}{v3}
\fmfforce{4/14w,1h}{v4}
\fmfforce{10/14w,0h}{v5}
\fmfforce{10/14w,1/2h}{u1}
\fmfforce{1w,0h}{w1}
\fmf{plain,right=1}{v3,v4,v3}
\fmf{plain}{u1,v5}
\fmf{plain}{v1,w1}
\fmf{plain}{v2,v3}
\fmf{plain}{v6,v5}
\fmfdot{v2,v3,v6,v5}
\fmfv{decor.shape=cross,decor.filled=shaded,decor.size=3thick}{u1}
\fmfv{decor.shape=cross,decor.filled=shaded,decor.size=3thick}{v1}
\fmfv{decor.shape=cross,decor.filled=shaded,decor.size=3thick}{w1}
\end{fmfgraph*}\end{center}}
\\ $3$ &$1$ & $0$ &
\begin{tabular}{@{}c}
$1/4$ \\ ${\scs ( 0, 1, 0 , 0 ; 2 )}$
\end{tabular}
\parbox{11mm}{\begin{center}
\begin{fmfgraph*}(8,8)
\setval
\fmfforce{0w,0h}{v1}
\fmfforce{1w,0h}{w1}
\fmfforce{1/2w,1h}{u1}
\fmfforce{1/2w,0h}{v2}
\fmfforce{1/2w,1/2h}{v3}
\fmf{plain,right=1}{v2,v3,v2}
\fmf{plain}{w1,v1}
\fmf{plain}{v3,u1}
\fmfdot{v2,v3}
\fmfv{decor.shape=cross,decor.filled=shaded,decor.size=3thick}{v1}
\fmfv{decor.shape=cross,decor.filled=shaded,decor.size=3thick}{w1}
\fmfv{decor.shape=cross,decor.filled=shaded,decor.size=3thick}{u1}
\end{fmfgraph*}\end{center}}
\hspace*{1mm}
\begin{tabular}{@{}c}
$1/4$ \\ ${\scs ( 1, 0, 0 , 0 ; 2 )}$
\end{tabular}
\parbox{11mm}{\begin{center}
\begin{fmfgraph*}(8,8)
\setval
\fmfforce{0w,0h}{v1}
\fmfforce{1w,0h}{w1}
\fmfforce{1/2w,1h}{u1}
\fmfforce{1/2w,0h}{v2}
\fmfforce{1/2w,1/2h}{v3}
\fmfforce{0w,1/2h}{v4}
\fmf{plain,right=1}{v4,v3,v4}
\fmf{plain}{w1,v1}
\fmf{plain}{v2,u1}
\fmfdot{v2,v3}
\fmfv{decor.shape=cross,decor.filled=shaded,decor.size=3thick}{v1}
\fmfv{decor.shape=cross,decor.filled=shaded,decor.size=3thick}{w1}
\fmfv{decor.shape=cross,decor.filled=shaded,decor.size=3thick}{u1}
\end{fmfgraph*}\end{center}}
\hspace*{1mm}
\begin{tabular}{@{}c}
$1/12$ \\ ${\scs ( 1, 0, 0 , 0 ; 6 )}$
\end{tabular}
\parbox{11mm}{\begin{center}
\begin{fmfgraph*}(8,12)
\setval
\fmfforce{0w,1/3h}{v1}
\fmfforce{1w,1/3h}{w1}
\fmfforce{1/2w,0h}{u1}
\fmfforce{1/2w,1/3h}{v2}
\fmfforce{1/2w,2/3h}{v3}
\fmfforce{1/2w,1h}{v4}
\fmf{plain,right=1}{v4,v3,v4}
\fmf{plain}{w1,v1}
\fmf{plain}{v3,u1}
\fmfdot{v2,v3}
\fmfv{decor.shape=cross,decor.filled=shaded,decor.size=3thick}{v1}
\fmfv{decor.shape=cross,decor.filled=shaded,decor.size=3thick}{w1}
\fmfv{decor.shape=cross,decor.filled=shaded,decor.size=3thick}{u1}
\end{fmfgraph*}\end{center}} \\ \hline \hline
$4$ &$0$ & $1$ &
\hspace{-10pt}
\rule[-10pt]{0pt}{26pt}
\begin{tabular}{@{}c}
$1/24$ \\ ${\scs ( 0, 0, 0 , 0 ; 24 )}$
\end{tabular}
\parbox{11mm}{\begin{center}
\begin{fmfgraph*}(8,12)
\setval
\fmfforce{0w,1/2h}{v1}
\fmfforce{1w,1/2h}{w1}
\fmfforce{1/2w,1/6h}{u1}
\fmfforce{1/2w,5/6h}{x1}
\fmfforce{1/2w,1/2h}{v2}
\fmf{plain}{w1,v1}
\fmf{plain}{x1,u1}
\fmfdot{v2}
\fmfv{decor.shape=cross,decor.filled=shaded,decor.size=3thick}{u1}
\fmfv{decor.shape=cross,decor.filled=shaded,decor.size=3thick}{v1}
\fmfv{decor.shape=cross,decor.filled=shaded,decor.size=3thick}{w1}
\fmfv{decor.shape=cross,decor.filled=shaded,decor.size=3thick}{x1}
\end{fmfgraph*}\end{center}}
\\ $4$ &$0$ & $0$ & 
\begin{tabular}{@{}c}
$1/8$ \\ ${\scs ( 0, 0, 0 , 0 ; 8 )}$
\end{tabular}
\parbox{11mm}{\begin{center}
\begin{fmfgraph*}(8,4)
\setval
\fmfforce{0w,0h}{v1}
\fmfforce{1w,0h}{w1}
\fmfforce{0w,1h}{u1}
\fmfforce{1w,1h}{x1}
\fmfforce{1/2w,0h}{v2}
\fmfforce{1/2w,1h}{v3}
\fmf{plain}{w1,v1}
\fmf{plain}{v2,v3}
\fmf{plain}{x1,u1}
\fmfdot{v2,v3}
\fmfv{decor.shape=cross,decor.filled=shaded,decor.size=3thick}{u1}
\fmfv{decor.shape=cross,decor.filled=shaded,decor.size=3thick}{v1}
\fmfv{decor.shape=cross,decor.filled=shaded,decor.size=3thick}{w1}
\fmfv{decor.shape=cross,decor.filled=shaded,decor.size=3thick}{x1}
\end{fmfgraph*}\end{center}}
\\ \hline $4$ &$1$ & $2$ &
\hspace{-10pt}
\rule[-10pt]{0pt}{26pt}
\begin{tabular}{@{}c}
$1/16$ \\ ${\scs ( 0, 1, 0 , 0 ; 8 )}$
\end{tabular}
\parbox{11mm}{\begin{center}
\begin{fmfgraph*}(8,4)
\setval
\fmfforce{0w,0h}{v1}
\fmfforce{1w,0h}{w1}
\fmfforce{0w,1h}{u1}
\fmfforce{1w,1h}{x1}
\fmfforce{1/2w,0h}{v2}
\fmfforce{1/2w,1h}{v3}
\fmf{plain,right=1}{v2,v3,v2}
\fmf{plain}{w1,v1}
\fmf{plain}{x1,u1}
\fmfdot{v2,v3}
\fmfv{decor.shape=cross,decor.filled=shaded,decor.size=3thick}{u1}
\fmfv{decor.shape=cross,decor.filled=shaded,decor.size=3thick}{v1}
\fmfv{decor.shape=cross,decor.filled=shaded,decor.size=3thick}{w1}
\fmfv{decor.shape=cross,decor.filled=shaded,decor.size=3thick}{x1}
\end{fmfgraph*}\end{center}}
\hspace*{1mm}
\begin{tabular}{@{}c}
$1/12$ \\ ${\scs ( 1, 0, 0 , 0 ; 6 )}$
\end{tabular}
\parbox{15mm}{\begin{center}
\begin{fmfgraph*}(12,8)
\setval
\fmfforce{0w,1/2h}{v1}
\fmfforce{1w,1/2h}{w1}
\fmfforce{1/3w,0h}{u1}
\fmfforce{1/3w,1h}{x1}
\fmfforce{1/3w,1/2h}{v2}
\fmfforce{2/3w,1/2h}{v3}
\fmfforce{2/3w,1h}{v4}
\fmf{plain,right=1}{v4,v3,v4}
\fmf{plain}{w1,v1}
\fmf{plain}{x1,u1}
\fmfdot{v2,v3}
\fmfv{decor.shape=cross,decor.filled=shaded,decor.size=3thick}{u1}
\fmfv{decor.shape=cross,decor.filled=shaded,decor.size=3thick}{v1}
\fmfv{decor.shape=cross,decor.filled=shaded,decor.size=3thick}{w1}
\fmfv{decor.shape=cross,decor.filled=shaded,decor.size=3thick}{x1}
\end{fmfgraph*}\end{center}}
\\ $4$ &$1$ & $1$ & 
\begin{tabular}{@{}c}
$1/4$ \\ ${\scs ( 0,0, 0 , 0 ; 4 )}$
\end{tabular}
\parbox{14.5mm}{\begin{center}
\begin{fmfgraph*}(11.464,3)
\setval
\fmfforce{0w,0h}{v1}
\fmfforce{4/11.464w,0h}{v2}
\fmfforce{7.464/11.464w,0h}{v3}
\fmfforce{1/2w,1h}{v4}
\fmfforce{1/2w,-1/3h}{v7}
\fmfforce{1w,0h}{w1}
\fmfforce{1.732/11.464w,1h}{u1}
\fmfforce{9.732/11.464w,1h}{x1}
\fmf{plain,right=1}{v4,v7,v4}
\fmf{plain}{u1,x1}
\fmf{plain}{v1,v2}
\fmf{plain}{v3,w1}
\fmfdot{v2,v3,v4}
\fmfv{decor.shape=cross,decor.filled=shaded,decor.size=3thick}{u1}
\fmfv{decor.shape=cross,decor.filled=shaded,decor.size=3thick}{v1}
\fmfv{decor.shape=cross,decor.filled=shaded,decor.size=3thick}{w1}
\fmfv{decor.shape=cross,decor.filled=shaded,decor.size=3thick}{x1}
\end{fmfgraph*}\end{center}}
\hspace*{1mm}
\begin{tabular}{@{}c}
$1/12$ \\ ${\scs ( 0,1, 0 , 0 ; 6 )}$
\end{tabular}
\parbox{11mm}{\begin{center}
\begin{fmfgraph*}(8,16)
\setval
\fmfforce{0w,1/4h}{v1}
\fmfforce{1w,1/4h}{w1}
\fmfforce{1/2w,0h}{u1}
\fmfforce{1/2w,1h}{x1}
\fmfforce{1/2w,1/4h}{v2}
\fmfforce{1/2w,1/2h}{v3}
\fmfforce{1/2w,3/4h}{v4}
\fmf{plain,right=1}{v4,v3,v4}
\fmf{plain}{v2,v3}
\fmf{plain}{w1,v1}
\fmf{plain}{v3,u1}
\fmf{plain}{v4,x1}
\fmfdot{v2,v3,v4}
\fmfv{decor.shape=cross,decor.filled=shaded,decor.size=3thick}{u1}
\fmfv{decor.shape=cross,decor.filled=shaded,decor.size=3thick}{v1}
\fmfv{decor.shape=cross,decor.filled=shaded,decor.size=3thick}{w1}
\fmfv{decor.shape=cross,decor.filled=shaded,decor.size=3thick}{x1}
\end{fmfgraph*}\end{center}}
\hspace*{1mm}
\begin{tabular}{@{}c}
$1/4$ \\ ${\scs ( 0,1, 0 , 0 ; 2 )}$
\end{tabular}
\parbox{11mm}{\begin{center}
\begin{fmfgraph*}(8,12)
\setval
\fmfforce{0w,0h}{v1}
\fmfforce{1w,0h}{w1}
\fmfforce{0w,1/3h}{u1}
\fmfforce{1/2w,1h}{x1}
\fmfforce{1/2w,0h}{v2}
\fmfforce{1/2w,1/3h}{v3}
\fmfforce{1/2w,2/3h}{v4}
\fmf{plain,right=1}{v4,v3,v4}
\fmf{plain}{v2,v3}
\fmf{plain}{w1,v1}
\fmf{plain}{v3,u1}
\fmf{plain}{v4,x1}
\fmfdot{v2,v3,v4}
\fmfv{decor.shape=cross,decor.filled=shaded,decor.size=3thick}{u1}
\fmfv{decor.shape=cross,decor.filled=shaded,decor.size=3thick}{v1}
\fmfv{decor.shape=cross,decor.filled=shaded,decor.size=3thick}{w1}
\fmfv{decor.shape=cross,decor.filled=shaded,decor.size=3thick}{x1}
\end{fmfgraph*}\end{center}}
\hspace*{1mm}
\begin{tabular}{@{}c}
$1/8$ \\ ${\scs ( 0,1, 0 , 0 ; 4 )}$
\end{tabular}
\parbox{11mm}{\begin{center}
\begin{fmfgraph*}(8,8)
\setval
\fmfforce{0w,0h}{v1}
\fmfforce{1w,0h}{w1}
\fmfforce{0w,1h}{u1}
\fmfforce{1w,1h}{x1}
\fmfforce{1/2w,0h}{v2}
\fmfforce{1/2w,1/2h}{v3}
\fmfforce{1/2w,1h}{v4}
\fmf{plain,right=1}{v4,v3,v4}
\fmf{plain}{v2,v3}
\fmf{plain}{u1,x1}
\fmf{plain}{v1,w1}
\fmf{plain}{v4,x1}
\fmfdot{v2,v3,v4}
\fmfv{decor.shape=cross,decor.filled=shaded,decor.size=3thick}{u1}
\fmfv{decor.shape=cross,decor.filled=shaded,decor.size=3thick}{v1}
\fmfv{decor.shape=cross,decor.filled=shaded,decor.size=3thick}{w1}
\fmfv{decor.shape=cross,decor.filled=shaded,decor.size=3thick}{x1}
\end{fmfgraph*}\end{center}}
\\ &&&
\begin{tabular}{@{}c}
$1/4$ \\ ${\scs ( 1,0, 0 , 0 ; 2 )}$
\end{tabular}
\parbox{11mm}{\begin{center}
\begin{fmfgraph*}(8,12)
\setval
\fmfforce{0w,0h}{v1}
\fmfforce{1w,0h}{w1}
\fmfforce{0w,1/3h}{u1}
\fmfforce{1/2w,1h}{x1}
\fmfforce{1/2w,0h}{v2}
\fmfforce{1/2w,1/3h}{v3}
\fmfforce{1/2w,2/3h}{v4}
\fmfforce{0w,2/3h}{v5}
\fmf{plain,right=1}{v4,v5,v4}
\fmf{plain}{v2,v3}
\fmf{plain}{w1,v1}
\fmf{plain}{v3,u1}
\fmf{plain}{v2,x1}
\fmfdot{v2,v3,v4}
\fmfv{decor.shape=cross,decor.filled=shaded,decor.size=3thick}{u1}
\fmfv{decor.shape=cross,decor.filled=shaded,decor.size=3thick}{v1}
\fmfv{decor.shape=cross,decor.filled=shaded,decor.size=3thick}{w1}
\fmfv{decor.shape=cross,decor.filled=shaded,decor.size=3thick}{x1}
\end{fmfgraph*}\end{center}}
\hspace*{1mm}
\begin{tabular}{@{}c}
$1/16$ \\ ${\scs ( 1,0, 0 , 0 ; 8 )}$
\end{tabular}
\parbox{11mm}{\begin{center}
\begin{fmfgraph*}(8,8)
\setval
\fmfforce{0w,0h}{v1}
\fmfforce{1w,0h}{w1}
\fmfforce{0w,1h}{u1}
\fmfforce{1w,1h}{x1}
\fmfforce{1/2w,0h}{v2}
\fmfforce{1/2w,1/2h}{v3}
\fmfforce{1/2w,1h}{v4}
\fmfforce{0w,1/2h}{v5}
\fmf{plain,right=1}{v3,v5,v3}
\fmf{plain}{w1,v1}
\fmf{plain}{x1,u1}
\fmf{plain}{v2,v4}
\fmfdot{v2,v3,v4}
\fmfv{decor.shape=cross,decor.filled=shaded,decor.size=3thick}{u1}
\fmfv{decor.shape=cross,decor.filled=shaded,decor.size=3thick}{v1}
\fmfv{decor.shape=cross,decor.filled=shaded,decor.size=3thick}{w1}
\fmfv{decor.shape=cross,decor.filled=shaded,decor.size=3thick}{x1}
\end{fmfgraph*}\end{center}}
\hspace*{1mm}
\begin{tabular}{@{}c}
$1/12$ \\ ${\scs ( 1,0, 0 , 0 ; 6 )}$
\end{tabular}
\parbox{15mm}{\begin{center}
\begin{fmfgraph*}(12,12)
\setval
\fmfforce{0w,1/3h}{v1}
\fmfforce{1w,1/3h}{w1}
\fmfforce{1/3w,0h}{u1}
\fmfforce{1/3w,2/3h}{x1}
\fmfforce{1/3w,1/3h}{v2}
\fmfforce{2/3w,1/3h}{v3}
\fmfforce{2/3w,2/3h}{v4}
\fmfforce{2/3w,1h}{v5}
\fmf{plain,right=1}{v4,v5,v4}
\fmf{plain}{v3,v4}
\fmf{plain}{w1,v1}
\fmf{plain}{x1,u1}
\fmfdot{v2,v3,v4}
\fmfv{decor.shape=cross,decor.filled=shaded,decor.size=3thick}{u1}
\fmfv{decor.shape=cross,decor.filled=shaded,decor.size=3thick}{v1}
\fmfv{decor.shape=cross,decor.filled=shaded,decor.size=3thick}{w1}
\fmfv{decor.shape=cross,decor.filled=shaded,decor.size=3thick}{x1}
\end{fmfgraph*}\end{center}}
\hspace*{1mm}
\begin{tabular}{@{}c}
$1/8$ \\ ${\scs ( 1,0, 0 , 0 ; 4 )}$
\end{tabular}
\parbox{11mm}{\begin{center}
\begin{fmfgraph*}(8,12)
\setval
\fmfforce{0w,0h}{v1}
\fmfforce{1w,0h}{w1}
\fmfforce{0w,1/3h}{u1}
\fmfforce{1w,1/3h}{x1}
\fmfforce{1/2w,0h}{v2}
\fmfforce{1/2w,1/3h}{v3}
\fmfforce{1/2w,2/3h}{v4}
\fmfforce{1/2w,1h}{v5}
\fmf{plain,right=1}{v4,v5,v4}
\fmf{plain}{v2,v4}
\fmf{plain}{w1,v1}
\fmf{plain}{x1,u1}
\fmfdot{v2,v3,v4}
\fmfv{decor.shape=cross,decor.filled=shaded,decor.size=3thick}{u1}
\fmfv{decor.shape=cross,decor.filled=shaded,decor.size=3thick}{v1}
\fmfv{decor.shape=cross,decor.filled=shaded,decor.size=3thick}{w1}
\fmfv{decor.shape=cross,decor.filled=shaded,decor.size=3thick}{x1}
\end{fmfgraph*}\end{center}}
\\ $4$ &$1$ & $0$ & 
\begin{tabular}{@{}c}
$1/8$ \\ ${\scs ( 0,0, 0 , 0 ; 8 )}$
\end{tabular}
\parbox{15mm}{\begin{center}
\begin{fmfgraph*}(12,12)
\setval
\fmfforce{0w,1/2h}{v1}
\fmfforce{1w,1/2h}{w1}
\fmfforce{1/2w,0h}{u1}
\fmfforce{1/2w,1h}{x1}
\fmfforce{1/3w,1/2h}{v2}
\fmfforce{1/2w,1/3h}{v3}
\fmfforce{2/3w,1/2h}{v4}
\fmfforce{1/2w,2/3h}{v5}
\fmf{plain,right=1}{v4,v2,v4}
\fmf{plain}{v1,v2}
\fmf{plain}{u1,v3}
\fmf{plain}{w1,v4}
\fmf{plain}{x1,v5}
\fmfdot{v2,v3,v4,v5}
\fmfv{decor.shape=cross,decor.filled=shaded,decor.size=3thick}{u1}
\fmfv{decor.shape=cross,decor.filled=shaded,decor.size=3thick}{v1}
\fmfv{decor.shape=cross,decor.filled=shaded,decor.size=3thick}{w1}
\fmfv{decor.shape=cross,decor.filled=shaded,decor.size=3thick}{x1}
\end{fmfgraph*}\end{center}}
\hspace*{1mm}
\begin{tabular}{@{}c}
$1/16$ \\ ${\scs ( 0,1, 0 , 0 ; 8 )}$
\end{tabular}
\parbox{11mm}{\begin{center}
\begin{fmfgraph*}(8,12)
\setval
\fmfforce{0w,0h}{v1}
\fmfforce{1w,0h}{w1}
\fmfforce{0w,1h}{u1}
\fmfforce{1w,1h}{x1}
\fmfforce{1/2w,0h}{v2}
\fmfforce{1/2w,1/3h}{v3}
\fmfforce{1/2w,2/3h}{v4}
\fmfforce{1/2w,1h}{v5}
\fmf{plain,right=1}{v4,v3,v4}
\fmf{plain}{v2,v3}
\fmf{plain}{u1,x1}
\fmf{plain}{v1,w1}
\fmf{plain}{v4,v5}
\fmfdot{v2,v3,v4,v5}
\fmfv{decor.shape=cross,decor.filled=shaded,decor.size=3thick}{u1}
\fmfv{decor.shape=cross,decor.filled=shaded,decor.size=3thick}{v1}
\fmfv{decor.shape=cross,decor.filled=shaded,decor.size=3thick}{w1}
\fmfv{decor.shape=cross,decor.filled=shaded,decor.size=3thick}{x1}
\end{fmfgraph*}\end{center}}
\hspace*{1mm}
\begin{tabular}{@{}c}
$1/4$ \\ ${\scs ( 0,0, 0 , 0 , 4 )}$
\end{tabular}
\parbox{14.5mm}{\begin{center}
\begin{fmfgraph*}(11.464,7)
\setval
\fmfforce{0w,0h}{v1}
\fmfforce{4/11.464w,0h}{v2}
\fmfforce{7.464/11.464w,0h}{v3}
\fmfforce{1/2w,3/7h}{v4}
\fmfforce{1/2w,1h}{v5}
\fmfforce{1/2w,-1/7h}{v7}
\fmfforce{1w,0h}{w1}
\fmfforce{1.732/11.464w,1h}{u1}
\fmfforce{9.732/11.464w,1h}{x1}
\fmf{plain,right=1}{v4,v7,v4}
\fmf{plain}{u1,x1}
\fmf{plain}{v1,v2}
\fmf{plain}{v3,w1}
\fmf{plain}{v4,v5}
\fmfdot{v2,v3,v4,v5}
\fmfv{decor.shape=cross,decor.filled=shaded,decor.size=3thick}{u1}
\fmfv{decor.shape=cross,decor.filled=shaded,decor.size=3thick}{v1}
\fmfv{decor.shape=cross,decor.filled=shaded,decor.size=3thick}{w1}
\fmfv{decor.shape=cross,decor.filled=shaded,decor.size=3thick}{x1}
\end{fmfgraph*}\end{center}}
\\ && &
\begin{tabular}{@{}c}
$1/4$ \\ ${\scs ( 0,1, 0 , 0 , 2 )}$
\end{tabular}
\parbox{11mm}{\begin{center}
\begin{fmfgraph*}(8,16)
\setval
\fmfforce{0w,0h}{v1}
\fmfforce{1w,0h}{w1}
\fmfforce{0w,1/4h}{u1}
\fmfforce{1/2w,1h}{x1}
\fmfforce{1/2w,0h}{v2}
\fmfforce{1/2w,1/4h}{v3}
\fmfforce{1/2w,1/2h}{v4}
\fmfforce{1/2w,3/4h}{v5}
\fmf{plain,right=1}{v4,v5,v4}
\fmf{plain}{v1,w1}
\fmf{plain}{v2,v4}
\fmf{plain}{x1,v5}
\fmf{plain}{u1,v3}
\fmfdot{v2,v3,v4,v5}
\fmfv{decor.shape=cross,decor.filled=shaded,decor.size=3thick}{u1}
\fmfv{decor.shape=cross,decor.filled=shaded,decor.size=3thick}{v1}
\fmfv{decor.shape=cross,decor.filled=shaded,decor.size=3thick}{w1}
\fmfv{decor.shape=cross,decor.filled=shaded,decor.size=3thick}{x1}
\end{fmfgraph*}\end{center}}
\hspace*{1mm}
\begin{tabular}{@{}c}
$1/4$ \\ ${\scs ( 1,0, 0 , 0 , 2 )}$
\end{tabular}
\parbox{11mm}{\begin{center}
\begin{fmfgraph*}(8,16)
\setval
\fmfforce{0w,0h}{v1}
\fmfforce{1w,0h}{w1}
\fmfforce{0w,1/4h}{u1}
\fmfforce{1w,1/2h}{x1}
\fmfforce{1/2w,0h}{v2}
\fmfforce{1/2w,1/4h}{v3}
\fmfforce{1/2w,1/2h}{v4}
\fmfforce{1/2w,3/4h}{v5}
\fmfforce{1/2w,1h}{v6}
\fmf{plain,right=1}{v5,v6,v5}
\fmf{plain}{v1,w1}
\fmf{plain}{v2,v5}
\fmf{plain}{x1,v4}
\fmf{plain}{u1,v3}
\fmfdot{v2,v3,v4,v5}
\fmfv{decor.shape=cross,decor.filled=shaded,decor.size=3thick}{u1}
\fmfv{decor.shape=cross,decor.filled=shaded,decor.size=3thick}{v1}
\fmfv{decor.shape=cross,decor.filled=shaded,decor.size=3thick}{w1}
\fmfv{decor.shape=cross,decor.filled=shaded,decor.size=3thick}{x1}
\end{fmfgraph*}\end{center}}
\hspace*{1mm}
\begin{tabular}{@{}c}
$1/16$ \\ ${\scs ( 1,0, 0 , 0 ; 8 )}$
\end{tabular}
\parbox{15mm}{\begin{center}
\begin{fmfgraph*}(12,8)
\setval
\fmfforce{0w,0h}{v1}
\fmfforce{2/3w,0h}{w1}
\fmfforce{0w,1h}{u1}
\fmfforce{2/3w,1h}{x1}
\fmfforce{1/3w,0h}{v2}
\fmfforce{1/3w,1/2h}{v3}
\fmfforce{1/3w,1h}{v4}
\fmfforce{2/3w,1/2h}{v5}
\fmfforce{1w,1/2h}{v6}
\fmf{plain,right=1}{v5,v6,v5}
\fmf{plain}{v1,w1}
\fmf{plain}{v2,v4}
\fmf{plain}{v3,v5}
\fmf{plain}{x1,u1}
\fmfdot{v2,v3,v4,v5}
\fmfv{decor.shape=cross,decor.filled=shaded,decor.size=3thick}{u1}
\fmfv{decor.shape=cross,decor.filled=shaded,decor.size=3thick}{v1}
\fmfv{decor.shape=cross,decor.filled=shaded,decor.size=3thick}{w1}
\fmfv{decor.shape=cross,decor.filled=shaded,decor.size=3thick}{x1}
\end{fmfgraph*}\end{center}}
\end{tabular}
\end{center}
\caption{Connected vacuum diagrams and their weights for the 
$\phi^3$-$\phi^4$-theory with $n=1,2,3,4$ currents for the respective
first two loop orders. Withing each loop order $l$ the diagrams are
distinguished with respect to the number $p$ of $4$-vertices.
Each diagram is characterized by the
vector $(S,D,T,F;N$) whose components specify the number of self, double,
triple, fourfold connections, and of the identical vertex permutations,
respectively.}
\end{table}
\end{fmffile}

\begin{fmffile}{fg14}
\begin{table}[t]
\begin{center}
\begin{tabular}{|cc|c|}
\,\,\,$l$\,\,\,
& \,\,\,$p$\,\,\, &
$\Gamma^{(0,l,p)}$
\\
\hline
$2$ & $0$ &
\hspace{-10pt}
\rule[-10pt]{0pt}{26pt}
\begin{tabular}{@{}c}
$\mbox{}$ \\
$1/12$ \\ ${\scs ( 0, 0, 1 , 0 ; 2 )}$\\
$\mbox{}$ 
\end{tabular}
\parbox{7mm}{\begin{center}
\begin{fmfgraph}(4,4)
\setval
\fmfforce{0w,0.5h}{v1}
\fmfforce{1w,0.5h}{v2}
\fmf{plain,left=1}{v1,v2,v1}
\fmf{plain}{v1,v2}
\fmfdot{v1,v2}
\end{fmfgraph}\end{center}}
\\ 
$2$ & $1$ &
\hspace{-10pt}
\rule[-10pt]{0pt}{26pt}
\begin{tabular}{@{}c}
$\mbox{}$ \\
$1/8$ \\ ${\scs ( 2, 1, 0 , 0 ; 1 )}$\\
$\mbox{}$ 
\end{tabular}
\parbox{11mm}{\begin{center}
\begin{fmfgraph}(8,4)
\setval
\fmfleft{i1}
\fmfright{o1}
\fmf{plain,left=1}{i1,v1,i1}
\fmf{plain,left=1}{o1,v1,o1}
\fmfdot{v1}
\end{fmfgraph}\end{center}}
\\
\hline
\hspace{-10pt}
\rule[-10pt]{0pt}{26pt}
$3$ & $0$  & 
\begin{tabular}{@{}c}
$\mbox{}$\\
$1/24$ \\ ${\scs ( 0, 0, 0 , 0 ; 24 )}$\\
$\mbox{}$
\end{tabular}
\parbox{9mm}{\begin{center}
\begin{fmfgraph}(4,4)
\setval
\fmfforce{0w,0h}{v1}
\fmfforce{1w,0h}{v2}
\fmfforce{1w,1h}{v3}
\fmfforce{0w,1h}{v4}
\fmf{plain,right=1}{v1,v3,v1}
\fmf{plain}{v1,v3}
\fmf{plain}{v2,v4}
\fmfdot{v1,v2,v3,v4}
\end{fmfgraph}\end{center}} 
\hspace*{1mm}
\begin{tabular}{@{}c}
$1/16$ \\ ${\scs ( 0, 2, 0 , 0 ; 4 )}$
\end{tabular}
\parbox{9mm}{\begin{center}
\begin{fmfgraph}(4,4)
\setval
\fmfforce{0w,0h}{v1}
\fmfforce{1w,0h}{v2}
\fmfforce{1w,1h}{v3}
\fmfforce{0w,1h}{v4}
\fmf{plain,right=1}{v1,v3,v1}
\fmf{plain,right=0.4}{v1,v4}
\fmf{plain,left=0.4}{v2,v3}
\fmfdot{v1,v2,v3,v4}
\end{fmfgraph}\end{center}} 
\\ 
$3$ & $1$ & 
\begin{tabular}{@{}c}
$\mbox{}$ \\
$1/8$ \\ ${\scs ( 0, 2, 0 , 0 ; 2 )}$\\
$\mbox{}$
\end{tabular}
\parbox{9mm}{\begin{center}
\begin{fmfgraph}(6,6)
\setval
\fmfforce{0w,1/2h}{v1}
\fmfforce{1w,1/2h}{v2}
\fmfforce{1/2w,1h}{v3}
\fmf{plain,left=1}{v1,v2,v1}
\fmf{plain,left=0.4}{v3,v1}
\fmf{plain,left=0.4}{v2,v3}
\fmfdot{v1,v2,v3}
\end{fmfgraph}\end{center}}
\hspace*{1mm}
\begin{tabular}{@{}c}
$1/8$ \\ ${\scs ( 1, 1, 0 , 0 ; 2 )}$
\end{tabular}
\parbox{11mm}{\begin{center}
\begin{fmfgraph}(4,8)
\setval
\fmfforce{0w,1/4h}{v1}
\fmfforce{1w,1/4h}{v2}
\fmfforce{1/2w,1/2h}{v3}
\fmfforce{1/2w,1h}{v4}
\fmf{plain,left=1}{v1,v2,v1}
\fmf{plain,left=1}{v3,v4,v3}
\fmf{plain}{v1,v2}
\fmfdot{v2,v3,v1}
\end{fmfgraph}\end{center}} 
\\
$3$ &  $2$ &
\begin{tabular}{@{}c}
$\mbox{}$\\
$1/48$ \\ ${\scs ( 0, 0, 0 , 1 ; 2 )}$\\
$\mbox{}$
\end{tabular}
\parbox{9mm}{\begin{center}
\begin{fmfgraph}(6,4)
\setval
\fmfforce{0w,0.5h}{v1}
\fmfforce{1w,0.5h}{v2}
\fmf{plain,left=1}{v1,v2,v1}
\fmf{plain,left=0.4}{v1,v2,v1}
\fmfdot{v1,v2}
\end{fmfgraph}\end{center}} 
\hspace*{1mm}
\begin{tabular}{@{}c}
$1/16$ \\ ${\scs ( 2, 1, 0 , 0 ; 2 )}$
\end{tabular}
\parbox{15mm}{\begin{center}
\begin{fmfgraph}(12,4)
\setval
\fmfleft{i1}
\fmfright{o1}
\fmf{plain,left=1}{i1,v1,i1}
\fmf{plain,left=1}{v1,v2,v1}
\fmf{plain,left=1}{o1,v2,o1}
\fmfdot{v1,v2}
\end{fmfgraph}\end{center}}
\\ \hline 
\hspace{-10pt}
\rule[-10pt]{0pt}{26pt}
$4$ & $0$ &
\begin{tabular}{@{}c}
$\mbox{}$\\
$1/72$ \\ ${\scs ( 0, 0, 0 , 0 ; 72 )}$\\
$\mbox{}$
\end{tabular}
\parbox{9mm}{\begin{center}
\begin{fmfgraph}(6,6)
\setval
\fmfforce{0w,0.5h}{v1}
\fmfforce{0.25w,0.933h}{v2}
\fmfforce{0.75w,0.933h}{v3}
\fmfforce{1w,0.5h}{v4}
\fmfforce{0.75w,0.067h}{v5}
\fmfforce{0.25w,0.067h}{v6}
\fmf{plain,right=1}{v1,v4,v1}
\fmf{plain}{v1,v4}
\fmf{plain}{v2,v5}
\fmf{plain}{v3,v6}
\fmfdot{v1,v2,v3,v4,v5,v6}
\end{fmfgraph}
\end{center}}
\hspace*{2mm}
\begin{tabular}{@{}c}
$1/12$ \\ ${\scs ( 0, 0, 0 , 0 ; 12 )}$
\end{tabular}
\parbox{9mm}{\begin{center}
\begin{fmfgraph}(6,6)
\setval
\fmfforce{0w,0.5h}{v1}
\fmfforce{0.25w,0.933h}{v2}
\fmfforce{0.75w,0.933h}{v3}
\fmfforce{1w,0.5h}{v4}
\fmfforce{0.75w,0.067h}{v5}
\fmfforce{0.25w,0.067h}{v6}
\fmf{plain,right=1}{v1,v4,v1}
\fmf{plain}{v1,v4}
\fmf{plain}{v2,v6}
\fmf{plain}{v3,v5}
\fmfdot{v1,v2,v3,v4,v5,v6}
\end{fmfgraph}
\end{center}}
\hspace*{2mm}
\begin{tabular}{@{}c}
$1/48$ \\ ${\scs ( 0, 3, 0 , 0 ; 6 )}$
\end{tabular}
\parbox{9mm}{\begin{center}
\begin{fmfgraph}(6,6)
\setval
\fmfforce{0w,0.5h}{v1}
\fmfforce{0.25w,0.933h}{v2}
\fmfforce{0.75w,0.933h}{v3}
\fmfforce{1w,0.5h}{v4}
\fmfforce{0.75w,0.067h}{v5}
\fmfforce{0.25w,0.067h}{v6}
\fmf{plain,right=1}{v1,v4,v1}
\fmf{plain,right=0.7}{v2,v3}
\fmf{plain,right=0.7}{v4,v5}
\fmf{plain,right=0.7}{v6,v1}
\fmfdot{v1,v2,v3,v4,v5,v6}
\end{fmfgraph}
\end{center}} 
\hspace*{2mm}
\begin{tabular}{@{}c}
$1/16$ \\ ${\scs ( 0, 2, 0 , 0 ; 4 )}$
\end{tabular}
\parbox{9mm}{\begin{center}
\begin{fmfgraph}(6,6)
\setval
\fmfforce{0w,0.5h}{v1}
\fmfforce{0.25w,0.933h}{v2}
\fmfforce{0.75w,0.933h}{v3}
\fmfforce{1w,0.5h}{v4}
\fmfforce{0.75w,0.067h}{v5}
\fmfforce{0.25w,0.067h}{v6}
\fmf{plain,right=1}{v1,v4,v1}
\fmf{plain,right=0.7}{v2,v3}
\fmf{plain}{v1,v4}
\fmf{plain,right=0.7}{v5,v6}
\fmfdot{v1,v2,v3,v4,v5,v6}
\end{fmfgraph}
\end{center}}
\hspace*{2mm}
\begin{tabular}{@{}c}
$1/8$ \\ ${\scs ( 0, 1, 0 , 0 ; 4 )}$
\end{tabular}
\parbox{9mm}{\begin{center}
\begin{fmfgraph}(6,6)
\setval
\fmfforce{0w,0.5h}{v1}
\fmfforce{0.25w,0.933h}{v2}
\fmfforce{0.75w,0.933h}{v3}
\fmfforce{1w,0.5h}{v4}
\fmfforce{0.75w,0.067h}{v5}
\fmfforce{0.25w,0.067h}{v6}
\fmf{plain,right=1}{v1,v4,v1}
\fmf{plain,right=0.7}{v2,v3}
\fmf{plain,right=0.2}{v4,v6}
\fmf{plain,right=0.2}{v5,v1}
\fmfdot{v1,v2,v3,v4,v5,v6}
\end{fmfgraph}
\end{center}}
\\ $4$ & $1$ &
\begin{tabular}{@{}c}
$\mbox{}$ \\
$1/8$ \\ ${\scs ( 0, 0, 0 , 0 ; 8 )}$\\
$\mbox{}$
\end{tabular}
\parbox{9mm}{\begin{center}
\begin{fmfgraph}(4,4)
\setval
\fmfforce{0w,0h}{v1}
\fmfforce{1w,0h}{v2}
\fmfforce{1w,1h}{v3}
\fmfforce{0w,1h}{v4}
\fmfforce{1/2w,1/2h}{v5}
\fmf{plain,right=1}{v1,v3,v1}
\fmf{plain}{v1,v3}
\fmf{plain}{v2,v4}
\fmfdot{v1,v2,v3,v4,v5}
\end{fmfgraph}\end{center}} 
\hspace*{1mm}
\begin{tabular}{@{}c}
$1/4$ \\ ${\scs ( 0, 2, 0 , 0 ; 2 )}$
\end{tabular}
\parbox{13mm}{\begin{center}
\begin{fmfgraph}(8,8)
\setval
\fmfforce{1/2w,1h}{v1}
\fmfforce{1/2w,0h}{v2}
\fmfforce{0.85w,0.85h}{v3}
\fmfforce{0.15w,0.85h}{v4}
\fmfforce{1/2w,0.7h}{v5}
\fmfforce{1/2w,0.3h}{v6}
\fmf{plain,left=1}{v1,v2,v1}
\fmf{plain,left=1}{v5,v6,v5}
\fmf{plain}{v6,v2}
\fmf{plain,right=0.4}{v5,v4}
\fmf{plain,right=0.4}{v3,v5}
\fmfdot{v2,v3,v4,v5,v6}
\end{fmfgraph}\end{center}} 
\hspace*{1mm}
\begin{tabular}{@{}c}
$1/16$ \\ ${\scs ( 0, 3, 0 , 0 ; 2 )}$
\end{tabular}
\parbox{9mm}{\begin{center}
\begin{fmfgraph}(6,6)
\setval
\fmfforce{1/2w,1h}{v1}
\fmfforce{1/2w,0h}{v2}
\fmfforce{0.25w,0.933h}{v3}
\fmfforce{0.75w,0.933h}{v4}
\fmfforce{0w,0.5h}{v5}
\fmfforce{1w,0.5h}{v6}
\fmf{plain,right=1}{v1,v2,v1}
\fmf{plain,right=0.7}{v3,v4}
\fmf{plain,right=0.4}{v2,v5}
\fmf{plain,right=0.4}{v6,v2}
\fmfdot{v2,v3,v4,v5,v6}
\end{fmfgraph}
\end{center}}
\hspace*{1mm}
\begin{tabular}{@{}c}
$1/32$ \\ ${\scs ( 0, 2, 0 , 0 ; 8 )}$
\end{tabular}
\parbox{9mm}{\begin{center}
\begin{fmfgraph}(4,8)
\setval
\fmfforce{1/2w,1h}{v1}
\fmfforce{1/2w,1/2h}{v2}
\fmfforce{1/2w,0h}{v3}
\fmfforce{1w,3/4h}{v4}
\fmfforce{0w,3/4h}{v5}
\fmfforce{1w,1/4h}{v6}
\fmfforce{0w,1/4h}{v7}
\fmf{plain,right=1}{v1,v2,v1}
\fmf{plain,right=1}{v3,v2,v3}
\fmf{plain}{v4,v5}
\fmf{plain}{v6,v7}
\fmfdot{v2,v4,v5,v6,v7}
\end{fmfgraph}
\end{center}}
\hspace*{1mm}
\begin{tabular}{@{}c}
$1/4$ \\ ${\scs ( 0, 2, 0 , 0 ; 1 )}$
\end{tabular}
\parbox{9mm}{\begin{center}
\begin{fmfgraph}(6,6)
\setval
\fmfforce{0w,1/2h}{v1}
\fmfforce{1/2w,1h}{v2}
\fmfforce{1/2w,0h}{v3}
\fmfforce{0.9w,0.75h}{v4}
\fmfforce{0.9w,0.25h}{v5}
\fmf{plain,right=1}{v2,v3,v2}
\fmf{plain}{v2,v3}
\fmf{plain,right=0.4}{v1,v2}
\fmf{plain,right=0.4}{v4,v5}
\fmfdot{v1,v3,v2,v4,v5}
\end{fmfgraph}
\end{center}}
\hspace*{1mm}
\begin{tabular}{@{}c}
$1/16$\\ 
${\scs ( 1, 2, 0 , 0 ; 2 )}$
\end{tabular}
\parbox{9mm}{\begin{center}
\begin{fmfgraph}(6,10)
\setval
\fmfforce{0w,0.3h}{v1}
\fmfforce{1w,0.3h}{v2}
\fmfforce{0.5w,0.6h}{v3}
\fmfforce{0.5w,1h}{v4}
\fmfforce{0.15w,0.5h}{v5}
\fmfforce{0.15w,0.1h}{v6}
\fmfforce{0.85w,0.5h}{v7}
\fmfforce{0.85w,0.1h}{v8}
\fmf{plain,left=1}{v1,v2,v1}
\fmf{plain,left=0.5}{v5,v6}
\fmf{plain,left=0.5}{v8,v7}
\fmf{plain,left=1}{v3,v4,v3}
\fmfdot{v3,v5,v6,v7,v8}
\end{fmfgraph}\end{center}} 
\\ & &
\begin{tabular}{@{}c}
$\mbox{}$\\
$1/8$\\ 
${\scs ( 1, 1, 0 , 0 ; 2 )}$\\
$\mbox{}$
\end{tabular}
\parbox{9mm}{\begin{center}
\begin{fmfgraph}(6,10)
\setval
\fmfforce{0w,0.3h}{v1}
\fmfforce{1w,0.3h}{v2}
\fmfforce{0.5w,0.6h}{v3}
\fmfforce{0.5w,1h}{v4}
\fmfforce{0.15w,0.5h}{v5}
\fmfforce{0.15w,0.1h}{v6}
\fmfforce{0.85w,0.5h}{v7}
\fmfforce{0.85w,0.1h}{v8}
\fmf{plain,left=1}{v1,v2,v1}
\fmf{plain,right=0.5}{v5,v7}
\fmf{plain,right=0.5}{v8,v6}
\fmf{plain,left=1}{v3,v4,v3}
\fmfdot{v3,v5,v6,v7,v8}
\end{fmfgraph}\end{center}} 
\hspace*{1mm}
\begin{tabular}{@{}c}
$1/8$\\ 
${\scs ( 1, 0, 0 , 0 ; 4 )}$
\end{tabular}
\parbox{9mm}{\begin{center}
\begin{fmfgraph}(6,10)
\setval
\fmfforce{0w,0.3h}{v1}
\fmfforce{1w,0.3h}{v2}
\fmfforce{0.5w,0.6h}{v3}
\fmfforce{0.5w,1h}{v4}
\fmfforce{0.15w,0.5h}{v5}
\fmfforce{0.15w,0.1h}{v6}
\fmfforce{0.85w,0.5h}{v7}
\fmfforce{0.85w,0.1h}{v8}
\fmf{plain,left=1}{v1,v2,v1}
\fmf{plain}{v5,v8}
\fmf{plain}{v7,v6}
\fmf{plain,left=1}{v3,v4,v3}
\fmfdot{v3,v5,v6,v7,v8}
\end{fmfgraph}\end{center}} 
\\ $4$ & $2$ &
\hspace*{1mm}
\begin{tabular}{@{}c}
$\mbox{}$ \\
$1/8$ \\ ${\scs ( 0, 1, 0 , 0 ; 4 )}$\\
$\mbox{}$
\end{tabular}
\parbox{9mm}{\begin{center}
\begin{fmfgraph}(6,6)
\setval
\fmfforce{0w,0.5h}{v1}
\fmfforce{1w,0.5h}{v2}
\fmfforce{1/2w,0.7h}{v3}
\fmfforce{1/2w,0.3h}{v4}
\fmf{plain,left=1}{v1,v2,v1}
\fmf{plain,left=0.4}{v1,v2,v1}
\fmf{plain}{v4,v3}
\fmfdot{v1,v2,v3,v4}
\end{fmfgraph}\end{center}} 
\hspace*{1mm}
\begin{tabular}{@{}c}
$1/16$ \\ ${\scs ( 0, 3, 0 , 0 ; 2 )}$
\end{tabular}
\parbox{9mm}{\begin{center}
\begin{fmfgraph}(4.5,4.5)
\setval
\fmfforce{0w,0h}{v1}
\fmfforce{1w,0h}{v2}
\fmfforce{1w,1h}{v3}
\fmfforce{0w,1h}{v4}
\fmf{plain,left=1}{v1,v3,v1}
\fmf{plain,right=0.4}{v3,v2}
\fmf{plain,right=0.4}{v4,v3}
\fmf{plain,right=0.4}{v1,v4}
\fmfdot{v1,v2,v3,v4}
\end{fmfgraph}\end{center}} 
\hspace*{1mm}
\begin{tabular}{@{}c}
$1/8$ \\ ${\scs ( 0, 2, 0 , 0 ; 2 )}$
\end{tabular}
\parbox{9mm}{\begin{center}
\begin{fmfgraph}(4.5,4.5)
\setval
\fmfforce{0w,0h}{v1}
\fmfforce{1w,0h}{v2}
\fmfforce{1w,1h}{v3}
\fmfforce{0w,1h}{v4}
\fmf{plain,left=1}{v1,v3,v1}
\fmf{plain,right=0.4}{v3,v2}
\fmf{plain}{v1,v3}
\fmf{plain,right=0.4}{v1,v4}
\fmfdot{v1,v2,v3,v4}
\end{fmfgraph}\end{center}} 
\hspace*{1mm}
\begin{tabular}{@{}c}
$1/24$ \\ ${\scs ( 0, 1, 1 , 0 ; 2 )}$
\end{tabular}
\parbox{9mm}{\begin{center}
\begin{fmfgraph}(4,10)
\setval
\fmfforce{0w,1/5h}{v1}
\fmfforce{1w,1/5h}{v2}
\fmfforce{0w,4/5h}{v3}
\fmfforce{1w,4/5h}{v4}
\fmf{plain,left=1}{v4,v3,v4}
\fmf{plain,left=1}{v1,v2,v1}
\fmf{plain}{v4,v3}
\fmf{plain,right=0.3}{v3,v1}
\fmf{plain,right=0.3}{v2,v4}
\fmfdot{v1,v2,v3,v4}
\end{fmfgraph}\end{center}} 
\hspace*{1mm}
\begin{tabular}{@{}c}
$1/4$ \\ ${\scs ( 1, 1, 0 , 0 ; 1 )}$
\end{tabular}
\parbox{9mm}{\begin{center}
\begin{fmfgraph}(6,10)
\setval
\fmfforce{0w,3/10h}{v1}
\fmfforce{1/2w,6/10h}{v2}
\fmfforce{1w,3/10h}{v4}
\fmfforce{1/2w,0h}{v5}
\fmfforce{1/2w,1h}{v6}
\fmf{plain,right=1}{v2,v5,v2}
\fmf{plain,right=1}{v2,v6,v2}
\fmf{plain,right=0.4}{v4,v5,v4}
\fmf{plain}{v1,v4}
\fmfdot{v1,v2,v4,v5}
\end{fmfgraph}
\end{center}}
\hspace*{1mm}
\begin{tabular}{@{}c}
$1/16$\\ 
${\scs ( 1, 2, 0 , 0 ; 2 )}$
\end{tabular}
\parbox{9mm}{\begin{center}
\begin{fmfgraph}(6,10)
\setval
\fmfforce{0w,0.3h}{v1}
\fmfforce{1w,0.3h}{v2}
\fmfforce{0.5w,0.6h}{v3}
\fmfforce{0.5w,1h}{v4}
\fmfforce{0.5w,0h}{v5}
\fmf{plain,left=1}{v1,v2,v1}
\fmf{plain,left=0.4}{v1,v5}
\fmf{plain,left=0.4}{v5,v2}
\fmf{plain,left=1}{v3,v4,v3}
\fmfdot{v1,v2,v3,v5}
\end{fmfgraph}\end{center}} 
\hspace*{1mm}
\\ &&
\begin{tabular}{@{}c}
$\mbox{}$\\
$1/16$ \\ ${\scs ( 1, 2, 0 , 0 ; 2 )}$\\
$\mbox{}$
\end{tabular}
\parbox{7mm}{\begin{center}
\begin{fmfgraph}(4,12)
\setval
\fmfforce{0w,1/6h}{v1}
\fmfforce{1w,1/6h}{v2}
\fmfforce{1/2w,1/3h}{v3}
\fmfforce{1/2w,2/3h}{v4}
\fmfforce{1/2w,1h}{v5}
\fmf{plain,left=1}{v1,v2,v1}
\fmf{plain,left=1}{v3,v4,v3}
\fmf{plain,left=1}{v4,v5,v4}
\fmf{plain}{v1,v2}
\fmfdot{v1,v2,v3,v4}
\end{fmfgraph}\end{center}} 
\hspace*{1mm}
\begin{tabular}{@{}c}
$1/16$ \\ ${\scs ( 2, 0, 0 , 0 ; 4 )}$
\end{tabular}
\parbox{15mm}{\begin{center}
\begin{fmfgraph}(12,4)
\setval
\fmfforce{0w,1/2h}{v1}
\fmfforce{1/3w,1/2h}{v2}
\fmfforce{1/2w,0h}{v3}
\fmfforce{1/2w,1h}{v4}
\fmfforce{2/3w,1/2h}{v5}
\fmfforce{1w,1/2h}{v6}
\fmf{plain,left=1}{v1,v2,v1}
\fmf{plain,left=1}{v2,v5,v2}
\fmf{plain,left=1}{v5,v6,v5}
\fmf{plain}{v3,v4}
\fmfdot{v2,v3,v4,v5}
\end{fmfgraph}\end{center}} 
\hspace*{1mm}
\begin{tabular}{@{}c}
$1/16$ \\ ${\scs ( 2, 1, 0 , 0 ; 2 )}$
\end{tabular}
\parbox{15mm}{\begin{center}
\begin{fmfgraph}(6,9.75)
\setval
\fmfforce{0w,0.3h}{v1}
\fmfforce{1w,0.3h}{v2}
\fmfforce{0.2w,0.55h}{v3}
\fmfforce{0.8w,0.55h}{v4}
\fmf{plain,left=1}{v1,v2,v1}
\fmf{plain}{v1,v2}
\fmfi{plain}{reverse fullcircle scaled 0.7w shifted (1.02w,0.72h)}
\fmfi{plain}{reverse fullcircle scaled 0.7w shifted (-0.02w,0.72h)}
\fmfdot{v1,v2,v3,v4}
\end{fmfgraph}\end{center}}
\\
$4$& $3$ &
\begin{tabular}{@{}c}
$\mbox{}$ \\
$1/48$ \\ 
${\scs ( 0, 3, 0 , 0 ; 6 )}$ \\
$\mbox{}$
\end{tabular}
\parbox{9mm}{\begin{center}
\begin{fmfgraph}(6,6)
\setval
\fmfforce{0.5w,0h}{v1}
\fmfforce{0.5w,1h}{v2}
\fmfforce{0.066987w,0.25h}{v3}
\fmfforce{0.93301w,0.25h}{v4}
\fmf{plain,left=1}{v1,v2,v1}
\fmf{plain}{v2,v3}
\fmf{plain}{v3,v4}
\fmf{plain}{v2,v4}
\fmfdot{v2,v3,v4}
\end{fmfgraph}
\end{center}} 
\hspace*{1mm}
\begin{tabular}{@{}c}
$1/24$\\ 
${\scs ( 1, 0, 1 , 0 ; 2 )}$
\end{tabular}
\parbox{9mm}{\begin{center}
\begin{fmfgraph}(6,10)
\setval
\fmfforce{0w,0.3h}{v1}
\fmfforce{1w,0.3h}{v2}
\fmfforce{0.5w,0.6h}{v3}
\fmfforce{0.5w,1h}{v4}
\fmf{plain,left=1}{v1,v2,v1}
\fmf{plain,left=0.4}{v1,v2,v1}
\fmf{plain,left=1}{v3,v4,v3}
\fmfdot{v1,v2,v3}
\end{fmfgraph}\end{center}} 
\hspace*{1mm}
\begin{tabular}{@{}c}
$1/48$\\ 
${\scs ( 3, 0, 0 , 0 ; 6 )}$
\end{tabular}
\parbox{15mm}{\begin{center}
\begin{fmfgraph}(12,12)
\setval
\fmfforce{1/2w,1/3h}{v1}
\fmfforce{1/2w,2/3h}{v2}
\fmfforce{1/2w,1h}{v3}
\fmfforce{0.355662432w,0.416666666h}{v4}
\fmfforce{0.64433568w,0.416666666h}{v5}
\fmfforce{0.067w,1/4h}{v6}
\fmfforce{0.933w,1/4h}{v7}
\fmf{plain,left=1}{v1,v2,v1}
\fmf{plain,left=1}{v2,v3,v2}
\fmf{plain,left=1}{v4,v6,v4}
\fmf{plain,left=1}{v5,v7,v5}
\fmfdot{v2,v4,v5}
\end{fmfgraph}\end{center}} 
\hspace*{1mm}
\begin{tabular}{@{}c}
$1/32$\\ 
${\scs ( 2, 2, 0 , 0 ; 2 )}$
\end{tabular}
\parbox{19mm}{\begin{center}
\begin{fmfgraph}(16,4)
\setval
\fmfleft{i1}
\fmfright{o1}
\fmf{plain,left=1}{i1,v1,i1}
\fmf{plain,left=1}{v1,v2,v1}
\fmf{plain,left=1}{v2,v3,v2}
\fmf{plain,left=1}{o1,v3,o1}
\fmfdot{v1,v2,v3}
\end{fmfgraph}\end{center}}
\hspace*{1mm}
\end{tabular}
\end{center}
\caption{One-particle irreducible vacuum diagrams and their 
weights of the $\phi^3$-$\phi^4$-theory without field expectation values
up to four loops. Within each loop order $l$ the diagrams are 
distinguished with respect to the number $p$ of $4$-vertices.
Each diagram is characterized by the
vector $(S,D,T,F;N$) whose components specify the number of self-, double,
triple, fourfold connections, and of the identical vertex permutations,
respectively.}
\end{table}
\end{fmffile}
\newpage
\hspace*{-2cm}
\begin{fmffile}{fg15}

\begin{table}[t]
\begin{center}
\begin{tabular}{|ccc|c|}
\,\,\,$n$\,\,\,
&\,\,\,$l$\,\,\,
& \,\,\,$p$\,\,\, &
$\Gamma^{(n ,l,p)}$
\\
\hline
$1$ & $1$ & $0$ &
\hspace{-10pt}
\rule[-10pt]{0pt}{26pt}
\begin{tabular}{@{}c}
$1/2$ \\ ${\scs ( 1, 0, 0 , 0 , 1 )}$
\end{tabular}
\parbox{11mm}{\begin{center}
\begin{fmfgraph*}(8,4)
\setval
\fmfforce{0w,1/2h}{v1}
\fmfforce{1/2w,1/2h}{v2}
\fmfforce{1w,1/2h}{v3}
\fmf{plain,left=1}{v2,v3,v2}
\fmf{boson}{v1,v2}
\fmfdot{v1,v2}
\end{fmfgraph*}\end{center}}
\\ \hline 
$1$ &$2$ & $1$ &
\hspace{-10pt}
\rule[-10pt]{0pt}{26pt}
\begin{tabular}{@{}c}
$1/6$ \\ ${\scs ( 0, 0, 1 , 0 ; 1 )}$
\end{tabular}
\parbox{11mm}{\begin{center}
\begin{fmfgraph*}(8,4)
\setval
\fmfforce{0w,1/2h}{v1}
\fmfforce{1/2w,1/2h}{v2}
\fmfforce{1w,1/2h}{v3}
\fmf{plain,left=1}{v2,v3,v2}
\fmf{boson}{v1,v2}
\fmf{plain}{v2,v3}
\fmfdot{v1,v2,v3}
\end{fmfgraph*}\end{center}}
\hspace*{1mm}
\begin{tabular}{@{}c}
$1/4$ \\ ${\scs ( 1, 1, 0 , 0 ; 1 )}$
\end{tabular}
\parbox{15mm}{\begin{center}
\begin{fmfgraph*}(12,4)
\setval
\fmfforce{0w,1/2h}{v1}
\fmfforce{1/3w,1/2h}{v2}
\fmfforce{2/3w,1/2h}{v3}
\fmfforce{1w,1/2h}{v4}
\fmf{plain,left=1}{v2,v3,v2}
\fmf{plain,left=1}{v3,v4,v3}
\fmf{boson}{v1,v2}
\fmfdot{v1,v2,v3}
\end{fmfgraph*}\end{center}}
\\ $1$ & $2$ & $0$ &
\begin{tabular}{@{}c}
$1/4$ \\ ${\scs ( 0, 1, 0 , 0 ; 2 )}$
\end{tabular}
\parbox{11mm}{\begin{center}
\begin{fmfgraph*}(8,4)
\setval
\fmfforce{0w,1/2h}{v1}
\fmfforce{1/2w,1/2h}{v2}
\fmfforce{1w,1/2h}{v3}
\fmfforce{3/4w,0h}{v4}
\fmfforce{3/4w,1h}{v5}
\fmf{plain,left=1}{v2,v3,v2}
\fmf{boson}{v1,v2}
\fmf{plain}{v4,v5}
\fmfdot{v1,v2,v4,v5}
\end{fmfgraph*}\end{center}}\\ \hline\hline
$2$ & $1$ & $1$ &
\hspace{-10pt}
\rule[-10pt]{0pt}{26pt}
\begin{tabular}{@{}c}
$1/4$ \\ ${\scs ( 1, 0, 0 , 0 ; 2 )}$
\end{tabular}
\parbox{11mm}{\begin{center}
\begin{fmfgraph*}(8,4)
\setval
\fmfforce{0w,0h}{v1}
\fmfforce{1/2w,0h}{v2}
\fmfforce{1/2w,1h}{v3}
\fmfforce{1w,0h}{w1}
\fmf{plain,left=1}{v2,v3,v2}
\fmf{boson}{v1,w1}
\fmfdot{v1,w1,v2}
\end{fmfgraph*}\end{center}}
\\ $2$ &$1$ & $0$ &
\begin{tabular}{@{}c}
$1/4$ \\ ${\scs ( 0, 1, 0 , 0 ; 2 )}$
\end{tabular}
\parbox{15mm}{\begin{center}
\begin{fmfgraph*}(12,4)
\setval
\fmfforce{0w,1/2h}{v1}
\fmfforce{1/3w,1/2h}{v2}
\fmfforce{2/3w,1/2h}{v3}
\fmfforce{1w,1/2h}{w1}
\fmf{plain,left=1}{v2,v3,v2}
\fmf{boson}{v1,v2}
\fmf{boson}{v3,w1}
\fmfdot{v1,w1,v2,v3}
\end{fmfgraph*}\end{center}}
\\ \hline $2$ &$2$ & $2$ &
\hspace{-10pt}
\rule[-10pt]{0pt}{26pt}
\begin{tabular}{@{}c}
$1/12$ \\ ${\scs ( 0, 0, 1 , 0 ; 2 )}$
\end{tabular}
\parbox{15mm}{\begin{center}
\begin{fmfgraph*}(12,4)
\setval
\fmfforce{0w,1/2h}{v1}
\fmfforce{1/3w,1/2h}{v2}
\fmfforce{2/3w,1/2h}{v3}
\fmfforce{1w,1/2h}{w1}
\fmf{plain,left=1}{v2,v3,v2}
\fmf{plain}{v2,v3}
\fmf{boson}{v1,v2}
\fmf{boson}{w1,v3}
\fmfdot{v1,w1,v2,v3}
\end{fmfgraph*}\end{center}}
\hspace*{1mm}
\begin{tabular}{@{}c}
$1/8$ \\ ${\scs ( 1, 1, 0 , 0 ; 2 )}$
\end{tabular}
\parbox{11mm}{\begin{center}
\begin{fmfgraph*}(8,8)
\setval
\fmfforce{0w,0h}{v1}
\fmfforce{1/2w,0h}{v2}
\fmfforce{1/2w,1/2h}{v3}
\fmfforce{1/2w,1h}{v4}
\fmfforce{1w,0h}{w1}
\fmf{plain,left=1}{v2,v3,v2}
\fmf{plain,left=1}{v3,v4,v3}
\fmf{boson}{v1,w1}
\fmfdot{v1,w1,v2,v3}
\end{fmfgraph*}\end{center}}
\\ $2$ &$2$ & $1$ &
\hspace{-10pt}
\rule[-10pt]{0pt}{26pt}
\begin{tabular}{@{}c}
$1/8$ \\ ${\scs ( 0, 2, 0 , 0 ; 2 )}$
\end{tabular}
\parbox{19mm}{\begin{center}
\begin{fmfgraph*}(16,4)
\setval
\fmfforce{0w,1/2h}{v1}
\fmfforce{1/4w,1/2h}{v2}
\fmfforce{1/2w,1/2h}{v3}
\fmfforce{3/4w,1/2h}{v4}
\fmfforce{1w,1/2h}{w1}
\fmf{plain,left=1}{v2,v3,v2}
\fmf{plain,left=1}{v3,v4,v3}
\fmf{boson}{v1,v2}
\fmf{boson}{v4,w1}
\fmfdot{v1,w1,v2,v3,v4}
\end{fmfgraph*}\end{center}}
\hspace*{1mm}
\begin{tabular}{@{}c}
$1/8$ \\ ${\scs ( 0, 1, 0 , 0 ; 4 )}$
\end{tabular}
\parbox{15mm}{\begin{center}
\begin{fmfgraph*}(8,4)
\setval
\fmfforce{0w,0h}{v1}
\fmfforce{1/2w,0h}{v2}
\fmfforce{1w,0h}{w1}
\fmfforce{1/2w,1h}{v3}
\fmfforce{1/4w,1/2h}{v4}
\fmfforce{3/4w,1/2h}{v5}
\fmf{plain,left=1}{v2,v3,v2}
\fmf{boson}{v1,w1}
\fmf{plain}{v4,v5}
\fmfdot{v1,w1,v2,v4,v5}
\end{fmfgraph*}\end{center}}
\hspace*{1mm}
\begin{tabular}{@{}c}
$1/2$ \\ ${\scs ( 0, 1, 0 , 0 ; 1 )}$
\end{tabular}
\parbox{14.5mm}{\begin{center}
\begin{fmfgraph*}(11.464,3)
\setval
\fmfforce{0w,0h}{v1}
\fmfforce{4/11.464w,0h}{v2}
\fmfforce{7.464/11.464w,0h}{v3}
\fmfforce{1/2w,1h}{v4}
\fmfforce{1/2w,-1/3h}{v7}
\fmfforce{1w,0h}{w1}
\fmf{plain,right=1}{v4,v7,v4}
\fmf{boson}{v1,v2}
\fmf{plain}{v4,v2}
\fmf{boson}{v3,w1}
\fmfdot{v1,w1,v2,v3,v4}
\end{fmfgraph*}\end{center}}
\hspace*{1mm}
\begin{tabular}{@{}c}
$1/4$ \\ ${\scs ( 1, 0, 0 , 0 ; 2 )}$
\end{tabular}
\parbox{14.5mm}{\begin{center}
\begin{fmfgraph*}(11.464,7)
\setval
\fmfforce{0w,0h}{v1}
\fmfforce{4/11.464w,0h}{v2}
\fmfforce{7.464/11.464w,0h}{v3}
\fmfforce{1/2w,3/7h}{v4}
\fmfforce{1/2w,1h}{v5}
\fmfforce{1/2w,-1/7h}{v7}
\fmfforce{1w,0h}{w1}
\fmf{plain,right=1}{v4,v7,v4}
\fmf{plain,left=1}{v4,v5,v4}
\fmf{boson}{v1,v2}
\fmf{boson}{v3,w1}
\fmfdot{v1,w1,v2,v3,v4}
\end{fmfgraph*}\end{center}}
\\ $2$ &$2$ & $0$ &
\begin{tabular}{@{}c}
$1/4$ \\ ${\scs ( 0, 0, 0 , 0 ; 4 )}$
\end{tabular}
\parbox{15mm}{\begin{center}
\begin{fmfgraph*}(12,4)
\setval
\fmfforce{0w,1/2h}{v1}
\fmfforce{1/3w,1/2h}{v2}
\fmfforce{2/3w,1/2h}{v3}
\fmfforce{1w,1/2h}{w1}
\fmfforce{1/2w,0h}{v4}
\fmfforce{1/2w,1h}{v5}
\fmf{plain,left=1}{v2,v3,v2}
\fmf{boson}{v1,v2}
\fmf{boson}{v3,w1}
\fmf{plain}{v4,v5}
\fmfdot{v1,w1,v2,v3,v4,v5}
\end{fmfgraph*}\end{center}}
\hspace*{1mm}
\begin{tabular}{@{}c}
$1/4$ \\ ${\scs ( 0, 1, 0 , 0 ; 2 )}$
\end{tabular}
\parbox{15mm}{\begin{center}
\begin{fmfgraph}(12,4)
\setval
\fmfforce{1/3w,0h}{v1}
\fmfforce{2/3w,0h}{v2}
\fmfforce{2/3w,1h}{v3}
\fmfforce{1/3w,1h}{v4}
\fmfforce{0w,0h}{v5}
\fmfforce{1w,0h}{v6}
\fmf{plain,right=1}{v1,v3,v1}
\fmf{boson}{v1,v5}
\fmf{boson}{v2,v6}
\fmf{plain,right=0.5}{v4,v3}
\fmfdot{v1,v2,v3,v4,v5,v6}
\end{fmfgraph}\end{center}} 
\\ \hline\hline
$3$ &$0$ & $0$ &
\hspace{-10pt}
\rule[-10pt]{0pt}{26pt}
\begin{tabular}{@{}c}
$1/6$ \\ ${\scs ( 0, 0, 0 , 0 ; 6 )}$
\end{tabular}
\parbox{11mm}{\begin{center}
\begin{fmfgraph*}(6.928,12)
\setval
\fmfforce{1/2w,5/6h}{v1}
\fmfforce{1w,1/4h}{w1}
\fmfforce{0w,1/4h}{u1}
\fmfforce{1/2w,1/2h}{v2}
\fmf{boson}{v2,v1}
\fmf{boson}{v2,w1}
\fmf{boson}{v2,u1}
\fmfdot{v1,w1,u1,v2}
\end{fmfgraph*}\end{center}}
\\ \hline $3$ &$1$ & $1$ &
\hspace{-10pt}
\rule[-10pt]{0pt}{26pt}
\begin{tabular}{@{}c}
$1/6$ \\ ${\scs ( 0, 0, 0 , 0 ; 6 )}$
\end{tabular}
\parbox{15mm}{\begin{center}
\begin{fmfgraph}(12,12)
\setval
\fmfforce{1/2w,1/3h}{v1}
\fmfforce{1/2w,2/3h}{v2}
\fmfforce{1/2w,1h}{v3}
\fmfforce{0.355662432w,0.416666666h}{v4}
\fmfforce{0.64433568w,0.416666666h}{v5}
\fmfforce{0.067w,1/4h}{v6}
\fmfforce{0.933w,1/4h}{v7}
\fmf{plain,left=1}{v1,v2,v1}
\fmf{boson}{v2,v3}
\fmf{boson}{v4,v6}
\fmf{boson}{v5,v7}
\fmfdot{v2,v3,v4,v5,v6,v7}
\end{fmfgraph}\end{center}} 
\\ $3$ &$1$ & $0$ &
\begin{tabular}{@{}c}
$1/4$ \\ ${\scs ( 0, 1, 0 , 0 ; 2 )}$
\end{tabular}
\parbox{11mm}{\begin{center}
\begin{fmfgraph*}(8,8)
\setval
\fmfforce{0w,0h}{v1}
\fmfforce{1w,0h}{w1}
\fmfforce{1/2w,1h}{u1}
\fmfforce{1/2w,0h}{v2}
\fmfforce{1/2w,1/2h}{v3}
\fmf{plain,right=1}{v2,v3,v2}
\fmf{boson}{w1,v1}
\fmf{boson}{v3,u1}
\fmfdot{u1,v1,w1,v2,v3}
\end{fmfgraph*}\end{center}}  \\ \hline \hline
$4$ &$0$ & $1$ &
\hspace{-10pt}
\rule[-10pt]{0pt}{26pt}
\begin{tabular}{@{}c}
$1/24$ \\ ${\scs ( 0, 0, 0 , 0 ; 24 )}$
\end{tabular}
\parbox{11mm}{\begin{center}
\begin{fmfgraph*}(8,12)
\setval
\fmfforce{0w,1/2h}{v1}
\fmfforce{1w,1/2h}{w1}
\fmfforce{1/2w,1/6h}{u1}
\fmfforce{1/2w,5/6h}{x1}
\fmfforce{1/2w,1/2h}{v2}
\fmf{boson}{w1,v1}
\fmf{boson}{x1,u1}
\fmfdot{u1,v1,w1,x1,v2}
\end{fmfgraph*}\end{center}}
\\ \hline $4$ &$1$ & $2$ &
\hspace{-10pt}
\rule[-10pt]{0pt}{26pt}
\begin{tabular}{@{}c}
$1/16$ \\ ${\scs ( 0, 1, 0 , 0 ; 8 )}$
\end{tabular}
\parbox{11mm}{\begin{center}
\begin{fmfgraph*}(8,4)
\setval
\fmfforce{0w,0h}{v1}
\fmfforce{1w,0h}{w1}
\fmfforce{0w,1h}{u1}
\fmfforce{1w,1h}{x1}
\fmfforce{1/2w,0h}{v2}
\fmfforce{1/2w,1h}{v3}
\fmf{plain,right=1}{v2,v3,v2}
\fmf{boson}{w1,v1}
\fmf{boson}{x1,u1}
\fmfdot{u1,v1,w1,x1,v2,v3}
\end{fmfgraph*}\end{center}}
\\ $4$ &$1$ & $1$ & 
\begin{tabular}{@{}c}
$1/4$ \\ ${\scs ( 0,0, 0 , 0 ; 4 )}$
\end{tabular}
\parbox{14.5mm}{\begin{center}
\begin{fmfgraph*}(11.464,3)
\setval
\fmfforce{0w,0h}{v1}
\fmfforce{4/11.464w,0h}{v2}
\fmfforce{7.464/11.464w,0h}{v3}
\fmfforce{1/2w,1h}{v4}
\fmfforce{1/2w,-1/3h}{v7}
\fmfforce{1w,0h}{w1}
\fmfforce{1.732/11.464w,1h}{u1}
\fmfforce{9.732/11.464w,1h}{x1}
\fmf{plain,right=1}{v4,v7,v4}
\fmf{boson}{u1,x1}
\fmf{boson}{v1,v2}
\fmf{boson}{v3,w1}
\fmfdot{u1,v1,w1,x1,v2,v3,v4}
\end{fmfgraph*}\end{center}}
\\ $4$ &$1$ & $0$ & 
\begin{tabular}{@{}c}
$1/8$ \\ ${\scs ( 0,0, 0 , 0 ; 8 )}$
\end{tabular}
\parbox{15mm}{\begin{center}
\begin{fmfgraph*}(12,12)
\setval
\fmfforce{0w,1/2h}{v1}
\fmfforce{1w,1/2h}{w1}
\fmfforce{1/2w,0h}{u1}
\fmfforce{1/2w,1h}{x1}
\fmfforce{1/3w,1/2h}{v2}
\fmfforce{1/2w,1/3h}{v3}
\fmfforce{2/3w,1/2h}{v4}
\fmfforce{1/2w,2/3h}{v5}
\fmf{plain,right=1}{v4,v2,v4}
\fmf{boson}{v1,v2}
\fmf{boson}{u1,v3}
\fmf{boson}{w1,v4}
\fmf{boson}{x1,v5}
\fmfdot{u1,v1,w1,x1,v2,v3,v4,v5}
\end{fmfgraph*}\end{center}}
\end{tabular}
\end{center}
\caption{One-particle irreducible 
vacuum diagrams and their weights for the 
$\phi^3$-$\phi^4$-theory with $n=1,2,3,4$ field expectation values
for the respective first two loop orders. Within each loop order $l$ the
diagrams are distinguished with respect to the number $p$ of $4$-vertices.
Each diagram is characterized 
by the
vector $(S,D,T,F;N$) whose components specify the number of self, double,
triple, fourfold connections, and of the identical vertex permutations,
respectively.}
\end{table}

\end{fmffile}


\begin{thebibliography}{199}
%
\bibitem{Kleinert1} H. Kleinert, Fortschr. Phys. {\bf 30}, 187 (1982).
%
\bibitem{Kleinert2} H. Kleinert, Fortschr. Phys. {\bf 30}, 351 (1982).
%
\bibitem{QED} M. Bachmann, H. Kleinert, A. Pelster, Phys. Rev. {\bf D 61},
085017 (2000).
%
\bibitem{PHI4} H. Kleinert, A. Pelster, B. Kastening, M. Bachmann,
Phys. Rev. {\bf E} (in press), hep-th/9907168.
%
\bibitem{Boris} B. Kastening, Phys. Rev. {E 61}, 3501 (2000).
%
\bibitem{Kleinert3} H. Kleinert, {\it Gauge Fields in Condensed Matter,
Vol. I, Superflow and Vortex Lines}, World Scientific (1989).
%
\bibitem{Neu}
J. Neu, MS Thesis (in German), FU-Berlin (1990).
%
\bibitem{Verena}
H. Kleinert, V. Schulte-Frohlinde, {\it Critical Properties
of $\phi^4$-Theories}, World Scientific, in press.
%
\bibitem{Amit} D.J. Amit, {\it Field Theory, the Renormalization Group
and Critical Phenomena}, McGraw-Hill (1978).
%
\bibitem{Zuber} C. Itzykson, J.-B. Zuber, {\it Quantum Field Theory},
McGraw-Hill (1985).
%
\bibitem{Zinn} J. Zinn-Justin, {\it Quantum Field Theory and Critical
Phenomena}, Third Edition, Oxford  (1996).
%
\bibitem{CODE}
http://www.physik.fu-berlin.de/\~{}kleinert/294/programs.
%
\bibitem{Muenster} C. Gutsfeld, J. K\"uster, G. M\"unster
Nucl. Phys. {\bf B 479}, 654 (1996).
%
\bibitem{Dohm} S.A. Larin, M. M\"onnigmann, M. Str\"osser, V. Dohm,
Phys. Rev. {\bf B 58}, 3394 (1998). 
%
\bibitem{Kastening1} B. Kastening, Phys. Rev. {\bf D 54}, 3965 (1996).
%
\bibitem{Kastening2} B. Kastening, Phys. Rev. {\bf D 57}, 3567 (1998).
%
\end{thebibliography}
\end{document}